\documentclass[12pt]{article}
\pdfoutput=1
\usepackage{amsmath}
\usepackage{multirow}
\usepackage{amsfonts}
\usepackage{amssymb,graphics,psfrag}
\usepackage{array,epsfig,multirow,graphicx}
\usepackage{comment}
\usepackage[all,cmtip]{xy}
\usepackage{hyperref}

\numberwithin{equation}{section}
\numberwithin{table}{section}

\def\hybrid{\topmargin -20pt   \oddsidemargin 0pt
        \headheight 0pt \headsep 0pt
        \textwidth 6.25in      
        \textheight 9 in      
        \marginparwidth .875in
        \parskip 5pt plus 1pt
          \jot = 1.5ex
  }
\hybrid

\newcommand{\be}{\begin{equation}}
\newcommand{\ee}{\end{equation}}
\newcommand{\bea}{\begin{eqnarray}}
\newcommand{\eea}{\end{eqnarray}}

\newcommand{\nn}{\nonumber}

\newcommand{\ov}{\overline}

\newcommand{\beq}{\begin{equation}}
\newcommand{\eeq}{\end{equation}}

\newcommand{\cO}{\mathcal{O}}

\newcommand{\cC}{\mathcal{C}}

\newcommand{\cS}{\mathcal{S}}

\newcommand{\cref}{{\bf [check ref]}}

\newcommand{\Xsu}{\hat{X}_{\text{SU}(5)}}
\newcommand{\Ssu}{\cS_{\text{SU}(5)}}

\def\blfootnote{\xdef\@thefnmark{}\@footnotetext}
\long\def\symbolfootnote[#1]#2{\begingroup%
\def\thefootnote{\fnsymbol{footnote}}\footnote[#1]{#2}\endgroup}

\begin{document}

\baselineskip=15pt

\begin{titlepage}
\begin{flushright}
\parbox[t]{1.08in}{UPR-1250-T}
\end{flushright}

\begin{center}

\vspace*{ 0.0cm}

{\Large \bf \text{Chiral Four-Dimensional F-Theory Compactifications}\\ With SU(5) and Multiple U(1)-Factors}

\vskip 1cm

\renewcommand{\thefootnote}{}
\begin{center}
 {Mirjam Cveti\v{c}$^{1,2}$,  Antonella Grassi$^{3}$, Denis Klevers$^1$, Hernan Piragua$^1$}
\end{center}
\vskip .2cm
\renewcommand{\thefootnote}{\arabic{footnote}}

$\,^1$ {Department of Physics and Astronomy,\\
University of Pennsylvania, Philadelphia, PA 19104-6396, USA} \\[.3cm]

$\,^2$ {Center for Applied Mathematics and Theoretical Physics,\\ University of Maribor, Maribor, Slovenia}\\[.3cm]

$\,^3$ {Department of Mathematics,\\
University of Pennsylvania, Philadelphia, PA 19104-6396, USA} \\[.3cm]

{cvetic\ \textsf{at}\ cvetic.hep.upenn.edu, klevers\ \textsf{at}\ sas.upenn.edu, hpiragua\ \textsf{at}\ sas.upenn.edu, }

 \vspace*{0.2cm}

\end{center}

\vskip 0.0cm

\begin{center} {\bf ABSTRACT } \end{center}

We develop geometric techniques to determine the spectrum  and the chiral
indices of  matter multiplets for four-dimensional
F-theory compactifications on elliptic Calabi-Yau fourfolds with rank
two Mordell-Weil group.
The general elliptic fiber is the Calabi-Yau onefold in $dP_2$. We
classify its resolved elliptic  fibrations over a general base
$B$. The study of
singularities of these fibrations leads to explicit matter representations,
that we determine both for U$(1)\times $U(1) and
SU(5)$\times \text{U}(1)\times \text{U}(1)$ constructions.
We determine for the first time certain matter curves and surfaces
using techniques involving  prime ideals.
The vertical cohomology ring of these fourfolds is calculated for both
cases and general formulas for the Euler numbers are derived. Explicit
calculations are presented for a specific base $B=\mathbb{P}^3$.
We  determine the general $G_4$-flux that belongs to $H^{(2,2)}_V$  of
the resolved Calabi-Yau fourfolds.
As a by-product, we derive for the first time all conditions on
$G_4$-flux in general F-theory compactifications with a non-holomorphic
zero section.
These conditions have to be formulated after a circle reduction
in terms of Chern-Simons terms on the 3D Coulomb branch and invoke
M-theory/F-theory duality.
New Chern-Simons terms are generated by Kaluza-Klein
states of the circle compactification. We explicitly perform
the relevant field theory computations, that yield non-vanishing
results precisely for fourfolds with a non-holomorphic zero section.
Taking into account the new Chern-Simons terms, all 4D matter
chiralities are determined via  3D M-theory/F-theory duality.
We independently  check these chiralities using the subset of
matter surfaces we determined.  The presented techniques
are general and do not rely on toric data.
\begin{flushright}
\parbox[t]{1.2in}{June, 2013}
\end{flushright}
\end{titlepage}

\tableofcontents


\newpage
%
\section{Introduction and Summary of Results}
\label{sec:intro}

Four-dimensional F-theory  compactifications  provide a  broad domain of
the string theory  landscape with the potential to derive promising
particle physics consequences of string theory.  It has a number of
advantages, such as encoding  geometric structures of string theory
compactifications  at a finite string couplings.  The main focus in
recent years has been on studies  of F-theory GUT models with SU(5) as
well as SO(10) gauge groups, initiated in
\cite{Donagi:2008ca, Beasley:2008dc,Beasley:2008kw,Donagi:2008kj} within
local model building. Techniques for constructions of global models were
developed in
\cite{Blumenhagen:2009yv,Marsano:2009wr,Chen:2010ts,Grimm:2009yu,Knapp:2011ip},
and efforts to embed local models into global ones have been pursued,
see e.g.~\cite{Heckman:2010bq,Weigand:2010wm,Maharana:2012tu} for
reviews.

The origin of non-Abelian gauge symmetries in four-dimensional F-theory
compactifications is well understood since the origins of F-theory
\cite{Vafa:1996xn,Morrison:1996na,Morrison:1996pp}, and it is due to
the full classification of codimension one singularities of elliptically
fibered Calabi-Yau fourfolds with a section in the Weierstrass or Tate
model \cite{kodaira1963compact,tate1975algorithm,Bershadsky:1996nh}.
Matter multiplets appear at co-dimension two  singularities and were
studied originally in \cite{Vafa:1996xn,Morrison:1996na,Katz:1996xe} and
more recently in
\cite{Knapp:2011ip,Heckman:2010bq,Weigand:2010wm,Maharana:2012tu}.\footnote{Recent 
efforts clarified and filled the gaps  in
classifications  singularities at higher co-dimensions
\cite{Esole:2011sm,Marsano:2011hv,Lawrie:2012gg}.  For most recent
complementary advances, employing deformations of singularities, see
\cite{Grassi:2013kha}.}
Constructions  with chiral matter in four-dimensions require the addition
of $G_4$-flux, which belongs to the middle  (vertical) cohomology on the
resolved Calabi-Yau fourfold.  Specific  fluxes of this type were
constructed in  \cite{Krause:2012yh,Marsano:2011hv}.

On the other hand Abelian gauge symmetries in F-theory are less
understood and studied. This  is primarily due to the
fact that the classification of Abelian gauge symmetries  and their
matter spectrum depends crucially on  the global geometry of the elliptic
fibration.  Nevertheless, some  aspects of Abelian gauge symmetries  have
been studied in local F-theory models employing spectral cover methods
\cite{Donagi:2009ra,Marsano:2009gv,Marsano:2009wr,Dudas:2009hu,Cvetic:2010rq,Dudas:2010zb,Dolan:2011iu,Marsano:2012yc}.
The study of  higher rank Abelian sectors in  four-dimensional F-theory
compactifications is also of phenomenological interest,  since it  can
play an important r\^ole in model building beyond the Standard Model
model.

Abelian gauge theory sectors appear in compactifications  of F-theory on
elliptic Calabi-Yau varieties  with a general  fiber being an elliptic
curve with rational points. Elliptic curves with a so-called non-trivial
Mordell-Weil group of rational points are a classical subject in
mathematics
\cite{neron1964modeles,shioda1990mordell,shioda1989,Wazir:2001,silverman2009arithmetic}.
These rational points lift to rational sections of the  elliptically
fibered Calabi-Yau manifold, which contribute new harmonic
two-forms to the cohomology that support Abelian gauge fields in the F-
theory effective action. The number of Abelian gauge  fields is set by
the rank of the Mordell-Weil group of the elliptic curve and its torsion
subgroup gives rise to non-simply connected groups
\cite{Morrison:1996pp,Aspinwall:1998xj,Aspinwall:2000kf}. Rank one
Mordell-Weil groups in compact elliptic  fibrations have been studied
recently in the F-theory literature in a variety of  contexts
\cite{Grimm:2010ez,Braun:2011zm,Krause:2011xj,Grimm:2011fx,Morrison:2012ei,Cvetic:2012xn,Cvetic:2012xn,Braun:2013yti,Borchmann:2013jwa, Braun:2013nqa}. Elliptic fibrations with elliptic fiber of $D_5$-type have been constructed in the context of counting BPS states in \cite{Klemm:2004km}. The resolutions of their most general fibrations, which have a Mordell-Weil group of rank three, as well as the complete induced F-theory matter spectrum have been analyzed in \cite{Cvetic:2013qsa}.

While a classification of  possible Abelian gauge sectors in
F-theory, analogous to the well-studied non-Abelian sector, is lacking
(see, however \cite{Grassi:2012qw}
for a systematic study of rational sections on toric K3-surfaces),
significant progress has been made recently for the systematic study of
U(1) $\times$ U(1) gauge symmetry  on elliptically fibered Calabi-Yau
manifolds with a rank two Mordell-Weil group
\cite{Borchmann:2013jwa,Cvetic:2013nia}.\footnote{The elliptic curves in $\mathbb{P}^{2}(1,1,2)$ and 
$\mathbb{P}^2$ have also appeared  in \cite{Aldazabal:1996du} and independently in \cite{Klemm:1996hh}, cf.~also \cite{Klemm:2004km}, as F-theory duals
of heterotic backgrounds with U(1)-Wilson lines. In the former,  elliptic threefolds
with these elliptic fibers over $\mathbb{F}_n$ are constructed and the non-Abelian gauge groups are matched between the dual theories in 
dependence on $n$ as in \cite{Morrison:1996pp}. In the latter, the Kaehler classes of these elliptic threefolds, which are dual to the Wilson lines breaking 
$E_8\rightarrow E_{8-k}\times \text{U}(1)^k$, $k=0,1,2$, in the heterotic string, are identified by matching certain Gromov-Witten invariants with BPS-states of non-critical strings from $E_{8-k}$ small instantons 
computed on the heterotic side. However, the detailed analysis and classification of the Abelian sectors on the F-theory side, by studying the 
singularities of these elliptic fibrations, has first been carried out in \cite{Cvetic:2013nia} and in this work using the resolved $dP_2$-elliptic fibrations.}  The
analysis found that the
natural presentation of an elliptic curve with two rational points and a
zero point is the generic Calabi-Yau one-fold in $dP_2$ and the
birational map to its Tate and Weierstrass form  was  derived
\cite{Cvetic:2013nia}.  While the
discussion of its resolved elliptic  fibrations was done for a Calabi-Yau
variety  over a general base $B$, their classification  was first performed for
the base $B=\mathbb{P}^2$ and later for any two-dimensional base $B$ in \cite{Cvetic:2013jta}. 
One key finding of this classification, which 
we further elaborate on also in this work, was the identification of all 
topological degrees of freedom in the construction of a fibrations of a 
fixed elliptic curve over a fixed base $B$. These are encoded in the choice of two divisor 
classes $\cS_7$, $\cS_9$ \cite{Cvetic:2013nia}. Allowing for all their 
possible values yields elliptic fibrations with novel properties, most
notably with a non-holomorphic zero section, that were not constructed before because the existence
of a holomorphic zero section was enforced. Next, a 
thorough analysis of the generic codimension two
singularities of this most general family of elliptic Calabi-Yau threefolds was given. 
This determines geometrically all the matter representations under
U(1) $\times$ U(1) and their multiplicities,  that were shown  to
be consistent with anomaly cancellations in the six-dimensional
compactified theory.  Explicit toric examples were constructed, both with
U(1) $\times$ U(1) and SU(5)$\times $ U(1)$\times $ U(1) gauge symmetries.

It is the purpose of this paper to develop explicit techniques for the
calculation of the matter spectrum and their chiralities  for
four-dimensional F-theory compactifications with two U(1)-factors. This
involves  now F-theory compactifications on elliptically fibered
Calabi-Yau fourfolds with rank two Mordell-Weil groups. Furthermore the
appearance of the chiral matter requires an explicit  construction of
$G_4$-flux. We would like to emphasize that our calculations are performed for
general globally defined elliptic Calabi-Yau fourfolds  and are {\it not}
restricted to geometries described by toric reflexive polytopes.
In particular, we find closed formulas for the basis of the vertical cohomology 
groups, the consistent $G_4$-flux and the chiral indices for entire discrete families
of Calabi-Yau fourfolds which depend explicitly on the aforementioned degrees of freedom  
$\cS_7$, $\cS_9$ of these families. We underline that even if each
member of this family could be realized torically, it would be hard to 
obtain this dependence on $\cS_7$, $\cS_9$ by considering reflexive polytopes.
In recent works \cite{Braun:2013yti,Grimm:2013oga,Braun:2013nqa}  a
classification of  toric Calabi-Yau fourfolds with  $SU(5)$ and allowed U(1) factors was given (See also \cite{Borchmann:2013jwa}).
Note however that  techniques employed have, at least to the knowledge of
the authors, not  culminated in the determination of  the matter
representations and their 4D chiralities yet which are the main
results of this paper.

We advance the program in several important ways:

$\bullet$  The general elliptic fiber is the generic Calabi-Yau one-fold
in $dP_2$ with two rational points and a zero point  (as analyzed in
detail  in \cite{Cvetic:2013nia}). The chosen fiber determines the
representations of the matter multiplets under U(1) $\times$ U(1) over
any base $B$.   The spectrum for U(1) $\times$ U(1) and SU(5)$ \times $
U(1) $ \times $ U(1)  (for a specific SU(5)) is derived and summarized in
table \ref{tab:allReps}. However,  the analysis is performed now over a
general three-dimensional base and the matter appears over codimension
two Riemann surfaces in the base, denoted as matter curves. We employ
explicit algebraic geometry techniques using prime ideals\footnote{Note that in the case of Calabi-Yau threefolds these geometric
techniques were sufficient to determine multiplicities  of all the matter
multiplets. \cite{Cvetic:2013nia}.} to represent
these matter curves and classify the singularities of the elliptic
fibration over them using the  Calabi-Yau fourfold with the resolved elliptic
fibration (See sections \ref{sec:ellipticCurvedP2},
\ref{sec:SingularFibU1U1} and \ref{sec:SingularitiesSU5U1U1}).
At present these techniques allow
us, however, to determine only a subset of matter surfaces (three out of six in the U(1) $
\times$ U(1) and six out of twelve in the SU(5) $\times $ U(1) $ \times $ U(1)
case). 

$\bullet$ As a preparation for the construction of $G_4$-flux on these
Calabi-Yau fourfolds, we find for the first time consistent conditions on
the $G_4$-flux in F-theory compactifications with a non-holomorphic zero
section. These conditions have to be formulated in three dimensions after
the compactification on a circle and require the use of M-theory/F-theory
duality. We show the connections between holomorphicity of the zero
section and Kaluza Klein-states via Chern-Simons (CS) terms on the 3D
Coulomb branch. In particular, there are new CS-terms for a
non-holomorphic zero section.

$\bullet$ We develop techniques to calculate explicitly  the most general
$G_4$-flux that belongs to the vertical cohomology $H^{(2,2)}_V$
resolved Calabi-Yau fourfolds (See Section \ref{sec:CYFFCohomologyRing}
and \ref{sec:G4review}).  To this end we algebraically calculate the full
vertical cohomology ring of these Calabi-Yau fourfolds. These cohomology
calculations allow us to compute the general expression for the Euler
number and the Chern classes
of these fourfolds for an arbitrary base $B$, both in the U(1)$\times$U(1) and
SU(5)$\times$U(1)$\times$U(1) cases. As an application of these
techniques we derive an explicit basis of the cohomology group for
all elliptically fibered Calabi-Yau fourfolds with fiber in $dP_2$ and
base $B=\mathbb{P}^3$.\footnote{These
techniques have been used in the context of mirror symmetry on Calabi-Yau
fourfolds in \cite{Mayr:1996sh,Klemm:1996ts,Grimm:2009ef}.}
Again these techniques are general and not restricted to toric examples. In particular 
the dependence on the divisors $\cS_7$, $\cS_9$ is manifest.
When  the $G_4$-flux is integrated over matter surfaces (determined via
the geometric techniques mentioned above) we obtain the chiralities of
three  matter representations (second set in Table \ref{tab:allReps}).
Chiralities of the remaining matter representations are determined by a
subset of the 3D CS-terms of the dual M-theory invoking M-/F-theory
duality. 4D anomalies are found to be cancelled. We note that the rest of
the 3D CS-terms, in particular those for the
Kaluza-Klein vector, provide an independent check for  chiralities
of matter multiplets obtained via geometric techniques.
It is important to note that, given the list of representations that are
realized,  all  CS-terms taken together are sufficient
to determine chiralities of all the matter  multiplets. Our geometric
techniques allow us to have an independent determination for a subset of
them (See Section \ref{sec:G4+chiralitiesU1xU1}).  We also perform an
independent check that  with the obtained  spectrum the four-dimensional
anomalies are cancelled.  Explicit results are  presented for the most
general $G_4$-flux for all generic
resolved elliptic Calabi-Yau fourfolds over the  base
$B=\mathbb{P}^3$, both for $\text{U}(1)\times \text{U}(1)$  and
$\text{SU}(5)\times \text{U}(1)\times \text{U}(1)$ (for a specific
embedding of  $\text{SU}(5)$).

\begin{table}[ht!]
\begin{center}
\begin{tabular}{|c|c|}
\hline & \\ $U(1)\times U(1)$ & $SU(5)\times U(1)\times U(1)$ \\  & \\ \hline\hline
& \\ $(1,0)$ $(0,1)$ $(1,1)$ & $(\mathbf{ 5}, -\textstyle{\frac{2}{5}},0)$  $(\mathbf{ 5}, \textstyle{\frac{3}{5}},0)$ $(\mathbf{ 5},-\textstyle{\frac{2}{5}},-1)$  \\  &  \\ \hline
& \\
$(-1,1)$ $(0,2)$ $(-1,-2)$& $(\mathbf{ 5}, -\textstyle{\frac{2}{5}},1)$  $(\mathbf{ 5}, \textstyle{\frac{3}{5}},1)$ $(\ov{\mathbf{10}},\textstyle{-\frac{1}{5}},0)$ \\  &  \\ \hline
\end{tabular}
\caption{Matter representation for F-theory  compactifications with rank-
two Mordell Weil  group.  While   U(1)$\times$ U(1)  charges for
$\text{SU}(5)$ singlets are general,  U(1)$\times$ U(1)  charges for
the additional non-singlet matter representations of $\text{SU}(5)$
depend on a specific realization of the SU(5) gauge symmetry.}
\label{tab:allReps}
\end{center}
\end{table}

We note that the fibrations
over the chosen base $B=\mathbb{P}^3$ are generally non-flat at a single
codimension three locus. This can be
circumvented in two ways. Either, one can forbid the existence of the
non-flat fiber geometrically, or restrict the allowed $G_4$-flux by
requiring a vanishing integral of it over the non-flat fiber. Both
approaches yield an anomaly-free 4D spectrum with no chiral excess of
additional light states.

The paper is organized in the following way:
In Section \ref{sec:ellipticCurvedP2} we summarize the geometry of the
general elliptic curve with rank two Mordell Weil group in $dP_2$ and
classify its fibrations. Section \ref{sec:SingularFibU1U1} is devoted
to the analysis of the  codimension two and three singularities of the
fibrations  specifying, respectively, the matter content, the matter
curves and surfaces as well as the Yukawa couplings. In Section
\ref{sec:CYFFCohomologyRing} we determine the cohomology ring and Chern
classes for resolved elliptic Calabi-Yau fourfolds with rank two
Mordell-Weil group.
Section \ref{sec:G4review} is devoted to  a detailed discussion of
$G_4$-flux and conditions imposed on them in general F-theory
compactifications with a non-holomorphic zero section. This study is
based on relations between 3D Chern-Simons terms under F-theory/M-theory
duality.  A special emphasis is on quantum loop corrections due to
Kaluza-Klein states, that are computed explicitly.  In section
\ref{sec:G4+chiralitiesU1xU1} the general $G_4$-flux is explicitly
calculated  (Section \ref{sec:G4U1xU1}),  and matter chiralities both
geometrically and via 3D Chern-Simons terms of dual the M-theory
are derived (Section \ref{sec:4DChiralityU1U1}).  In Section
\ref{sec:anomalyCancellation} anomaly cancellation of the four-
dimensional field theory is checked and in section \ref{sec:ToricExsU1U1}
an explicit toric example is presented.
A  generalization to F-theory compactifications with an additional
SU(5) non-Abelian gauge symmetry is spelled out in sections
\ref{sec:FFEllipticFibSU5U12} and \ref{sec:G4FluxChiralitiesSU5}. In
section \ref{sec:SingularitiesSU5U1U1} the singularities at codimension
one, two (matter representations) and three (Yukawa points) are
presented. A non-flat fiber at codimension three is discovered. The
cohomology ring and the general Euler number are calculated in
\ref{sec:CohomologyFFSU5}. In section \ref{sec:G4FluxChiralitiesSU5} the
$G_4$-flux and the 4D chiralities are computed and anomaly cancellation
is checked in detail for all elliptic fibrations with $dP_2$-fiber over
$\mathbb{P}^3$. Section \ref{sec:G4U1xU1xSU5} demonstrates the general
construction of $G_4$-flux in F-theory compactifications with rank two
Mordell-Weil group and a resolved SU(5)-singularity with
$B=\mathbf{P}^3$. In section \ref{sec:ChiralitiesU1xU1xSU5} the
4D chiralities are evaluated and anomaly cancellation is checked.  We
finish the section with one concrete toric example in section
\ref{sec:ExamplesSU5U1U1}. Conclusions and future directions can
be found in section \ref{sec:conclusions}.

This work has seven appendices. Appendix \ref{app:CohomologyRing}
contains formulas for all Chern classes of $dP_2$-elliptic fibrations over an arbitrary base
$B$ along with the Chern classes and Euler numbers for their resolved
Calabi-Yau two-, three- and fourfolds. In appendix \ref{app:intsP3}
we present the cohomology ring of the generic fourfolds with
$B=\mathbb{P}^3$ and $dP_2$-elliptic fiber. Appendix
\ref{app:ChernOfdP2BU} contains  the Chern classes
of  $dP_2$-fibrations with resolved SU(5)-singularities, in appendices
\ref{app:fluxP3SU5} and \ref{app:H22BasisSU5} the intersections  and
vertical cohomology ring of fourfolds with $B=\mathbb{P}^3$ are
computed. Appendix \ref{app:fluxesNChiralities} contains all 3D CS-terms
and chiralities. Appendix \ref{app:tuning-s7-s9} concludes by
describing how to systematically construct the toric polytopes for the example of Calabi-Yau fourfolds with 
base $B=\mathbb{P}^3$.

\section{The Elliptic Curve in \texorpdfstring{$dP_2$}{Lg} And Its Fibrations}
\label{sec:ellipticCurvedP2}

In this section we review the construction
of the elliptic curve $\mathcal{E}$ in $dP_2$ and its  Calabi-Yau
elliptic fibrations over a general $B$.
These Calabi-Yau manifolds have a rank two Mordell-Weil group,
that gives rise to U$(1)\times$U(1) gauge symmetry in F-theory.

In section \ref{sec:EllCurveMWrk2} we review the geometry of
$\mathcal{E}$ as the generic Calabi-Yau onefold in the
del Pezzo surface $dP_2$ following the conventions and notations of
\cite{Cvetic:2013nia}, to which
we also refer for more details. The reader familiar with the geometry
of this elliptic curve $\mathcal{E}$ can safely skip the first part of
the section.
Then, in section \ref{sec:GeneraldP2fibrations} we construct resolved
elliptically fibered Calabi-Yau manifolds
$\hat{\pi}:\hat{X}\rightarrow X$ over an arbitrary
base $B$ with this elliptic curve $\mathcal{E}$ as the general fiber.
The singular Calabi-Yau manifold is denoted by $X$
We show that
these Calabi-Yau manifolds $\hat{X}$ are classified by the choice of two
divisors $\cS_7$, $\cS_9$ in the base $B$. In particular, we work out all
the line bundles that are relevant to formulate the Calabi-Yau constraint
of $\hat{X}$, which is the analog of the Tate model for elliptic
fibrations with $dP_2$-elliptic fiber.

The content of section
\ref{sec:GeneraldP2fibrations} is a direct extension of the discussion
in \cite{Cvetic:2013nia}, where the possibility of a full classification
of all Calabi-Yau elliptic fibrations with general fiber $\mathcal{E}$
was pointed out, but demonstrated explicitly only for $B=\mathbb{P}^2$.

\subsection{The Elliptic Curve with Rank Two Mordell-Weil Group}
\label{sec:EllCurveMWrk2}

\begin{figure}[ht!]
\centering
 \includegraphics[scale=0.35]{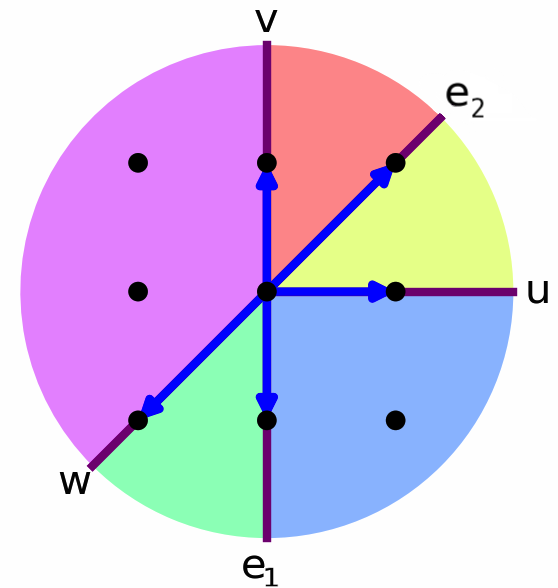}
 \caption{Fan of $dP_2$. The coordinates corresponding to its rays are indicated.}
 \label{fig:dP2poly}
\end{figure}

The hypersurface description of the elliptic curve $\mathcal{E}$ with a
zero point $P$ and two rational points $Q$ and $R$ has been derived in
\cite{Cvetic:2013nia} from the existence of a degree three ample line
bundle $M=\mathcal{O}(P+Q+R)$ on $\mathcal{E}$.
The result of this analysis is that such an elliptic curve is naturally
represented as  the generic Calabi-Yau hypersurface in the del Pezzo
$dP_2$. The Calabi-Yau constraint in $dP_2$ takes the form
\beq \label{eq:CYindP2}
	p\!\!=\! u (s_1u^2e_1^2e_2^2 + s_2 u v e_1 e_2^2 + s_3  v^2e_2^2
	+ s_5 u w e_1^2e_2+ s_6 vwe_1e_2 + s_8 w^2 e_1^2)
	+ s_7 v^2 we_2 + s_9 v w^2e_1\,,
\eeq
where we introduced the homogeneous coordinates $[u:v:w:e_1:e_2]$ on
$dP_2$.\footnote{We
deviate here from the conventions of \cite{Cvetic:2013nia} by denoting
coordinates on $dP_2$ by $[u:v:w:e_1:e_2]$ and those on $\mathbb{P}^2$
by $[u':v':w']$.} One readily checks
that $p$ is the most general section of the anti-canonical bundle
$K^{-1}_{dP_2}=\mathcal{O}(3H-E_1-E_2)$, by noting the following
divisor classes of the homogeneous coordinates,
\beq \label{eq:dP2divs}
	\begin{array}{|c|c||rrr|}
	\hline
	 & \text{divisor class}& \multicolumn{3}{c|}{\mathbb{C}^*\text{-actions}}\\
	\hline
	  u& D_u=H-E_1-E_2& 1 & 1 &1\\
	  v& D_v=H-E_2& 1 & 0 & 1\\
      w& D_w=H-E_1 & 1& 1& 0\\
     e_1& E_1&0 &-1 &0\\
     e_2& E_2&0 & 0 &-1\\
		\hline
	\end{array}
\eeq
Here we also introduced the divisors $D_{u}$, $D_{v}$ and $D_{w}$ obtained by setting $u$, $v$, $w$ to zero,
respectively. The group of divisors is generated by $H$, $E_1$ and $E_2$ whose geometric interpretation we give momentarily. We note that $dP_2$ is a toric variety with the $(\mathbb{C}^*)^3$-action on the
homogeneous coordinates\footnote{Denoting one column vector in the last three columns of \eqref{eq:dP2divs} by
$\ell$ and the homogeneous coordinates collectively as $x_i$, than the corresponding
$\mathbb{C}^*$-action is defined as $x_i\mapsto \lambda^{\ell_i} x_i$ with $\lambda\in\mathbb{C}^*$.} specified by
the last three columns of \eqref{eq:dP2divs}. Its reflexive two-dimensional polytope is given in figure
\ref{fig:dP2poly}.

It is useful for our purposes to recall that a general del Pezzo surface
$dP_n$ is
obtained from $\mathbb{P}^2$ by a blow-up at $n$ generic points. Thus,
the del Pezzo surface
$dP_2$ is the blow-up of $\mathbb{P}^2$ at two generic points. In terms
of the homogeneous coordinates $[u':v':w']$ on $\mathbb{P}^2$, this
blow-up is performed in our case at $u'=w'=0$ and $u'=v'=0$,
so that the blow-down map takes the form
\beq \label{eq:blowdownmap}
	u'=u e_1 e_2\,,\qquad v'=ve_2\,,\qquad w'=w e_1\,,
\eeq
with the two sections $e_i$ associated to
the exceptional divisors $E_i$. We note that one can use this map to represent the elliptic curve $\mathcal{E}$ in
\eqref{eq:CYindP2} as a \textit{non-generic} Calabi-Yau onefold in $\mathbb{P}^2$. As discussed in
\cite{Cvetic:2013nia} this presentation of $\mathcal{E}$ suffices to obtain its Weierstrass model and discriminant.
However, for the understanding of elliptic fibrations the completely
resolved curve $\mathcal{E}$ in $dP_2$ with is inevitable.

The divisors classes in \eqref{eq:dP2divs} are the classes of the  $e_i$, i.e.~the exceptional divisors $E_i$
of the blow-up \eqref{eq:blowdownmap}, and the divisor class $H$, that is the pullback of the hyperplane class on $\mathbb{P}^2$.
We note the intersections on $dP_2$ are given as
\beq \label{eq:dP2ints}
	H^2=1\,,\qquad H\cdot E_i=0\,,\qquad E_i\cdot E_j=-\delta_{ij}\,.
\eeq
These intersections immediately follow on the one hand from the rule that
two divisors corresponding to rays of the same
two-dimensional cone have intersection number one and on the other hand
from the exceptional set, the Stanley-Reissner ideal $SR$. The latter
encodes all rays which do not share a two dimensional cone and have intersections number zero. From figure \ref{fig:dP2poly}
one readily obtains
\beq \label{eq:SRidealdP2}
	SR=\{uv,\,uw,\,e_1 e_2,\,e_1 v,\,e_2 w\}\,.
\eeq

The rational points $Q$ and $R$ as well as the zero point $P$ on
$\mathcal{E}$ are simply given
by the intersections of the three independent divisors in
\eqref{eq:dP2divs} with the elliptic curve
\eqref{eq:CYindP2}. We have chosen the three different points on
$\mathcal{E}$ that are obtained by intersecting the three
divisors $D_u$, $E_1$ and $E_2$ on $dP_2$ with $p=0$ in
\eqref{eq:CYindP2}. Upon setting all coordinates to one that can not
vanish simultaneously with the divisor under
consideration due to the exceptional set \eqref{eq:SRidealdP2} we obtain
\bea \label{eq:coordsPQRdP2}
	&P:\,\,E_2\cap p=[-s_9:s_8:1:1:0]\,,\quad Q:\,\,E_1\cap p=[-s_7:1:s_3:0:1]\,,\nn&\\
	&R:\,\,D_{u}\cap p=[0:1:1:-s_7:s_9]\,&\,.
\eea
The point $P$ is considered as the zero point on $\mathcal{E}$ and the points $Q$, $R$ are
the rational points.  They are the two generators of the rank two Mordell-Weil group of rational points, with
the group law given by the addition of points on $\mathcal{E}$.
We note that these three points are generically distinct. $P$ is always distinct from $Q$, however, we
observe $P=R$ for $s_9=0$  and $Q=R$ for $s_7=0$.

When considering elliptically fibered Calabi-Yau fourfolds $\hat{X}$ with
the curve \eqref{eq:CYindP2} as the general elliptic fiber, the points
$P$, $Q$ and $R$ lift to rational sections of the fibration. We note that
the coefficients $s_i$ in \eqref{eq:CYindP2} are then base-dependent
functions, that can vanish on $B$. In particular, we see from \eqref{eq:coordsPQRdP2} that the
points $P$, $Q$, and $R$, respectively, are ill-defined when $s_8=s_9=0$,
$s_3=s_8=0$ and $s_7=s_9=0$, respectively. This behavior is typical for
rational sections and further discussed in section
\ref{sec:SingularFibU1U1}.
In F-theory compactifications on such a fourfold $\hat{X}$
each of the rational sections gives rise to an Abelian gauge symmetry.
Thus, Calabi-Yau fourfolds with general elliptic fiber in $dP_2$
generically have a rank two Abelian gauge group, i.e.~an U$(1)\times$U(1)
gauge symmetry.

\subsection{General Calabi-Yau Fibrations with $dP_2$-Elliptic Fiber}
\label{sec:GeneraldP2fibrations}

In this section we discuss the construction of resolved elliptically
fibered Calabi-Yau manifolds $\hat{X}$ with general elliptic fiber in
$dP_2$. The following results hold for general complex dimension
of $\hat{X}$, in particular for Calabi-Yau three- and fourfolds.
We end this section with the concrete example of $B=\mathbb{P}^3$.

\subsubsection*{Classifying $dP_2$-fibrations and their Calabi-Yau hypersurfaces $\hat{X}$}
In general an elliptically fibered Calabi-Yau manifold
$\mathcal{E} \rightarrow\hat{X}\stackrel{\pi}{\rightarrow} B$ with $\pi$
denoting the projection to the base $B$ is
constructed by first considering
the defining equation for the desired elliptic curve $\mathcal{E}$ alone
and then by lifting the coefficients
in this equation to sections over the base $B$. In the case at hand, the
elliptic curve is described by
\eqref{eq:CYindP2}. Thus, all we have to do to obtain an elliptic
fibration is to promote the coefficients $s_i$
to sections of line bundles on the base $B$. Finally, the Calabi-Yau
condition for \eqref{eq:CYindP2} fixes the respective line bundles for
the sections $s_i$.

The procedure of lifting the $s_i$ to sections of $B$ is described as
follows. First, we have to define the ambient space in which the
elliptically fibered manifold $\hat{X}\rightarrow B$ is embedded. Since
the constraint \eqref{eq:CYindP2} merely cuts the elliptic curve
$\mathcal{E}$  out of $dP_2$, the  ambient space is simply
a $dP_2$-fibration over the base $B$ of $\hat{X}$. It takes the form
\beq \label{eq:dP2fibration}
	\xymatrix{
	dP_2 \ar[r] & 	dP^B_2(\cS_7,\cS_9) \ar[d]\\
	& B\,
	}
\eeq
which can be viewed as a generalization of a projective bundle.
Here $\cS_7$ and $\cS_9$ are two divisors on $B$ associated to the
vanishing loci of the sections $s_7$ and $s_9$
in \eqref{eq:CYindP2}. The total space is denoted $dP^B_2(\cS_7,\cS_9)$
since it is uniquely determined by these divisors $\cS_7$ and
$\cS_9$ if we demand that the constraint \eqref{eq:CYindP2} defines
a Calabi-Yau manifold $\hat{X}$. In fact, we first note that
any the  $dP_2$-fibration is specified by only two divisors on $B$.
This can be seen by noting that in  a general such fibrations the
homogeneous coordinates $[u:v:w:e_1:e_2]$ on $dP_2$ are sections of
five  different line bundles on the base $B$, respectively.
However, we can always use the three $\mathbb{C}^*$-actions in
\eqref{eq:dP2divs} to eliminate three of these line
bundles, so that only  two of the five coordinates on $dP_2$ take values
in non-trivial line bundles. We make the
following assignment of line bundles on $B$ to the coordinates,
\beq \label{eq:LBassignment}
	u\in \mathcal{O}_B(\cS_9+ [K_{B}])\,,\qquad v\in \mathcal{O}_B(\cS_9-\cS_7)\,,
\eeq
where $K_B$ denotes the canonical bundle on $B$ and $[K_B]$ the
associated divisor. All other coordinates on $dP_2$
transform as the trivial bundle on $B$. We note that this parametrization
of the two line bundles for $u$ and $v$ is completely general, because
$\cS_7$ and $\cS_9$ are completely general divisors on $B$ at the moment.

Next, we use these results to readily calculate the total Chern class of
$dP^B_2(\cS_7,\cS_9)$ from adjunction,
see \eqref{eq:c(dP2B)} in appendix \ref{app:CohomologyRing},  from which
we obtain its anti-canonical bundle
\beq \label{eq:antiKdP2}
	K_{dP^B_2}^{-1}=\mathcal{O}(3H-E_1-E_2+2\cS_9-\cS_7)\,,
\eeq
where we suppressed the dependence on $\cS_7$, $\cS_9$ for brevity
of our notation.
Then the Calabi-Yau condition implies that the constraint
\eqref{eq:CYindP2} has to be a section of
$K_{dP^B_2}^{-1}$. This immediately fixes the line bundles
of all the sections $s_i$ on $B$. We summarize the
sections defining the elliptically fibered Calabi-Yau manifold $\hat{X}$
as follows
\beq \label{eq:sectionsFibration}
\text{
\begin{tabular}{c|c}
\text{section} & \text{bundle}\\
\hline
	$u$&$\mathcal{O}(H-E_1-E_2+\cS_9+[K_B])$\rule{0pt}{13pt} \\
	$v$&$\mathcal{O}(H-E_2+\cS_9-\cS_7)$\rule{0pt}{12pt} \\
	$w$&$\mathcal{O}(H-E_1)$\rule{0pt}{12pt} \\
	$e_1$&$\mathcal{O}(E_1)$\rule{0pt}{12pt} \\
	$e_2$&$\mathcal{O}(E_2)$\rule{0pt}{12pt} \vspace{1.73cm}\\
\end{tabular}
}\qquad \text{
\begin{tabular}{c|c}
\text{section} & \text{bundle}\\
\hline
	$s_1$&$\mathcal{O}(3[K_B^{-1}]-\cS_7-\cS_9)$\rule{0pt}{13pt} \\
	$s_2$&$\mathcal{O}(2[K_B^{-1}]-\cS_9)$\rule{0pt}{12pt} \\
	$s_3$&$\mathcal{O}([K_B^{-1}]+\cS_7-\cS_9)$\rule{0pt}{12pt} \\
	$s_5$&$\mathcal{O}(2[K_B^{-1}]-\cS_7)$\rule{0pt}{12pt} \\
	$s_6$&$\mathcal{O}([K_B^{-1}])$\rule{0pt}{12pt} \\
	$s_7$&$\mathcal{O}(\cS_7)$\rule{0pt}{12pt} \\
	$s_8$&$\mathcal{O}([K_B^{-1}]+\cS_9-\cS_7)$\rule{0pt}{12pt} \\
	$s_9$&$\mathcal{O}(\cS_9)$ \rule{0pt}{12pt}
\end{tabular}
}
\eeq
In particular we see that with the parametrization
\eqref{eq:LBassignment} the divisors $\cS_7$ and $\cS_9$ are indeed
associated to $s_7$ and $s_9$ as claimed at the beginning.

\subsubsection*{Basic geometry of Calabi-Yau manifolds with $dP_2$-elliptic
fiber}
Having constructed the general elliptically fibered Calabi-Yau manifolds
$\hat{X}$ over $B$, we discuss next the group of divisors on $\hat{X}$.
By construction, the basis of divisors on a generic\footnote{By generic
we mean the absence of Cartan divisors $D_i$ from resolutions of
codimension one singularities of the fibration of $\hat{X}$. We will
briefly discuss the geometry of $\hat{X}$ in the presence of $D_i$ at the
end of this section. We refer to section \ref{sec:FFEllipticFibSU5U12}
for more details.} $\hat{X}$ is induced
by a basis of divisors on the ambient space $dP_2^B(\cS_7,\cS_9)$, which
consists of divisors of the base $B$ and the fiber $dP_2$.
The divisors induced from a basis of divisors $D_\alpha^b$ of
the base $B$ are the vertical divisors
$D_\alpha=\pi^*(D_\alpha^b)$ of the elliptic fibration
$\pi:\,\hat{X}\rightarrow B$ . Similarly, the classes $H$, $E_1$, $E_2$
of the fiber $dP_2$ in \eqref{eq:dP2divs} become divisors on $\hat{X}$.
Then, the points $P$, $Q$ and $R$ in \eqref{eq:coordsPQRdP2} lift to,
in general, \textit{rational sections} of the fibration of
$\pi:\,\hat{X}\rightarrow B$, denoted
$\hat{s}_P$, $\hat{s}_Q$ and $\hat{s}_R$, with $\hat{s}_P$ the zero
section.
We denote the homology classes of the associated divisors
by capital letters,
\beq \label{eq:divSPSQSR}
	S_P=E_2\,,\qquad S_Q=E_1\,,\qquad S_R=H-E_1-E_2+\cS_9+[K_B]\,.
\eeq

In general, a rational section is a
non-holomorphic map  of the base $B$ into $\hat{X}$, such as
$\hat{s}_P:\,B \rightarrow \hat{X}$ for example. A rational section
$B\rightarrow \hat{X}$ is ill-defined over codimension two loci to the
effect that it wraps entire
fiber components over these loci. From a given rational section, one can
easily obtain a \textit{holomorphic section}, i.e.~a holomorphic map
$\hat{B}\rightarrow \hat{X}$, by a birational transformation,
namely a blow-up $\hat{B}\rightarrow B$ at those codimension two loci of
$B$. Usually the zero section $\hat{s}_P$ has been assumed to be
holomorphic in F-theory. Only lately, the possibility of a
non-holomorphic zero section  $\hat{s}_P$ in F-theory has
been studied \cite{Braun:2013yti,Cvetic:2013nia,Grimm:2013oga}.
The group of sections excluding the zero section $\hat{s}_P$
is the \textit{Mordell-Weil group} of rational sections on $\hat{X}$,
which in the case at hand is rank two and generated by $\hat{s}_Q$,
$\hat{s}_R$. For brevity of our notation, we will occasionally denote
the generators of the Mordell-Weil group and their divisor classes
collectively as
\beq \label{eq:defS_m}
	\hat{s}_m=(\hat{s}_Q,\hat{s}_R)\,,\qquad S_m=(S_Q,S_R)\,.
\eeq

There are some characteristic intersections involving the divisors $S_P$,
$S_Q$ and $S_R$ in \eqref{eq:divSPSQSR} that immediately follow from
the defining properties of a section.
We list them in the following and refer to
\cite{Park:2011ji,Morrison:2012ei,Cvetic:2012xn,Cvetic:2013nia} for a
more thorough discussion. We also give a simple criterion to distinguish
between rational and holomorphic sections.  A
more detailed account on intersections in the
presence of a rational zero section can be found in \cite{Grimm:2013oga}.
Here, we content ourselves with noting that
$\hat{s}_P$ is holomorphic if $\cS_9=0$ or $\cS_8=0$, $\hat{s}_Q$ is
holomorphic if $\cS_3=0$ or $\cS_7=0$ and $\hat{s}_R$ is holomorphic if
$\cS_7=0$ or $\cS_9=0$, cf.~the paragraph following
\eqref{eq:coordsPQRdP2} and \cite{Cvetic:2013nia}.

The following intersections and definitions will be crucial in
the rest of this work:
\\
\fbox{
\begin{minipage}{0.3\textwidth}
~\vspace{0.3cm}\\
\textbf{Universal intersection:}\\
~\vspace{0.5cm}\\
\textbf{\phantom{....\,...}Rational sections:}\\
~\vspace{0.6cm}\\
\textbf{\phantom{.}Holomorphic sections:}\\
~\vspace{0.4cm}\\
\phantom{\ .}\textbf{\phantom{.......... .}Shioda maps:}\\
~\vspace{0.2cm}\\
~\\
\textbf{\phantom{............}\text{Height pairing:}}\\
~\vspace{0.4cm}
\end{minipage}}
\hspace{-0.1cm}\fbox{
\begin{minipage}{0.7\textwidth}
\beq\label{eq:intSecFiber}
 \,\,\,\,\phantom{....}S_P\cdot F=S_m\cdot F=1\text{ with general fiber
$F\cong \mathcal{E}$}\,, \!\!
\eeq
\bea\label{eq:SP^2}
&\!\!\pi(S_P^2+[K_B^{-1}]\cdot S_P)=\pi(S_m^2+[K_B^{-1}]\cdot S_m)=0\,,&\!\\
	\label{eq:S7S9}
	& \quad\cS_7=\pi(S_P\cdot S_R)\,,\qquad \cS_9=\pi(S_Q\cdot S_R)\,,&
\eea
\beq\label{eq:holomorphicSec}
 S_P^2+[K_B^{-1}]\cdot S_P=S_m^2+[K_B^{-1}]\cdot S_m=0\,,
\eeq
\bea \label{eq:ShiodaMapSQSR}
\sigma(\hat{s}_Q)\!\!&\!\!=\!\!&\!\!S_Q-S_P-[K_B^{-1}]\,,\\
 \sigma(\hat{s}_R)\!\!&\!\!=\!\!&\!\!S_R-S_P-[K_B^{-1}]-\cS_9\,,\nn
\eea
\beq \label{eq:height_pairing}
\pi(\sigma(\hat{s}_m)\cdot \sigma(\hat{s}_n))\!=\!\begin{pmatrix}
	2[K_B]& [K_B]-\cS_7+\cS_9\\
	[K_B]+\cS_7-\cS_9 &2[K_B]-2\cS_9
\end{pmatrix}_{mn}
\eeq
\end{minipage}}
~\\

Let us briefly comment on these intersections in the order
of their appearance. The intersection \eqref{eq:intSecFiber} is an
immediate consequence of the definition of
a section: its divisor class intersects the general
class of the fiber $F\cong \mathcal{E}$ at a point.
The relation \eqref{eq:SP^2} can be shown by an adjunction argument, see
section \ref{sec:CYFFCohomologyRing} for direct cohomology
computations.
Here we have defined the a projection onto the homology $H^4(B)$ of the
base as
\beq \label{eq:pi}
	\pi(\mathcal{C})=(\mathcal{C}\cdot \Sigma^\alpha)D^b_\alpha\,,\qquad\quad \Sigma_b^\alpha\cdot D_\beta^b=\delta^\alpha_\beta\,
\eeq
for every complex surface $\mathcal{C}$ in $\hat{X}$. The intersection
pairings on $\hat{X}$, respectively, $B$ are denoted $\cdot$   and the
$\Sigma^\alpha=\pi^*(\Sigma^\alpha_b)$ arise from a basis of curves
$\Sigma_b^\alpha$ dual to the divisors $D_\alpha^b$ on $B$ as indicated
in the last equation in \eqref{eq:pi}.
We emphasize that in the case of a holomorphic section, the
relations \eqref{eq:SP^2} hold in the full homology of $\hat{X}$ as
indicated in \eqref{eq:holomorphicSec}.
The divisors $\cS_7$, $\cS_9$ are the codimension one
loci where the sections collide in the fiber $\mathcal{E}$, as discussed
below \eqref{eq:coordsPQRdP2}. They are encoded in the
intersections \eqref{eq:S7S9}.
Next, we introduce the divisors $\sigma(\hat{s}_Q)$, $\sigma(\hat{s}_R)$
in \eqref{eq:ShiodaMapSQSR}.
The map $\sigma$ is the Shioda map that takes here the form
\beq \label{eq:ShiodaMapU1only}	
	\sigma(\hat{s}_m) := S_m-\tilde{S}_P-\pi(S_m\cdot\tilde{S}_P)\ ,
\eeq
where we introduced the combination  \cite{Grimm:2011sk,Bonetti:2011mw}
\beq \label{eq:SPtilde}
	\tilde{S}_P=S_P+\frac{1}{2}[K_B^{-1}]\,.
\eeq
We refer to
\cite{shioda1990mordell,Park:2011ji,Morrison:2012ei,Cvetic:2012xn,Cvetic:2013nia}
for more details on the Shioda map and to section
\ref{sec:FFEllipticFibSU5U12} for the inclusion of an SU$(5)$-sector.
We note that the divisors \eqref{eq:ShiodaMapSQSR}
support U(1)-gauge fields in F-theory due to their vanishing
intersections with vertical divisors $D_\alpha$ and the zero-section, as
well as potential Cartan divisors $D_i$ of non-Abelian groups.
Finally, we have calculated the intersection matrix of the Shioda map
of $\hat{s}_Q$, $\hat{s}_R$ in \eqref{eq:height_pairing}.

We finish this section by some concluding definitions and remarks
on the general structure of the fibrations \eqref{eq:dP2fibration} and
$\hat{X}$. First, we
summarize the basis of divisors on $\hat{X}$ as
\beq \label{eq:basisH11X}
	D_A=(\tilde{S}_P,D_\alpha,D_i,\sigma(\hat{s}_m))\,, \quad A=0,1,\ldots,h^{(1,1)}(\hat{B})+\text{rk}(G)+3\,,
\eeq
where we have collectively denoted the basis \eqref{eq:ShiodaMapSQSR} as
$\sigma(\hat{s}_m)$. We have also introduced one set of Cartan divisors
$D_i$ with $i=1,\ldots,\text{rk}(G)$ in order to prepare
for the presence of a non-Abelian group $G$, as in section
\ref{sec:FFEllipticFibSU5U12} with $G=$SU(5). These divisors $D_i$ are
present for non-generic $\hat{X}$ with a resolved singularity of type $G$
of the elliptic fibration over codimension one in $B$. The $D_i$ admit a
fibration
\beq \label{eq:S_G}
	\xymatrix{
	c_{-\alpha_i} \ar[r] & 	D_i \ar[d]\\
	&\mathcal{S}^b_{G}
	}
\eeq
where the general fiber is a rational curve
$c_{-\alpha_i}\cong \mathbb{P}^1$ that corresponds to the simple root
$-\alpha_i$ of $G$. The divisor $\mathcal{S}_G^b$ in $B$
physically  supports  7-branes that give rise to the non-Abelian gauge
symmetry $G$ in F-theory
\cite{Morrison:1996na,Morrison:1996pp,Bershadsky:1996nh}.

Next, we expand the canonical bundle $K_B$ of the base $B$ in terms
of the vertical divisors $D_\alpha$ as
\beq \label{eq:KBexpansion}
	[K_B]=K^\alpha D_\alpha\,
\eeq
with coefficients $K^\alpha$. Similarly, we expand the divisors
\beq \label{eq:S7S9exp}
	\cS_7=n_7^\alpha D^b_\alpha\,,\qquad \cS_9=n_9^\alpha D^b_\alpha\,,
\eeq
with general positive integral coefficients $n_7^\alpha$, $n_9^\alpha$,
$\alpha=1,\ldots, h^{(1,1)}(B)$.
It is important to emphasize that the
coefficients $n_7^\alpha$, $n_9^\alpha$ are in general further bounded
from above by the requirement
that all sections $s_i$ in \eqref{eq:sectionsFibration} are generic,
i.e.~that the line bundle of $s_i$ admits sufficiently many holomorphic
sections. If this is not
the case we expect additional singularities in $\hat{X}$, potentially
corresponding to a minimal (non-Abelian) gauge symmetry in F-theory.
For this reason, we will in the rest of this work assume that $\hat{X}$
can be constructed with generic $s_i$.

Despite these restrictions on the integers $n_7^\alpha$ and $n_9^\alpha$
we would like to point out that the constructions of the fibration
\eqref{eq:dP2fibration} and of $\hat{X}$
hold in general for an arbitrary base $B$ and arbitrary complex
dimension. In particular this analysis applies to an
arbitrary choice of divisors $\cS_7$ and $\cS_9$ within these bounds.
In particular the general construction here reproduce immediately the
classification in \cite{Cvetic:2013nia} with $B=\mathbb{P}^2$ as a
special case.

\subsubsection*{$dP_2$-fibrations over $B=\mathbb{P}^3$ with generic Calabi-Yau hypersurfaces $\hat{X}$}

We conclude with the discussion of the special case $B=\mathbb{P}^3$,
which will be considered in later sections of this work. In this case,
there is only one divisor in the base, the
hyperplane $H_B$, so that the $dP_2$-fibration
\eqref{eq:dP2fibration} is specified only by two integers
$n_7\equiv n_7^1$, $n_9\equiv n_9^1$.
In this case we use the notation
\beq \label{eq:dP2fibrationP3}
	\xymatrix{
	dP_2 \ar[r] & 	dP_2(n_7,n_9) \ar[d]\\
	& \mathbb{P}^3\,
	}
\eeq
where we suppress the base $B=\mathbb{P}^3$ when denoting the total space
\eqref{eq:dP2fibration} of the fibration if the context is clear.

\begin{figure}[ht!]
\centering
\includegraphics[scale=0.5]{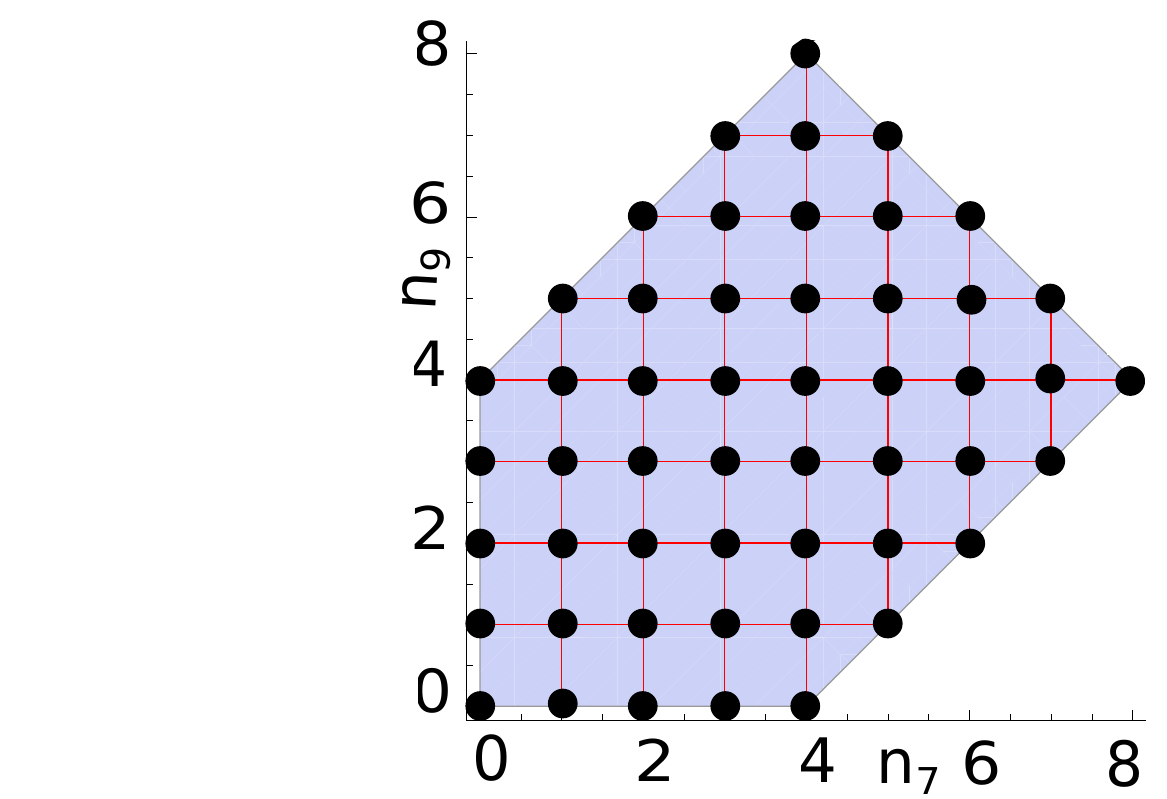} \qquad \qquad
 \caption{Each dot corresponds to a $dP_2$-fibration over
 $\mathbb{P}^3$ with
 generic Calabi-Yau $\hat{X}$.}
 \label{fig:XP3n7n9region}
\end{figure}

We note that
$K_{\mathbb{P}^3}^{-1}=\mathcal{O}_{\mathbb{P}}(4)$.
In this case all sections $s_i$ exist iff all bundles in
the second table in \eqref{eq:sectionsFibration} have
non-negative degree. This puts  the following conditions on
the integers $n_7$, $n_9$,
\beq \label{eq:XP3condn7n9}
		0\leq n_7,n_9\leq 8\,,\quad n_7+n_9\leq 12\,,\quad 0\leq 4+n_7-n_9\,,\quad 0\leq 4+n_9-n_7\,.
\eeq
The domain of allowed valued for $n_7$ and $n_9$ are displayed in figure
\ref{fig:XP3n7n9region}. As we will see in section \ref{sec:ToricExsU1U1}
we can torically construct the Calabi-Yau fourfolds $\hat{X}$
for a wide range of values of $n_7$, $n_9$. The general
strategy to build the corresponding reflexive polytopes is outlined in
appendix \ref{app:tuning-s7-s9}. It is satisfying,  that in the toric
context the conditions \eqref{eq:XP3condn7n9} are enforced by
reflexivity of the toric polytope, i.e.~for values $n_7$, $n_9$ exceeding
the bounds \eqref{eq:XP3condn7n9} the toric polytope is no longer
reflexive.

\section{Calabi-Yau Fourfolds with Rank Two Mordell-Weil}
\label{sec:FFEllipticFib}

In this section we analyze F-theory compactifications to four dimensions
on a generic elliptically fibered Calabi-Yau fourfold $\hat{X}$ over
a base $B$ with general fiber $\mathcal{E}$ in $dP_2$. These
compactifications have a gauge theory with U$(1)\times$U(1) gauge group
and a number of chiral matter fields in representations
$\mathbf{1}_{(q_1,q_2)}$. The possible U$(1)\times$U(1)-charges
$(q_1,q_2)$ have been determined
recently in \cite{Borchmann:2013jwa,Cvetic:2013nia} and
the full 6D anomaly-free spectrum including matter multiplicities has
been derived for $B=\mathbb{P}^2$ in \cite{Cvetic:2013nia}.

Here we extend this geometric analysis to fourfolds.
The main difference to the 6D case is that matter is not localized
anymore at points in $B$, but on in general rather complicated matter
curves. The determination of these matter curves and some of their
associated matter surfaces, along with the Yukawa points, is presented in
section \ref{sec:SingularFibU1U1}. Then, in section
\ref{sec:CYFFCohomologyRing} we present a method to determine the
cohomology ring of the fourfold $\hat{X}$. We use these
techniques to derive general expressions for the Euler number of
$\hat{X}$ and its second Chern class. For the example of $B=\mathbb{P}^3$
we finally compute  the full vertical cohomology group. These
calculations serve as a preparation for the computation of 4D chiralities
in section \ref{sec:G4+chiralitiesU1xU1}, which requires both
the knowledge of matter surfaces and the construction of $G_4$-flux.

\subsection{Singularities of the Fibration: Matter Surfaces \& Yukawa
Points}
\label{sec:SingularFibU1U1}

We organize this section into a detailed discussion of codimension
two singularities in section \ref{sec:codim2U1xU1} and a very brief
account on codimension three singularities in  section \ref{sec:codim3U1xU1}.

\subsubsection{Matter: Codimension Two}
\label{sec:codim2U1xU1}

In general, the determination of the matter sector in F-theory vacua with
general gauge group requires a detailed analysis of singularities of the
elliptic fibration of the Calabi-Yau fourfold at codimension two
in the base $B$, where the elliptic fiber $\mathcal{E}$ becomes
reducible. Then one has
to identify the isolated rational curve $c_\mathbf{w}$ in the fiber over
these loci, since these correspond in F-theory to matter in a
representation $\mathbf{R}$ from wrapped M2-brane states. These curves
are in one-to-one correspondence to the weights $\mathbf{w}$ of
the representations $\mathbf{R}$ and accordingly labeled. In the case of
elliptically fibered Calabi-Yau fourfolds, the codimension two matter
loci are Riemann surfaces of genus $g$, the so-called matter curves
$\Sigma_\mathbf{R}$ in $B$ conveniently labeled by the corresponding
matter representation $\mathbb{R}$.  In addition, for the determination
of four-dimensional chirality, compare section
\ref{sec:G4+chiralitiesU1xU1}, we have to know the homology classes of
the associated matter surfaces
\beq \label{eq:matterSurfs}
	\xymatrix{
	c_\mathbf{w} \ar[r] & 	\cC^\mathbf{w}_\mathbf{R} \ar[d]\\
	& \Sigma_\mathbf{R}
	}
\eeq
which are constructed as the fibration of the rational curve
$c_\mathbf{w}$ corresponding to a given weight $\mathbf{w}$
of the representation $\mathbf{R}$ fibered over $\Sigma_\mathbf{R}$.

In this section we determine the matter curves $\Sigma_\mathbf{R}$ and
the matter surfaces $\cC^\mathbf{w}_\mathbf{R}$
for the six representations occurring in the Calabi-Yau fourfold
$\hat{X}$. As we demonstrate, their determination  is complicated
by the fact that three of the six the codimension two loci in the base
$B$ where the elliptic fiber $\mathcal{E}$ becomes reducible
are themselves reducible curves. Their irreducible components are
multiple different matter curves $\Sigma_{\mathbf{R}}$.
Some of these matter curves, denoted $\Sigma_{\mathbf{R}'}$,
fail to be complete intersection and can
only be described in terms of their prime ideals. These
prime ideals are straightforwardly constructed from the two equations of
the original reducible codimension two locus.
However, the isolation of rational curves $c_\mathbf{w}$ over
those matter curves $\Sigma_{\mathbf{R}'}$ is very involved.
Thus, in these cases we can not determine the corresponding matter
surfaces \eqref{eq:matterSurfs} explicitly. Fortunately, we can obtain
the other three matter surfaces straightforwardly, and are still
able to determine the full F-theory matter spectrum for the fourfold
$\hat{X}$, as outlined in section \ref{sec:G4+chiralitiesU1xU1}.
It would be desirable, however, to reproduce the results obtained there
invoking M-/F-theory duality by direct geometric computation
based on a better understanding of the matter surfaces
$\cC_{\mathbf{R}'}$ in general.

In any case, we can qualitatively describe all the matter surfaces
$\cC^{\mathbf{w}}_{\mathbf{R}}$  by recalling the
construction of the resolved fourfold $\hat{X}$.
The smooth fourfold $\hat{X}$ is formed by two consecutive blow-ups of
a singular Weierstrass model $X$. We depict this
schematically as
\beq \label{eq:TwoResolutions}
	\xymatrix{{ \begin{array}{c}
	\hat{X} \subset dP_2^B(\cS_7,\cS_9) \\
	\text{generic CY}
	\end{array}}
	 \ar[rr]^{\hat{\pi}}\ar[rd]_{\pi_2}& & {\begin{array}{c}
	X\subset (\mathbb{P}^2(1,2,3)\rightarrow B) \\
	\text{non-generic WSF}
	\end{array}}\\
	& {\begin{array}{c}
	\tilde{X}\subset (\mathbb{P}^2\rightarrow B) \\
	\text{non-generic cubic}
	\end{array}}\ar[ru]_{\pi_1}&	}
\eeq
where the full blow-down map $\hat{\pi}:\,\hat{X}\rightarrow X$
is consequently a composition $\hat{\pi}=\pi_1\circ\pi_2$.
On the left we have the smooth geometry with elliptic fiber constructed
in section \ref{sec:ellipticCurvedP2}. It can be understood
as a toric blow-up $\pi_2:\,\hat{X}\rightarrow\tilde{X}$ from a non-
generic cubic in $\mathbb{P}^2$, with corresponding fourfold denoted by
$\hat{X}$. A final blow-down $\pi_1$ yields the singular Weierstrass form (WSF) $X$
with $\mathbb{P}^2(1,2,3)$-fiber. The birational map $\pi_1$
is derived in detail in \cite{Cvetic:2013nia}, see its defining equations Eqs.~(3.18) and (3.20) therein.  

Having the diagram \eqref{eq:TwoResolutions} in mind, the three
matter  surfaces $\cC_{\mathbf{R}}$ which
have a simple description are those generated in the blow-up $\pi_2$.
There are three simple codimension two singularities in $\tilde{X}$,
which are precisely the three simple matter curves $\Sigma_{\mathbf{R}}$.
Their pull-backs under $\pi_2$ are precisely the matter surfaces
$\cC_{\mathbf{R}}=\pi_2^*(\Sigma_{\mathbf{R}})$. Because of the
simplicity of both $\Sigma_{\mathbf{R}}$ and the blow-up $\pi_2$,
these surfaces have a description as a simple complete intersection in
the ambient space $dP_2(\cS_7,\cS_9)$. In contrast,
the other three matter curves $\Sigma_{\mathbf{R}'}$ are the loci of
codimension two singularities in the WSF $X$, which are
resolved by the map $\pi_1$. However, these curves $\Sigma_\mathbf{R}'$
have a description only in terms of prime ideals and the map $\pi_1$ is
not a simple toric blow-up but a fully-fledged birational map \cite{Cvetic:2013nia}. These two
complications make an explicit determination of the surfaces
$\cC^{\mathbf{w}}_{\Sigma_\mathbf{R}'}$ hard. Nevertheless,  the matter
surfaces are again abstractly given by
$\cC_{\mathbf{R}'}=\pi_1^*(\Sigma_\mathbf{R}')$, which are ruled surfaces over $\Sigma_\mathbf{R}'$.
Thus, the determination of the exceptional loci of the map $\pi_1$ might
be a first step towards an understanding of these matter surfaces.

\subsubsection*{Summary of Matter Representations \& Their Matter Curves}

Before going into technical calculations of matter curves and surfaces,
let us briefly summarize the matter content as it has been determined in
\cite{Borchmann:2013jwa,Cvetic:2013nia}.

There are six different matter representations
$\mathbf{R}=\mathbf{1}_{(q_1,q_2)}$ in the F-theory
compactification on the fourfold $\hat{X}$. The list of realized
U$(1)\times$U(1)-charges, together with the cohomology class of the
corresponding matter curves $\Sigma_{\mathbf{R}}$ determined below, reads
\begin{equation}
\label{eq:U1spectrum}
	\text{\begin{tabular}{|c||c|}
\hline
	Matter& Homology class of $\Sigma_\mathbf{R}$ in $B$\rule{0pt}{13pt} \\\hline  \hline
	 $\mathbf{1}_{(1,0)}$& $6[K_B^{-1}]^2+4[K_B^{-1}] \cdot\cS_7-5[K_B^{-1}]\cdot \cS_9+\cS_9^2+\cS_7\cdot \cS_9-2 \cS_7^2$\rule{0pt}{14pt} \\ \hline
	 $\mathbf{1}_{(0,1)}$& $6[K_B^{-1}]^2+4 [K_B^{-1}]\cdot (\cS_7+\cS_9)-2\cS_7^2-2\cS_9^2$\rule{0pt}{14pt}\\\hline
	 $\mathbf{1}_{(1,1)}$&$6[K_B^{-1}]^2+4[K_B^{-1}]\cdot \cS_9-5[K_B^{-1}]\cdot \cS_7+\cS_7^2+\cS_7\cdot \cS_9-2\cS_9^2$\rule{0pt}{14pt}\\\hline
	 $\mathbf{1}_{(-1,1)}$&$\left([K_B^{-1}]+\cS_7-\cS_9\right)\cdot\cS_7 $\rule{0pt}{14pt}\\\hline
	 $\mathbf{1}_{(0,2)}$&$ \cS_7\cdot \cS_9$\rule{0pt}{14pt}\\\hline
	 $\mathbf{1}_{(-1,-2)}$&$ \cS_9 \cdot\left([K_B^{-1}]+\cS_9-\cS_7\right)$ \rule{0pt}{14pt}\\\hline
\end{tabular}}
\end{equation}
Here we used as before the notation $[K_B^{-1}]$ for the anti-canonical
divisor of the base and denoted the intersection on $B$ as '$\cdot$'.
These representations of matter fields are model-independent and in
particular do not depend on the choice of base $B$. The last three
matter representations arise from rational curves created in
the blow-up $\pi_2^{-1}$ in \eqref{eq:TwoResolutions}. Their matter
curves are simply described by  $s_3=s_7=0$, $s_7=s_9=0$ and $s_8=s_9=0$ in
the order of their appearance in \eqref{eq:U1spectrum}. The first three
representations arise from rational curves from the blow-up
$\pi_1^{-1}$ in \eqref{eq:TwoResolutions}. The determination of their
matter curves is more involved and presented below.

\begin{figure}[ht!]
\centering
 \includegraphics[scale=0.4]{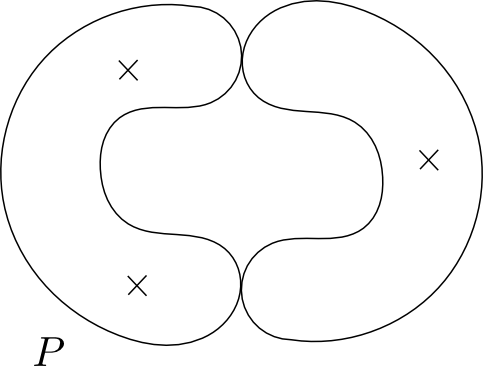}
 \caption{$I_2$-fiber from resolving a codimension two
 singularity of the fibration of $\hat{X}$.}
 \label{fig:I2fiber}
\end{figure}

All the matter representations in \eqref{eq:TwoResolutions} arise
from M2-branes on rational curves $c_{\mathbf{w}}$ with
wight $\mathbf{w}=(q_1,q_2)$. These charges are calculated by the
intersection of the curve $c_\mathbf{w}$  with
the Shioda maps $\sigma(\hat{s}_Q)$, $\sigma(\hat{s}_R)$ defined in
\eqref{eq:ShiodaMapSQSR} as
\beq \label{eq:U1chargeGen}
	q_m\equiv \sigma(\hat{s}_m)\cdot c_\mathbf{w}=(S_{m}\cdot c_\mathbf{w})-(S_P\cdot c_\mathbf{w})\,,
\eeq
All curves $c_{\mathbf{w}}$  are part of an $I_2$-fiber.
Along the matter surfaces in \eqref{eq:U1spectrum} the general
elliptic fiber $\mathcal{E}$ splits into two
rational curves $c_1,\,c_2\cong \mathbb{P}^1$ intersecting in two points
with one curve, say $c_1$, the original singular fiber and the other
curve $c_2\equiv c_{\mathbf{w}}$. We write this as
\beq \label{eq:I2split}
	I_2\text{-fiber}\,:\qquad \mathcal{E}=c_1+c_2\,, \quad c_1\cdot c_2=2\,.
\eeq
A cartoon of such a reducible fiber together with possible locations
of the points $P$, $Q$ and $R$ is depicted in figure \ref{fig:I2fiber}.
In terms of the Calabi-Yau constraint \eqref{eq:CYindP2} the split of
$\mathcal{E}$ into an $I_2$-fiber is visible as a factorization at a
point $pt\in \Sigma_{\mathbf{R}}$ as
\beq \label{eq:CYfactorized}
	p\vert_{pt}=p_1\cdot p_2\,.
\eeq
Here the two rational curves in \eqref{eq:I2split} are described by one
of these two factors, for example $c_1=\{p_1=0,pt\in \Sigma_{\mathbf{R}}\}$ and $c_2=\{p_2=0\,,pt\in \Sigma_{\mathbf{R}}\}$.

\subsubsection*{Matter Surfaces $\cC_{(-1,1)}$, $\cC_{(0,2)}$, $\cC_{(-1,-2)}$ and their Homology Classes}

Next, we determine the matter surfaces
$\mathcal{C}_{\mathbf{R}}^{\mathbf{w}}$ for the last three
representations in \eqref{eq:U1spectrum}. As we will see, this is
straightforward since the matter curves $\Sigma_\mathbf{R}$ in
these cases are irreducible varieties of the simple form $s_i=s_j=0$ for
appropriate $i$, $j$. This
implies that the factorization of the elliptic fiber
$\mathcal{E}=c_1+c_2$ is manifest over the entire matter
curve $\Sigma_\mathbf{R}$ and the matter
surfaces $\cC^{\mathbf{w}}_\mathbf{R}$ can be described by a complete
intersection of three constraints in the ambient space
$dP_2^B(\cS_7,\cS_9)$. Then its homology class is given simply by the
product of the divisor classes of each of these
constraints.

The resulting homology classes of matter surfaces read
\beq	
	\label{eq:MatSurfaces1}
	\text{\begin{tabular}{|c|c|}
\hline
	Matter surface& Homology class \rule{0pt}{13pt} \\\hline  \hline
	 $\cC_{\mathbf{1}_{(-1,1)}}$& $([K_B^{-1}]+\cS_7-\cS_9)\cdot \cS_7\cdot E_1$\rule{0pt}{14pt} \\ \hline
	 $\cC_{\mathbf{1}_{(0,2)}}$& $\cS_7\cdot \cS_9 \cdot ([K_B^{-1}]+\cS_9-\cS_7+2H)$\rule{0pt}{14pt}\\\hline
	 $\cC_{\mathbf{1}_{(-1,-2)}}$&$([K_B^{-1}]+\cS_9-\cS_7)\cdot \cS_9\cdot(3H-E_1-2E_2+2\cS_9-\cS_7) $\rule{0pt}{14pt}\\\hline
\end{tabular}}
\eeq
Here we suppressed the weight $\mathbf{w}$ since it is identical to
the charges $(q_1,q_2)$.
We obtain the homology class of the first matter surface
$\cC_{\mathbf{1}_{(-1,1)}}$ by noting its description as the complete
intersection
\beq
\label{eq:charge-11}
		\cC_{\mathbf{1}_{(-1,1)}}\,=\{ s_3=s_7=0\,, \,\,e_1=0\}\,
\eeq
in the ambient space $dP_2^B(\cS_7,\cS_9)$.
Here the first two equations describe the matter curve
$\Sigma_{\mathbf{1}_{(-1,1)}}=\{s_3=s_7=0\}$ over which the Calabi-Yau
constraint \eqref{eq:CYindP2} factorizes as $p=p_1 p_2$ with one
factor given by $e_1$, cf.~section 4.3 of
\cite{Cvetic:2013nia}. Thus, the rational isolated curve is described
as $c_{(-1,1)}=\{e_1=0,\, pt\in \Sigma_{(-1,1)}\}$ over all the points of
$\Sigma_{(-1,1)}$.
The homology class in the first line of \eqref{eq:MatSurfaces1} for
$\cC_{\mathbf{1}_{(-1,1)}}$ in the ambient
space $dP_2^B(\cS_7,\cS_9)$ is then easily obtained
from \eqref{eq:charge-11}  employing the assignments
\eqref{eq:sectionsFibration} of line bundles to $s_3$, $s_7$ and $e_1$.

Similarly we obtain the homology classes of the matter surfaces
$\cC_{\mathbf{1}_{(0,2)}}$ and $\cC_{\mathbf{1}_{(-1,-2)}}$. In the
former case the matter curve is
$\Sigma_{\mathbf{1}_{(0,2)}}=\{s_7=s_9=0\}$ over which
the Calabi-Yau constraint \eqref{eq:CYindP2} factorizes globally with
the isolated rational curve given by \cite{Cvetic:2013nia}
\beq \label{eq:c02}
c_{(0,2)}=\{ s_1e_1^2 e_2^2  u^2 + s_2e_1 e_2^2  u v + s_3e_2^2  v^2 + s_5e_1^2 e_2  u w +s_6
   e_1 e_2 v w +  s_8 e_1^2w^2=0\}\,.
\eeq
Here it is understood that the sections $s_i$ are evaluated on
$\Sigma_{\mathbf{1}_{(0,2)}}$. The homology class of this complete
intersection is the product of the class of $c_{(0,2)}$ and of
$\Sigma_{\mathbf{1}_{(0,2)}}$ and we immediately  reproduce  the second
line in \eqref{eq:MatSurfaces1} using \eqref{eq:sectionsFibration}.
Finally, the matter curve for $\mathbf{1}_{(-1,-2)}$ is given by
$\Sigma_{\mathbf{1}_{(-1,-2)}}=\{s_8=s_9=0\}$ and the isolated rational
curve is \cite{Cvetic:2013nia}
\beq \label{eq:cm1m2}
	c_{(-1,-2)}=\{s_1 e_1^2 e_2  u^3 + s_2 e_1 e_2 u^2 v + s_3 e_2 u v^2 +s_5  e_1^2 u^2 w +
  s_6e_1 u v w + s_7 v^2 w=0\}\,,
\eeq
where as before the $s_i$ are evaluated on
$\Sigma_{\mathbf{1}_{(-1,-2)}}$. Then, the matter surface is
again  a complete intersection in the ambient space
\eqref{eq:dP2fibration} and its homology class, employing the line
bundles \eqref{eq:sectionsFibration}, is indeed given by the third line
in \eqref{eq:MatSurfaces1}.

\subsubsection*{Matter Curves $\Sigma_{(1,0)}$, $\Sigma_{(0,1)}$, $\Sigma_{(1,1)}$ and their Prime Ideals}

As mentioned before, the three remaining  matter curves
$\Sigma_{\mathbf{R}'}$ are themselves no simple complete intersections,
but  contained in a  reducible codimension two
subvarieties in $B$ that are complete intersections. To isolate the
component $\Sigma_{\mathbf{R}'}$ of interest we have to determine its
prime ideal. This prime ideal is generated by more than two constraints,
but still describes a codimension two variety in $B$. In
addition, the factorization \eqref{eq:CYfactorized}  describing
the split $\mathcal{E}=c_1+c_2$ of the elliptic curve does not occur
globally over the matter curves $\Sigma_{\mathbf{R}'}$, but is manifest
only at generic points of $\Sigma_{\mathbf{R}'}$.
These combined effects render the determination of the homology
class of the matter surfaces $\cC_{\mathbf{R}'}$ unfeasible.
However, we can obtain the homology class of the matter
curves $\Sigma_{\mathbf{R}'}$ as shown next. For completeness
we will also present the prime ideal for one illustrative example.
In general the prime ideals are needed for a thorough analysis
of codimension three singularities presented in section
\ref{sec:codim3U1xU1}.

We begin with the determination of the homology classes of the matter
curves $\Sigma_{\mathbf{R}'}$.
They can be obtained by first determining the homology class of the two
equations for the reducible codimension two locus in $B$ and by then
subtracting the classes of those components $\Sigma_{\mathbf{R}}$ we
are not interested in. As in the six-dimensional case
\cite{Cvetic:2013nia} we have to subtract the components
$\Sigma_{\mathbf{R}}$ with the right multiplicity, which is computed by
the resultant\footnote{In general, the resultant gives the order of a
root of two polynomials in two variables.} of the two equations at the
root corresponding to $\Sigma_{\mathbf{R}}$. The resulting homology
classes of this computations give the first
three lines of \eqref{eq:U1spectrum}. We work out these homology classes
in detail in the remainder of this section.

First we present the equations for the reducible codimension two loci in
$B$ that contain the three matter
curves $\Sigma_{\mathbf{R}}$ that we are interested in as
irreducible  components.  These codimension two loci read
\cite{Cvetic:2013nia}
\bea \label{eq:charge11reducible}
  \text{loc}_1\!\!&\!\!=\!\!&\!\!\{s_7 s_8^2 + s_9 ( s_5 s_9-s_6 s_8)=s_3 s_8^2 - s_2 s_8 s_9 + s_1 s_9^2=0\}\,,\\
  \text{loc}_2\!\!&\!\!=\!\!&\!\!\{s_3 s_6 s_8 - s_2 s_7 s_8 - s_3 s_5 s_9 + s_1 s_7 s_9=s_3^2 s_8^2+s_7 \left(s_1 s_7 s_8+s_2 s_5 s_9-s_2 s_6 s_8\right)\nn\\&&+s_3 \left(s_6^2 s_8-s_5 s_7 s_8-s_5 s_6 s_9-s_2 s_8 s_9+s_1 s_9^2\right) =0\}\,,\nn\\
  \text{loc}_3\!\!&\!\!=\!\!&\!\!\{2 s_7^3 s_8^3 + s_3 s_9^3 ( s_5 s_9-s_6 s_8) +
 s_7^2 s_8 s_9 ( 2 s_5 s_9-3 s_6 s_8)- s_7 s_9^2 ( s_5 s_6 s_9 + s_2 s_8 s_9-s_6^2 s_8 \nn\\&&- 2 s_3 s_8^2 - s_1 s_9^2)=s_7^4 s_8^4+2 s_7^3 s_8^2 s_9 (s_5 s_9-s_6 s_8)+s_7 s_9^3 (2 s_3 s_8-s_2 s_9) (s_5 s_9-s_6 s_8)\nn\\
 &&-s_7^2 s_9^2 (2 s_5 s_6 s_8 s_9+s_2 s_8^2 s_9\!-\!s_6^2 s_8^2\!-\!2 s_3 s_8^3\!-\!s_5^2 s_9^2)+s_3 s_9^4(s_3 s_8^2+s_9 (s_1 s_9\!-\!s_2 s_8))=0\}\nn
\eea

By calculating the
associated prime ideals\footnote{The ideal generated by $\text{loc}_1$ is
the intersection of its associated primary ideals. An ideal $I$ is
primary ideal if $a b\in I$ implies $a\in I$ or $b^n\in I$ for some
$n>0$. If $n=1$, the ideal $I$ is a prime ideal.} of $\text{loc}_1$ we
see that it has two irreducible components. One is obviously the matter
curve $\Sigma_{\mathbf{1}_{(-1,-2)}}=\{s_8=s_9=0\}$ and the other one is
the matter curve $\Sigma_{\mathbf{1}_{(1,1)}}$. Then we
determine the homology class of the reducible variety $\text{loc}_1$.
We further recall from \cite{Cvetic:2013nia} that the order
of the root $(s_8,s_9)=(0,0)$ of the two polynomials in $\text{loc}_1$ is
$4$. Thus, decompose the class of $\text{loc}_1$ as
\bea \label{eq:Sigma11Homology}
		&[\text{loc}_1]&\!\!=(2[K_B^{-1}]+2\cS_9-\cS_7)\cdot (3[K_B^{-1}]+\cS_9-\cS_7)\cong \Sigma_{(1,1)}+4\Sigma_{(-1,-2)}\nn\\
		&\Rightarrow \!&\!\! \Sigma_{(1,1)}\cong 6[K_B^{-1}]^2+4[K_B^{-1}]\cdot \cS_9-5[K_B^{-1}]\cdot \cS_7+\cS_7^2+\cS_7\cdot \cS_9-2\cS_9^2\,,
\eea
where we used \eqref{eq:sectionsFibration} and denote the equivalence
relation in homology  as '$\cong$'. Here
we also used that the homology class of $\Sigma_{\mathbf{1}_{(-1,-2)}}$
as given in the third line of \eqref{eq:MatSurfaces1} by
the first two factors. Thus, we have obtained
the homology class of $\Sigma_{\mathbf{1}_{(1,1)}}$ as in
\eqref{eq:U1spectrum}.

Similarly, we obtain the homology class of $\Sigma_{\mathbf{1}_{(1,0)}}$.
We calculate four associated prime ideals of $\text{loc}_2$ in
\eqref{eq:charge11reducible} that correspond to four different
irreducible components. These components are the curves
$\Sigma_{\mathbf{1}_{(-1,1)}}=\{s_3=s_7=0\}$,
$\Sigma_{\mathbf{1}_{(-1,-2)}}=\{s_8=s_9=0\}$,
$\Sigma_{\mathbf{1}_{(1,1)}}$ and finally
$\Sigma_{\mathbf{1}_{(1,0)}}$. By calculating  the
resultants of  $\text{loc}_2$ at the relevant roots, we obtain
multiplicities one for all irreducible
components. Thus, we obtain the homology class of $\Sigma_{(1,0)}$ from
decomposition of the class of $\text{loc}_2$ as
\bea 	&[\text{loc}_2]&=12[K_B^{-1}]^2=\Sigma_{(-1,1)}+\Sigma_{(-1,-2)}+\Sigma_{(1,1)}+\Sigma_{(1,0)}\nn\\
		&\Rightarrow &\!\! \Sigma_{(1,0)}\cong 6[K_B^{-1}]^2+4[K_B^{-1}] \cdot\cS_7-5[K_B^{-1}]\cdot \cS_9+\cS_9^2+\cS_7\cdot \cS_9-2 \cS_7^2\,,
\eea
where we have used \eqref{eq:sectionsFibration}, the homology classes of
matter curves in \eqref{eq:MatSurfaces1} as well
as in \eqref{eq:Sigma11Homology}. This is the result in
\eqref{eq:U1spectrum}.

Finally, we determine the homology class of the matter curve
$\Sigma_{\mathbf{1}_{(0,1)}}$.
The ideal $\text{loc}_3$ in  \eqref{eq:charge11reducible} has five prime
ideals corresponding to the matter curves
$\Sigma_{\mathbf{1}_{(-1,1)}}$, $\Sigma_{\mathbf{1}_{(0,2)}}$,
$\Sigma_{\mathbf{1}_{(-1,-2)}}$, $\Sigma_{\mathbf{1}_{(1,1)}}$
and the matter curve $\Sigma_{\mathbf{1}_{(0,1)}}$ we are interested in.
The  multiplicities of the irreducible components we are not interested
are calculated as one, 16, 16, one, respectively, by the corresponding
resultants of $\text{loc}_3$.
Thus, we calculate the homology class of the curve
$\Sigma_{\mathbf{1}_{(0,1)}}$ from the homology class of
$\text{loc}_3$ as
\bea
		&\text{loc}_3&=12[K_B^{-1}]^2=\Sigma_{(-1,1)}+16\cdot \Sigma_{(0,2)}+16\cdot\Sigma_{(-1,-2)}+\Sigma_{(1,1)}+\Sigma_{(0,1)}\nn\\
		&\Rightarrow &\!\! \Sigma_{(0,1)}=6[K_B^{-1}]^2+4 [K_B^{-1}]\cdot (\cS_7+\cS_9)-2\cS_7^2-2\cS_9^2\,,
\eea
where we used the homology class of the matter curves in
\eqref{eq:MatSurfaces1} and \eqref{eq:Sigma11Homology}. This is the
homology class in \eqref{eq:U1spectrum}.

We conclude this discussion by presenting the associated
prime ideal of selected matter surfaces as an instructive preparation
of section \ref{sec:codim3U1xU1}.
The prime ideal of $\Sigma_{\mathbf{1}_{(1,1)}}$ reads
\bea  \label{eq:charge11PI}
	\mathcal{P}&=&\left\{s_3^2 s_5^2+s_7 \left(s_2^2 s_5-s_1 s_2 s_6+s_1^2 s_7\right)+s_3 \left(-s_2 s_5 s_6+s_1 \left(s_6^2-2 s_5 s_7\right)\right),\right.\nn\\
	&\phantom{=}&\left. s_3 s_5^2 s_9+s_2 s_5 \left(s_7 s_8-s_6 s_9\right)+s_1 \left(-s_6 s_7 s_8+s_6^2 s_9-s_5 s_7 s_9\right),\right.\nn\\&\phantom{=}&\left. s_3 s_5 s_8-s_1 s_7 s_8-s_2 s_5 s_9+s_1 s_6 s_9,s_3 s_6 s_8-s_2 s_7 s_8-s_3 s_5 s_9+s_1 s_7 s_9,\right.\nn \\ &\phantom{=}&\left.s_3 s_8^2+s_9 \left(-s_2 s_8+s_1 s_9\right),  s_7 s_8^2+s_9 \left(-s_6 s_8+s_5 s_9\right)\right\}.
\eea
The dimension of $\mathcal{P}$ is calculated to be six in the ring
generated by the $s_i$ which confirms that the
irreducible variety  described by it is codimension two in $B$ as
expected. It is evident from \eqref{eq:charge11PI} that the two
irreducible components
$\Sigma_{\mathbf{1}_{(1,1)}}$ and
$\Sigma_{\mathbf{1}_{(-1,-2)}}=\{s_8=s_9=0\}$ of $\text{loc}_1$
intersect at points in
$B$. These points may correspond to Yukawa points in F-theory
since the fiber in the resolved space $\hat{X}$ splits further
into three components, as can be seen by a prime ideal analysis.
We also determine the prime ideal of the matter curve
$\Sigma_{\mathbf{1}_{(1,0)}}$ as
\bea \label{eq:charge10PI}
	\mathcal{P}&=&\left\{-s_2^2 s_8^2+s_2 \left(s_5 s_6 s_8-s_5^2 s_9+2 s_1 s_8 s_9\right)-s_1 \left(s_6^2 s_8-s_5 s_6 s_9+s_1 s_9^2\right),\right.\nn\\
	&\phantom{=}&\left. s_2 s_5 s_7-s_1 s_6 s_7-s_2 s_3 s_8+s_1 s_3 s_9,s_3 s_5 s_7-s_1 s_7^2-s_3^2 s_8,\right.\nn\\
	&\phantom{=}&\left. s_3 s_6 s_7-s_2 s_7^2-s_3^2 s_9,s_3 s_6 s_8-s_2 s_7 s_8-s_3 s_5 s_9+s_1 s_7 s_9\right\}\,.
\eea
From this ideal we see that $\Sigma_{\mathbf{1}_{(1,0)}}$ intersects the
matter curve $\Sigma_{\mathbf{1}_{(-1,1)}}$, where again there is an
additional split of the fiber.

\subsubsection{Yukawa Couplings: Codimension Three}
\label{sec:codim3U1xU1}

At codimension three the singularities of the fibration enhance further
signaling the presence of a  Yukawa point. In the case at hand we find an
enhancement to an $I_3$-singularity, which is resolved into three
intersecting $\mathbb{P}^1$'s in $\hat{X}$. The loci of $I_3$-fibers
are determined by looking at zeros of higher order of the discriminant
and by checking whether the fiber in the resolution $\hat{X}$ splits
further. We find the following loci,
\beq
\text{
\begin{tabular}{|c|c|} \hline
Loci & Yukawa \rule{0pt}{14pt}\\ \hline
$s_8=s_9=s_7=0$ & $\mathbf{1}_{(-1,-2)}\times \mathbf{1}_{(0,2)}\times \mathbf{1}_{(1,0)}$ \rule{0pt}{13pt}\\
$s_3=s_7=s_9=0$ & $\mathbf{1}_{(0,2)} \times\overline{{\mathbf{1}}}_{(1,-1)}\times \overline{\mathbf{1}}_{(-1,-1)}$ \rule{0pt}{13pt}\\
$\Sigma_{1_{(1,0)}}\cap\Sigma_{1_{(1,0)}}\cap\Sigma_{1_{(1,0)}}$ & $\overline{\mathbf{1}}_{(-1,-1)} \times\mathbf{1}_{(1,0)} \times\mathbf{1}_{(0,1)}$\rule{0pt}{13pt}\\ \hline
$\Sigma_{1_{(1,0)}}\cap\lbrace s_3=s_7=0\rbrace$ & $\overline{\mathbf{1}}_{(1,-1)}\times\overline{\mathbf{1}}_{(-1,0)} \times \mathbf{1}_{(0,1)}$ \rule{0pt}{13pt}\\
$\Sigma_{1_{(1,1)}}\cap\lbrace s_8=s_9=0 \rbrace$ & $\mathbf{1}_{(1,1)} \times{\mathbf{1}}_{(-1,-2)}\times \mathbf{1}_{(0,1)}$\rule{0pt}{13pt}\\ \hline
\end{tabular}
}
\eeq
We note that the first three agree with earlier results, see
\cite{Borchmann:2013jwa}. The last two loci also produce reducible fibers
with three irreducible components that can be described in terms of the
prime ideals. The study of these new Yukawa points, along
with a more thorough discussion of the use of prime ideals, will be
postponed to future work.

\subsection{The Cohomology Ring and the Chern Classes of $\hat{X}$}
\label{sec:CYFFCohomologyRing}

In this section we abstractly calculate the cohomology ring of the
fourfold $\hat{X}$. The central result of these computations
is the basis of surfaces or dual $(2,2)$-forms in $H^{(2,2)}(\hat{X})$,
which is relevant for the construction of $G_4$-flux, see section
\ref{sec:G4U1xU1}. Furthermore, for the calculation of the D3-brane
tadpole and the quantization of the $G_4$-flux, we use these techniques
to calculate  the general expression for the Euler number and the second
Chern class of $\hat{X}$ for  a general
base $B$. In addition, we derive the full cohomology
ring explicitly for $B=\mathbb{P}^3$, leaving the straightforward
generalization to other bases for future works.

We note that the presentation of the cohomology of $\hat{X}$ used
here has been employed in the context of toric mirror symmetry for a long
time and is in this sense not new. For an F-theory context see
e.g.~\cite{Grimm:2009ef,Klevers:2011xs} and references therein.
We refer also to \cite{Marsano:2011hv} for cohomology calculations in the
same spirit. However, we emphasize that, except for the language
that we borrow from toric geometry, the following discussion is based
only on reasonable assumptions on the intersections of $\hat{X}$.
Thus, we expect the following procedure to work also in the non-toric
case. \text{In particular, not all fourfolds considered here
have, to our} \text{knowledge, a description in terms of a reflexive polytope,
which does, however, not keep} us from using them for F-theory and
computing their  full chiral 4D spectrum.

The basic idea to calculate the cohomology ring $H^{(*,*)}_V(\hat{X})$ of
a general elliptically fibered Calabi-Yau
fourfold $\hat{X}$ over a base $B$ with general fiber the elliptic
curve in $dP_2$ is to exploit the
Stanley-Reissner (SR) ideal\footnote{We merely borrow this term from
toric geometry. In general, $SR$ can  be any ideal containing all
vanishing intersections of divisors on $dP_2^B(\cS_7,\cS_9)$, which not
necessarily has to be a toric variety.} $SR$ of the ambient space
$dP_2^B(\cS_7,\cS_9)$
together with the linear equivalences of divisors.
After dividing out the linear equivalences, the cohomology ring
$H^{(*,*)}_V(\hat{X})$ can be represented as the quotient
ring $R$ of the form\footnote{We note that this polynomial ring
is only the primary vertical cohomology
$H_V^{(*,*)}(\hat{X})$ \cite{Greene:1993vm}. This is the subspace of
$H^{(*,*)}(\hat{X})$ relevant for $G_4$-flux inducing
chirality in F-theory. Its complement in $H^{(2,2)}(\hat{X})$ is the
horizontal cohomology $H^{(2,2)}_H(\hat{X})$ that
encodes complex structure moduli of $\hat{X}$. See
e.g.~\cite{Grimm:2009ef} for an analysis of $G_4$-flux in
$H^{(2,2)}_H(\hat{X})$ in F-theory.}
\beq \label{eq:abstractCohomRing}
	H^{(*,*)}_V(\hat{X})\cong
	\frac{\mathbb{C}[D_\alpha,S_P,S_Q,S_R]\cdot \big[\hat{X}\big]}{SR}\,,
\eeq
where the basis \eqref{eq:divSPSQSR} together with the vertical
divisors $D_\alpha$ are the variables of the free polynomial
ring $\mathbb{C}[D_\alpha,S_P,S_Q,S_R]$ and
$SR$ is considered as an ideal in this ring. For this purpose, the ideal
$SR$ has to be translated  into
intersection relations of those divisors. Note that we have to
multiply by the homology class of $\hat{X}$ in $dP_2^B(\cS_7,\cS_9)$ in
\eqref{eq:antiKdP2} to restrict the
intersections on the ambient space $dP_2^{B}(\cS_7,\cS_9)$ to $\hat{X}$.\footnote{Generally, not all divisors
on $\hat{X}$ arise as restrictions of divisors on the
ambient space. However, for generic $\hat{X}$ with elliptic fiber
in $dP_2$, only divisors in $B$ can potentially miss this assumption. We
exclude those $B$ in the following. Note that non-generic $\hat{X}$ can have additional divisors, see
the footnote \ref{fn1}.}  By the Calabi-Yau condition
this class is precisely given as $\mathcal{O}(\hat{X})=K^{-1}_{dP_2^B}$.

The quotient  ring \eqref{eq:abstractCohomRing} is graded with each
graded piece being finitely generated by monomials in the divisors
$D_\alpha$, $S_P$ and $S_m$ of appropriate degree. We denote this ring by
$R$. The $k$-th graded piece is then identified with
\beq
	H^{(k,k)}_V(\hat{X})=R^{(k)}\,,
\eeq
after restriction to $\hat{X}$, i.e.~after dropping the overall factor
$K^{-1}_{dP_2^B}$ in \eqref{eq:abstractCohomRing}.
More precisely, at grade zero we obtain
$H^{(0,0)}(\hat{X})=\langle 1\rangle$, at
grade one we have $H^{(1,1)}(\hat{X})
=\langle D_\alpha,S_P,S_Q,S_R\rangle$. At higher grade
we obtain naively as many generators as homogeneous monomials of
appropriate degree in the divisors in \eqref{eq:abstractCohomRing}.
However, due to  equivalence relation in $R$ the number of independent
monomials is in general
smaller. In fact, by Poincar\'e duality the rings $R^{(3)}$, and
$R^{(4)}$ are fixed, i.e.~the corresponding Hodge
numbers are related as  $h^{(3,3)}(\hat{X})
\stackrel{!}{=}h^{(1,1)}(\hat{X})=3+h^{(1,1)}(B)$ and
$h^{(4,4)}(\hat{X})\stackrel{!}{=}h^{(0,0)}(\hat{X})=1$.
At degree $k\geq 5$ we trivially have $R^{(k)}=\{0\}$
by reasons of dimensionality.
In this sense
the only non-trivial piece is $R^{(2)}\cong H^{(2,2)}_V(\hat{X})$, and
the corresponding  Hodge-number $h^{(2,2)}_V(\hat{X})$. Furthermore, it
is precisely
the elements $H^{(2,2)}_V(\hat{X})$
of independent surfaces on $\hat{X}$ into which the
general $G_4$-flux on $\hat{X}$ has to be expanded, as discussed
in section \ref{sec:G4U1xU1}.

The main advantage of the representation \eqref{eq:abstractCohomRing}
compared to concrete toric models is that it allows us to
determine the cohomology ring for all fibrations $dP_2^B(\cS_7,\cS_9)$
in \eqref{eq:dP2fibration} over a given base $B$ with general
divisors $\cS_7$, $\cS_9$. Of course the relevant computations depend
on the geometry of the base $B$ since the ideal $SR$ in
\eqref{eq:abstractCohomRing} in general is generated by the
SR-ideal \eqref{eq:SRidealdP2} of the fiber $dP_2$ and
of the base $B$, which differs from case to case. Nevertheless, as we
demonstrate next, it is possible to calculate the total Chern class
$c(\hat{X})$ and Euler number $\chi(\hat{X})$ of $\hat{X}$ for any
base $B$ using minimal assumptions.

\subsubsection{Second Chern Class and Euler Number of $\hat{X}$: General Formulas}

For the purpose of finding the general expression for $c(\hat{X})$ and, thus, the Euler number $\chi(\hat{X})$, it
suffices to know, that the intersections of more than three vertical divisors $D_\alpha$ in both $dP_2^B(\cS_7,\cS_9)$
and $\hat{X}$ are zero. The latter is true because of the properties of fibrations.
Thus, we are working in the following with the ideal of vanishing intersections on $dP_2^B(\cS_7,\cS_9)$ generated by the
ideal \eqref{eq:dP2divs} of the fiber $dP_2$ supplemented by the vanishing of quartic intersection of vertical divisors,
\bea \label{eq:SRintsdP2B}
		SR'\!\!&\!\!=\!\!&\!\!\left\{S_R\cdot (S_R+S_Q-\cS_7 -[K_B])\,,\,\,S_R\!\cdot\! (S_R+S_P-\cS_9-[K_B]),\,\,S_Q\!\cdot\! S_P\,,
		\right.\\
		\!\!&\!\!\!\!&\!\!\left.\,\,\,S_Q\cdot (S_R+S_Q-\cS_7 -[K_B])\,,\,S_P\!\cdot\!
		(S_R+S_P-\cS_9-[K_B])\,,\,\,
		D_\alpha\!\cdot\! D_\beta\!\cdot\!  D_\gamma\!\cdot\! D_\delta\right\}\,.\nn
\eea
Here we have employed \eqref{eq:sectionsFibration} in combination with
\eqref{eq:divSPSQSR} to translate \eqref{eq:SRidealdP2} into intersection
relations. The
prime in $SR'$ reminds us that we are not working with the full SR-ideal
of the base $B$, but
just assume vanishing quartic intersections. As before '$\cdot$' denotes
the intersections product in $dP_2^B(\cS_7,\cS_9)$.

Next we can perform the calculation of the total Chern class of $\hat{X}$. For this purpose we first compute the formal
expression of the Chern class $c(\hat{X})$ by adjunction,
\beq \label{eq:cXadjunction}
	c(\hat{X})=\frac{c(dP_2^B)}{1+c_1(\mathcal{O}(\hat{X}))}\,.
\eeq
The numerator denotes the total Chern class of $dP_2^B(\cS_7,\cS_9)$
and the denominator is the Chern
class of its anti-canonical bundle  \eqref{eq:antiKdP2}, which is the
class of $\hat{X}$ as mentioned above.

Then, we
reduce this expression in the quotient ring
\eqref{eq:abstractCohomRing} with $SR$  replaced by the reduced
ideal $SR'$ in \eqref{eq:SRintsdP2B}. We refer to appendix
\ref{app:CohomologyRing} for the detailed calculations leading to the
following results, as well as for the general expression of the total
Chern classes $c(dP_2^B)$ and $c(\hat{X})$ for Calabi-Yau two-,
three- and fourfolds.
We obtain for the second Chern class $c_2(\hat{X})$ of $\hat{X}$
the expression
\bea \label{eq:c2XgenB}
	c_2(\hat{X})\!\!&\!\!=\!\!&\!\! 3 c_1^2 + c_2 - 2 S_Q^2 - 3 S_P^2 + c_1 (2 S_Q + S_P+4 S_R - 2 (\cS_7 + \cS_9)) \nn\\
\! \!&\!&\!\! +2\cS_7(S_Q  -  S_P)+\cS_9(3 S_P- 2 S_Q  -S_R
  +  \cS_7) \,,
\eea
where we have expressed all cohomology classes in terms of the
basis of divisors \eqref{eq:divSPSQSR} on the fiber $dP_2$ and
the first and second Chern classes $c_1\equiv c_1(B)$, respectively,
$c_2\equiv c_2(B)$ of the base $B$. By abuse of notation we denote
a divisor and its Poincar\'e dual $(1,1)$-form by the same symbol.

As a first sanity check we note that \eqref{eq:c2XgenB} is
consistent with the formula for the second
Chern class of a fourfold with a generic $E_6$-elliptic fiber,
i.e.~with the elliptic curve in $\mathbb{P}^2$. In fact,
in the limit $\cS_7=\cS_9=0$, the total space
$dP_2^B(\cS_7,\cS_9)$ formally turns into
$\mathbb{P}(\mathcal{O}_B\oplus K_{B}^{-1}\oplus K_{B}^{-1})$, the
sections $S_P$, $S_Q$ and $S_R$ become indistinguishable and fuse into
a single holomorphic three-section $\sigma$, following conventions
in the literature. Then the second line in \eqref{eq:c2XgenB}
vanishes and we use the relation \eqref{eq:holomorphicSec} for $\sigma^2$
to rewrite the first line as
\beq
	c_2(\hat{X})\,\,\rightarrow\,\, c_2(X_{E_6})=3c_1^2+c_2
	+12\sigma c_1\,,
\eeq
where we denote by $X_{E_6}$ the fourfold with $E_6$-elliptic fiber.
This expression is in line with the results obtained in
\cite{Klemm:1996hh}.

Similarly, the Euler number of $\hat{X}$ is calculated from the
integration of the fourth Chern class $c_4(\hat{X})$ as
\beq \label{eq:EulerNumberX}
	\chi(\hat{X})=3\int_B\left[24 c_1^3 + 4c_1 c_2 - 16 c_1^2 (\cS_7+
	\cS_9)+  c_1 (8\cS_7^2  + \cS_7 \cS_9 + 8 \cS_9^2)
	-  \cS_7 \cS_9(\cS_7  +  \cS_9)\right]\,.
\eeq
Here the integral over $\hat{X}$ has been reduced to an integral over
the base $B$ by first consecutive application of the relation
\eqref{eq:SP^2} and then by employing \eqref{eq:intSecFiber}, which
can be rewritten for Calabi-Yau fourfolds as the intersection
relation
\beq \label{eq:intSecFiber1}
	S_P\cdot D_\alpha\cdot D_\beta\cdot D_\gamma= S_m\cdot D_\alpha\cdot
	D_\beta\cdot D_\gamma=(D_\alpha\cdot D_\beta\cdot
	D_\gamma)\vert_B\,.
\eeq	
Here $S_m$ collectively denotes the divisors $S_Q$, $S_R$ of the
sections $\hat{s}_Q$, $\hat{s}_R$ and $D_\alpha$, $D_\beta$,
$D_\gamma$ are general vertical divisors. We emphasize that our
expression of the Euler number \eqref{eq:EulerNumberX} reproduces
the Euler number of \cite{Klemm:1996hh} as the special case
$\cS_7=\cS_9=0$.  As before $S_P$, $S_Q$ and $S_R$ become homologous
and we obtain
\beq
	\chi(\hat{X}_{E_6})=72 \int_B c_1^3 + 12\int_B c_1 c_2\,.
\eeq

As another consistency check, and also for the sake of the
discussion of general flux quantization and the D3-brane tadpole in
section \ref{sec:G4U1xU1}, we calculate the arithmetic genus
$\chi_0(\hat{X})$ on $\hat{X}$. It is calculated from the Todd class
$\text{Td}_4(\hat{X})$ by the Hirzebruch-Riemann-Roch index theorem.
Since $\hat{X}$ is a simply-connected Calabi-Yau fourfold, its
arithmetic genus has to be two,
\beq \label{eq:chi_0hp0}
	\chi_0(\hat{X}):=\sum_p (-1)^p h^{(p,0)}(\hat{X})\stackrel{!}{=}2\,.
\eeq
This immediately follows from $h^{(0,0)}(\hat{X})=h^{(4,0)}(\hat{X})=1$
and $h^{(p,0)}(\hat{X})=0$ otherwise. From index theory, however, we
obtain
\beq \label{eq:chi0X_Euler}
	 \chi_0(\hat{X})=\int_{\hat{X}}\text{Td}_4(\hat{X})=\frac{1}{720}\int_{\hat{X}}(3c_2(\hat{X})^2-c_4(\hat{X}))=\frac{1}{720}\left(3\int_{\hat{X}}c_2(\hat{X})^2-\chi(\hat{X})\right)\,.
\eeq
Evaluating this integral using our expressions \eqref{eq:c2XgenB},
\eqref{eq:EulerNumberX} for the second Chern class and Euler number on
$\hat{X}$, cf.~appendix \ref{app:CohomologyRing} for details, we obtain
\beq \label{eq:chi0X_chi0B}
	\chi_0(\hat{X})=\frac{1}{12}\int_B c_1c_2=2\chi_0(B)\stackrel{!}{=}2\,.
\eeq
Here the first equality is due to a remarkable cancellation
of all terms containing the divisors $\cS_7$, $\cS_9$ in the second
Chern class and Euler number of $\hat{X}$. In the second
equality we used the index theorem for the arithmetic genus of the base
\beq
	\chi_0(B)=\frac{1}{24}\int_B c_1 c_2\,,
\eeq
and the last equality follows from the constraint \eqref{eq:chi_0hp0}.
Thus we see, that the arithmetic genus $\chi_0(\hat{X})=2$ precisely
iff $\chi_0(B)=1$. In general, one demands the stronger conditions
$h^{(1,0)}(B)=h^{(2,0)}(B)=h^{(3,0)}(B)=0$ since non-trivial
$(p,0)$-forms of the base $B$ would pull back to $(p,0)$-forms
on $\hat{X}$ under the projection $\pi:\,\hat{X}\rightarrow B$,
which we excluded by assumption.

We note that our result \eqref{eq:chi0X_chi0B} for the arithmetic genus is
in line with the computations in \cite{Sethi:1996es,Klemm:1996ts}, whose
analysis we followed. We also refer to \cite{Marsano:2011hv} for an
application of these techniques to F-theory with SU$(5)$  gauge group.

\subsubsection{The Full Cohomology Ring of $\hat{X}$: Base $B=\mathbb{P}^3$}

As we demonstrate next, the representation
\eqref{eq:abstractCohomRing} for a concrete base $B$ allows us to
calculate the full cohomology ring for a general Calabi-Yau fourfold
$\hat{X}$ in $dP_2^B(\cS_7,\cS_9)$ with
general divisors $\cS_7$, $\cS_9$. We exemplify this in the following
for the base $B=\mathbb{P}^3$, but note that this analysis can
be generalized to other bases. For all details of the intersection
calculations as well as the quartic intersections,
we refer the reader to appendix \ref{app:intsP3}.

In the case $B=\mathbb{P}^3$  the cohomology $H^{(1,1)}(\hat{X})$ is
generated according to \eqref{eq:basisH11X} by the divisors $D_A$.
We choose the following basis,
\beq \label{eq:divsdP2P3}
	H^{(1,1)}(\hat{X})=\langle H_B,S_P,S_Q,S_R\rangle\,,
\eeq
where $H_B$ is the only vertical divisor of the fibration, which is
pullback of the hyperplane of $\mathbb{P}^3$
to $\hat{X}$. The three other divisors are related to the
in general rational sections $\hat{s}_P$, $\hat{s}_Q$
and $\hat{s_R}$. Employing \eqref{eq:sectionsFibration} and
\eqref{eq:divSPSQSR}, their associated divisor classes are
\beq \label{eq:SPSQSRP3}
	S_P=E_2\,,\qquad S_Q=E_1\,,\qquad S_R=H-E_1-E_2+(n_9-4)H_B\,,
\eeq
where we have used $c_1(K_{\mathbb{P}^3})=-c_1(\mathbb{P}^3)=-4H_B$.
We recall that the divisors $\cS_7$, $\cS_9$ on
$\mathbb{P}^3$ are specified by integers $n_7$, $n_9$ in the region in
figure \ref{fig:XP3n7n9region} specifying the total space
\eqref{eq:dP2fibrationP3} of the $dP_2$-fibration over $\mathbb{P}^3$.

We set up the construction of the cohomology ring of
$\pi:\,\hat{X}\rightarrow \mathbb{P}^3$ via \eqref{eq:abstractCohomRing}
by specifying the ideal $SR$. We note that the SR-ideal  in
the case of $B=\mathbb{P}^3$ is generated
by the Stanley-Reissner ideal \eqref{eq:SRidealdP2} of the fiber
$dP_2$ and the base, which is just $H_B^4=0$.
Thus, using the divisor classes \eqref{eq:divsdP2P3}, the resulting ideal is identical to \eqref{eq:SRintsdP2B} with
all vertical divisors equal to $H_B$ and with
$K_{\mathbb{P}^3}^{-1}=\cO_{\mathbb{P}^3}(4)$.
Then, we need the anti-canonical bundle $K^{-1}_{dP_2(n_7,n_9)}$ of
$dP_2(n_7,n_9)$. It is given in general in \eqref{eq:antiKdP2} and
easily specialized to $B=\mathbb{P}^3$ using $\cS_7=n_7H_B$ and
$\cS_9=n_9H_B$ as well as expressed in the basis \eqref{eq:SPSQSRP3}.
Now we are equipped with all the necessary quantities to construct
the quotient ring representation \eqref{eq:abstractCohomRing} of
the cohomology ring $H_V^{(*,*)}(\hat{X})$.

We begin by summarizing the Hodge numbers of the vertical cohomology of
$\hat{X}$ as
\beq \label{eq:XP3_Hodgenums}
	h^{(0,0)}(\hat{X})=h^{(4,4)}(\hat{X})=1\,,\quad h^{(1,1)}(\hat{X})=h^{(3,3)}(\hat{X})=4\,,\quad h^{(2,2)}_V(\hat{X})=5\,\, (4)\,,
\eeq
where the subscript $V$ indicates that we are considering the vertical
subspace, and the number in the bracket denotes the non-generic case
with $(n_7,n_9)$ on the boundary\footnote{\label{fn1}We note that for the two special values
$(n_7,n_9)=(4,8),\,(8,4)$ there is one additional divisor on $\hat{X}$ that is not induced from the ambient space. For these special
values we see from \eqref{eq:sectionsFibration} that $\hat{X}$ is not generic since $s_1$, $s_2$, $s_3$, respectively, $s_1$, $s_5$, $s_8$ are 
constants. Then, we can perform a variable transformation on the fiber coordinates to achieve $s_1=0$, i.e.~the elliptic curve will have an 
additional section at $u=1$, $v=w=0$. The elliptic fiber can then be embedded into $dP_3$ with all sections toric. We thank Jan Keitel for pointing
out the existence of a non-toric divisor for non-generic $\hat{X}\rightarrow \mathbb{P}^3$.} of the allowed region in
figure \ref{fig:XP3n7n9region}. These are the lines $n_7=0$, $n_9=0$,
$n_9=4+n_7$ for $n_7\leq 4$, $n_9=n_7-4$  and $n_9=12-n_7$, both of
the latter two for $4<n_7$.

Indeed, we obtain these Hodge numbers as
follows from the ring \eqref{eq:abstractCohomRing}. At
degree zero, which is $H^{(0,0)}(\hat{X})$, the only
generator is the trivial element $1$. The graded piece
$R^{(1)}\cong H^{(1,1)}(\hat{X})$  is
generated by the four divisors $D_A$.
At degree two, i.e.~$H_V^{(2,2)}(\hat{X})$, there are ten different
combinations $D_A\cdot_{\hat{X}} D_B$, of which, however, only five
are generically inequivalent. As outlined in appendix \ref{app:intsP3}
a choice of basis, denoted in general by $\mathcal{C}_r$ with $r=1,
\ldots,h^{(2,2)}_V(\hat{X})$, for $H^{(2,2)}_V(\hat{X})$ is given by
\beq \label{eq:XP3_H22V}
	H^{(2,2)}_V(\hat{X})=\langle H_B^2, H_B\cdot S_P,
	H_B\cdot \sigma(\hat{s}_Q), H_B \cdot\sigma(\hat{s}_R),S_P^2\rangle\,.
\eeq
Here $\sigma(\hat{s}_Q)$, respectively, $\sigma(\hat{s}_R)$ are the Shioda
maps \eqref{eq:ShiodaMapSQSR} of the sections $\hat{s}_Q$, $\hat{s}_R$. In
the case at hand these take the form
\beq \label{eq:XP3_Shioda}
	\sigma(\hat{s}_Q)=S_Q-S_P-4H_B\,,\quad \sigma(\hat{s}_R)
	=S_R-S_P-(4+n_9)H_B
\eeq	
We can evaluate the $5\times 5$-intersection matrix $\eta^{(2)}$
in the basis \eqref{eq:XP3_H22V} using the quartic intersections in
\eqref{eq:XP3_C0} as
\beq  \label{eq:XP3_eta2_1}
	 \eta^{(2)}=
\left(
\begin{array}{ccccc}
 0 & 1 & 0 & 0 & -4 \\
 1 & -4 & 0 & 0 & 16+\left(n_7-n_9-4\right) n_9 \\
 0 & 0 & -8 & n_7-n_9-4 & n_9 \left(4-n_7+n_9\right) \\
 0 & 0 & \eta^{(2)}_{34} & -2 \left(4+n_9\right) & 2 n_9 \left(4-n_7+n_9\right) \\
 -4 & \eta^{(2)}_{25} & \eta^{(2)}_{35} & \eta^{(2)}_{45} & -64-\left(8+n_7-2 n_9\right) \left(n_7-n_9-4\right) n_9 \\
\end{array}
\right)\,.
\eeq
Here entries $\eta^{(2)}_{rs}$ that are determined by symmetry
are omitted and denoted by  $\eta^{(2)}_{sr}$.
We note that for values of $(n_7,n_9)$ on the boundary
of figure \ref{fig:XP3n7n9region}, there are only four inequivalent
such surfaces. A quick way
to see this is by calculating the rank of the matrix
\eqref{eq:XP3_eta2_1}
which is generically five, but decreases to four in these cases.
In all these cases we can drop the basis element $S_P^2$ in
\eqref{eq:XP3_H22V}
since it becomes homologous to the other four basis elements,
cf.~\eqref{eq:SP^2n7=0}, \eqref{eq:SP^2n9_1}, \eqref{eq:SP^2n9_2} and
\eqref{eq:SP^2calculated} in appendix \ref{app:intsP3}. The corresponding
intersection
matrix $\eta^{(2)}$ is then obtained from \eqref{eq:XP3_eta2_1}
by deleting the last row and column. We note that both the
knowledge of the basis \eqref{eq:XP3_H22V} as well as of the
intersections \eqref{eq:XP3_eta2_1} is essential for the construction of
$G_4$-flux in section \ref{sec:G4U1xU1}.

At degree
three, there are 20 combinations of three divisors $D_A$, however,
there are only four inequivalent ones, which is expected by duality of
$H^{(3,3)}(\hat{X})$ and $H^{(1,1)}(\hat{X})$. Finally, at degree
four, which is $H^{(4,4)}(\hat{X})$, there are 35 different quartic
monomials in the $D_A$, of which there is only one inequivalent
combination. This combination is precisely the quartic intersections
on $\hat{X}$. The higher graded pieces of $H^{(k,k)}(\hat{X})$, $k>4$
vanish, which is intuitively clear since there are at most quartic
intersections on a Calabi-Yau fourfold.

We conclude by summarizing some key intersections on $\hat{X}$ which
are discussed in section \ref{sec:GeneraldP2fibrations} as general
properties of the fibrations, that can, however, be proven explicitly
using the representation \eqref{eq:abstractCohomRing}. Of the complete
quartic intersections summarized in \ref{eq:XP3_C0} of appendix
\ref{app:intsP3} we highlight the following intersections,
\beq
S_P\cdot S_R\cdot H_B^2=n_{9} S_{*}\cdot H_B^3 \,,\qquad S_Q\cdot S_R \cdot H_B^2=n_7 S_{*}\cdot H_B^3\,,\quad S_{*}\cdot H_B^3=1\,,\quad
S_{*}^2 \cdot H_B^2=-4\,,
\eeq
where $S_{*}$ collectively denotes all the divisor classes $S_P$, $S_Q$
and $S_R$ of the sections. Here the first two relations are the versions
of \eqref{eq:S7S9}, respectively, on $B=\mathbb{P}^3$.
The third relation implies that a section of the elliptic fibration of
$\hat{X}$ intersects the generic fiber $F=\pi^*(pt)$ for a generic point
$pt$ in $B$ precisely at one point, cf.~\eqref{eq:intSecFiber}. Finally,
the last relation is the analog of \eqref{eq:SP^2} on $B=\mathbb{P}^3$.
We note that for a holomorphic section, this
relation holds without intersection with $H_B^2$. Indeed, this is
confirmed by the concrete cohomology calculation in appendix
\ref{app:intsP3} for the zero-section $S_P$ in \eqref{eq:SP^2calculated}.


\section{$G_4$-Flux Conditions in F-Theory from CS-Terms: Kaluza-Klein States on the 3D Coulomb Branch}
\label{sec:G4review}

In this section we discuss the construction of $G_4$-flux in F-theory
compactifications on general elliptically fibered Calabi-Yau fourfolds
$\hat{X}$ with a non-trivial Mordell-Weil group and a non-holomorphic
zero section.

We define $G_4$-flux in F-theory through
the  \textit{M-theory} compactification on the resolved fourfold
$\hat{X}$, that is dual to F-theory  reduced on a circle to
3D. The general constraints on $G_4$-flux in
M-theory compactifications are reviewed in section \ref{sec:G4Mtheory}.
In addition to these conditions, $G_4$-flux that is admissible for
an \textit{F-theory} compactification has to obey additional constraints.
The form of $G_4$-flux that yields a consistent F-theory
has been derived first in \cite{Dasgupta:1999ss} by requiring a 
Lorentz-invariant uplift to four dimensions.
Here we discuss a different logic to obtain constraints on the $G_4$-flux. 
As we point out in section \ref{sec:G4constraintsF}, these constraints
are appropriately formulated as the requirement of the vanishing of
certain Chern-Simons (CS) terms on the Coulomb branch of the effective
three-dimensional theory\footnote{See also
\cite{Closset:2012vg,Closset:2012vp} for recent related studies of
connections between CS-terms and contact terms in 3D effective field
theories with background fields in the context of F-maximization.}.
In particular, consistent conditions on the $G_4$-flux are obtained
only  if one-loop corrections of both massive states on the 3D Coulomb
branch as well as \textit{Kaluza-Klein} (KK) states are taken into
account. Most importantly, the presence of a non-holomorphic zero section
is linked to the existence of new CS-terms for the KK-vector,
that are generated by KK-states, whereas other CS-terms receive
additional shifts. 

We present for the first time a consistent
set of conditions on $G_4$-flux in F-theory compactifications with a
non-holomorphic zero section.
We also evaluate explicitly the corrections of massive states to 3D 
CS-levels for the F-theory/M-theory compactification on the fourfold 
$\hat{X}$ with $dP_2$-elliptic fiber.

The following discussion is an extension of \cite{Cvetic:2012xn}, where
KK-states have first been discussed in the context of 4D anomaly
cancellation, and inspired by the analogous six-dimensional
analysis in \cite{Grimm:2013oga}\footnote{We are grateful to Thomas W.
Grimm for explanations and comments on the importance of $\Theta_{00}$.},
see also \cite{Bonetti:2012fn} for the relevance of KK-states in the
description of self-dual two-forms in 6D/5D.

\subsection{A Brief Portrait of $G_4$-Flux in M-Theory}
\label{sec:G4Mtheory}

Let $\hat{X}$ denote an arbitrary smooth Calabi-Yau
fourfold.
In general, $G_4$-flux in M-theory can only be defined on such a smooth
manifold, that in the context of F-theory typically arises from
resolutions of both codimension one singularities from non-Abelian
gauge groups or, as in the case considered here, from higher
codimension singularities in the presence of a non-trivial Mordell-Weil
group.

Then, $G_4$-flux in F-theory is defined as $G_4$-flux in M-theory with as
set of additional F-theoretic restrictions discussed in
the next section \ref{sec:G4constraintsF}.  $G_4$-flux in M-theory
is consistent if it obeys two basic conditions. First, $G_4$  has to be
quantized as \cite{Witten:1996md}
\begin{equation}
G_4 + \frac{c_2(\hat{X})}{2} \in H^4(\hat{X},\mathbb{Z})\,,
\label{eq:G4quantization}
\end{equation}
which depends on the second Chern class $c_2(\hat{X})$ of $\hat{X}$.
In addition, the M2-brane tadpole has to be cancelled
\cite{Sethi:1996es,Gukov:1999ya},\footnote{We are working
here in the carefully checked conventions of \cite{Bilal:2003es},
where also comparison with other, inconsistent sign choices in the
literature can be found.}
\begin{equation}
\frac{\chi(\hat{X})}{24}=n_3+\frac{1}{2}\int_{\hat{X}}G_4\wedge G_4\,,
\label{eq:D3tadpole}
\end{equation}
where $\chi(\hat{X})$ is the Euler characteristic of $\hat{X}$ and $n_3$
the number of spacetime-filling M2-branes. This tadpole lifts to the
D3-brane tadpole in F-theory with $n_{3}$ denoting the number
of D3-branes.

In addition, one can distinguish $G_4$-flux further by decomposing
$H^{4}(\hat{X})$ into its primary vertical and horizontal subspaces
$H^{4}(\hat{X})=H^{4}_V(\hat{X})\oplus H^{4}_H(\hat{X})$
\cite{Greene:1993vm}. As mentioned
before, only $G_4$-flux in the vertical homology induces 4D chirality
as well as gaugings of axions in F-theory and is considered here.
A general $G_4$-flux in the vertical cohomology
$H^{(2,2)}_V(\hat{X})$ than has an expansion as
\beq \label{eq:G4expansion}
	G_4=m^r\cC_r\,,
\eeq
where $\cC_r$ with $r=1,\ldots,h^{(2,2)}_V(\hat{X})$ denotes an
integral basis
of $H^{(2,2)}_V(\hat{X})$ and $m^r$ are the flux-quanta with integrality
fixed by the quantization condition \eqref{eq:G4quantization}.\footnote{In 
general, $c_2(\hat{X})$ has to be decomposed into the
integral basis of $H^{(2,2)}_V(\hat{X})$. However, the determination of
this basis is very involved and requires more sophisticated
techniques that would exceed the scope of this work. We refer to
\cite{Grimm:2009ef} for the application of mirror symmetry to fix the
integral basis. See also \cite{Intriligator:2012ue} for a discussion of
potential conflicts between the split of $H^4(\hat{X})$ into vertical and
horizontal subspace and the choice of an integral basis.}
Such a basis can be constructed explicitly, as demonstrated in sections
\ref{sec:CYFFCohomologyRing} and \ref{sec:CohomologyFFSU5}, for concrete
examples.

\subsection{Deriving Conditions on $G_4$-Flux in F-theory}
\label{sec:G4constraintsF}

The additional constraints on the $G_4$-flux in F-theory
compactifications are
most conveniently formulated in the three-dimensional theory
obtained after compactification of the 4D $\mathcal{N}=1$
effective action of F-theory  on $S^1$.
Then we can use the basic duality between three-dimensional F-theory on
$\hat{X}\times S^1$ and M-theory on $\hat{X}$.
Consideration of the resolved fourfold $\hat{X}$ means in terms of
the 3D $\mathcal{N}=2$ effective theory obtained in the circle reduction
to go to the 3D Coulomb. The corresponding fields acquiring  a VEV
are the adjoint valued scalars $\zeta^{A}$, $A=0,\ldots,
h^{(1,1)}(B)+\text{rk}(G)+n_{\text{U}(1)}$, along the Cartan directions
of the 4D gauge group $G\times \text{U}(1)^{n_{U(1)}}$. The zeroth
component denotes the scalar in the multiplet of the KK-vector.
Then the 3D gauge group is broken in an ordinary Higgs effect to the
maximal torus $\text{U}(1)^{h^{(1,1)}(B)+\text{rk}(G)+n_{\text{U}(1)}+1}$
and in addition the charged fermions, in particular those from the 4D
massless matter multiplets, obtain  a  mass  (shift) as
$m=q_A\cdot \zeta^A$, where $q_A$ denotes the full 3D charge vector.

This Coulomb branch then describes the IR dynamics of the dual
M-theory compactification on the fourfold $\hat{X}$ in the supergravity
approximation. It is the key point of the following analysis that
the matching of the two dual descriptions works only
on the level of the quantum effective action after massive degrees
of freedom have been integrated out on the F-theory side. As we
emphasize in the  following, also corrections due to KK-states have to be
considered. In fact, consistent conditions on
the $G_4$-flux in the presence of a non-holomorphic zero section
are only obtained if new CS-terms for the KK-vector are taken into
account.

The general approach of matching F- and M-theory in 3D initiated
in \cite{Grimm:2010ks} has been exploited recently in
\cite{Grimm:2011fx,Grimm:2012rg,Cvetic:2012xn,Braun:2013yti}
to study various aspects of the F-theory effective action. We refer
to these references for the background of the following discussion.

\subsubsection{F-Theory Conditions from KK-States Corrected CS-Terms}

First we recall that $G_4$-flux in M-theory
induces CS-terms for the U(1)-gauge fields $A^A$ on the 3D Coulomb
branch that read
\beq \label{eq:3DCSterms}
	S^{(3)}_{CS}=-\frac{1}{2}\int \Theta^M_{AB}A^A\wedge F^{B}\,\qquad \Theta^M_{AB}=\frac{1}{2}\int_{\hat{X}} G_4\wedge \omega_A\wedge\omega_B\,.
\eeq
This can be shown by reducing the M-theory three-form $C_3$ along
$(1,1)$-forms $\omega_A$ on the fourfold $\hat{X}$ that are dual to the
basis of divisors $D_A$,
$A=0,1,\ldots,h^{(1,1)}(\hat{X})-1$.
We recall, cf.~\eqref{eq:basisH11X} for the case $n_{\text{U}(1)}=2$,
that this basis is given by $h^{(1,1)}(B)$ vertical divisors $D_\alpha$,
the divisor $\tilde{S}_P=S_P+\frac{1}{2}[K_B^{-1}]$, see
\eqref{eq:SPtilde}, associated to the zero section, the Shioda maps
$\sigma(\hat{s}_m)$ for a rank $n_{\text{U}(1)}$ Mordell-Weil
group and $\text{rk}(G)$ Cartan divisors $D_i$ in the presence of a four-
dimensional non-Abelian gauge group $G$.

Now the strategy to formulate conditions on the $G_4$-flux for a
valid F-theory compactification is as follows.
We require that the CS-levels on the F-theory side, denoted
$\Theta_{AB}^F$, have to agree with the M-theory CS-levels, denoted
$\Theta_{AB}^M$ and given by the flux integral in \eqref{eq:3DCSterms},
\beq \label{eq:FluxCondMF}
	\Theta_{AB}^F\stackrel{!}{=}\Theta_{AB}^M\,.
\eeq
This means, whenever a non-vanishing CS-level
$\Theta^F_{AB}$ is there on the F-theory side,
as a classical CS-level or generated by loops of massive matter on
the 3D Coulomb branch, then the same CS-level has to be there on the
M-theory side,  i.e.~it has to be generated by the $G_4$-flux.
In contrast, when a CS-term is not there on the F-theory effective field
theory side, the corresponding flux integral in \eqref{eq:3DCSterms} for
the same CS-term in M-theory has to vanish. However, the critical point
is  to allow in the corresponding F-theory loop-computation also for
loops with an infinite tower of KK-states. If KK-states are not included,
the $G_4$-flux is in general over-constrained, in particular in the
presence of a non-holomorphic zero section. In addition, certain
CS-levels get shifted by KK-states and a consistent match of CS-terms
in F- and M-theory is only possible if these corrections are included.

The general form for the correction to the classical
CS-level on the F-theory side, denoted by $\Theta_{\text{cl},\, AB}^{F}$,
has been worked out in \cite{Niemi:1983rq,Redlich:1983dv,Aharony:1997bx}.
The correction is one-loop exact with all 3D massive fermions
contributing in the loop. Assuming a fermion with charge vector $q_A$,
the loop corrected CS-term takes the simple form
\beq \label{eq:CSGenCorr}
\Theta^F_{AB}=
\Theta^F_{\text{cl},\, AB}
+ \frac12\sum_{\underline{q}}n(\underline{q})q_A q_B \ \text{sign}(q_A\zeta^A)\,,
\eeq
where $n(\underline{q})$ is the number of fermions with charge vector
$\underline{q}$ and the sum runs over all these charge vectors. We note
that since real masses can be negative in 3D, the sign-function is
non-trivial. We can rewrite this expression further by noting
the general form of thee charge vector
\beq
	q_A=(n,q_\alpha,q_i,q_m)\,,
\eeq
where we recall the 3D gauge group
$\text{U}(1)^{h^{(1,1)}(B)+\text{rk}(G)+n_{\text{U}(1)}+1}$ on the
Coulomb branch and that the charge under the KK-vector $A^0$ is just
the KK-label of KK-states, $q_0\equiv q_{KK}=n$.

Next, we assume that in a theory obtained by circle reduction from 4D,
there are no states with charge under the $A^\alpha$, since these do not
correspond to a 4D gauge symmetry, i.e.~we assume
$q_\alpha=0$.\footnote{One might wonder whether these states correspond,
on the M-theory side, to M2-branes wrapping curves in the base $B$,
e.g.~generated from resolved conifolds in $B$.} We consider in
the following the massive charged states obtained from the reduction
of four-dimensional massless chiral matter to 3D, along with their
KK-states.
Denoting their charge vectors as $q_A=(n,0,q_i,q_m)$ with $q_i$ the
Dynkin labels of their non-Abelian representations $\mathbf{R}$
under $G$, we write the loop-correction \eqref{eq:CSGenCorr} as
\bea \label{eq:CSloops}
	\Theta^F_{00}=\frac{1}{2}
	\sum_{\mathbf{R}_{q_m}}\sum_{\underline{q}\in \mathbf{R}_{q_m}} \chi(\mathbf{R}_{q_m})\sum_{n=-\infty}^\infty n^2
	\text{sign}(m_{CB}+n\cdot m_{KK})\,,\nn\\
	\Theta^F_{0\Lambda}=\frac{1}{2}
	\sum_{\mathbf{R}_{q_m}}\sum_{\underline{q}\in \mathbf{R}_{q_m}}
	q_\Lambda\,\chi(\mathbf{R}_{q_m})\sum_{n=-\infty}^\infty n\,
	\text{sign}(m_{CB}+n\cdot m_{KK})\,,\nn\\
	\Theta^F_{\Sigma\Lambda}=\frac{1}{2}
	\sum_{\mathbf{R}_{q_m}}\sum_{\underline{q}\in \mathbf{R}_{q_m}} q_{\Sigma}\,q_{\Lambda}\,\chi(\mathbf{R}_{q_m})\sum_{n=-\infty}^\infty
	\text{sign}(m_{CB}+n\cdot m_{KK})\,,
\eea
where we invoked the absence of classical CS-terms. Here we have
suppressed a labeling of the KK-charges $n$ by the weights $\mathbf{w}$ 
of the representation
$\mathbf{R}_{q_m}$ and unified the labels $i$ and $m$ as $\Lambda=(i,m)$ and $\chi(\mathbf{R}_{q_m})$
denote the \textit{4D chiralities} of chiral matter fields.
We have also defined the Coulomb branch mass $m_{CB}=q_\Sigma
\zeta^\Sigma$ and the KK-mass $m_{KK}=\frac{1}{R_{KK}}$ with $R_{KK}$ the
radius of the $S^1$.
The terms \eqref{eq:CSloops} are, bearing our assumptions on the spectrum
of massive fermions in mind, the only CS-terms receiving loop-corrections
via \eqref{eq:CSGenCorr}.
Other CS-terms are classically generated, either in 4D or in the 3D
reduction.

We summarize all CS-terms, the presence of classical terms,
the potential correction via loop-effects, their physical
interpretation and a related reference in the following,
\beq \label{eq:CStermslist}
\text{\begin{tabular}{|c|c|c|c|c|} \hline
 $\Theta^F_{AB}$   &     $\Theta_{\text{cl},\,AB}^F$  &  loop-corr.& $G_4$-condition&
 interpretation  \rule{0pt}{15pt}\\ \hline \hline
$\Theta^F_{00}$ &     -    & yes& - & 4D chiralities  \rule{0pt}{14pt}\\ \hline
$\Theta^F_{0\alpha}$ & yes & - &$ \stackrel{!}{=}0$& $S^1$-circle fluxes \cite{Grimm:2011sk}\rule{0pt}{14pt}\\ \hline
 $\Theta^F_{0i}$  &  -   & yes &- & 4D chiralities\rule{0pt}{14pt}\\ \hline
 $\Theta^F_{0m}$  &   -& yes & -&4D  anomaly cancellation \cite{Cvetic:2012xn} \rule{0pt}{14pt}\\
 & & && (for holomorphic $\hat{s}_P$)\\ \hline
 $\Theta^F_{\alpha\beta}$  & - (?) & - &$ \stackrel{!}{=}0$& non-geometric flux? \rule{0pt}{14pt}\\ \hline
 $\Theta^F_{\alpha i}$ & yes & - & $ \stackrel{!}{=}0$&4D gaugings by GUT Cartans \rule{0pt}{14pt}\\ \hline
 $\Theta^F_{\alpha m}$ & yes & - & -& 4D gaugings by U$(1)_m$\rule{0pt}{14pt}\\
 & & && (4D GS-mechanism)\\ \hline
 $\Theta^F_{\Sigma\Lambda}$ & -& yes &-& 4D chiralities \rule{0pt}{14pt}\\
 \hline
\end{tabular}}
\eeq
Here we also mention in one column whether the corresponding
CS-term is used to impose a condition on the $G_4$-flux. Indeed,
the effect of CS-terms $\Theta^F_{0\alpha }$
is the induction of circle fluxes along the $S^1$ in compactification
from 4D and, thus, not a physical effect in 4D. Therefore,
we impose these CS-terms to vanish. Then, the CS-terms
$\Theta^F_{\alpha\beta}$  obstruct the lift back to 4D
\cite{Grimm:2010ks}, potentially by non-geometric effects,  and are
required to vanish. Finally, the CS-terms $\Theta_{\alpha i}$
correspond to 4D gaugings of axions via the maximal torus of the GUT
and are thus set to zero. Thus, using the M-/F-theory duality relation
\eqref{eq:FluxCondMF} we formulate the F-theory conditions on
the $G_4$-flux in terms of 3D CS-levels $\Theta_{AB}$ as
\beq \label{eq:CStermsConditions}
	\boxed{ \quad\mathbf{G_4}\text{\textbf{-flux
conditions:}}\qquad\Theta_{0\alpha}=\Theta_{\alpha\beta}=\Theta_{i\alpha}
	=0\quad\rule{0pt}{14pt}}
\eeq
We note that these conditions on the $G_4$-flux look much weaker than
the ones considered in the literature before. We claim that these
conditions are the appropriate ones, in particular in cases with a
non-holomorphic the zero section $\hat{s}_P$.
In contrast, however, in compactifications with a holomorphic zero
section, the vanishing of the CS-levels in \eqref{eq:CStermsConditions}
implies the vanishing of other, dependent CS-levels. We discuss
this in the following and contrast it to the situation with
a non-holomorphic $\hat{s}_P$.

\subsubsection{KK-Corrected 3D CS-Terms: Field Theory Computations}
\label{sec:KKcomputations}

In certain cases, the loop-corrections \eqref{eq:CSloops} can
vanish, leading to additional vanishing CS-terms. In particular, for
a holomorphic zero-section, the CS-term $\Theta^F_{00}$ in field theory
vanishes, and consistently  also the geometric CS-term $\Theta^M_{00}$ in
M-theory. The latter can be seen easily using
the conditions \eqref{eq:CStermsConditions} in the relation
\beq \label{eq:Theta00Relation}
	\Theta_{00}=\frac{1}{4}K^\alpha K^\beta\Theta_{\alpha\beta}\,,
\eeq	
which is derived employing the definition \eqref{eq:SPtilde}, the
intersection property \eqref{eq:holomorphicSec}, and $K^\alpha$
introduced in
\eqref{eq:KBexpansion}. In order to see the same from the field theory
side and, in general, to compute the loop-corrections to the CS-terms, we
first have to evaluate the sign-function in \eqref{eq:CSloops}.

We calculate the sign-function geometrically, recalling from the
discussion of section \ref{sec:SingularFibU1U1} that to every weight
$\mathbf{w}$ of a representation $\mathbf{R}_{q_m}$ realized in F-theory
there is a corresponding curve $c_{\mathbf{w}}$,
cf.~\eqref{eq:matterSurfs}. The sign-function in
\eqref{eq:CSloops} is then determined by testing whether
the curve $c_{\mathbf{w}}$ associated to a given weight $\mathbf{w}$
with Dynkin labels $q_\Lambda$ and KK-charge $n$ is in the Mori cone
$M(\hat{X})$ of effective curves on $\hat{X}$. We define
\beq \label{eq:signw}
	\text{sign}(q_A\zeta^A)=\left\{\begin{matrix}
	\phantom{-}\,\,\,1\,,\qquad c_{\mathbf{w}}\in M(\hat{X})\,,\\
	-1\,, \qquad \text{otherwise}\,.
	\end{matrix}\right.
\eeq
Here the KK-charge $q_0=n$ of a curve
$c_{\mathbf{w}}$ is obtained geometrically as
\beq \label{eq:KKcharge}
	n=c_{\mathbf{w}}\cdot \tilde{S}_P\,.
\eeq
In general, a curve is in the Mori cone if it is described by
holomorphic equations and  an analysis of the
geometry allows in general to find all holomorphic curves corresponding
to matter, cf.~sections \ref{sec:SingularFibU1U1} and
\ref{sec:SingularitiesSU5U1U1} as well as
\cite{Cvetic:2013nia}.\footnote{In general, the values of the
sign-function on a given representation $\mathbf{R}_{q_m}$ depend on the
phase of $\hat{X}$, respectively, of the 3D gauge theory. See
\cite{Hayashi:2013lra} for a detailed discussion of phases structure of
Calabi-Yau fourfolds and 3D SU(5) gauge theories. However, it can be
shown by a similar argument as in \cite{Witten:1996qb} that
4D observables like the chiralities $\chi(\mathbf{R}_{q_m})$ are not
expected to depend on the phase.}
In the toric context, the relevant parts of the Mori cone can be
constructed systematically as recently demonstrated in
\cite{Grimm:2013oga}.

Now, in the presence of a holomorphic zero section the sign-function
is centered around $0$ because no curve $c_{\mathbf{w}}$ has KK-charge.
This follows from the simple geometric fact that by definition any
rational curve
$c_{\mathbf{w}}$ does not intersect $S_P$, which always
goes through the original singular curve, cf.~figure \ref{fig:I2fiber}.
This implies that the loop-corrections in \eqref{eq:CSloops}
that are odd in the KK-level $n$ vanish
since KK-states with charge $-|n|$ cancel those with charge $|n|$.
In particular, $\Theta^F_{00}=0$, confirming the geometric result
\eqref{eq:Theta00Relation}. In addition,
the sum over KK-states in $\Theta^F_{\Lambda\Sigma}$ reduces to
\beq	\Theta^F_{\Lambda\Sigma}=\frac{1}{2}
	\sum_{\mathbf{R}_{q_m}}\sum_{\underline{q}\in \mathbf{R}_{q_m}}
	q_{\Sigma}\,q_{\Lambda}\,
	\chi(\mathbf{R}_{q_m})\,\text{sign}(m_{CB})\,,
\eeq
which has been used in \cite{Grimm:2011fx,Cvetic:2012xn}.

In contrast, the CS-levels $\Theta_{0\Lambda}$ receive an infinite
loop-correction that
has to be regularized by zeta-function regularization. In
\cite{Cvetic:2012xn} this zeta function regularization has been performed
and it was shown that the field theory result for $\Theta_{0m}$ agrees
with the 4D mixed Abelian-gravitational anomaly,
\beq \label{eq:Theta0m}
	\Theta_{0m}^F=-\frac{1}{12}\sum_{\underline{q}}n(\underline{q})q_m\,,
\eeq
with $n(\underline{q})$ denoting the number of fermions in 4D with
U(1)-charge vector $\underline{q}$. In particular, anomaly
cancellation follows then using the M-/F-theory relation
\eqref{eq:FluxCondMF}  and the geometric result
\beq \label{eq:Theta0Lambda}
	\Theta^M_{0\Lambda}=\frac{1}{2}K^\alpha \Theta^M_{\alpha \Lambda}\,,
\eeq
where the right side is immediately identified with the 4D Green-Schwarz
term for $\Lambda=m$, cf.~section \ref{sec:anomalyCancellation}. Here we
used the general result $S_P\cdot D_\Lambda=0$ for a holomorphic
zero section. We also infer from \eqref{eq:Theta0Lambda}
that the CS-terms $\Theta_{0i}$ are set to zero recalling
\eqref{eq:CStermsConditions}.

In the presence of a non-holomorphic section $\hat{s}_P$ the only thing
that changes on the field theory side is a shift of the sign function
in \eqref{eq:signw}.
It is no longer centered symmetrically around the origin of the
sum over KK-labels $n$. This is geometrically clear because
there are now rational curves $c_{\mathbf{w}}$ in the Mori cone
that have non-zero intersection with the rational zero  section
$\hat{s}_P$, i.e.~that have non-zero KK-charge
\eqref{eq:KKcharge}. These curves have to be located precisely at the
loci where the rational zero section is ill-defined and wraps a
whole $\mathbb{P}^1$ in the fiber, which is the original
singular curve. Everywhere else in the base $B$ the zero section
is only a point on the original singular fiber and does not intersect
curves $c_{\mathbf{w}}$.

The effect of the shift of the sign-function
is dramatic because  now  KK-states with positive
and negative KK-label $|n|$, respectively, $-|n|$ do no longer
cancels in corrections \eqref{eq:CSloops}  that are odd in $n$.
The infinite parts of the sums over KK-states still cancel, but with
a non-zero remainder. Thus, the CS-term $\Theta_{00}^F$ that was zero for
a holomorphic zero section is now generated by a contribution of a finite
number of KK-states. For example, assuming, with $k$ denoting an integer,
a sign-function of the form
\beq \label{eq:signshifted}
	\text{sign}(m_{CB}+n\cdot m_{KK})=\left\{\begin{matrix}
	\phantom{-}\,\,\,1\,,\qquad\text{for } n\geq -k\,,\\
	-1\,, \qquad\text{for } n<-k\,,
	\end{matrix}\right.
\eeq
for only one weight  of a single matter multiplet $\mathbf{R}_{q_m}$
the loop induced CS-term in \eqref{eq:CSloops} reads
\beq \label{eq:Theta001rep}
	\Theta_{00}^F=\frac{k(k+1)(2k+1)}{6}\chi(\mathbf{R}_{q_m})\,.
\eeq
See e.g.~\cite{Grimm:2013oga} for a formal derivation or the following
for an intuitive understanding,
\beq \label{eq:sign2}
	\text{\begin{tabular}{|c||c|c|c|c|c|c||c|} \hline
 Integers   &     $-k-1$  &  $-k$ & $-k+1$&  $\ldots$  & $0$ & $1$ & $\sum_{n} n^2 \text{sign}$\rule{0pt}{14pt} \\ \hline \hline
 $ n^2 \text{sign}(n)$ & $-(k+1)^2$ & $-k^2$ & $-(k-1)^2$ & $\ldots$ &$0$ & $1$ &
 $0$\rule{0pt}{14pt}
 \\ \hline
 $ n^2 \text{sign}(n+k)$ & $-(k+1)^2$ & $k^2$ & $(k-1)^2$ & $\ldots$ &$0$ & $1$ &
 $\frac{k(k+1)(2k+1)}{3}$\rule{0pt}{22pt}
 \\ \hline
 \end{tabular}}
\eeq
Here the unshifted sum was normalized to zero, then the shifted
sum differs only by the amount obtained as the finite
sum over the differences between the first and second row in
\eqref{eq:sign2} for each column.

Thus we see that $\Theta_{00}^F$ in \eqref{eq:Theta001rep}
is directly proportional to one chirality. Consequently, imposing
$\Theta_{00}^M=0$, as done in the literature with holomorphic zero
sections, also in the non-holomorphic case would unnecessarily set this
chirality to zero. We will see that the
loop-correction to $\Theta_{00}^F$ precisely takes the form
\eqref{eq:Theta001rep} for $dP_2$-elliptic fibrations, both without
and with an additional SU(5) gauge group, cf.~sections
\ref{sec:4DChiralityU1U1} and \ref{sec:SingularitiesSU5U1U1}, since
only one singlet  has a non-trivial KK-charge $n=2$ and, thus, a shifted
sign-function \eqref{eq:signshifted} with $k=-2$.

We conclude by  mentioning that the CS-terms
$\Theta_{\Lambda\Sigma}$ are shifted in a similar way,
where as in \eqref{eq:sign2} the shift originates from a finite
number of KK-states. We note that this shift has to be taken into
account when we determine certain 4D chiralitites via 3D CS-terms
in sections \ref{sec:4DChiralityU1U1} and \ref{sec:ChiralitiesU1xU1xSU5}.
In addition, also the relation \eqref{eq:Theta0m} is modified
by a finite correction due to KK-states.  Assuming again that only
one weight in the  representation $\mathbf{R}_{q_m}$ has a
non-trivial KK-charge with shifted sign-function \eqref{eq:signshifted},
we obtain the corrected expression
\beq \label{eq:Theta0mshifted}
	\Theta_{0m}^F=-\frac{1}{12}\sum_{\underline{q}}n(\underline{q})q_m-\frac{k(k+1)}{2}
	\chi(\mathbf{R}_{q_m})q_m\,.
\eeq
Thus, we see that these CS-levels are no longer directly related to the
4D mixed Abelian-gravitational anomaly as in \eqref{eq:Theta0m} with a
holomorphic zero section. Geometrically  this is also clear since
in general $S_P\cdot \sigma(\hat{s}_m)\neq 0$ in compactifications
with rational zero sections. Thus, \eqref{eq:Theta0Lambda} does not
hold and the cancellation of 4D mixed Abelian-gravitational anomaly
is not geometrically implied. However, after subtracting
the contribution of KK-states in \eqref{eq:Theta0mshifted}, 4D anomaly
cancellation requires the remaining piece to again equal
$\frac{1}{2}K^\alpha\Theta_{\alpha m}$. In other words, we can formulate
the relation
\beq \label{eq:sPG4}
	\frac{1}{4}\int_{\hat{X}}S_P\cdot \sigma(\hat{s}_m)\cdot G_4
	=\frac{k(k+1)}{2}q_m\chi(\mathbf{R}_{q_m})
\eeq
on the fourfold $\hat{X}$ as a necessary and sufficient condition for
4D anomaly cancellation. It would be nice to proof this
relation purely geometrically.
Finally, we note that also the CS-terms $\Theta^F_{0i}$ need no longer
vanish, since geometrically $\Theta^M_{0i}\neq
\frac{1}{2}K^\alpha\Theta^M_{\alpha i}$. However, in the geometry
$\hat{X}$  at hand, cf.~section \ref{sec:SingularitiesSU5U1U1},
this relation still holds since the singularity of the zero section
happens away from the resolved SU(5) singularity and, thus,
no SU(5)-matter has non-trivial intersections with $S_P$. Moreover, we
obtain $\tilde{S}_P\cdotp D_i=0$ for all SU(5)-Cartan divisors $D_i$
in section \ref{sec:CohomologyFFSU5}.

\section{$G_4$-Flux \& Chiralities on Fourfolds with Two U(1)s}
\label{sec:G4+chiralitiesU1xU1}

In this section we analyze chirality-inducing $G_4$-flux in F-theory
on the fourfolds $\hat{X}$ with $dP_2$-elliptic fiber and a
non-holomorphic zero section.  In section \ref{sec:G4U1xU1} we first
construct the general $G_4$-flux for the fourfold
$\pi:\,\hat{X}\rightarrow \mathbb{P}^3$, where we also comment
on the general D3-brane tadpole.
Then in section \ref{sec:4DChiralityU1U1} we  outline first our general
strategy to obtain chiralities on Calabi-Yau fourfolds with
higher rank Mordell-Weil group and a rational zero section, before
applying it again to the fourfold $\hat{X}$ with $B=\mathbb{P}^3$.
Then in section \ref{sec:anomalyCancellation} we first review anomaly
cancellation conditions in general 4D effective field theories.
Then we show that the general spectrum obtained for
the Calabi-Yau fourfold $\hat{X}$ is anomaly-free for the entire
allowed region of figure \ref{fig:XP3n7n9region}. We conclude in
section \ref{sec:ToricExsU1U1} with a concrete toric example to
exemplify the general findings.

The case of an un-Higgsed SU$(5)$ GUT-group
is considered separately in section \ref{sec:FFEllipticFibSU5U12}.

\subsection{$G_4$-Flux on Fourfolds with Two Rational Sections}
\label{sec:G4U1xU1}

In this section
we first show that on a general fourfold $\hat{X}$ with $dP_2$-elliptic
fiber the D3-brane tadpole can always be solved  by adding a sufficient
amount of $n_{D3}$ of integral D3-brane charge.  Then we
construct viable $G_4$-flux for F-theory on a general fourfold $\hat{X}$
with base $B=\mathbb{P}^3$. We show that for different elliptic
fibrations, i.e.~for different values of the integers $n_7$, $n_9$,
the number of independent $G_4$-flux quanta changes.

\subsubsection*{Integral D3-Tadpole on General Fourfolds with U$(1)\times$U(1)}

In the following we prove the necessary condition
for D3-tadpole cancellation on $\hat{X}$, namely that the induced
D3-brane charge from the combination of the Euler number and the quantized
$G_4$-flux in the tadpole equation \eqref{eq:D3tadpole} is always
integral. The following discussion is an application of the arguments
in \cite{Sethi:1996es,Witten:1996md,Klemm:1996ts}, that immediately
carry over to elliptic fibrations $\hat{X}$ with $dP_2$-elliptic curve.

To this end, we use the relation \eqref{eq:chi0X_Euler} between the
arithmetic genus $\chi_0(\hat{X})$ and the Euler number $\chi(\hat{X})$
to rewrite the tadpole \eqref{eq:D3tadpole} as
\beq
	 n_{D3}=\frac{\chi(\hat{X})}{24}-\frac{1}{2}\int_{\hat{X}}G_4^2=-60+\frac{1}{2}\int_{\hat{X}}\left(\frac{1}{4}c_2(\hat{X})^2-G_4^2\right)\,.
\eeq
Here we have also employed that $\chi_0(\hat{X})=2$,
cf.~\eqref{eq:chi0X_chi0B}.
Using the flux quantization condition \eqref{eq:G4quantization} this
can be written as
\beq \label{eq:nD3induced}
	n_{D3}=-60-\frac{1}{2}\int_{\hat{X}}\left(x^2-x\wedge c_2(\hat{X})\right)\,,
\eeq
where we used $x=G_4+\tfrac{1}{2}c_2(\hat{X})$. By flux quantization
\eqref{eq:G4quantization} we know that $x$ is integral, i.e.~an
element in $H^{4}(\hat{X},\mathbb{Z})$. This implies
by Wu's theorem that $x^2\cong c_2(\hat{X})\wedge x\,\,\text{mod}\,2 $
\cite{Witten:1996md}, so that the integrand in \eqref{eq:nD3induced} is
divisible by two. Thus, the number  $n_{D3}$ of D3-branes is integral for
every elliptically Calabi-Yau fourfold $\hat{X}$ with general elliptic
fiber in $dP_2$.

\subsubsection*{The $G_4$-Flux on $\hat{X}$ with $B=\mathbb{P}^3$}

Next we explicitly determine the $G_4$-flux on $\hat{X}$ for a general
elliptic fibration over the base $B=\mathbb{P}^3$, i.e.~for all
integers $n_7$, $n_9$ in the allowed region in figure
\ref{fig:XP3n7n9region}.

We begin by expanding the $G_4$-flux according to \eqref{eq:G4expansion}
into the basis of $H^{(2,2)}_V(\hat{X})$ determined in
\eqref{eq:XP3_H22V},
\beq \label{eq:G4expansionP3}
	G_4=a_1 H_B^2+a_2 H_B\cdot S_P+a_3 H_B\cdot \sigma(\hat{s}_Q)
	+a_4 H_B\cdot \sigma(\hat{s}_R)+a_5 S_P^2\,,
\eeq
for general coefficients $a_i$, where as before the application
of Poincar\'e duality is understood. Then we calculate the CS-levels
\eqref{eq:3DCSterms} employing the intersection ring
\eqref{eq:XP3_C0} in the basis of divisors \eqref{eq:divsdP2P3},
but with $\tilde{S}_P=S_P+2H_B$ as defined in \eqref{eq:SPtilde} replacing
the zero section $S_P$.
The generic solution is a three-parameter family of $G_4$-flux
given by
\beq \label{eq:G4solP3}
	G_4=a_5 n_9 \left(4-n_7+n_9\right)H_B^2+4 a_5H_B\cdot S_P+a_3H_B \cdot\sigma(\hat{s}_Q)+a_4 H_B\cdot\sigma(\hat{s}_R)+a_5S_P^2\,,
\eeq
which is valid for all values of $n_7$ and $n_9$ in the allowed region
figure \ref{fig:XP3n7n9region}.

This generic three-parameter solution for the $G_4$-flux is expected
since there are generically five different surfaces in
\eqref{eq:G4expansionP3}
and two independent conditions \eqref{eq:CStermsConditions}, namely
$\Theta_{0\alpha}=\Theta_{\alpha\beta}=0$ with $\alpha,\,\beta=1$.
However, the situation becomes more interesting at special values for
$(n_7,n_9)$.
First, we recall that for $(n_7,n_9)$ on the boundary of the
region in figure \ref{fig:XP3n7n9region}, the dimensionality of
$H^{(2,2)}_V(\hat{X})$ decreases to $4$. The surface $S_P^2$ becomes
linearly dependent in homology on the four other surfaces, as
noted below \eqref{eq:XP3_eta2_1}. At the same time, the number of
independent conditions on the $G_4$-flux remains two. Thus, we find  two
independent $G_4$-fluxes on the boundary.
We have depicted this situation in figure \ref{fig:I2fiber}
\begin{figure}[ht!]
\centering
\includegraphics[scale=0.5]{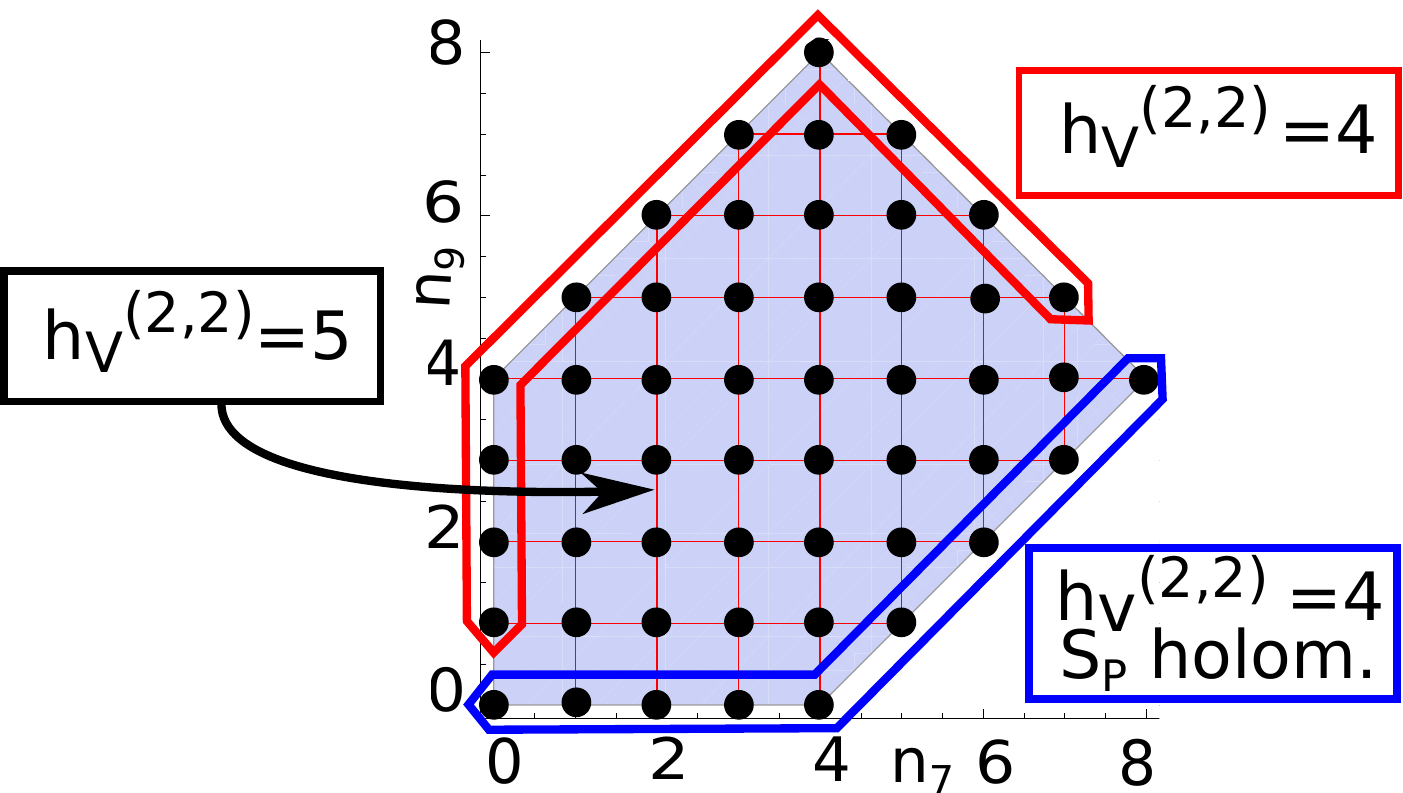} \qquad \qquad
 \caption{The region of allowed values for $(n_7,n_9)$ from figure
 \ref{fig:XP3n7n9region}. On the entire region, there are two conditions
 on the flux. In the interior
 of this region, \eqref{eq:G4solP3} holds. On the red and the blue
 boundary, there are only four independent
 $(2,2)$-forms in the expansion \eqref{eq:G4expansionP3}. On the
 blue boundary, $\hat{s}_P$ is holomorphic.}
 \label{fig:G4FluxRegionU1}
\end{figure}

In all these cases we can obtain the expression for the $G_4$-flux by
specializing \eqref{eq:G4solP3}.
This ensures that all quantities, in particular the chiralities of
4D charged matter, that are calculated from the most general
$G_4$-flux specialize correctly for non-generic values of
$(n_7,n_9)$. We discuss this specialization at the end of this
subsection, but note
that the reader may want to skip these details on a first read and
proceed with the chirality formulas in section \ref{sec:4DChiralityU1U1}.

Before delving into the details of this analysis, we evaluate the
D3-brane tadpole \eqref{eq:D3tadpole}  for the Calabi-Yau fourfold
$\hat{X}\rightarrow \mathbb{P}^3$ and the $G_4$-flux \eqref{eq:G4solP3}.
First, we calculate the individual terms in the D3-brane tadpole.
We obtain the Euler number for $\hat{X}$ with  $B=\mathbb{P}^3$ from the
general formula \eqref{eq:EulerNumberX}  as
\beq \label{eq:XP3_Eul}
	\chi(\hat{X})=4896+3 \left[-256(n_7+n_9)+4(8n_7^2+n_7n_9+8n_9^2)
	-n_7n_9(n_7+n_9)\right]\,,
\eeq
where we employed $c_1(\mathbb{P}^3)=4H_B$, $c_2(\mathbb{P}^3)=6H_B^2$.
We note that this expression is manifestly positive due to the bounds
on $n_7$ and $n_9$ in \eqref{eq:XP3condn7n9}, respectively, figure
\ref{fig:G4FluxRegionU1}.
Then we calculate the contribution of the $G_4$-flux \eqref{eq:G4solP3}
to the D3-tadpole as
\bea
	\int_{\hat{X}} G_4^2&=&-8 a_3^2+2 a_3 a_4 \left(n_7-n_9-4\right)
	+a_5^2 n_9\left(2 n_9-n_7\right) \left(n_7-n_9-4\right)
	-2 a_4^2 \left(4+n_9\right)\nn\\&&+2 a_5\left(a_3+2 a_4\right)  n_9 \left(4-n_7+n_9\right)\,.
\eea
Finally, for the purpose of $G_4$-flux quantization via
\eqref{eq:G4quantization} we note the following expression for
the second Chern-class of $\hat{X}$,
\bea
	c_2(\hat{X})&=&(182 + 3 n_9( n_7 - n_9-4) )H_B^2+
	28 H_B\cdot S_P+2 (8-n_9)H_B\cdot \sigma(\hat{s}_Q)\nn\\
	&&+(16 + 2 n_7 - 3 n_9)H_B\cdot \sigma(\hat{s}_R)-5S_P^2\,,
\eea
where we expanded the general expression \eqref{eq:c2XgenB} for
a better comparison with the $G_4$-flux \eqref{eq:G4solP3} in the basis
\eqref{eq:XP3_H22V} of the vertical cohomology. Also for completeness
we calculate the square of the second Chern-class from the general
formula \eqref{eq:c2X^2} as
\beq
\int_{\hat{X}}c_2(\hat{X})^2=2112-256(n_7+n_9)+4(8n_7^2+n_7n_9+8n_9^2)
	-n_7n_9(n_7+n_9)\,.
\eeq
Comparing this with the Euler number \eqref{eq:XP3_Eul} we
reconfirm the general relation \eqref{eq:chi0X_Euler} between
the arithmetic genus and the Euler number of a Calabi-Yau fourfold.

Finally, we conclude by a discussion of \eqref{eq:G4solP3} for
non-generic values of $(n_7,n_9)$.
Along each component of the boundary we have to use the homology
relations between $S_P^2$ and the four remaining basis elements
\eqref{eq:XP3_H22V} of $H^{(2,2)}_V(\hat{X})$. The relevant relations are
\eqref{eq:SP^2n7=0}, \eqref{eq:SP^2n9_1}, \eqref{eq:SP^2n9_2} and
\eqref{eq:SP^2calculated} worked out in the appendix. In all these cases
the general formula
\eqref{eq:G4solP3} for the $G_4$-flux can be shown to reduce to the
two-parameter
\beq \label{eq:G4solP3n7=0}
	G_4=\tilde{a}_3H_B\cdot \sigma(\hat{s}_Q)
	+\tilde{a}_4 H_B\cdot \sigma(\hat{s}_R)\,.
\eeq
Here the coefficients $\tilde{a}_3$ and $\tilde{a}_4$ are the following
linear combination of $a_3$, $a_4$ and $a_5$ on the three boundary
components,
\bea
	&\{n_7=0\}:\,\, \tilde{a}_3=a_3\,,\,\,\tilde{a}_4=a_4-n_9 a_5\,,\quad \{n_9=n_7+4\}:\,\, \tilde{a}_3=a_3-n_7 a_5\,,\,\,
	\tilde{a}_4=a_4-4a_5\,,\!\!\!\!\!\!&\nn\\
	&\{n_9=12-n_7\}:\,\,\,\tilde{a}_3=a_3+(n_7-8)a_5\,,\,\,
	\tilde{a}_4=a_4+(n_7-8) a_5\,,&\nn\\
	&\{n_9=n_7-4\}\cup \{n_9=0\}:\,\,\, \tilde{a}_3=a_3\,,\,\, \tilde{a}_4=a_4\,.&
\eea
It is satisfying to see that the parameters $\tilde{a}_3$, $\tilde{a}_4$
are continuous at the intersection points of boundary components.
We note that the blue boundaries in figure
\ref{fig:G4FluxRegionU1}, $n_9=0$ and $n_9=n_7-4$, are
special because $\hat{s}_P$ is holomorphic.

We conclude with one remark on the form of the $G_4$-flux obtained here.
In some cases,  Type IIB seven-brane gauge
fluxes $F^{(1)}$, $F^{(2)}$ of two single D7-branes can be lifted into
F-theory by considering a $G_4$-flux  of the type
\beq
	G_4=F^{(1)}\cdot\sigma(\hat{s}_Q)+F^{(2)}\cdot\sigma(\hat{s}_R)\,.
\eeq	
We note that the flux \eqref{eq:G4solP3n7=0} is precisely of this form.
In contrast,t he interpretation of the
general $G_4$-flux in \eqref{eq:G4solP3} in terms of Type IIB quantities,
if possible, is less clear and would be very interesting to investigate.

\subsection{4D Chiralities from Matter Surfaces \& 3D CS-Terms}
\label{sec:4DChiralityU1U1}

Finally we are prepared to calculate the chirality of
matter in four-dimensional F-theory compactifications on $\hat{X}$
with U$(1)\times$U(1) gauge group. We demonstrate that the chirality
of all six matter representations discussed in section
\ref{sec:FFEllipticFib} can be determined uniquely. Half of the
chiralities are determined by integration of the $G_4$-flux over matter
surfaces $\cC_{\mathbf{R}}^{\mathbf{w}}$, the other half from
the 3D CS-terms $\Theta_{\Sigma\Lambda}$. As we see explicitly,
the 3D CS-terms $\Theta_{\Lambda\Sigma}$ are
not sufficient to fix all chiralities, in contrast to earlier works
with a holomorphic zero section. However, supplemented
by new CS-terms present only for non-holomorphic zero sections,
see section \ref{sec:G4constraintsF},
the chiralities can be obtained also exclusively via 3D CS-terms.
We emphasize that this analysis requires the inclusion of
KK-charges for all curves $c_{\mathbf{w}}$. In the fourfold
$\hat{X}$ only the representation $\mathbf{1}_{(-1,-2)}$ has non-trivial
KK-charge $q_{KK}=2$.

We outline the general strategy to obtain 4D chiralities in section
\ref{sec:ChiralitiesStrat} before we determine them
explicitly in section \ref{sec:XP3chiralities}
The concrete calculations are performed for the Calabi-Yau fourfold
$\hat{X}$ with base $B=\mathbb{P}^3$ for  the entire
allowed region in figure \eqref{fig:XP3n7n9region}.

\subsubsection{General Strategy to Determine 4D Chiralities}
\label{sec:ChiralitiesStrat}

We begin our discussion by recalling how to extract the chiral index
of 4D matter in a representation $\mathbf{R}$ under a gauge group $G$
in a global F-theory compactification on an arbitrary resolved
elliptically fibered Calabi-Yau fourfold $\hat{X}$.

As explained in section \ref{sec:FFEllipticFib}, each matter
representation $\mathbf{R}$ possesses an associated ruled surface,
the matter surface $\mathcal{C}_{\mathbf{R}}^{\mathbf{w}}$, where
$\mathbf{w}$ denotes a
weight of $\mathbf{R}$. Then, the chiral index of charged matter in this
representation $\mathbf{R}$ is given by the flux integral\footnote{The
factor $-\frac{1}{4}$ has been introduced to be consistency with
the conventions of \cite{Cvetic:2012xn}. It can be reabsorbed into the
$G_4$-flux.}
\cite{Donagi:2008ca,Hayashi:2008ba,Braun:2011zm,Marsano:2011hv}
\beq \label{eq:chi(R)}
	\chi(\mathbf{R})= -\frac{1}{4}\int_{\cC_{\mathbf{R}}^\mathbf{w}}G_4\,,
\eeq
where the $G_4$-flux is a quantized M-theory flux subject to the
conditions \eqref{eq:CStermsConditions}.
It is important to note that these conditions, more precisely the
conditions $\Theta_{i\beta}=0$, imply that the integral \eqref{eq:chi(R)}
is independent on the choice of a particular weight $\mathbf{w}$, since
different weights $\mathbf{w}$, $\mathbf{w}'$ of the same representation
$\mathbf{R}$ are related by a root $\alpha_i$,
$\mathbf{w}-\mathbf{w}'=  \alpha_i$.

In cases where all the matter surfaces $\cC_{\mathbb{R}}^{\mathbf{w}}$
are known,
the integral \eqref{eq:chi(R)} is the most direct and geometric way of
calculating the chirality $\chi(\mathbf{R})$. However, for the fourfold
$\hat{X}$ at hand, but also for the fourfold $\Xsu$ with an additional
resolved SU(5) singularity constructed in section
\ref{sec:FFEllipticFibSU5U12}, the homology class of all matter
surfaces $\cC_{\mathbb{R}}^{\mathbf{w}}$ is not known. Fortunately,
since we know the expected matter spectrum from the geometric analysis
there is to determine the missing chiralities $\chi(\mathbf{R})$
indirectly via 3D CS-terms.

For this purpose we recall the three-dimensional  M-/F-duality
explained in section \ref{sec:G4constraintsF}.
There we have seen that the 3D CS-terms \eqref{eq:3DCSterms} for the
U(1)-vector fields $A^{\Lambda}=(A^i,A^m)$ on the Coulomb branch
can be calculated in two independent ways. On the one hand,
after solving the $G_4$-flux condition \eqref{eq:CStermsConditions},
one can calculate all CS-terms $\Theta^M_{\Lambda\Sigma}$
on the M-theory side by evaluating
the classical flux integrals in \eqref{eq:3DCSterms}.
On the other hand, the CS-levels
$\Theta^F_{AB}$ on the F-theory side are generated and corrected at
one-loop from integrating out charged matter and, thus, contain, as shown
in section \ref{sec:G4constraintsF}, cf.~\eqref{eq:CSloops}, the 4D
chiral indices $\chi(\mathbf{R}_{q_m})$ of all 4D matter.
Since we know which representations $\mathbf{R}_{q_m}$ are there from our
geometric analysis in section \ref{sec:SingularFibU1U1} and
the later section \ref{sec:SingularitiesSU5U1U1} for the SU(5) case, we
can solve the matching condition \eqref{eq:FluxCondMF} for the
chiralities $\chi(\mathbf{R}_{q_m})$.

We note that in both  fourfolds $\hat{X}$ and $\Xsu$ we know the homology
classes for a number of matter surfaces $\cC_{\mathbb{R}}^{\mathbf{w}}$,
but not for all. For these matter surfaces, we can evaluate the index
\eqref{eq:chi(R)} directly. The remaining chiralities can be fixed,
as demonstrated in sections \ref{sec:4DChiralityU1U1} and
\ref{sec:ChiralitiesU1xU1xSU5} by the matching of the CS-terms
$\Theta_{\Lambda\Sigma}$, that are given on the F-theory
side by the  loop-expression given in \eqref{eq:CSloops}.
However, we can also obtain \textit{all} 4D chiralities by taking
into account the CS-terms $\Theta_{00}$ and $\Theta_{0m}$ in
\eqref{eq:CSloops}, respectively, in \eqref{eq:Theta001rep}
and \eqref{eq:Theta0mshifted}. As we will see concretely,
the obtained results agree with the direct computations via
\eqref{eq:chi(R)}, confirming the validity of application
of the M-/F-theory duality \eqref{eq:FluxCondMF} to the determination
of 4D chiralities.

We conclude by noting that the matching \eqref{eq:FluxCondMF} of CS-terms
$\Theta_{\Sigma\Lambda}$ has been used in
\cite{Grimm:2011fx,Cvetic:2012xn,Braun:2013yti}
to calculate successfully
the chiralities of F-theory compactifications with SU$(5)$ and
SU$(5)\times$U(1) gauge symmetry.
In this work, however,
we encounter the novel situation that the conditions
arising from $\Theta_{\Sigma\Lambda}$
are not sufficient to determine the chiralities of the full spectrum
with both U$(1)\times$U(1) as well as SU$(5)\times$U$(1)\times$U$(1)$
gauge symmetry as outlined in the following and in section
\ref{sec:FFEllipticFibSU5U12}. The reason for this is precisely the
existence of a non-holomorphic zero section. However, either supplemented
by the chiralities that can be directly determined  by the flux integrals
\eqref{eq:chi(R)} or by the CS-terms $\Theta_{00}$ and $\Theta_{0m}$,
we obtain all chiralities. For an application of the latter resolution,
see also the analogous analysis of \cite{Grimm:2013oga}  in 6D.

One might wonder whether the CS-terms
always provide enough conditions to solve for the chiralities of
a known F-theory spectrum. By a simple counting argument assuming
a rank $n_{\text{U}(1)}$ Mordell-Weil group, we enumerate the number
of conditions arising from the matching of these CS-terms as
\beq
	\#(\text{CS-terms})=\frac{(n_{\text{U}(1)}+2)(n_{\text{U}(1)}+1)}{2}\,.
\eeq
If all these conditions remain independent, the CS-terms might indeed
be sufficient to determine the chiralities in F-theory compactifications
with more U(1)-symmetries. This can be seen by a similar estimate on the
growth of the number of different representations as a function of the
rank $n_{\text{U}(1)}$ of the Mordell-Weil group.

\subsubsection{Chiralities on $\hat{X}$ with $B=\mathbb{P}^3$: Matter
Surfaces \& CS-Terms}
\label{sec:XP3chiralities}

In the following we calculate for the first time 4D chiralities
$\chi(\mathbf{R})$ of an F-theory compactification on a general
elliptically fibered Calabi-Yau fourfold
$\hat{X}$ with rank two Mordell-Weil group and with the full
matter spectrum analyzed in section \ref{sec:FFEllipticFib}.
The following is a direct extension of the six-dimensional analysis in
\cite{Cvetic:2013nia} to a 4D chiral theory.

As mentioned before, the 4D chirality of a given representation
$\mathbf{R}$ is computed by the flux integrals
\eqref{eq:chi(R)} given the $G_4$-flux and the corresponding
matter surfaces $\cC_{\mathbf{R}}^{\mathbf{w}}$.
The matter surfaces for the matter fields $\mathbf{1}_{(q_1,q_2)}$
with the three different
U$(1)^2$-charges $(q_1,q_2)=(-1,1)$, $(0,2)$, $(-1,-2)$ have been
determined in \eqref{eq:MatSurfaces1}. Using the general
$G_4$-flux \eqref{eq:G4solP3} on $\hat{X}\rightarrow \mathbb{P}^3$
we obtain the following chiralities
\bea \label{eq:XP3_chi(R)1}
	\chi(\mathbf{1}_{(-1,1)})&=&
	-\tfrac{1}{4}\int_{\cC_{(-1,1)}}G_4
	=\tfrac{1}{4} \left(a_3-a_4\right) n_7 \left(4+n_7-n_9\right)\,,\nn\\
\chi(\mathbf{1}_{(0,2)})&=&
	-\tfrac{1}{4}\int_{\cC_{(0,2)}}G_4
	=\tfrac{1}{4} n_7 n_9 (-2 a_4+a_5 (4-n_7+n_9))\,,
	\nn \\
	\chi(\mathbf{1}_{(-1,-2)})&=&
	-\tfrac{1}{4}\int_{\cC_{(-1,-2)}}G_4
	=\tfrac{1}{4} n_9(4-n_7+n_9) (a_3+2 a_4+a_5 (n_7-2 n_9)) \,.
\eea
In order to evaluate the involved intersections we have made use of
the topological metric $\eta^{(2)}$ on the cohomology
$H^{(2,2)}_V(\hat{X})$ calculated in \eqref{eq:XP3_eta2_1}.

In order to obtain the chiralities for the matter fields
$\mathbf{1}_{(q_1,q_2)}$ with charges $(q_1,q_2)=(1,0)$, $(0,1)$,
$(1,1)$, we have to use the 3D CS-levels and the matching condition
\eqref{eq:FluxCondMF}. For this purpose we have to determine the
KK-charges of all six matter representations on $\hat{X}$.
As mentioned before, a non-trivial KK-charge is calculate by
the intersection \eqref{eq:KKcharge} of the curve $c_{\mathbf{w}}$
and the zero section $\tilde{S}_P$. Such a non-trivial intersection
can only occur at loci, where the zero section is ill-defined and
wraps fiber components. This is precisely the case at the loci
$s_8=s_9=0$, where $\mathbf{1}_{(-1,-2)}$ is supported. Since the
fiber is an $I_2$-fiber over all matter loci, we obtain a KK-charge
\beq \label{eq:KKm1m2}
	q_{KK}(\mathbf{1}_{(-1,-2)})=c_{(-1,-2)}\cdotp \tilde{S}_P=2\,,
\eeq
and zero for all other matter representations. We note
that \eqref{eq:KKm1m2} implies that the sign-function
\eqref{eq:signw} is given by the shifted sign-function
\eqref{eq:signshifted}  with $k=-2$ whereas the other
matter fields retain a point-symmetric sign-function.

We can immediately cross-check this result using the field theory
computations of sections \ref{sec:KKcomputations}. We first calculate
the CS-level $\Theta_{00}^F$ for the 3D KK-vector
on the field theory side. Using the
general expression \eqref{eq:Theta001rep} for $k=-2$ we obtain
\beq \label{eq:XP3_Theta00}
	\Theta_{00}^F=-\chi(\mathbf{1}_{(-1,-2)})\,,
\eeq
with the chirality $\chi(\mathbf{1}_{(-1,-2)})$ determined
in \eqref{eq:XP3_chi(R)1}. We readily calculate the corresponding
flux integral $\Theta_{00}^M$ via \eqref{eq:3DCSterms}  and immediately
reproduce \eqref{eq:XP3_Theta00}.
Next we check the relation
\eqref{eq:Theta0mshifted} for the CS-level $\Theta_{0m}$, respectively,
\eqref{eq:sPG4}. Again we
start with the field theory result for the right hand side of
\eqref{eq:sPG4} which requires
\beq \label{eq:XP3_Theta0m}
	\tfrac{1}{4}\int_{\hat{X}}S_P\cdot\sigma(\hat{s}_Q)\cdot G_4\stackrel{!}{=}-\chi(\mathbf{1}_{(-1,-2)})\,,\qquad
	\tfrac{1}{4}\int_{\hat{X}}S_P\cdot\sigma(\hat{s}_R)\cdot G_4\stackrel{!}{=}-2\chi(\mathbf{1}_{(-1,-2)}).
\eeq
We confirm this relation easily by calculating the intersections
on the left hand side directly from the $G_4$-flux
\eqref{eq:G4solP3}.
Thus, as we have just demonstrated the results for the chiralities in
\eqref{eq:XP3_chi(R)1} obtained from the matter surface integrals
can be employed as an independently check of the field theory expressions
for the CS-levels in \ref{sec:KKcomputations} and the M-/F-theory
duality relation \eqref{eq:FluxCondMF}.

Next we proceed with the computation of the other CS-levels
$\Theta_{mn}$, $m,n=1,2$, for the two U(1) gauge fields $A^m$
corresponding to the divisors $\sigma(\hat{s}_Q)$, respectively, $\sigma(\hat{s}_R)$. Beginning with the M-theory expressions, we
obtain using the Shioda maps \eqref{eq:XP3_Shioda},
\eqref{eq:3DCSterms} and the general $G_4$-flux \eqref{eq:G4solP3},
\bea \label{eq:XP3_CSM}
	\Theta^M_{11}\!\!&\!=\!&\!\!\frac{1}{2} \left[a_3 (96-n_7 (4+n_7)+n_9 (4+n_9))+a_5 (4-n_7+n_9) n_9 (n_7-12-2n_9))\right.\nn\\&&\left.+a_4 (n_7^2+2 (4+n_9) (6+n_9)-n_7 (8+3 n_9))\right]\,,
\nn\\
	\Theta^M_{12}\!\!&\!=\!&\!\!\tfrac{1}{2} \left[a_3 (n_7^2+2 (4+n_9) (6+n_9)-n_7 (8+3 n_9))+(4-n_7+n_9) (a_5 (-12+3 n_7-5 n_9) n_9\right.\nn\\
	 &&\left.+a_4 (12+5 n_9))\right]\,,\nn\\
	\Theta^M_{22}\!\!&\!=\!&\!\!\frac{1}{2} \left[96 a_4+a_3 (4-n_7+n_9) (12+5 n_9)+2 a_4 n_9 (32-4 n_7+5 n_9)\nn\right.\\
	&&\left.-2 a_5 n_9 (4-n_7+n_9) (12-2 n_7+5 n_9)\right]\,,
\eea
with $\Theta^M_{21}=\Theta^M_{12}$.
Here we have used the quartic intersections \eqref{eq:XP3_C0} in
appendix \ref{app:intsP3}.

Then we compute the one-loop CS-terms $\Theta_{mn}^{F}$ in
\eqref{eq:CSloops}  on the F-theory side. For the matter spectrum at
hand, cf.~\eqref{eq:U1spectrum}, we obtain
\bea \label{eq:XP3_CSF}
	\Theta_{11}^F&=&\tfrac{1}{2} (\chi (\mathbf{1}_{(1,0)})+\chi (\mathbf{1}_{(1,1)})+\chi (\mathbf{1}_{(-1,1)})-3\chi (\mathbf{1}_{(-1,-2)})))\,,\nn\\
	\Theta_{12}^F&=&\tfrac{1}{2} (\chi (\mathbf{1}_{(1,1)})-\chi (\mathbf{1}_{(-1,1)})-6\chi (\mathbf{1}_{(-1,-2)}))\,,\nn\\
	\Theta_{22}&=&\tfrac{1}{2} (\chi (\mathbf{1}_{(0,1)})+\chi (\mathbf{1}_{(1,1)})+\chi (\mathbf{1}_{(-1,1)})+4 (\chi (\mathbf{1}_{(0,2)})-3\chi (\mathbf{1}_{(-1,-2)})))\,.
\eea
We note that the factor of $-3$ in front of $\chi(\mathbf{1}_{(-1,-2)})$
occurs due to the shifted sign-function \eqref{eq:signshifted} with
$k=-2$. Indeed, the relevant sum over KK-states in this case yields
\beq
	\sum_n \text{sign}(n+k)=-3\,.
\eeq

We note that the matching of 3D CS-terms \eqref{eq:FluxCondMF} only using
the $\Theta_{mn}$ yields three conditions for the six a priori unknown
chiralities in \eqref{eq:XP3_CSF}. Thus, it is impossible to determine
the full matter
spectrum from these CS-terms alone, which is in contrast to
earlier studies in the literature with holomorphic zero sections. We
emphasize that the same
complication arises in the case of SU$(5)\times$U$(1)^2$ gauge group
considered in section \ref{sec:FFEllipticFibSU5U12}, since the
representations
in the SU$(5)$-singlet sector will be identical to the six
representations considered here. However, we can either
use the results \eqref{eq:XP3_chi(R)1} from the integral of the
$G_4$-flux over the matter surfaces or have to incorporate
the CS-terms $\Theta_{00}$ and $\Theta_{0m}$ to obtain three
further conditions and to fix all six chiralities.

Consequently, taking into account the results \eqref{eq:XP3_chi(R)1},
we apply the matching condition \eqref{eq:FluxCondMF}
for the dual CS-terms \eqref{eq:XP3_CSM} and \eqref{eq:XP3_CSF}
we obtain the remaining three chiralities as
\bea \label{eq:XP3_chi(R)2}
	\chi(\mathbf{1}_{(1,0)})\!\!&\!=\!&\!\!\tfrac{1}{4} \left[a_5 n_7 n_9
	\left(4-n_7+n_9\right)+a_3 \left(2 n_7^2-\left(12-n_9\right)
	\left(8-n_9\right)-n_7 \left(16+n_9\right)\right)\right]\,,\\
	\chi(\mathbf{1}_{(0,1)})\!\!&\!=\!&\!\!\tfrac{1}{2} \left[a_5
	n_9\left(4-n_7+n_9\right) \left(12-n_9\right)
	-a_4 \left(n_7\left(8-n_7\right) +\left(12-n_9\right)
	\left(4+n_9\right)\right)\right]\,,\nn\\
	\chi(\mathbf{1}_{(1,1)})\!\!&\!=\!&\!\!\tfrac{1}{4}
	\left[2 a_5 n_9(4\!-\!n_7\!+\!n_9) (12\!-\!n_9)\!-\!(a_3\!+\!a_4)
	(n_7^2\!+\!n_7 (n_9\!-\!20)
	+2 (12\!-\!n_9) (4\!+\!n_9))\right]\,.\nn
\eea

We conclude by noting that the chiralities we obtain include factors of
$\frac{1}{2}$ and $\frac{1}{4}$. These factors should disappear once the
$G_4$-flux has been quantized appropriately according to
\eqref{eq:G4quantization}. The precise quantization will, however,
depend on the
values of $n_7$, $n_9$ and has to be done in a case by case
analysis. In addition, for concrete $n_7$, $n_9$
the factors in the numerator \eqref{eq:XP3_chi(R)1} and
\eqref{eq:XP3_chi(R)2} have different divisibility properties and can
cancel the denominators. Therefore, in order to not obscure these
cancellation effects, we keep the normalization in
\eqref{eq:XP3_chi(R)1}, \eqref{eq:XP3_chi(R)2} and the mild
fractions of $\frac{1}{2}$ and $\frac{1}{4}$. In concrete toric
examples, they can be cancelled appropriately.

\subsection{4D Anomaly Cancellation: F-Theory with Multiple U(1)s}
\label{sec:anomalyCancellation}

Finally, after having calculated the matter spectrum of an F-theory
compactification, we check consistency of the obtained
low-energy effective physics. One check is anomaly cancellation.
In the following we introduce the necessary quantities to
analyze anomalies and refer to \cite{Cvetic:2012xn}
for more details on anomalies in general and, in particular, in F-theory.
Then we use these techniques to show that the general spectrum
found in section \ref{sec:4DChiralityU1U1} for F-theory compactifications 
on $\hat{X}\rightarrow \mathbf{P}^3$ with gauge group U$(1)^2$
is anomaly-free.

\subsubsection{4D Anomaly Cancellation: General Discussion}

In the following we assume a 4D gauge group $G$ with a single non-Abelian
factors, see for example \cite{Cvetic:2012xn} for the general case, and
with a number $n_{U(1)}$ of Abelian factors. Charges are summarized by a
charge vector $\underline{q}=(q_m)$ and
the number of left Weyl fermions in a matter representation
$\mathbf{R}_{\underline{q}}$ are denoted in general by
$n(\mathbf{R}_{\underline{q}})$. The number of fermions in a
representation $\mathbf{R}$ irrespective of their U(1)-charges is
denoted by $n(\mathbf{R})$, whereas $n(\underline{q})$ indicates
the number of fermions with charges $\underline{q}$ regardless
of their non-Abelian representation. All these numbers can be expressed
in terms of the chiralities $\chi(\mathbf{R}_{\underline{q}})$.

The
conditions for 4D anomaly cancellation via the generalized Green-Schwarz
(GS) mechanism \cite{Green:1984sg,Sagnotti:1992qw} yield a system of linear equations involving
the spectrum of the theory as well as parameters encoding the GS-counter
terms and the gaugings of axions.
These conditions for cancellation of 4D purely non-Abelian, purely
Abelian, mixed Abelian-non-Abelian
and mixed Abelian-gravitational anomalies read, in the same order,
\bea \label{eq:4Danomalies}
\text{purely non-Abelian anomaly} &:& \,\,\, \sum_{\mathbf{R}}n(\mathbf{R}) V(\mathbf{R})=0 \,,\nn\\
\text{U}(1)_k\times\text{U}(1)_l\times\text{U}(1)_m\text{-anomaly} &:&  \,\,\, \frac{1}{6}\sum_{\underline{q}} n(\underline{q}) q_{(m} q_n  q_{k)} = \frac{1}{4} b^\alpha_{(mn} \Theta_{k)\alpha}\,, \nn\\
U(1)_m\text{-non-Abelian anomaly} &:&\,\,\, \frac{1}{2}\sum_{\mathbf{R}} \sum_{\underline{q}} n(\mathbf{R}_{\underline{q}}) U(\mathbf{R})q_m = \frac{1}{4\lambda} b^\alpha \Theta_{\alpha m}\,,\nn \\
U(1)_m\text{-gravitational anomaly}&:& \,\,\, \frac{1}{48} \sum_q n(\underline{q}) q_m = -\frac{1}{16} a^\alpha \Theta_{m\alpha}\,.
\eea
Here $V(\mathbf{R})$ and $U(\mathbf{R})$ denote group theoretical
constants that arise when rewriting traces in the representation
$\mathbf{R}$  as traces in the fundamental representation $\mathbf{f}$.
Letting $F$ denote the non-Abelian field strength, we set
\beq \label{eq:traceRelations}
     \text{tr}_{\mathbf{R}} F^3 = V(\mathbf{R})  \text{tr}_{\mathbf{f}}
     F^3\,,\qquad
	\text{tr}_{\mathbf{R}}
	F^2=U(\mathbf{R})\text{tr}_{\mathbf{f}}F^2\,.
\eeq
Similarly, we note that $\lambda=2c_{G}/V(\mathbf{adj})$
with $c_{G}$ the dual Coxeter number of $G$ and $\mathbf{adj}$ its
adjoint representation. Note that $\lambda=1$ for $G=\text{SU}(N)$,
the case of interest here.

There are some remarks in order. The left hand sides of
\eqref{eq:4Danomalies} are the actual one-loop triangle anomalies
of the field theory. These are determined entirely by the spectrum.
The right hand sided of \eqref{eq:4Danomalies} are the contribution
of the anomalous tree-level GS-counter terms. In an F-theory
compactification
on a Calabi-Yau fourfold $\hat{X}$ over a base $B$
they are identified as follows \cite{Grimm:2012yq,Cvetic:2012xn}
\beq \label{eq:GScoefficients}
	\Theta_{m\alpha}=\Theta_{m\alpha}^M\,,\quad
	b^\alpha_{mn}=-\pi(\sigma(\hat{s}_m)\cdot\sigma(\hat{s}_n))\cdot \Sigma_b^\alpha\,,\quad b^\alpha=S^b_G\cdot \Sigma_b^\alpha\,,\quad a^\alpha=K^\alpha\,.
\eeq
Here the CS-terms $\Theta_{AB}^M$ are defined in
\eqref{eq:3DCSterms}, the N\'eron-Tate height pairing has been introduced
in \eqref{eq:height_pairing} for two sections and can
be straightforwardly generalized to an arbitrary number of sections,
$\Sigma_b^\alpha$ denotes a basis of curves
defined in \eqref{eq:pi}, $S_G^b$ is the divisor in the base $B$
supporting the gauge symmetry $G$, cf.~\ref{eq:S_G}, and $K^\alpha$
is the coefficient \eqref{eq:KBexpansion} in the expansion of $K_B$.

\subsubsection{Anomaly Cancellation in 4D F-Theory with a U$(1)^2$-Sector: $B=\mathbb{P}^3$}

The spectrum of the F-theory compactification to four dimensions on
the fourfold $\hat{X}\rightarrow \mathbb{P}^3$ has been calculated in
\eqref{eq:XP3_chi(R)1} and \eqref{eq:XP3_chi(R)2}. The various
anomalies for this spectrum read
\bea \label{eq:XP3_anomalies}
	A&\hspace{-0.87cm}\phantom{.}^{\text{U}(1)}_{111}\!:\,2 [a_3 (n_9-n_7-12)+ a_5 n_7 \left(4-n_7+n_9\right)]\,,\quad A^{\text{U}(1)}_{222}\!:\,n_7 \left(a_3+a_5 \left(4-n_9\right)\right) \left(4+n_9\right)\,,\nn\\
	A&\hspace{-0.45cm}\phantom{.}^{\text{U}(1)}_{112}\!\!:\tfrac{1}{6} \left[a_5 n_7 \left(48+n_7^2+n_9^2-2 n_7 \left(4+n_9\right)\right)+a_3 \left(n_7^2-2 n_7 \left(n_9-8\right)+\left(n_9-12\right) \left(4+n_9\right)\right)\right]\,,\nn\\
	A&\hspace{-1.15cm}\phantom{.}^{\text{U}(1)}_{122}\!:\tfrac{1}{6} \left(a_5 n_7 \left(n_7-n_9-4\right) \left(n_9-12\right)+a_3 \left(-2 n_7^2+n_7 \left(4+n_9\right)+\left(n_9-12\right) \left(4+n_9\right)\right)\right)\,,\nn\\
	A&\hspace{-1.4cm}\phantom{.}^{\text{U}(1)\text{-grav}}_{1}\!\!: a_3 \left(n_9\!-\!n_7\!-12\right)+a_5 n_7 \left(4\!-\!n_7+n_9\right)\,,\quad A^{\text{U}(1)\text{-grav}}_{2}\!\!:2 n_7 \left(a_3+a_5 \left(4-n_9\right)\right)\,,		
\eea
where we brought all numerical factors in \eqref{eq:4Danomalies} to the
left hand sides. We denoted by $A^{\text{U}(1)}_{klm}$
the $\text{U}(1)_k\times\text{U}(1)_l\times\text{U}(1)_m$- and
by $A^{\text{U}(1)-\text{grav}}_{k}$ the U$(1)$-gravitational
anomalies. Clearly, there are no non-Abelian anomalies due to the absence
of a non-Abelian group $G$, however, see section
\ref{sec:FFEllipticFibSU5U12} for the inclusion of
$G=\text{SU}(5)$.
Thus, since the anomalies \eqref{eq:XP3_anomalies}
are all non-vanishing, a non-trivial GS-mechanism is required for
consistency of the theory.

Therefore, all
that is left to check cancellation to prove anomaly  cancellation
of these F-theory compactifications  is to calculate the quantities
on the right of \eqref{eq:GScoefficients} that encode the GS-mechanism.
First, we obtain
\bea \label{eq:XP3_GScoeffs}
	\Theta_{\alpha m}&=&\left(\tfrac{1}{4} [(-12-n_7+n_9) a_3+n_7 (4-n_7+n_9) a_5],\tfrac{1}{2} n_7 [a_3+(4-n_9) a_5]\right)_m\,,\nn\\
	b_{mn}^\alpha&=&\begin{pmatrix}
		8 & 4-n_7+n_9\\4-n_7+n_9 & 8+2 n_9
	\end{pmatrix}\,,\qquad\qquad  a^\alpha=-4\,.
\eea
where we evaluated the CS-terms \eqref{eq:3DCSterms} for the flux
\eqref{eq:G4solP3}, computed the height pairing \eqref{eq:height_pairing}
and \eqref{eq:KBexpansion} for
$K_{\mathbb{P}^3}=\mathcal{O}_{\mathbb{P}^3}(-4)$.
We note that the index $\alpha=1$ since the
only vertical divisor is the hyperplane $H_B$ and $m=1,2$ for
the two rational sections $\hat{s}_Q$, $\hat{s}_R$. Equipped with
the coefficients in \eqref{eq:XP3_GScoeffs} we finally calculate
the GS-terms on the right side of \eqref{eq:4Danomalies}, which
precisely yield \eqref{eq:XP3_anomalies}. Thus, we see that all
anomalies are cancelled by the GS-mechanism.

We conclude note that the spectrum calculated in section
\ref{sec:4DChiralityU1U1}
is the uniquely determined anomaly-free spectrum if only one chirality
$\chi(\mathbf{1}_{(q_1,q_2)})$ is calculated independently. Anomaly
cancellation is not sufficient to fix the spectrum completely, since the
$U(1)_1^3$-anomaly is proportional to the $U(1)_1$-gravitational anomaly
as is evident from \eqref{eq:XP3_anomalies}.

\subsection{A Toric Example}
\label{sec:ToricExsU1U1}

We conclude the analysis of F-theory compactifications
with U(1)$\times$U(1) gauge group with explicit toric constructions
of the fourfolds $\hat{X}$ with $dP_2$-elliptic curve
$\mathcal{E}$. We focus on elliptically fibered Calabi-Yau fourfolds
$\hat{X}\rightarrow \mathbb{P}^3$ over $\mathbb{P}^3$.

First we note that we can construct using the algorithm of appendix
\ref{app:tuning-s7-s9} toric fourfolds realizing a wide variety of values
for $n_7$ and $n_9$ inside the allowed region in figure
\ref{fig:XP3n7n9region} of $dP_2$-fibrations over $\mathbb{P}^3$ with
generic Calabi-Yau hypersurface $\hat{X}$.
In the following, we present the case $n_7=4$ and $n_9=5$ to illustrate
key ideas of the toric construction.

The toric reflexive polytope
defining the toric variety $dP_2(4,5)$ takes the form
\beq \label{eq:ExampleU1U1_45}
\text{
 \begin{tabular}{|c||ccccc||c|c|c|c||c|} \hline
  variable  & \multicolumn{5}{c||}{vertices}	& \multicolumn{4}{c||}{$\mathbf{C}^*$-action}&divisor class   \rule{0pt}{13pt}\\  \hline
  $z_0$ & 1  & 1  & 1 & -1  &  -1 	&1 &0 &0 &0& $H_B$ 		 	\\
  $z_1$ & -1 & 0 & 0  & 0  &  0	&1 &0 &0 &0& $H_B$ 		 	\\
  $z_3$ & 0& -1 & 0  &  0  &  0	&1 &0 &0 &0& $H_B$ 		 	\\
  $z_2$ & 0 & 0  & -1 & 0  &  0	&1 &0 &0 &0& $H_B$ 			\\
\hline
  $u$  & 0&0  & 0  & 1  &  0 	&0 &1 &1 &-1& $H-E_1-E_2+H_B$ 		\\
  $v$ & 0&0  & 0  & 0  &  1   	&0 &1 &0 &0& $H-E_2+H_B$ 			\\
  $w$ & 0&0  & 0  & -1 & -1		&-1 &0 &1 &0& $H-E_1$ 		 	\\
  $e_1$ & 0&0  & 0  & 0  & -1 	&0 &0 &-1 &1& $E_1$			\\
  $e_2$ & 0&0  & 0  & 1  & 1 	&0 &-1 &0 &1& $E_2$ 			\\ \hline
 \end{tabular}
 }
\eeq
Here the first column denoted the projective coordinates on the
toric variety $dP_2(4,5)$, the next five columns are the vertices of
the polytope, that are given as five-dimensional row vectors. Next
we displayed the four $\mathbb{C}^*$-actions of $dP_2(4,5)$ along with
the divisor classes in the last column.

Comparing the divisor classes in \eqref{eq:ExampleU1U1_45} with the
general assignments in \eqref{eq:sectionsFibration} we immediately
confirm that the above polytope indeed describes a fibration with
divisors $\cS_7=4 H_B$ and $\cS_9=5 H_B$. This can also be checked
explicitly by torically calculating  the intersections \eqref{eq:S7S9} in
the toric variety associated to the single star triangulation of
the polytope specified in \eqref{eq:ExampleU1U1_45}.
The Euler number, as well as the Hodge numbers, of the fourfold $\hat{X}$
in $dP_2(4,5)$ read
\beq \label{eq:ex1X_P3U1U1}
	\chi(\hat{X})=1620\,,\qquad h^{(1,1)}(\hat{X})=4\,, (0)\,,\qquad  h^{(2,2)}_V(\hat{X})=5\,,\qquad h^{(3,1)}(\hat{X})=258\,,(0)\,,
\eeq	
where the zeros in parenthesis indicate the absence of any non-toric
divisors, respectively, complex structure moduli on $\hat{X}$ as expected
by construction. The Euler number $\chi(\hat{X})$ as well as the number
$h^{(2,2)}_V(\hat{X})$ of independent  surfaces in \eqref{eq:ex1X_P3U1U1}
agree with the general formula in \eqref{eq:XP3_Eul}, respectively,
the cohomology calculations leading to \eqref{eq:XP3_H22V}.
The topological metric $\eta^{(2)}$ of these five classes is read off
from  \eqref{eq:XP3_eta2_1} with $n_7=4$, $n_9=5$.

The codimension two singularities of this concrete Calabi-Yau fourfold
$\hat{X}$ yield the full spectrum \eqref{eq:U1spectrum} of six different
singlet representations $\mathbf{1}_{(q_1,q_2)}$. The cohomology classes
of the three sections are given by the toric divisors $S_P=E_2$,
$S_Q=E_1$ and $S_R=H-E_1-E_2+H_B$
in \eqref{eq:ExampleU1U1_45} and agree with the general expression
\eqref{eq:SPSQSRP3}. We obtain the $(1,1)$-forms inducing the
U$(1)\times$U(1)-gauge fields by evaluating the
Shioda map of the sections $\hat{s}_Q$, $\hat{s}_R$ following
\eqref{eq:XP3_Shioda} as
\beq
	\sigma(\hat{s}_Q)=S_Q-S_P-4H_B\,,\qquad \sigma(\hat{s}_Q)=S_R-S_P-9H_B\,.
\eeq
The Nero-Tate height pairing $b_{mn}^{\alpha}$ of these two classes
follows from \eqref{eq:XP3_GScoeffs} with $n_7=4$, $n_9=5$.

Finally, we calculate the general $G_4$-flux on $\hat{X}$ employing the
procedure of section \ref{sec:G4U1xU1}. We precisely obtain the
$G_4$-flux in \eqref{eq:G4solP3} in the case $n_7=4$, $n_9=5$.
With this result, we readily calculate the chiralities
for the matter representations, along with the
CS-terms \eqref{eq:3DCSterms},
again obtaining a perfect match with the results
\eqref{eq:XP3_chi(R)1}, \eqref{eq:XP3_chi(R)2} and \eqref{eq:XP3_CSM}
evaluated for $n_7=4$, $n_9=5$. Then we also calculate the GS-terms
\eqref{eq:XP3_GScoeffs}, most prominently the gaugings
$\Theta_{m\alpha}$, and show, following section
\ref{sec:anomalyCancellation}, that all anomalies are cancelled.

\section{Calabi-Yau Fourfolds with SU$(5)\times$U$(1)\times\text{U}(1)$}
\label{sec:FFEllipticFibSU5U12}

Building on  the results of previous sections, the aim of this section
is to develop tools for  the construction of phenomenologically appealing
F-theory compactifications with additional non-Abelian gauge symmetries.
For concreteness we consider in this section a Calabi-Yau fourfold $\Xsu$
with a resolved $SU(5)$-singularity,  while maintaining the rank two 
Mordell-Weil group. Compactifications of F-theory on $\Xsu$ give rise to a
four-dimensional GUT with $\text{SU}(5)\times\text{U}(1)\times
\text{U}(1)$ gauge group.

We consider here one particular geometric way of adding and resolving
the SU$(5)$-singularity. Once this non-Abelian sector is added, it is
rather straightforward to implement  the techniques developed in previous
sections for the rank-two Abelian symmetry.  As the first step
in section \ref{sec:SingularitiesSU5U1U1} we describe the resolution to
$\Xsu$, determine the matter representations as well as Yukawa points and
the matter surfaces  for a subset of matter multiplets. We note, that the
spectrum that we obtain in this analysis has not been found via the
toric classification of SU(5)-TOPs in \cite{Borchmann:2013jwa}.
As the next step
we calculate the vertical cohomology ring $H^{(*,*)}_V(\Xsu)$
in section \ref{sec:CohomologyFFSU5}. For a general base $B$ we derive
expressions for the Chern classes and Euler number of $\Xsu$, before
we calculate the full vertical cohomology for the fourfold with
$B=\mathbb{P}^3$ explicitly.

We emphasize one finding of our analysis. With the addition of an
SU(5)-singularity we encounter at one codimension three locus a non-flat
fiber (NFF),  i.e.~at this particular  codimension three locus the
fiber of $\Xsu$ is no longer a complex one-dimensional elliptic curve,
but a complex two-dimensional surface $\cC_{NFF}$.
We note that this was also
observed  in \cite{Borchmann:2013jwa}. The
physics of non-flat fibers has been studied in \cite{Candelas:2000nc}, to which we refer 
for more details. In summary, the complex surface in the fibration can be wrapped by an
M5-brane, that gives rise, in the blow-down of the resolved fibration of
$\Xsu$, to a charged tensionless string with an infinite tower of massless
excitations. In addition, further light states are contributed by
M2-branes wrapping holomorphic curves in the surface of the 
non-flat fiber.

Compactifications of F-theory with such a spectrum of  additional light
states are problematic for phenomenology. One obvious way out is to geometrically 
avoid the presence of the Yukawa couplings supporting the dimension two fiber. This 
can be achieved by considering a base $B$ where the dangerous Yukawa coupling does not
exist but all the other physical features of the compactification on
$\Xsu$ are retained. See also \cite{Braun:2013yti} for a toric
analysis. Such bases are easily constructed and realize
phenomenologically interesting F-theory compactifications
\cite{workinprogress1}.

Another way to deal with the non-flat fiber is to forbid at least a
chiral excess of the additional light states induced by $\cC_{NFF}$.
This requires the $G_4$-flux on $\Xsu$ to integrate to zero  over
$\cC_{NFF}$,
\beq \label{eq:NFFcond}
 \boxed{ \text{\textbf{Non-flat fiber
condition:}}\qquad\int_{\cC_{NFF}} G_4 = 0\,,\rule{0pt}{18pt}}
\eeq
Following this approach later in section
\ref{sec:G4FluxChiralitiesSU5}, we show explicitly that we can
obtain a physical and consistent low-energy spectrum with all 4D
field theory anomalies cancelled even in the presence of a non-flat
fiber at codimension three. It would be interesting to investigate the
microscopic meaning of \eqref{eq:NFFcond} in more detail.

\subsection{Singularities of the Fibration: Gauge Group, Matter \& Yukawa
Couplings}
\label{sec:SingularitiesSU5U1U1}

In this section we analyze the codimension one, two and three
singularities and their resolutions in $\Xsu$. We thoroughly determine
all matter representations and determine some of the associated matter
surfaces.

Before going into the details of this geometric analysis of $\Xsu$
it is instructive to introduce some terminology and the following
simple geometric picture.
We construct $\Xsu$ by first considering a non-generic Calabi-Yau
fourfold $\hat{X}$ with an SU(5)-singularity over codimension one in $B$.
Then we resolve all singularities and obtain the smooth  $\Xsu$.
Thus, the smooth Calabi-Yau  fourfold $\Xsu$ is
schematically obtained from a singular Weierstrass model $X$ as,
\beq \label{eq:buXSU5}
	\hat{X}_{\text{SU}(5)}\,\,\,
	\stackrel{\pi_{\text{SU}(5)}}{\longrightarrow}\,\,\,	\hat{X}\,\,\,
	\stackrel{\hat{\pi}}{\longrightarrow}\,\,\, X\,.
\eeq
Here $\pi_{\text{SU}(5)}$ blows down the four Cartan divisors of SU(5)
and $\hat{\pi}=\pi_1\circ \pi_2$ is the two-step blow-up
in \eqref{eq:TwoResolutions}. Thus the fourfold $\Xsu$ is constructed via
a combination of a total of six resolutions starting from the singular
Weierstrass model $X$.

\subsubsection{Explicit Resolution of a Codimension One
SU(5)-Singularity}

In the following we first engineer and then resolve an SU(5)-singularity
over codimension one in $B$.
We introduce the section $z \in \cO(\Ssu)$ that vanishes along the
divisor $\Ssu^b$ in the base $B$ of  SU$(5)$-singularities of the elliptic
fibration of $\hat{X}$.
Here we focus  on the interplay between the
non-Abelian and Abelian sector and choose one particular
SU(5)-singularity first constructed in \cite{Cvetic:2013nia}
that has  been missed in the literature before.

\begin{figure}[ht!]
\centering
\includegraphics[scale=0.5]{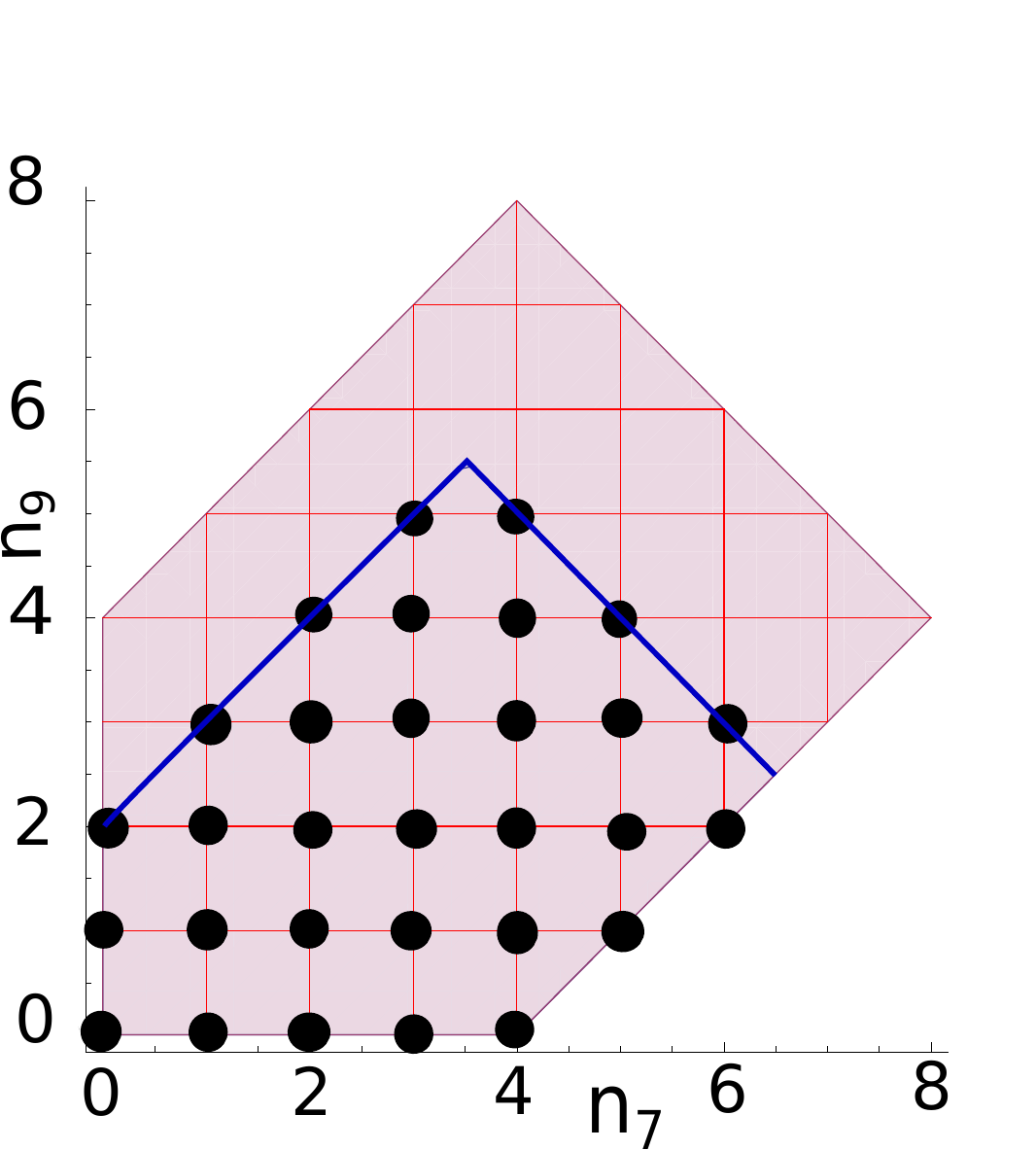} \qquad \qquad
 \caption{Each dot represents a $dP_2$-fibration over
 $\mathbb{P}^3$ with
 generic Calabi-Yau $\Xsu$. The red region is the set of $dP_2$-fibration with generic $\hat{X}$ without SU(5), cf.~figure \ref{fig:XP3n7n9region}.}
 \label{fig:XP3SU5n7n9region}
\end{figure}

The SU(5)-singularity of interest is obtained by considering
non-generic coefficients $s_i$ in the Calabi-Yau
equation \eqref{eq:CYindP2} of the form
\bea \label{eq:s'SU5}
s_1 = z^3 s'_1\,, \qquad  s_2 = z^2 s'_2\,, \qquad s_3 = z^2 s'_3\,, \qquad s_5 = z s'_5\,.
\eea
Since the $s_i$ are still required to be sections of the bundles
\eqref{eq:sectionsFibration} by the Calabi-Yau condition, this implies
that the $s_i'$  have to be sections of the following line bundles:
\beq \label{eq:sectionsSU50}
\text{
\begin{tabular}{c|c}
\text{section} & \text{bundle}\\
\hline
$s'_1$&$\mathcal{O}(3[K_B^{-1}]-\cS_7-\cS_9-3\Ssu)$\rule{0pt}{13pt} \\
	$s'_2$&$\mathcal{O}(2[K_B^{-1}]-\cS_9-2\Ssu)$\rule{0pt}{12pt} \\
	$s'_3$&$\mathcal{O}([K_B^{-1}]+\cS_7-\cS_9-2\Ssu)$\rule{0pt}{12pt} \\
	$s'_5$&$\mathcal{O}(2[K_B^{-1}]-\cS_7-\Ssu)$\rule{0pt}{12pt} \\
\end{tabular}
}
\eeq
We note that we assume in the following that we obtain a fourfold $\Xsu$
that is a generic besides the resolved SU(5)-singularity,
i.e.~that all the $s_i$ respectively $s_i'$  in
\eqref{eq:sectionsFibration} and \eqref{eq:sectionsSU5} exist.
As before in the U$(1)^2$-case, this poses upper and lower bounds on the
coefficients $n_7^\alpha$, $n_9^\alpha$ in the expansion
\eqref{eq:S7S9exp} of $\cS_7$, $\cS_9$, that depend on the base $B$.
For the case $B=\mathbb{P}^3$ that is of most interest in this work,
these bounds replace \eqref{eq:XP3condn7n9} and read, using
$K^{-1}_{\mathbb{P}^3}=\cO_{\mathbb{P}^3}(4)$ and $\Ssu=H_B$,
\beq \label{eq:XP3SU4condn7n9}
		0\leq n_7\leq 7\,,\quad 0\leq n_9\leq 6\,,\quad n_7+n_9\leq 9\,,\quad 0\leq 2+n_7-n_9\,,\quad 0\leq 4+n_9-n_7\,.
\eeq
The allowed region is displayed in figure \ref{fig:XP3SU5n7n9region}.

Next we confirm that the specialization \ref{eq:s'SU5} of the
coefficients indeed gives rise to an SU(5)-singularity.  The discriminant
takes the form
\be \label{eq:DeltaSU5}
 \Delta = -z^5 \left( \beta_5^4 P + z \beta^2_5 P_2 R + \mathcal{O}(z^2) \right)
\ee
where we defined
\be \label{eq:betaP}
	\beta_5=s_6\,, \quad P := P_1 P_2 P_3 P_4 P_5 =  (s'_2 s'_5 - s'_1 s_6) s_7 (s'_2 s_7-s'_3 s_6) s_8 (s_6 s_9-s_7 s_8 )\,,
\ee
and $R$ is a polynomial in $s_i$ with no common factors.

We note that the factor $z^5$ in \eqref{eq:DeltaSU5} signals an
$A_4$-singularity over the divisor $z=0$,  as desired. In
addition, we see that the singularity gets further enhanced at
$\beta_5=0$ to $D_5$ and at $P=0$ to $A_5$, indicating appearance of matter multiplets at codimension two singularities. Yukawa points are located at the codimension three loci.

In order to resolve all singularities, we perform four
blow-ups of the ambient space \eqref{eq:dP2fibration} that induce
two blow-ups and two small resolutions on the Calabi-Yau fourfold.
The blow-down map $\pi_{\text{SU(5)}}:\,\Xsu\rightarrow \hat{X}$ of this
resolution then reads \cite{Cvetic:2013nia},
\beq \label{eq:resMapSU5}
\pi_{\text{SU(5)}}\,:\quad w= d_1 d_2^2 d_3^3 d_4^2 \tilde{w}\,, \quad v=
d_1 d_2 d_3 d_4  \tilde{v}\,, \quad u =d_3 d_4  \tilde{u}\,, \quad z = d_1 d_2 d_3 d_4  \tilde{z}\,,
\eeq
where we have introduced new coordinates $ \tilde{w}$, $ \tilde{v}$,
$\tilde{u}$ and $ \tilde{z}$ as
well as the $d_i$ which are sections of the
line bundles $\mathcal{O}(D_i)$  associated with the exceptional
divisors $D_i$ for $i=1\dots4$. The latter are
the Cartan divisors of $\Xsu$ and admit the fibration structure
\eqref{eq:S_G} with $\Ssu=\pi(\cS_G^b)$.
In the following it will be convenient
to denote the new class of $ \tilde{z}$ as
$D_0$ and supplement the Cartan divisors as $D_I=(D_0,D_i)_I$ with
$I=0,1,\ldots, 4$. The divisor classes of all homogeneous coordinates
in \eqref{eq:resMapSU5} are then given  by
\beq \label{eq:sectionsSU5}
\text{
\begin{tabular}{c|c}
\text{section} & \text{bundle}\\
\hline
	$ \tilde{u}$&$\mathcal{O}(H-E_1-E_2+\cS_9+[K_B]-D_3-D_4)$\rule{0pt}{13pt} \\
	$ \tilde{v}$&$\mathcal{O}(H-E_2+\cS_9-\cS_7-D_1-D_2-D_3-D_4)$\rule{0pt}{12pt} \\
	$ \tilde{w}$&$\mathcal{O}(H-E_1-D_1-2D_2-3D_3-2D_4)$\rule{0pt}{12pt} \\
	$ \tilde{z}$&$\mathcal{O}(D_0)\equiv\mathcal{O}(\cS_{SU(5)}-D_1-D_2-D_3-D_4)$\rule{0pt}{12pt} \\
\end{tabular}
}
\eeq
The total transform of the Calabi-Yau hypersurface \eqref{eq:CYindP2}
then reads
\bea \label{eq:CYSU5}
 p &=& s_6 (e_1 e_2)  u v w + s_7 (d_1 d_2) e_2 v^2 w + s_8 (d_2 d_3^2 d_4) e_1^2  u w^2 +
 s_9 (d_1 d_4 d_3^2 d_2^2) e_1 v w^2  \\ &+& z_2 s_5 (d_4 d_3) e_1^2 e_2 u^2 w +
  z_2^2 s_2 (d_1 d_4) e_1 e_2^2 u^2 v  + z_2^2  s_3 (d_1^2 d_2 d_4) e_2^2 u v^2
 + z_2^3 s_1 (d_1 d_4^2 d_3) e_1^2 e_2^2  u^3\, ,\nn
\eea
where we dropped, by abuse of notation, all tildes of the coordinates in
\eqref{eq:resMapSU5} and all primes of  the sections in
\eqref{eq:sectionsSU5}.
The hypersurface \eqref{eq:CYSU5} is a section of the anti-canonical
bundle of the new ambient space, denoted $\widehat{dP}_2^B(\cS_7,\cS_9)$,
after the blow-up \eqref{eq:resMapSU5}. It reads
\beq \label{eq:KdP2SU5}
K_{\widehat{dP}_2^B}^{-1} = \pi_{\text{SU}(5)}^*\big(K^{-1}_{dP_2^B}\big)
- 2D_1-3D_2-5 D_3-4 D_4\,,
\eeq
where $K^{-1}_{dP_2^B}$ denotes the anti-canonical bundle
\eqref{eq:antiKdP2} of $dP_2^B(\cS_7,\cS_9)$ that is pulled back via
the blow-down map \eqref{eq:resMapSU5}.

There is one caveat when working with a particular resolution
such as \eqref{eq:resMapSU5}. It has been discussed in detail in
\cite{Esole:2011sm,Marsano:2011hv,Lawrie:2012gg},
but has also been long clear in the toric context, see
e.g.~\cite{Krause:2011xj,Grimm:2011fx,Cvetic:2012xn,Hayashi:2013lra} for
recent studies, that the resolution of an SU(5)-singularity over
codimension one in $B$ is not unique. The
different resolutions are related by mild birational transformations.
In the context of toric geometry, this is visible as
different phases of the toric ambient variety\footnote{In general, there
can also be phases of the toric variety and $\Xsu$ that do not correspond
to different resolutions but for example to different phases of $B$ or of
the general fiber $\mathcal{E}$. These might even change the
intersections  e.g.~of the fiber $\mathcal{E}$ from the ones considered
in section \ref{sec:EllCurveMWrk2} by going for instance into a phase
corresponding to $F_1$ blown up at a point. We exclude these phases here
and in particular in sections \ref{sec:ToricExsU1U1} and
\ref{sec:ExamplesSU5U1U1}.},
that descend to different
phases on the Calabi-Yau fourfold $\Xsu$. One way to specify these
different phases is by the Stanley-Reissner ideal $SR$,
which describes which divisors in \eqref{eq:sectionsSU5} do not
intersect.  However, we can always restrict ourselves to one particular
resolution
in order to extract physical quantities like the matter content,
Yukawa points and chiralities in F-theory. This is intuitively
clear since the resolved geometry $\Xsu$ is just a tool to analyze the
geometry and not physical in F-theory. This can be argued
more precisely in the dual M-theory compactification similar to the
discussion in \cite{Witten:1996qb}.

Thus, we focus here on one particular phase of the resolution
\eqref{eq:resMapSU5} to extract the effective physics of F-theory on
$\Xsu$.
Following the spirit of toric geometry, we specify
the precise resolution \eqref{eq:resMapSU5} by stating a specific
SR-ideal. Here we consider an ideal of the form
\bea \label{eq:SR-SU5}
SR &=& \{ v e_1, e_1 e_2, u v, u w, w e_2 \}\cup \{ z d_3, d_1 d_3, d_1 d_4\} \cup \{ d_1 e_1, z e_1, d_2 e_1, d_4 e_1, d_1 u, \nn \\ &&     d_2 u,
d_1 e_2, d_2 e_2, d_3 e_2, d_4 e_2,  z w, d_1 w, d_4 w, d_3 v, d_4 v,  z d_2 v\}\cup \pi_{\text{SU}(5)}^*(SR_B)\,.
\eea
We immediately recover the SR-ideal \eqref{eq:SRidealdP2} of the
$dP_2$-fiber as the first set in \eqref{eq:SR-SU5}, ensuring the
intersections \eqref{eq:dP2ints} of the fiber. The second set gives rise
to the correct intersections of the Cartan divisors to obtain the
intersection pattern of the $A_4$ Dynkin diagram. The third part encodes
the intersections of the Cartan divisors with the  sections
$\hat{s}_P$, $\hat{s}_m$. The last part in \eqref{eq:SR-SU5} is the
pull-back of the SR-ideal $SR_B$ of the base $B$ under the
blow-down map $\pi_{\text{SU}(5)}$  in \eqref{eq:resMapSU5}.
We will comment on the structure of this ideal in more detail in
section \ref{sec:CohomologyFFSU5}. For now we just state that it
contains the SR-ideal of the base and additional intersections involving
the $D_i$ and the vertical divisors, that take care of the split  of the
divisor $z$ into multiple components under the blow-up map
\eqref{eq:resMapSU5}.

\begin{figure}[ht!]
\centering
  \includegraphics[scale=0.6]{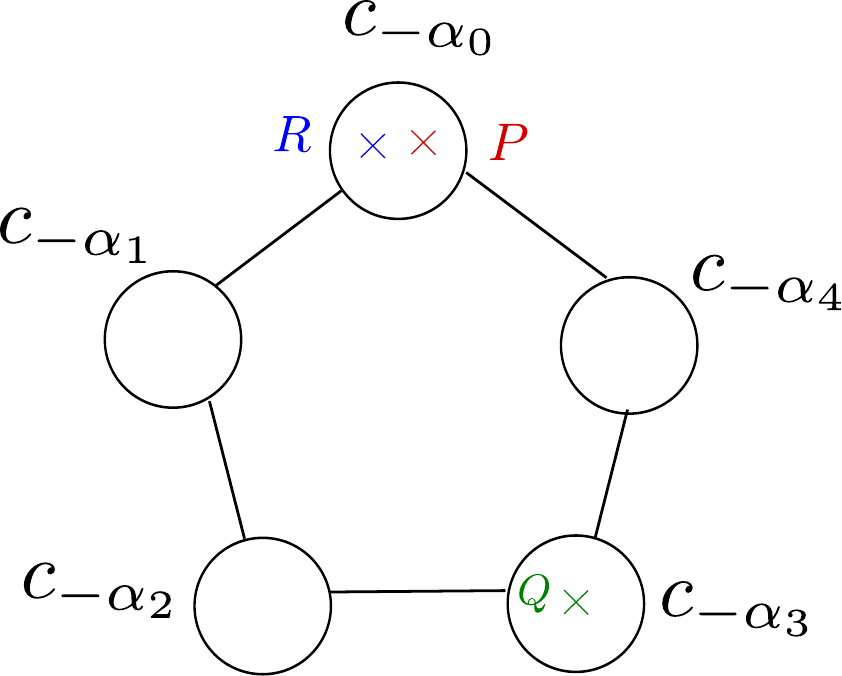}
  \caption{$I_4$-fiber over $\Ssu^b$ marked according to the
  intersections of the  nodes with  $S_P$, $S_Q$, $S_R$, corresponding to
  a $2-3$-split of a  spectral cover.}
  \label{fig:I4-Singularity}
\end{figure}

Let us next discuss the structure of the resolution $\Xsu$ at codimension
one in $B$.  The general fiber of $\Xsu$ is the $dP_2$-elliptic curve
$\mathcal{E}$. However, over $\Ssu^b$ the fiber becomes reducible.
One way to see this is by noting the following intersections of the
Cartan divisors $D_I$,
\beq \label{eq:intsCartansSU5}
  D_I \cdotp D_J \cdotp D_\alpha \cdotp D_\beta  = -C_{IJ}  \cS_{SU(5)} \cdotp S_P \cdotp D_\alpha, \cdotp D_\beta
\eeq
where  $D_\alpha$, $D_\beta$ are arbitrary vertical divisors and $C_{IJ}$
is the extended Cartan matrix of $SU(5)$ reading
\beq \label{eq:SU5Cartan}
(-C_{IJ}) = \left( \begin{array}{ccccc}
                 -2 &  1 &  0 &  0 &  1 \\
                 1  & -2 &  1 &  0 &  0 \\
                 0  & 1  & -2 &  1 &  0 \\
                 0  &  0 &  1 & -2 &  1 \\
                 1  &  0 &  0 &  1 & -2
                \end{array}
                \right).
\eeq
We note that \eqref{eq:intsCartansSU5} can be calculated via the abstract
presentation \eqref{eq:abstractCohomRing} of the cohomology ring of
$\Xsu$, see section \ref{sec:CohomologyFFSU5} and appendix
\ref{app:fluxP3SU5} for details.
The intersections \eqref{eq:intsCartansSU5} then immediately imply that
the divisors $D_I$ resolve a singularity of type $G=\text{SU}(5)$.
Indeed, we note  first that \eqref{eq:intsCartansSU5} implies
a fibration structure \eqref{eq:S_G} of the Cartan divisors
$\pi:\,D_I\rightarrow \Ssu^b$ with the rational fibers $c_{-\alpha_I}$
corresponding to the simple roots $-\alpha_i$ and the extended root
$-\alpha_0=\sum_i\alpha_i$ of SU(5). The latter conclusion can be shown
by representing the  individual curves
$c_{-\alpha_I}$ as intersections of the divisors $D_i$
with a curve $\Sigma_b$ dual to $\Ssu^b$ in the base, i.e.~assuming
$\Ssu^b \cdotp \Sigma_b = 1$ we obtain
\beq \label{eq:c_alphaI}
D_I \cdot \Sigma_b = c_{-\alpha_I}\,.
\eeq
Then, we immediately confirm from \eqref{eq:intsCartansSU5} and
\eqref{eq:SU5Cartan} that the $c_{-\alpha_I}$ in fact intersect as the
affine Dynkin diagram of SU(5), see figure \ref{fig:I4-Singularity},
justifying the assignment $c_{-\alpha_I}\leftrightarrow -\alpha_I$.

Next we turn to the behavior of the rational sections of $\Xsu$.
Generically, the sections are points in the fiber and only
intersect one of its irreducible components. As we have just seen
the fiber over $\Ssu^b$ splits into five rational curves.
The component which is pierced by the three sections $\hat{s}_P$,
$\hat{s}_Q$ and $\hat{s}_R$ can be determined by calculating the
intersections of their divisor classes $S_P$, $S_{Q,R}$ with the curves
\eqref{eq:c_alphaI}. We find that $S_P$ and $S_R$ intersect
$c_{-\alpha_0}$, and $S_Q$ passes through $c_{-\alpha_3}$, i.e.
\beq \label{eq:-alpha_IS_m}
	S_P\cdot c_{-\alpha_I} =S_R\cdot c_{-\alpha_I} =(1,0,0,0,0)_I\,,\quad  S_{Q} \cdot
	c_{-\alpha_I}=(0,0,0,1,0,0)_I\,.
\eeq
This situation is summarized in figure  \ref{fig:I4-Singularity}. We
note that this intersection pattern of the rational sections  would
correspond to a $2-3$ split in the language of spectral covers. Again,
the result \eqref{eq:-alpha_IS_m} can be obtained via the abstract
intersection computation outlined in section \ref{sec:CohomologyFFSU5}
and in appendix \ref{app:fluxP3SU5}.

Equipped with these intersections of the sections with the nodes
of the Dynkin diagram, we can calculate the Shioda maps of the
sections $\hat{s}_Q$, $\hat{s}_R$. We obtain
\beq \label{eq:ShiodaMapSQSRSU5}
\sigma(\hat s_Q)=  S_Q - S_P - [K_B^{-1}]+\frac{1}{5}(2 D_1+4 D_2+ 6 D_3+3 D_4)\,,\qquad
\sigma(\hat s_R) = \hat S_R - S_P - [K_B^{-1}] -  \cS_9\,.
\eeq
Here we have used the general formula for the Shioda map in the presence
of one set of Cartan divisors $D_i$ due to a codimension one singularity
of type $G$ of the elliptic fibration,
\beq \label{eq:ShiodaMapFull}	
	\sigma(\hat{s}_m) := S_m-\tilde{S}_P-\pi(S_m\cdot\tilde{S}_P)+\left(\cC^{-1}\right)^{ij}(S_m\cdot c_{-\alpha_i})D_j\,.
\eeq
The first two terms are identical to those in the Shioda
map \eqref{eq:ShiodaMapU1only}, whereas the last term incorporates
the presence of $G$ and involves the
fibral curves $c_{-\alpha_i}$ corresponding
to the simple roots of $G$ as well as the inverse of the normalized
coroot matrix $\cC_{ij}$ of $G$. For the case of SU(5) considered here
$\cC_{ij}$ is identical to the Cartan matrix $C_{ij}$.

\subsubsection{Matter: Codimension Two}

Now we proceed to analyze the resolved codimension two singularities in
$\Xsu$. There are five loci $z=P_i=0$, $i=1,\ldots,5$, in
\eqref{eq:betaP} where the order of vanishing of the discriminant
enhances, indicating the presence of an $I_5$-singularity and of
$\mathbf{5}$ representations in F-theory. In addition, there is one
codimension two loci $z=\beta_5=0$, where the singularity enhances to
type $D_5$ and where matter in the $\mathbf{10}$ representation is
located F-theory.

We analyze in the following the splitting of the
nodes of the $A_4$-fiber over all these codimension two loci, determine
the U$\times$U(1)-charges $(q_1,q_2)$ of all the non-Abelian
representations along with their KK-charges, which we find to be zero.
As in the U$(1)\times$U(1) case there will be only the SU(5)-singlet
$\mathbf{1}_{(-1,-2)}$ with non-trivial KK-charge.  We also determine
which weights of the respective representations are realized as
holomorphic curves and thus lie in the Mori cone of $\Xsu$. This fixes
the sign-function in \eqref{eq:signw} for each of these representation
and allows for the field theoretic computations outlined in section
\ref{sec:G4constraintsF}, that will be used in section
\ref{sec:G4FluxChiralitiesSU5}. We conclude with a determination
of some matter surfaces $\cC_{\mathbf{R}}^{\mathbf{w}}$.

The analysis of the splitting of nodes at codimension two presented in
this section is very similar to the ones performed
earlier in the literature based on the Tate model of elliptic fibrations.
Therefore, our discussion will be brief and we refer to
\cite{Marsano:2011hv,Lawrie:2012gg} for more background on the general
methodology.

\subsubsection*{The $\mathbf{5}$-representations and their
U$(1)\times\text{U}(1)$-charges}

In the following we study the codimension two loci $z=P_1=0$
in detail. The analysis of the other loci $z=P_i=0$ of
$\mathbf{5}$-representations is very similar. Therefore,
we will only summarize the results of our computations at the
end of this section.

\begin{figure}[ht!]
\centering
  \includegraphics[scale=0.6]{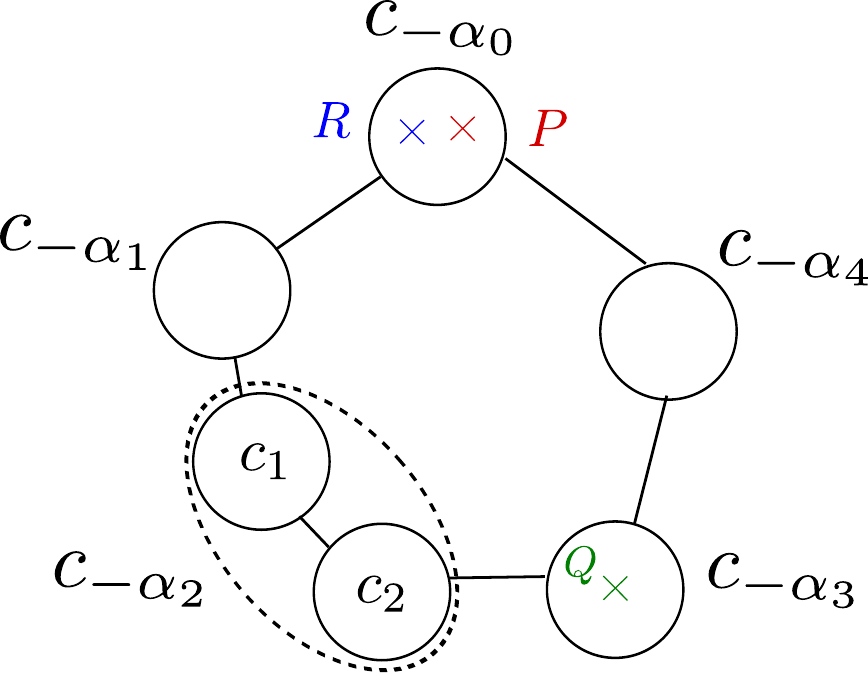}
  \caption{Splitting of the generic $I_4$-fiber over $P_1=0$.}
  \label{fig:I5fiber}
\end{figure}

At the loci $P_1=(s_2s_5-s_1s_6)=0$ defined \eqref{eq:betaP}
the second node in figure of the fiber becomes
reducible, as depicted  figure \ref{fig:I5fiber}.
This can be observed by intersecting the locus
$d_2=P_1=0$ with the Calabi-Yau equation \eqref{eq:CYSU5}. Since we are
only interested in the split of the fiber, we evaluate all coefficients
$s_i$ at generic points on $B$, cf.~\eqref{eq:c_alphaI}. Then, we locally
obtain
\beq \label{eq:splitD2}
 \left. p\right|_{d_2=P_1=0} = \frac{1}{s_1} \left( s_2 v + s_1 d_3 d_4 z \right) \left( s_5 w + s_1d_1 d_4  z^2\right).
\eeq
This means that the class of the curve $c_{-\alpha_2}$
splits into two curves $c_1$, $c_2$ when restricted to the non-generic
loci $P_1=0$ .
From the perspective of the ambient space $\widehat{dP}_2^B(\cS_7,\cS_9)$
we can write this splitting using \eqref{eq:splitD2} as
\beq \label{eq:splitD2class}
 c_{-\alpha_2} = c_1+c_2\,, \qquad [c_1]= H_B^3\cdot[v]\,,\quad
 [c_2]=H_B^3\cdot [w]\,,
\eeq
where we denoted the homology classes of the curve $c_i$ as $[c_i]$
and used the shorthand notation $[v]$ for the homology class
of $v$, cf.~\eqref{eq:sectionsSU5}.
In order to arrive at this result we first have to use the SR-ideal
\ref{eq:SR-SU5} and the fact that $H_B^4=0$.

The splitted curves $c_1$, $c_2$ correspond to a weight $\mathbf{w}$,
respectively, $-\mathbf{w}$ of the $\mathbf{5}$-representation according
to the general correspondence between isolated rational curves over
codimension two in $B$ and matter in F-theory explained before
\eqref{eq:matterSurfs}. The Dynkin  label\footnote{The Dynkin label
of $\mathbf{w}$ is the vector of `U(1)-charges' of $\mathbf{w}$ under
all Cartan generators.} $(q_i)$ of a weight
$\mathbf{w}$ is calculated in general geometrically as the intersection
\beq \label{eq:Dynkinlabel}
	q_i=D_i\cdot c_{\mathbf{w}}\,,
\eeq
which is in complete analogy with the formula \eqref{eq:U1chargeGen} for
the computation of U(1)-charges.
We apply \eqref{eq:Dynkinlabel} for the curves in \eqref{eq:splitD2class}
to obtain the Dynkin labels of $c_1$ and $c_2$ as
\beq \label{eq:DLc1c2}
  D_i\cdot c_1=(1,-1,0,0)_{i}\,, \qquad D_i\cdot c_2=(0,-1,1,0)_{i}\,,
\eeq
respectively. As an immediate consistency check we recover the Dynkin
label of $-\alpha_2$ as the sum of these Dynkin labels.
We also note, for completeness, that in this case, none of the curves
$c_1$, $c_2$ intersect the rational sections, as can be seen
by calculating intersections as in \eqref{eq:-alpha_IS_m}. This
is  again expected since the original unsplitted curve $c_{-\alpha_2}$
does not intersect the rational sections either, see figure
\ref{fig:I5fiber}.

The first set of Dynkin labels in \eqref{eq:DLc1c2} is recognized as
the Dynkin label of the weight $-(\mu_5-\alpha_1)$, where $-\mu_5$ is the
highest weight with Dynkin label $\Lambda_{\mathbf{5}}=(-1,0,0,0)$.
Since the Cartan divisors correspond to $-\alpha_i$, however, we
interpret this as $\mathbf 5$-representation
The second Dynkin label corresponds to
the  weight $(\mu_5-\alpha_1-\alpha_2)$ of the $\mathbf 5$
representation. All the other weights of $\mathbf5$ can then be obtained
by adding appropriately one of the curves $c_{1,2}$ and  a number of
curves $c_{-\alpha_i}$ of the simple roots $c_{-\alpha_i}$. We exemplify
this here to obtain the remaining weights of the $\overline{\mathbf5}$,
\beq \label{eq:signWeights5}
\text{
\begin{tabular}{|c|c|}\hline
Dynkin label  & Curves \\ \hline
(-1,0,0,0) & $c_1+c_{-\alpha_1}$\rule{0pt}{13pt} \\
(1,-1,0,0) & $c_1$ \\
(0,1,-1,0) & $-(c_2)$ \\
(0,0,1,-1) & $-(c_2+c_{-\alpha_3})$ \\
(0,0,0,1) & $-(c_2+c_{-\alpha_3}+c_{-\alpha_4})$ \\ \hline
\end{tabular}
}
\eeq
From this table we can directly see which curves are in the Mori cone
of $\Xsu$. The two curves $c_{1}$, $c_2$ are effective because we
explicitly know their holomorphic representation \eqref{eq:splitD2} .
Also by construction,  the curves associated to the simple roots
$c_{-\alpha_i}$ are effective and so are the sums of effective curves
with positive coefficients. This implies that we can now
evaluate the sign-function \eqref{eq:signw}. It is positive for
the first two representations in table \ref{eq:signWeights5}, but
negative for the other three curves, that are not in the Mori cone.
We summarize these findings, along with the signs of the other
representations in \eqref{eq:signsSignum5}.

Next, we calculate the U$(1)\times$U(1)-charge using the first equality
in the charge formula \eqref{eq:U1chargeGen}. For the U(1) associated
to the $S_Q$ section, we intersect the Shioda map $\sigma(\hat{s}_Q)$
in \eqref{eq:ShiodaMapSQSRSU5} with any of the curves in
\eqref{eq:signWeights5} of the representation. For example using
the locations in figure \ref{fig:I5fiber} of the sections on
the fiber, as well as the intersections \eqref{eq:intsCartansSU5} and
\eqref{eq:DLc1c2} we obtain
\beq
q_1=\sigma(\hat{s}_Q) \cdotp (c_1+c_{-\alpha_1}) = 0 - 0 + (-1,0,0,0) \,\,  \frac{1}{5} \left( \begin{array}{c c c c}
						  4 &  3 &  2 &  1   \\
						  3  &  6 &  4 & 2    \\
						  2  &  4 & 6 &  3  \\
						  1  & 2 &  3 &  4
                                                  \end{array} \right)
 \left( \begin{array}{c}
         0 \\ 0 \\ 1 \\ 0
        \end{array}
\right) = -\frac{2}{5},
\eeq
The second U(1)-charge related to the section  $S_R$ follows similarly as
\beq
  q_2=\sigma(\hat{s}_R) \cdotp (c_1+c_{-\alpha_1}) = 0.
\eeq
In summary, we have found the matter representation
$\mathbf{5}_{(-\frac{2}{5},0)}$ is realized at the codimension two loci
$P_1=z=0$.

A similar analysis can be performed at all other loci $z=P_i=0$,
$i=2,3,4,5$ of $\mathbf{5}$-representations. The computations are
completely analogous to the ones performed above, but not very
enlightening. In all cases we obtain a split of nodes similar
to the one in figure \ref{fig:I5fiber} and obtain an $I_5$-fiber.
The U$(1)\times$U(1)-charges can be straightforwardly calculated. We skip
these details and just summarize all charges of $\mathbf{5}$
representations we obtain:
\beq
\text{
\begin{tabular}{|c||c|c|c|c|c|} \hline
 Locus 		& $P_1=0$ & $P_2=0$ & $P_3=0$ & $P_4=0$ & $P_5=0$ \rule{0pt}{13pt}\\ \hline
 $(q_1, q_e)$ & $(-\tfrac{2}{5},0)$ & $(-\tfrac{2}{5},1)$ &$(\tfrac{3}{5},0)$ & $(\tfrac{3}{5},1)$ & $(-\tfrac{2}{5},-1)$ \rule{0pt}{13pt} \\ \hline
\end{tabular}
}
\eeq
We note that we are working in a normalization of charges which allows
for fraction. We can rescale our charges as usually done in
the literature in this case by a factor of $5$ to obtain integral
charges.

We also summarize which curves $c_{\mathbf{w}}$ of which weights
are in the Mori cone of $\Xsu$. As before we use this to determine
the values of the sign-function in \eqref{eq:signw}.
We obtain the following results:
\beq \label{eq:signsSignum5}
\text{
\begin{tabular}{|c||c|c|c|c|c|} \hline
 Weight		& $\mathbf{5}_{(-\tfrac{2}{5},0)}$  &
 $\mathbf{5}_{(-\tfrac{2}{5},1)}$ & $\mathbf{5}_{(\tfrac{3}{5},0)}$ &
 $\mathbf{5}_{(\tfrac{3}{5},1)}$ & $\mathbf{5}_{(-\tfrac{2}{5},-1)}$
 \rule{0pt}{13pt}\\ \hline
 $(-1,0,0,0)$ & + & + & + & + & - \rule{0pt}{13pt}\\ \hline
 $(1,-1,0,0)$ & + & + & + & + & -  \rule{0pt}{13pt}\\ \hline
 $(0,1,-1,0)$ & - & + & + & + & -  \rule{0pt}{13pt}\\ \hline
 $(0,0,1,-1)$ & - & - & - & + & -  \rule{0pt}{13pt}\\ \hline
 $(0,0,0,1)$  & - & - & - & + & -  \rule{0pt}{13pt}\\ \hline
\end{tabular}
}
\eeq

\subsubsection*{The $\mathbf{10}$-representation and its
U$(1)\times$U(1)-charges}

\begin{figure}[ht!]
\centering
  \includegraphics[scale=0.6]{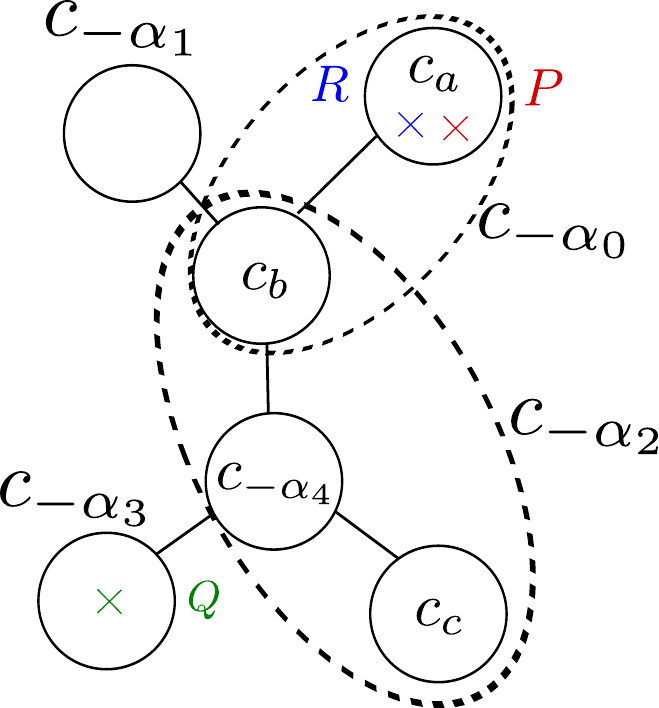}
  \caption{$D_5$-curve over $z=\beta_5=0$.}
  \label{fig:d6Fiber}
\end{figure}

For completeness, we briefly discuss the fiber at the loci $z=\beta_5=0$.
In this case more curves $c_{-\alpha_i}$ become reducible. Repeating the
same procedure as above, we insert $z=\beta_5=0$ and  $d_i=\beta_5=0$
into the Calabi-Yau equation \eqref{eq:CYSU5}, we obtain the following
splitting of nodes,
\beq \label{eq:splitSO10}
\text{
\begin{tabular}{|c|c|c|} \hline
  Node & splits into & Cartan charges \rule{0pt}{13pt}\\ \hline
 $z=0$  & $c_a: \,d_4s_8u+d_1 d_2 d_4 s_9 v + d_1 e_2 s_7 v^2=0$ & $(0,1,0,0)$ \rule{0pt}{13pt}\\
         & $c_b: \, d_2=0 $ & $(1,-1,0,1)$\rule{0pt}{13pt} \\ \hline
 $d_1=0$ &  $c_{-\alpha_1}: \,\, d_2 s_8+s_5 z_2=0$ & $(-2,1,0,0)$\rule{0pt}{13pt}\\  \hline
$d_2=0$ & $c_b: \,\, z_2=0$ & $(1,-1,0,-1)$\rule{0pt}{13pt}\\
        & $c_{-\alpha_4}: \,\, d_4=0$ & $(0,0,1,-2)$\rule{0pt}{13pt}\\
	& $c_c: \,\, d_1 d_3 d_4 s_1 z_2^2 + d_1 s_2 v z_2 + d_3 s_5 w=0 $ & $(0,-1,0,1)$\rule{0pt}{13pt}\\ \hline
  $d_3=0$ & $c_{-\alpha_3}: \, d_2 d_4 s_3 u+d_4 e_1 s_2 u^2 + d_2 s_7 w = 0 $ & $(0,1,-2,1)$\rule{0pt}{13pt}\\ \hline
$d_4=0$ & $ c_{-\alpha_4}: \,\, d_4=0$ & $(0,0,1,-2)$ \rule{0pt}{13pt}\\ \hline
\end{tabular}
}
\eeq
We observe that both the curves $c_{-\alpha_0}$ and $c_{-\alpha_2}$ split
into two, respectively, three components, denoted $c_a$, $c_b$,
respectively, $c_b$, $c_{-\alpha_4}$ and $c_c$, where we note that the
curve $c_a$ is the extended note, i.e.~the original singular fiber. The fact that the curves
$c_b$ and $c_{-\alpha_4}$ appear twice means that they have multiplicity
two. The intersection numbers between all the curves in
\eqref{eq:splitSO10} can be calculated taking into account the SR-ideal
of the ambient space. The intersection pattern of the curves reproduces
a $D_5$-curve, see figure \ref{fig:d6Fiber}.

The curves $c_{a,b,c,d}$ carry some weights of the $\mathbf{10}$ and
$\overline{\mathbf{10}}$ representations. All the other weights are
obtained as before by addition of curves $c_{-\alpha_i}$.  The
U$(1)\times$U(1)-charges of the $\overline{\mathbf{10}}$ representation
are determined by intersections with the Shioda maps $\sigma(\hat{s}_m)$
of the rational sections. We obtain
the representation
\beq \label{eq:10q1q2}
	\mathbf{10}_{(q_1,q_2)}=\mathbf{10}_{(\frac{1}{5},0)}\,.
\eeq
Finally, we enumerate which curves $c_{\mathbf{w}}$ of this
representation are in the Mori cone of $\Xsu$. As before we indicate
this by the values of the sign-function \eqref{eq:signw}, which reads
\beq
\text{
\begin{tabular}{|c||c|} \hline
 Weight		& Sign \\ \hline
 $(0,-1,0,0)$ & +  \\ \hline
 $(-1,1,-1,0)$ & +  \\ \hline
 $(1,0,-1,0)$ & +  \\ \hline
 $(-1,0,1,-1)$ & +  \\ \hline
 $(1,-1,1,-1)$  & +  \\ \hline
\end{tabular}
\begin{tabular}{|c||c|} \hline
 Weight		& Sign \\ \hline
 $(-1,0,0,1)$ & +  \\ \hline
 $(0,1,0,-1)$ & -  \\ \hline
 $(1,-1,0,1)$ & +  \\ \hline
 $(0,1,-1,1)$ & -  \\ \hline
 $(0,0,1,0)$  & -  \\ \hline
\end{tabular}
}
\eeq

\subsubsection*{KK-Charges}

We note that  at all codimension two singularities inside the divisor
$\cS_{SU(5)}$ the section $S_P$ is holomorphic. This means for these type
of matter the KK-charges vanish, $q_{KK}=0$. The only representation
with non-trivial KK-charge is, as before in section
\ref{sec:4DChiralityU1U1} the matter field $\mathbf{1}_{(-1,-2)}$.
Its KK-charge is $q_{KK}=2$.

\subsubsection*{Matter surfaces}

We conclude the discussion of codimension two singularities by
constructing the classes of certain matter surfaces
$\cC_{\mathbf{R}}^{\mathbf{w}}$. As before in section
\ref{sec:SingularFibU1U1} we are able to obtain these classes only for
a subset of the representations.

We first comment on the calculation of matter curves in the base.
A very convenient way to calculate all matter curves
$\Sigma_{\mathbf{R}}$ is  by projecting the intersections
$d^\star_i=P_l=0$, $l=1,\ldots,5$ or
$d^\star_i=\beta_5=0$ to the base, where $d^\star_i$ is the coordinates
associated to the Cartan divisor $D_i$ that has split at that codimension
two loci under consideration. The homology classes of the matter surfaces
are then given as
\beq
\Sigma_{\mathbf{R}}\cong[B]\cdotp[D^\star_i]\cdotp[P_i]=[B]\cdot[\Ssu]\cdotp[P_i]\,,
\eeq
where the divisor $D^\star_i$ has to correspond to the node that has
split at the location of the matter representation $\mathbf{R}$. Here
we have used the property  $\pi(D_I)=\cS_{SU(5)}$  in the second
equality.

As in the Abelian case, we can only specify some matter surfaces
completely. As before, this complication arises since we are only able to
isolate the irreducible components in the elliptic fiber locally, but can
not infer the total space of the fibration
$\cC_\mathbf{R}^{\mathbf{w}}\rightarrow \Sigma_{\mathbf{R}}$.
The subset of matter surfaces, that we can explicitly determine, along
with their homology classes, reads,
\beq \label{eq:matterSurfsSU5}
\text{
\begin{tabular}{|c|c|c|} \hline
 Rep.\!\!\! &  Matter curve \rule{0pt}{14pt} $\Sigma_{\mathbf{R}} $  &  Matter surface $\cC_{\mathbf{R}}$  \\ \hline
 \rule{0pt}{13pt}${\mathbf{5}}_{(-\frac{2}{5},0)}$\!\!\! & $\cS_{SU(5)}\cdotp ([s_2]+[s_5])$ & -  \\
 \rule{0pt}{13pt}${\mathbf{5}}_{(-\frac{2}{5},1)}$\!\!\! & $\cS_{SU(5)}\cdotp \cS_7$ & $( [K_B^{-1}]\!+\! \cS_9\! -\!\cS_7 \!+\! 2 H \!-\!2 D_1\! -\! 3 D_2\! -\! 4 D_3\! -\! 3 D_4 )\cdotp D_3\cdotp \cS_7$ \\
 \rule{0pt}{13pt}${\mathbf{5}}_{(\frac{3}{5},0)}$ \!\!\!& \!$\cS_{SU(5)}\cdotp ([s_3]+[K_B^{-1}])$\!\!\! & - \\
 \rule{0pt}{13pt}${\mathbf{5}}_{(\frac{3}{5},1)}$\!\!\! & \!\!$ \cS_{SU(5)}\cdotp [s_8]$ & \!\!$(\cS_9\!+\! 2 H\!-\!\! E_1\!-\!D_1 \!\!-\! 2 D_2\! -\!\! 4 D_3\! -\! 3 D_4 )\cdotp D_0\cdotp ([K_B^{-1}]\!+\!\cS_9\!-\!\cS_7)$\!\!\\
 \rule{0pt}{13pt}\!${\mathbf{5}}_{(-\frac{2}{5},-1)}$\!\!\! & $ \cS_{SU(5)}\cdotp (\cS_7+\cS_9)$ & - \\
 \rule{0pt}{13pt}${\mathbf{10}}_{(\frac{1}{5},0)}$\!\!\! & $ \cS_{SU(5)}\cdotp [K_B^{-1}]$ & $D_0\cdotp D_2\cdotp [K_B^{-1}]$ \\ \hline
\end{tabular}
}
\eeq
Here we made use of the SR-ideal \eqref{eq:SR-SU5},
the assignments of sections given in \eqref{eq:sectionsSU50},
\eqref{eq:sectionsSU5} and have abbreviated the divisors class of the
$s_i$ as $[s_i]$. In addition,
we have suppressed the intersection with the class of the base $B$
when denoting the matter curves $\Sigma_{\mathbf{R}}$ and by abuse of
notation used same symbol for vertical divisors and divisor in the base.

The matter surfaces for the singlets $\mathbf{1}_{(-1,1)}$,
$\mathbf{1}_{(0,2)}$ and $\mathbf{1}_{(-1,-2)}$ have been analyzed in
detail in section \ref{sec:SingularFibU1U1} in the case of a
U$(1)\times$U$(1)$ gauge group only. In the presence of an SU(5) at
codimension one, the matter surfaces determined in
\eqref{eq:MatSurfaces1} change. Taking into account the new classes
\eqref{eq:sectionsSU50}, \eqref{eq:sectionsSU5} and the new equation of
the Calabi-Yau fourfold
\eqref{eq:CYSU5} we obtain the matter surfaces as
\beq	
	\label{eq:MatSurfacesSingSU5}
	\text{\begin{tabular}{|c|c|}
\hline
	Matter surface\!\! &\! Homology class \rule{0pt}{13pt} \\\hline  \hline
	 $\cC_{\mathbf{1}_{(-1,1)}}$& $([K_B^{-1}]+\cS_7-\cS_9-2\Ssu)\cdot \cS_7\cdot E_1$\rule{0pt}{14pt} \\ \hline
	 $\cC_{\mathbf{1}_{(0,2)}}$& $\cS_7\cdot \cS_9 \cdot ([K_B^{-1}]+\cS_9-\cS_7+2H-2D_1-3D_2-4D_3-3D_4)$\rule{0pt}{14pt}\\\hline
	 $\cC_{\mathbf{1}_{(-1,-2)}}$&$([K_B^{-1}]\!+\!\cS_9\!-\!\cS_7)\cdotp \cS_9\cdotp(3 H \!-\! E_1\! -\! 2 E_2 \!+\!  2\cS_9\!-\!\cS_7\!-\!2 D_1\! -\! 3 D_2\! -\! 5 D_3\! -\! 4 D_4) \!$\rule{0pt}{14pt}\\\hline
\end{tabular}}
\eeq
This can be seen by specializing \eqref{eq:CYSU5} to the matter
curves $\Sigma_{\mathbf{1}_{(-1,1)}}=\{s_3'=s_7=0\}$,
$\Sigma_{\mathbf{1}_{(0,2)}}=\{s_7=s_9=0\}$, respectively,
$\Sigma_{\mathbf{1}_{(-1,-2)}}=\{s_8=s_9=0\}$ and, then, by
multiplying with the class of the isolated rational curves
over each matter curves.
We recall that the latter are given by $c_{(-1,1)}=\{e_1\}$ as well as
the total transforms of \eqref{eq:c02} and \eqref{eq:cm1m2}.

\subsubsection{Yukawa Couplings: Codimension Three}
\label{sec:codim3}

We focus in this section on the determination of those Yukawa points
that involve at least one the non-trivial representations under SU(5).
For an analysis of the Yukawa points of the singlets, we refer
to the discussion in section \ref{sec:codim3U1xU1}, that applies without
any changes.

Further enhancement of singularity type of the elliptic fibration
can be read directly from the discriminant \eqref{eq:DeltaSU5}
of $\Xsu$. Our strategy in determining these loci is to look for two
polynomials, typically describing two matter curves,
such that at their common vanishing locus the discriminant vanishes
with order $n$ greater than five, i.e.~$\Delta \sim z^n$ for $n>5$.
Then we check explicitly that the fiber of the resolution $\Xsu$ splits
further. We perform this analysis in the following for all
present matter representations of $\Xsu$.

We begin with the loci of the single $\mathbf{10}$
representation. We note that here, since $\beta_5=s_6=0$, the discriminant already
vanishes to degree seven and takes the form
\beq \label{eq:Delta10}
\Delta = -16 z^7  (s_5 s_7)^3 (s_2 s_8 s_7)^2 + \mathcal{O}(z^8)\,.
\eeq
Vanishing of any of these prefactors will enhance the zero of the
discriminant  to degree eight or higher.
First, we analyze the vanishing $s_6=s_5=0$ on the divisor $z=0$. The
polynomial $P_1$
vanishes but no other $P_i$ does, so this point belongs  both
to the matter curve of $\mathbf{5}_{2/5,0}$ and of
$\mathbf{10}$. Although the fiber does not degenerated
to an $E_6$-fiber the Yukawa coupling exists. The gauge invariant
coupling we read off is
\beq
s_5=s_6=z=0\,:\qquad  {\mathbf{10}}_{(\tfrac{1}{5},0)}\times{\mathbf{10}}_{(\tfrac{1}{5},0)}\times{\mathbf{5}}_{(-\tfrac{2}{5},0)}\,,
\eeq

Next, we study the vanishings of $s_2$ or $s_8$ with $s_6=z=0$.
In this case we obtain couplings of the type $5 \times 5 \times
\overline{10}$. We note that the fiber does not enhance to a $D_6$-fiber,
but the Yukawa is still realized geometrically.

Finally, we consider the last enhancement of the vanishing of
\eqref{eq:Delta10} with $s_7=s_6=0$. In this case the intersection of the
Cartan divisor $D_4$ with the Calabi Yau hypersurface is
automatically zero. Since we are now only specifying three conditions
$d_4=s_6=s_7=0$ in a five-dimensional ambient space, the fiber has to be
complex two-dimensional, namely the whole del Pezzo surface $dP_2$.
The elliptic fibration of $\Xsu$ is non-flat, with a complex 
two-dimensional fiber over codimension three. We note that this effect  
occurs also for other SU(5)-embeddings \cite{Borchmann:2013jwa}.

The physics of this
a non-flat fibration has been discussed in \cite{Candelas:2000nc}. As
mentioned before, an M5-brane can wrap the surface in the non-flat fiber,
and give rise to a string in a three-dimensional M-theory
compactification. In addition, M2-branes can wrap holomorphic curves
in $dP_2$. In  the blow-down of $\Xsu$, and in particular in the F-theory
limit,  the states associated to these degrees of freedom become
massless. For phenomenology these additional light states  are
usually undesired and can render the compactification unphysical,
since we have to deal with an infinite tower of charged excitations of
the string, which cannot easily be represented by a finite number of
fields.

As discussed in the introduction of the section, one attempt to deal
with the presence of the extra degrees of freedom from the non-flat
fiber is to forbid a chiral excess of states. As we will demonstrate
in section \ref{sec:G4FluxChiralitiesSU5} this can be achieved by tuning
the $G_4$-flux as in \eqref{eq:NFFcond}, so that it integrates to zero
over the non-flat fiber. Alternatively, we can simply forbid the Yukawa
point with the non-flat fiber geometrically. Recalling
$[s_6]=[K_B^{-1}]$, cf.~\eqref{eq:sectionsFibration}, and
$\pi(D_4)=\cS_{SU(5)}$, all we have to demand is that
\beq \label{eq:lociNFF}
 \cS_7 \cdotp [K_B^{-1}] \cdotp \cS_{\text{SU}(5)} = 0
\eeq
One obvious but drastic solution is $\cS_7=0$ since it deprives us from
some singlets and non-singlets representations.
A more sophisticated solution to \eqref{eq:lociNFF} is obtained as follows. We recall the basis expansion \eqref{eq:S7S9exp}
of the divisor $\cS_7$  and in addition expand
\bea
	\cS_{SU(5)}= n_{\text{SU}(5)}^\alpha D_\alpha\,.
\eea
Then, it is possible over an appropriate base $B$ to tune the integers
$n_{7}^\alpha$ and $n_{\text{SU}(5)}^\alpha$ in such a way
to only forbid the dangerous Yukawa point \eqref{eq:lociNFF} without
restricting the spectrum of $\Xsu$ and the other Yukawa points. As a concrete
simple example one can choose $B=\text{Bl}\mathbb{P}^3$, the blow-up of 
$\mathbb{P}^3$ along the curve $x_i=x_j=0$. Then \eqref{eq:lociNFF} yields 
$n_{\text{SU}(5)}^1 (4 n_7^1 + 3 n_7^2)+3 n_{\text{SU}(5)}^2 n_7^1 =0$ for $D_\alpha=\{H,H-E\}$ with $H$ the 
hyperplane and $E$ the exceptional divisor and $K^{-1}_{\text{Bl}_1\mathbb{P}^3}=4H-E$. Clearly, this can be solved
without removing other codimension two or three singularities.

We summarize all Yukawa couplings involving the $\mathbf{10}$-representation at $s_6=z=0$ in the following 
table,
\beq
\text{
\begin{tabular}{|c|c|} \hline
Loci & Yukawa coupling \rule{0pt}{13pt}\\ \hline
$s_5=0$ & ${\mathbf{5}}_{(-\tfrac{2}{5},0)} \times\mathbf{10}_{(\tfrac{1}{5},0)} \times\mathbf{10}_{(\tfrac{1}{5},0)}$ \rule{0pt}{13pt}\\
$s_2=0$ & ${\mathbf{5}}_{(-\tfrac{2}{5},0)} \times{\mathbf{5}}_{(\tfrac{3}{5},0)} \times\overline{\mathbf{10}}_{(-\tfrac{1}{5},0)}$ \rule{0pt}{13pt}\\
$s_8=0$ & ${\mathbf{5}}_{(\tfrac{3}{5},1)}\times {\mathbf{5}}_{(-\tfrac{2}{5},-1)}\times \overline{\mathbf{10}}_{(-\tfrac{1}{5},0)}$ \rule{0pt}{13pt}\\
$s_7=0$ & Non-flat fiber \rule{0pt}{13pt}\\ \hline
\end{tabular}
}
\eeq

Besides these Yukawa couplings there are additional Yukawa couplings
involving singlets and the $\mathbf{5}$-representations. All these Yukawa
couplings are localized at $z=s_7=0$. Evaluating the discriminant at
this locus we obtain  the first non-vanishing terms as
\beq
\Delta \cong -s_3^2 s_6^5 (s_2 s_5 - s_1 s_6) s_8 s_9^2.
\eeq
Setting each prefactor to zero we obtain Yukawa couplings, that  we
summarize in the following table,
\beq
\text{
\begin{tabular}{|c|c|} \hline
Loci & Yukawa coupling\rule{0pt}{13pt}\\ \hline
$s_3=0$ & ${\overline{\mathbf{5}}}_{(\tfrac{2}{5},-1)} \times{\mathbf{5}}_{(\tfrac{3}{5},0)} \times\mathbf{1}_{(-1,1)}$ \rule{0pt}{13pt}\\
$(s_2 s_5 - s_1 s_6)=0$ & $ {\mathbf{5}}_{(-\tfrac{2}{5},0)} \times{\ov{\mathbf{5}}}_{(\tfrac{2}{5},-1)} \times\mathbf{1}_{(0,1)}$ \rule{0pt}{13pt}\\
$s_8=0$ & ${\ov{\mathbf{5}}}_{(\tfrac{2}{5},-1)}\times {\mathbf{5}}_{(\tfrac{3}{5},1)} \times\ov{\mathbf{1}}_{(-1,0)}$ \rule{0pt}{13pt}\\
$s_9=0$ & ${\ov{\mathbf{5}}}_{(\tfrac{2}{5},-1)}\times \mathbf{5}_{(-\tfrac{2}{5},-1)}\times \mathbf{1}_{(0,2)}$ \rule{0pt}{13pt}\\ \hline
\end{tabular}
}
\eeq

\subsection{The Cohomology Ring and the Chern Classes of $\hat{X}_{\text{SU}(5)}$}
\label{sec:CohomologyFFSU5}

In this section we apply the techniques described in section
\ref{sec:CYFFCohomologyRing} to perform computations in the vertical
cohomology $H^{(*,*)}_V(\Xsu)$ of the Calabi-Yau fourfold $\Xsu$ with a
resolved SU$(5)$-singularity over $\cS_{\text{SU}(5)}$.
The computations are completely analogous to the ones performed
to obtain the cohomology of $\hat{X}$. All we have
to do is to add the Cartan divisors $D_i$, $i=1,\ldots,4$,
as additional variables to the polynomial ring $R$ in
\eqref{eq:abstractCohomRing}. We note
that the extended node $D_0=\Ssu-\sum_i D_i$ corresponding to $z$ is then
automatically included.
We also have to replace the ideal $SR$ in \eqref{eq:abstractCohomRing} by
the ideal \eqref{eq:SR-SU5} arising after the resolution process
\eqref{eq:resMapSU5}. Finally, we have to employ the anti-canonical
bundle \eqref{eq:KdP2SU5}, which is the class of $\Xsu$ in the blown-up
ambient space $\widehat{dP}_2^B(\cS_7,\cS_9)$.
We summarize the presentation of the vertical cohomology ring
of $\Xsu$ as
\beq \label{eq:abstractCohomRingSU5}
	H^{(*,*)}_V(\Xsu)\cong
	\frac{\mathbb{C}[D_\alpha,\Ssu,S_P,S_Q,S_R,D_i]\cdot \big[\Xsu\big]}{SR}\,,
\eeq
with the ideal $SR$ given in \eqref{eq:SR-SU5}. Here we have singled
out the vertical divisor $\Ssu$ of the original SU(5)-singularity.

We begin by using these techniques to calculate the Euler number and
Chern classes of $\Xsu$ for a general base $B$. Then
we  construct the full cohomology in the case of
$B=\mathbb{P}^3$. In the following discussion we will quote the main
results from this analysis and refer to
appendix \ref{app:fluxP3SU5} for more details on the calculation, in
particular the full quartic intersections of $\Xsu$ from which all
the following is derived.

\subsubsection{Second Chern Classes and Euler Number of $\Xsu$: General
Formulas}

The strategy for the computation of the Euler number is is a mixture
of the strategy in section \ref{sec:CYFFCohomologyRing} and the approach
of \cite{Marsano:2011hv} to compute Chern classes of fourfolds with a
resolved SU(5)-singularity. The key point is to find a SR-ideal that
contains sufficiently fine information to do concrete calculations,
but that is sufficiently coarse in order to not depend on the details of
$B$. The starting point is the SR-ideal \eqref{eq:SR-SU5}. The only
degree of freedom here is the pullback $\pi_{\text{SU}(5)}(SR_B)$ of the
SR-ideal of the base. Instead of using the full ideal that depends on the
details of $B$ we again work with a simplified version denoted $SR_B'$
that is based on some universal geometric properties of the fibration of
$\Xsu$.

As in \eqref{eq:SRintsdP2B} we  assume that $SR_B'$ contains
all quartic intersections of the vertical divisors $D_\alpha$. In
addition we assume as in \cite{Marsano:2011hv} that three vertical
$D_\alpha$ never intersect the Cartan divisors $D_i$ due to a too large
codimension in the base. Thus we use the ideal
\beq
	 SR'_B=\{D_\alpha\cdot D_\beta\cdot D_\gamma\cdot D_\delta,D_\alpha\cdot D_\beta\cdot D_\gamma\cdot D_i\}
\eeq
instead of $\pi_{\text{SU}(5)}(SR_B)$ in \eqref{eq:SR-SU5}. As we
demonstrate next  this  is sufficient to
calculate the Chern classes and Euler number as well as to check basic
intersections of $\Xsu$, as for the fourfold $\hat{X}$ considered in
section \ref{sec:CYFFCohomologyRing}.

Next we calculate the total Chern class of $\Xsu$. All we need is the
Chern class of the blown-up ambient space $\widehat{dP}_2^B(\cS_7,\cS_9)$
and to apply adjunction. The computation of the Chern class of the
ambient space is straightforward, but lengthy and little illuminating.
We simply use the result \eqref{eq:cdP2BU} and refer to appendix
\ref{app:ChernOfdP2BU} for details.
Then we use adjunction as in \eqref{eq:cXadjunction} to obtain the
total Chern class of $\Xsu$.
We present here the results of the second Chern class $c_2(\Xsu)$ that
reads
\bea \label{eq:c2XSU5genB}
	c_2(\Xsu)\!\!&\!\!=\!\!&\!\!c_2(\hat{X})- c_1 (4D_1+6 D_2+2 D_3+ D_4)+2 D_1 D_2+D_2^2+2 D_3 D_4\\
	&\!\!+\!\!&\!\! \Ssu(D_1+D_2 +D_4-2 S_Q )+\cS_7(D_1 +2 D_2-D_4)+\cS_9(D_1 +D_2 +D_4)\,,\nn
\eea
where $c_2(\hat{X})$ denotes the second Chern class \eqref{eq:c2XgenB}
on $\hat{X}$.
As a sanity check we  take the  blow-down limit
by formally setting $D_i\rightarrow 0$ and $\Ssu\rightarrow 0$ in which
we precisely recover the Chern class \eqref{eq:c2XgenB} before
resolution.

Next we calculate the Euler number on $\Xsu$. For this purpose we can
either bootstrap ourselves and use \eqref{eq:c2XSU5genB} in combination
with the relation \eqref{eq:chi0X_Euler} or calculate the fourth Chern
class on $\Xsu$ directly by expanding \eqref{eq:cdP2BU} to higher order
in the adjunction formula expression for the Chern class for $\Xsu$. In
either case we obtain
\bea \label{eq:EulerXSU5}
	\chi(\Xsu)&=&\chi(\hat{X})-3\int_B \Ssu\big(52 c_1^2
	-c_1 (41 \Ssu+17 \cS_7 +31 \cS_9 )+10 \Ssu^2
	\nn\\
	&+&\Ssu(5 \cS_7+13 \cS_9)+5 \cS_7^2 +2 \cS_7 \cS_9 +5 \cS_9^2 \big)\,,
\eea
where $\chi(\hat{X})$ denotes the Euler number \eqref{eq:EulerNumberX}
of $\hat{X}$ before the blow-up, to which our result \eqref{eq:EulerXSU5}
specializes correctly in the limit $D_i,\Ssu\rightarrow 0$.

\subsubsection*{The Full Cohomology Ring of $\Xsu$: $B=\mathbb{P}^3$}

Next we compute the full vertical cohomology of $\Xsu$ with base
$B=\mathbb{P}^3$.
In this case, there is only one divisor in the base, the hyperplane
$H_B$, thus, we have $D_\alpha=H_B$. Furthermore, the resolved
SU(5)-singularity of $\Xsu$ is located over $\mathcal{S}^b_{\text{SU}(5)}=H_B$.

\begin{figure}[ht!]
\centering
\includegraphics[scale=0.5]{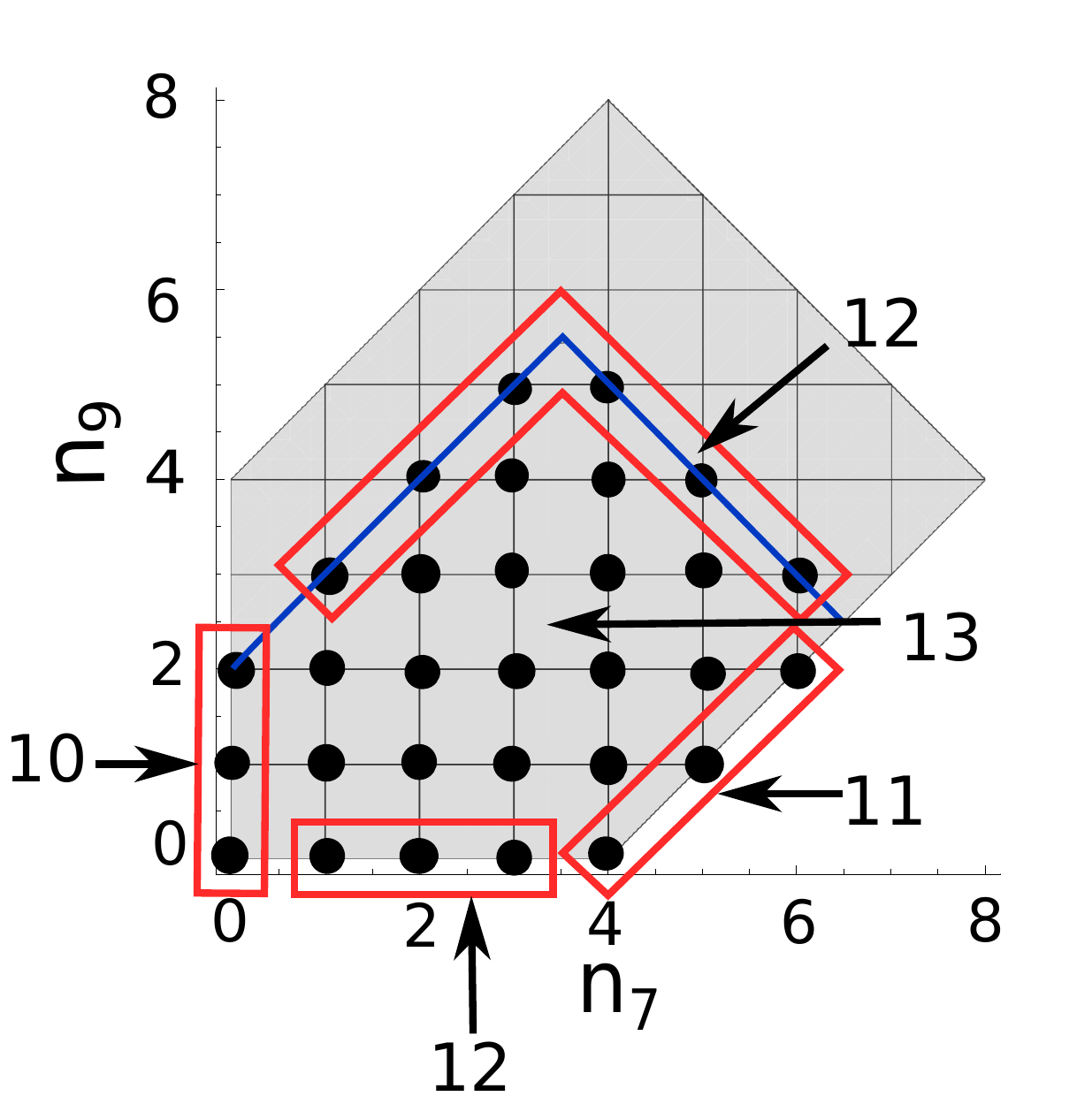} \qquad \qquad
 \caption{Number of independent surfaces $h^{(2,2)}_V(\Xsu)$ for the
 various values of allowed $n_7$, $n_9$. On the blue boundary and at
 $n_7=n_9=0$ we have a holomorphic zero section.}
 \label{fig:SU5n7n9-h22}
\end{figure}

The Hodge numbers are calculated to be
\beq
h^{(1,1)}(\Xsu)= h^{(3,3)}(\Xsu)= 8\,, \qquad h_V^{(2,2)}(\Xsu) \in [10,13]\,,
\eeq
where the number of independent surfaces as in the Abelian case depends
on the values of $n_7$, $n_9=0$, see figure \ref{fig:SU5n7n9-h22}. As
the basis of $H^{(1,1)}$, we use the following divisors
\beq \label{eq:basisH11XSU5}
H^{(1,1)}(\hat{X})= \langle H_B, S_P, S_Q, S_R, D_i\rangle\,,
\eeq
that fixes by Poincar\'e duality the basis of $H^{(3,3)}_V(\Xsu)$.
All quartic intersections of the divisors \eqref{eq:basisH11XSU5}
can be found in \eqref{eq:QuarticIntsSU5}.

Then we compute the grade two piece $R^{(2)}$ of the ring
\eqref{eq:abstractCohomRingSU5}.
We obtain 64 different monomials in two divisors, but due to equivalence
relations in \eqref{eq:abstractCohomRingSU5} at most 13 of them are
independent, with jumps along the boundary of the allowed region of
$n_7$, $n_9$ in figure \ref{fig:SU5n7n9-h22}.
 A carefully chosen basis
of $H_V^{(2,2)}(\Xsu)$, thus, has to have always the correct number of
surfaces for the entire allowed region. The analysis of appendix
\ref{app:H22BasisSU5} proves that one  convenient  basis is given by
\bea \label{eq:basisH22Su5}
H_V^{(2,2)}(\Xsu) &=& \langle H_B^2,\, H_B \cdotp S_P,\, H_B \cdotp \sigma(\hat{s}_Q),\, H_B \cdotp \sigma(\hat{s}_R),\, S_P \cdotp \sigma(\hat{s}_R),\,  \sigma(\hat{s}_Q) \cdotp \sigma(\hat{s}_R),\, \nn \\ && \sigma(\hat{s}_Q)^2,\,  H_B \cdotp D_1,\, H_B \cdotp D_2,\,
H_B \cdotp D_3,\, H_B \cdotp D_4,\, D_2 \cdotp D_4,\, D_1^2 \rangle\,.
\eea

At the boundaries of figure \ref{fig:SU5n7n9-h22}, one or more
elements in $H^{(2,2)}_V(\Xsu)$ become linearly dependent.
Thus, they can be expressed in terms of as smaller basis
that one obtains by dropping the linearly dependent elements from
\eqref{eq:basisH22Su5}. The expansions of the linearly dependent elements
in terms of this reduced basis for the various  boundary components of
figure \ref{fig:SU5n7n9-h22} can be found in appendix
\ref{app:H22BasisSU5}. The result of this analysis is that we
can obtain a basis on all boundaries
from the basis \ref{eq:basisH22Su5} by successively dropping the elements
$S_P \cdotp \sigma(\hat{s}_R)$, $H_B \cdotp \sigma(\hat{s}_Q)$ and $D_2 \cdotp D_4$.
This way we obtain, as required, in steps twelve, eleven and ten
independent surfaces on the relevant boundary components.

\section{$G_4$-Flux \& Chiralities on Fourfolds with
SU$(5)\times$U$(1)^2$}
\label{sec:G4FluxChiralitiesSU5}

With the cohomology ring at hand we finally calculate the general
$G_4$-flux, see section \ref{sec:G4U1xU1xSU5}, and 4D chiralities in this
section. As before, the lack of
knowledge of all the matter surfaces forces us to employ the 3D
Chern-Simons terms and M-/F-theory duality to determine the remaining
chiralities, cf.~section \ref{sec:ChiralitiesU1xU1xSU5}.
It turns out again, that the Chern-Simons terms by
themselves are sufficient to determine all the chiralities. The
chiralities determined by geometric techniques for a subset of matter
therefore provide an independent check.   We also check that for the
obtained spectrum all four-dimensional anomalies are cancelled.  Full
explicit results are again presented for the base $B=\mathbb{P}^3$.  We
also present one  concrete toric examples in section
\ref{sec:ExamplesSU5U1U1}.

There is one accompanying appendix
\ref{app:fluxesNChiralities} with further details on the $G_4$-flux and
the full expressions for the 4D chiralities.

We recall that there can be non-flat fibers at codimension three
in $\Xsu$. As mentioned before, these can be avoided for
general bases $B$, but are generically present for $B=\mathbb{P}^3$.\footnote{As 
mentioned above, we can forbid these points by setting
$\cS_7=0$, however, at the cost of losing many representations
of the general spectrum obtained from $\Xsu$, too.}
We deal with these complications here by forbidding a chiral excess
of the additional light degrees of freedom associated to the non-flat
fiber. This is achieved by requiring one  additional condition
on the $G_4$-flux, namely the vanishing of the integral
\eqref{eq:NFFcond}. As we demonstrate here, this ensures that the  4D
chiralities  can successfully be  determined both via geometric
techniques and  M-/F-theory  Chern-Simons terms, and that the resulting
chiral spectrum cancels all anomalies of the
four-dimensional quantum field theory.

\subsection{$G_4$-Flux on Fourfolds with Two Rational Sections \&  SU$(5)$}
\label{sec:G4U1xU1xSU5}

The construction of $G_4$-flux presented in this section
is very similar to the one of section \ref{sec:G4+chiralitiesU1xU1}.
Therefore, we will keep the following discussion as brief as possible
and choose, as before, the base $B=\mathbb{P}^3$ to demonstrate our
techniques.

We begin by expanding the $G_4$-flux in the generically 13-dimensional
basis \eqref{eq:basisH22Su5} of the cohomology group $H_V^{(2,2)}(\Xsu)$,
see \eqref{eq:G4expansion}. We note that the dimensionality
of $H^{(2,2)}_V(\Xsu)$ jumps at the boundaries of figure
\ref{fig:SU5n7n9-h22}, however, our general formulas for the $G_4$-flux
derived in this section will remain valid.
Then we enforce the generically six independent flux-conditions
\eqref{eq:CStermsConditions}.  Again we emphasize, that unlike in
compactifications with a holomorphic zero section, the CS-terms
$\Theta^M_{00}$, $\Theta^M_{0i}$ must not be required to vanish, nor
vanish automatically.

Since we are imposing generically six independent conditions
\ref{eq:CStermsConditions} on the 13-parameter $G_4$-flux, we obtain
a seven-dimensional $G_4$-flux,
\beq \label{eq:G4genSU5}
G_4 =  a_3 G_4^{(3)}+a_4 G_4^{(4)}+a_5 G_4^{(5)}+a_6 G_4^{(6)}+a_{7} G_4^{(7)}+a_{12} G_4^{(12)}+a_{13} G_4^{(13)},
\eeq
where the $a_i$ denote the free parameters of the $G_4$-flux.
The independent fluxes $G_4^{(i)}$ read
\bea
G_4^{(3)} \!\!&\!\!=\!\!&\!\! H_B \cdot \sigma(\hat s_Q), \qquad \quad
G_4^{(4)} = H_B \cdot \sigma(\hat s_R)\,, \qquad\quad
G_4^{(12)} = D_4 \cdot (D_2 - 4 H_B)\,,\nn\\
G_4^{(5)} \!\!&\!\!=\!\!&\!\!-2 H_B^2 n_9 (4 - n_7 + n_9) + S_P \cdot \sigma(\hat s_R)\,,\nn\\
G_4^{(6)}\!\! &\!\!=\!\!&\!\! \tfrac{1}{25} \big[4 n_7  D_1 \cdot H_B + 8 n_7  D_2 \cdot H_B + 12 n_7 D_3 \cdot H_B  + 6 D_4 n_7  \cdot H_B \nn \\ && +
   50 \cdot H_B^2 (4 - n_7 + n_9) (2 + n_9) + 25  H_B \cdot S_P (4 - n_7 + n_9) +
   25 \sigma(\hat s_Q) \cdot \sigma(\hat s_R)\big]\,, \nn \nn\\
G_4^{(7)} \!\! &\!\!=\!\!&\!\! \tfrac{1}{125} \big[-2 D_3 \cdot  H_B (282 + 31 n_7 - 35 n_9) - 2 D_1 \cdot H_B (144 + 7 n_7 - 15 n_9) \nn \\ && -
  4 D_2 \cdot H_B (144 + 7 n_7 - 15 n_9) + D_4 \cdot H_B (-282 - 31 n_7 + 35 n_9) \nn \\ &&+
  125 \cdot H_B^2 \left(\frac{136}{5} + n_9 (4 - n_7 + n_9)\right) + 850 \cdot H_B \cdot S_P + 125 \sigma(\hat s_Q)^2\Big] \,,\nn\\
G_4^{(13)} \!\! &\!\!=\!\!&\!\! \tfrac{1}{5} \big[5 D_1^2 + 18 D_1 \cdot H_B - 24 D_2 \cdot H_B - 16 D_3 \cdot H_B - 8 D_4 \cdot H_B + 40 \cdot H_B^2 \nn \\ && +
   3 n_9 D_1 \cdot H_B  + 6 n_9 D_2 \cdot H_B  + 4 n_9  D_3 \cdot H_B + 2 n_9 D_4 \cdot H_B  + 10 H_B \cdot S_P\big] \nn \\
\eea

The general $G_4$-flux \eqref{eq:G4genSU5} specializes correctly
at the boundaries of figure \ref{fig:SU5n7n9-h22} and
the concrete expressions for it can be obtained at every boundary
component, completely analogous to the discussion of section
\ref{sec:G4U1xU1}. The necessary  analysis of the behavior of
the basis \eqref{eq:basisH22Su5} along with the homology relations
between linearly dependent elements on the respective boundaries can
be found in appendix \ref{app:H22BasisSU5}.

\subsection{4D Chiralities from Matter Surfaces \& 3D CS-Terms}
\label{sec:ChiralitiesU1xU1xSU5}

In this section we finally calculate the chiralities of the total matter
content of the F-theory compactification on $\Xsu$, that we
have determined in section \ref{sec:SingularitiesSU5U1U1}. We recall
that there are six singlets, five different $\mathbf{5}$-representations
and one $\mathbf{10}$, cf.~section \ref{sec:SingularitiesSU5U1U1}.
After having determined these chiralities, we check cancellation of 4D
anomalies at the end of this section.  We find that all anomalies are
cancelled in general.

As mentioned before, all chiralities
can in principle be calculated from the $G_4$-flux in \eqref{eq:G4genSU5}
and the matter surfaces by evaluating the index \eqref{eq:chi(R)}.
However, as in the discussion of section \eqref{sec:4DChiralityU1U1},
we only have the explicit homology classes of the six matter surfaces
\eqref{eq:MatSurfaces1} and \eqref{eq:matterSurfsSU5}. Again, we can
obtain all chiralities by combining this information with the matching of
3D Chern-Simons terms as outline in section \ref{sec:ChiralitiesStrat}.
We first calculate the chiralities from the six matter surfaces
for the general $G_4$-flux \eqref{eq:G4genSU5} and then impose the
non-flat fiber condition \eqref{eq:NFFcond}. Only then it is possible
to obtain the remaining chiralities from the matching of 3D CS-terms.

The chiralities for the six representations whose surfaces
$\cC_{\mathbf{R}}^{\mathbf{w}}$ we know,
cf.~\eqref{eq:MatSurfacesSingSU5} and \eqref{eq:matterSurfsSU5}, are
calculated for the general $G_4$-flux \eqref{eq:G4genSU5} employing
\eqref{eq:chi(R)} as
\bea \label{eq:directChis}
\chi\big(\mathbf{5}_{(-\tfrac{2}{5},1)}\big)\!\! &\!\!=\!\!&\!\! -\tfrac{1}{500} n_7 \big[ -50 a_3 + 125 a_4 +5 (21 n_7 - 5 (8 + 3 n_9)) a_6 \nn \\ &&
   + (158 - 111 n_7 + 115 n_9) a_7  + 500 a_{12} + 50 ( n_9-4) a_{13}\big]\,,\nn\\
\chi\big(\mathbf{5}_{(\tfrac{3}{5},0)}\big) \!\! &\!\!=\!\!&\!\! -\tfrac{1}{500} ( 4-n_7 + n_9) \big[ 75 a_3 + 125 a_4 +  5 (  21 n_7 - 65 n_9-160) a_6 \nn \\ && + (  14 n_7 - 135 n_9-742) a_7 +
 50 (  n_9-24) a_{13}\big]\,,\nn\\
\chi\big(\mathbf{10}_{(\tfrac{1}{5},0)}\big) \!\! &\!\!=\!\!&\!\! \tfrac{1}{125} \big[-25 a_3 - 5 (2 n_7 - 5 (4 + n_9)) a_6 + (134 + 7 n_7 + 20 n_9) a_7  \nn \\ && + 125 n_7 a_{12}  + 50 (-27 + 5 n_7 - 2 n_9) a_{13} \big]\,, \nn\\
\chi(\mathbf{1}_{(-1,1)}) \!\! &\!\!=\!\!&\!\! -\tfrac{1}{20} n_7 (2 + n_7 - n_9) \big[-5 a_3 +
   5 a_4 + (  5 n_7-4) a_6 \nn  + (58 - 10 n_7 + 5 n_9) a_7 \big]\,,\nn\\
\chi(\mathbf{1}_{(0,2)}) \!\! &\!\!=\!\!&\!\! \tfrac{1}{100} n_7 n_9 \big[-50 a_4 + 50 (  n_7 \!-\! n_9-4) a_5   +
   5 (64 \!- \!15 n_7 + 15 n_9) a_6 \!+\! 162 a_7 \!+\! 50 a_{13}\big]\,,\nn\\
\chi(\mathbf{1}_{(-1,-2)}) \!\! &\!\!=\!\!&\!\! -\tfrac{1}{20} ( 4-n_7 + n_9) n_9 \big[ -5 a_3 - 10 a_4 +(10 n_7  - 20 n_9 ) a_5\nn \\ && + ( 60 - 15 n_7 a_6 + 25 n_9 ) a_6   +
 (54  - 5 n_7  + 10 n_9) a_7 + 10 a_{13}\big]\,,
\eea

We then obtain the other chiralities by using the identification
\eqref{eq:FluxCondMF} of 3D CS-levels for
$\Theta_{\Sigma\Lambda}$. Both the CS-levels
$\Theta_{\Sigma\Lambda}^M$ on the M-theory side and
the loop-generated CS-levels $\Theta_{\Sigma\Lambda}^F$
on the F-theory side are readily computed.
All we have to know for the computation of the CS-terms
$\Theta_{\Sigma\Lambda}^F$ is the KK-charge of the matter
representations. We recall from section
\ref{sec:SingularitiesSU5U1U1} that  only the singlet
$\mathbf{1}_{(-1,-2)}$ has a non-trivial KK-charge $q_{KK}=2$
and thus a shifted sign-function \eqref{eq:signshifted} with
$k=-2$. All the other matter representations have $q_{KK}=0$
and a symmetric sign-function.
Then, all the computations are straightforward but lengthy.
Thus, we refer the interested reader to appendix
\ref{app:fluxesNChiralities}, where the results of these calculations
are presented.

We make a short stop to show some interesting results/checks of the
CS-levels and the discussion in section \ref{sec:KKcomputations}. First,
we calculate the CS-term $\Theta_{00}$ for the KK-vector. On the  F-
theory side, we have to evaluate  the loop-corrections of KK-states as in
\eqref{eq:Theta001rep}. Since only the singlet $\mathbf{1}_{(-1,-2)}$ has
a non-trivial KK-charge, we again obtain the result
\eqref{eq:XP3_Theta00}.  On the M-theory
side, $\Theta^M_{00}$ is calculated directly from $G_4$-flux and we
obtain
\bea
\Theta^M_{00}&=&-\tfrac{1}{20} (4 - n_7 + n_9) n_9 \big[ -5 a_3 - 10 a_4 +(10 n_7  - 20 n_9 ) a_5 \nn \\ && + ( 60 - 15 n_7 a_6 + 25 n_9 ) a_6 +
 (54  - 5 n_7  + 10 n_9) a_7 + 10 a_{13}\big] \nn \\
 &=& \chi(\mathbf{1}_{(-1,-2)})\,,
\eea
where we used \eqref{eq:directChis} in the last equality.
Thus, we confirm the result \eqref{eq:XP3_Theta00}.

Similarly, we check
the matching of the CS-levels $\Theta_{0m}$. As before, a matching
of the M- and F-theory expression for these CS-levels requires
the relation \eqref{eq:XP3_Theta0m} to hold,
\beq
	\tfrac{1}{4}\int_{\hat{X}}S_P\cdot\sigma(\hat{s}_Q)\cdot G_4\stackrel{!}{=}-\chi(\mathbf{1}_{(-1,-2)})\,,\qquad
	\tfrac{1}{4}\int_{\hat{X}}S_P\cdot\sigma(\hat{s}_R)\cdot G_4\stackrel{!}{=}-2\chi(\mathbf{1}_{(-1,-2)})\,.
\eeq
Indeed, we evaluate the flux integral on the left to confirm this
equality.

Finally, we comment on the CS-terms $\Theta_{0i}$. From the direct
computation of the relevant integrals of the $G_4$-flux,
cf.~\eqref{eq:3DCSterms}, we obtain $\Theta_{0i}^M=0$, since
$S_P\cdot D_i=0$ in the homology of $\Xsu$, cf.~\eqref{eq:SPDi} in
appendix \ref{app:fluxP3SU5}. This immediately agrees with the field
theory result $\Theta_{0i}^F$ in \eqref{eq:CSloops} since there the only
charged matter states in 3D with non-trivial KK-charge is the singlet
$\mathbf{1}_{(-1,-2)}$, which however has $q_i=0$ for all $i$.

Coming back to the calculation of chiralities, we recall that for
$n_7>0$ there is a non-flat fiber, since the codimension three point
$s_6=s_7=d_4=0$ exists, cf.~section \ref{sec:codim3}. If we ignore this
fact and try to solve for the chiralities using the matching condition
\ref{eq:FluxCondMF} we find
\beq \label{eq:CSwithNFF}
\begin{aligned}[c]
 \Theta^F_{i=2,j=4} &= \Theta^M_{i=2,j=4}\,,\nn \\
 0 &= \frac{2}{5} n_7 a_{12}\,, \nn
\end{aligned}\quad\qquad
\begin{aligned}[c]
  \Theta^F_{i=4,m=1} &= \Theta^M_{i=4,m=1}\,, \nn\\
 0 &= 2 n_7 a_{12}\,, \nn
\end{aligned}\quad\qquad
\begin{aligned}[c]
  \Theta^F_{i=4,m=2} &= \Theta^M_{i=4,m=2}\,, \nn \\
 0 &= n_7 a_{12}\,, \nn
\end{aligned}\quad
\eeq
where the left hand sides are functions of chiralities. These equations
are obviously inconsistent on the F-theory side. However, this is not
surprising since the existence of a non-flat fiber implies the
presence of additional light states, that we have not taken
into account on the field theory side.  Since these states are probably
unwanted in setups aimed at the construction of phenomenologically
appealing F-theory compactifications, we require the absence of a
\textit{chiral excess} of these light states. As we see from
\eqref{eq:CSwithNFF} the appropriate condition to impose is to
demand the vanishing of the second line.

It is satisfying to see that the vanishing of this second line is
precisely the non-flat fiber condition \eqref{eq:NFFcond} on the
$G_4$-flux. Indeed, using the general $G_4$-flux \eqref{eq:G4genSU5} and
the homology class of the loci of the non-flat fiber on the left of
equation \eqref{eq:lociNFF}, we obtain
\bea \label{eq:nff-cond}
\int_{X_5} G_4 D_4 \cdotp \cS_7 \cdotp [s_6] &=& -8 n_7 a_{12}\stackrel{!}{=}0\,,
\eea
which implies exactly a vanishing of  the second line in
\eqref{eq:CSwithNFF}. Thus,
we see that we obtain a consistent set of equations in
\eqref{eq:CSwithNFF} if we impose the non-flat fiber condition
\eqref{eq:NFFcond}, as claimed.

\begin{figure}[ht!]
\centering
\includegraphics[scale=0.6]{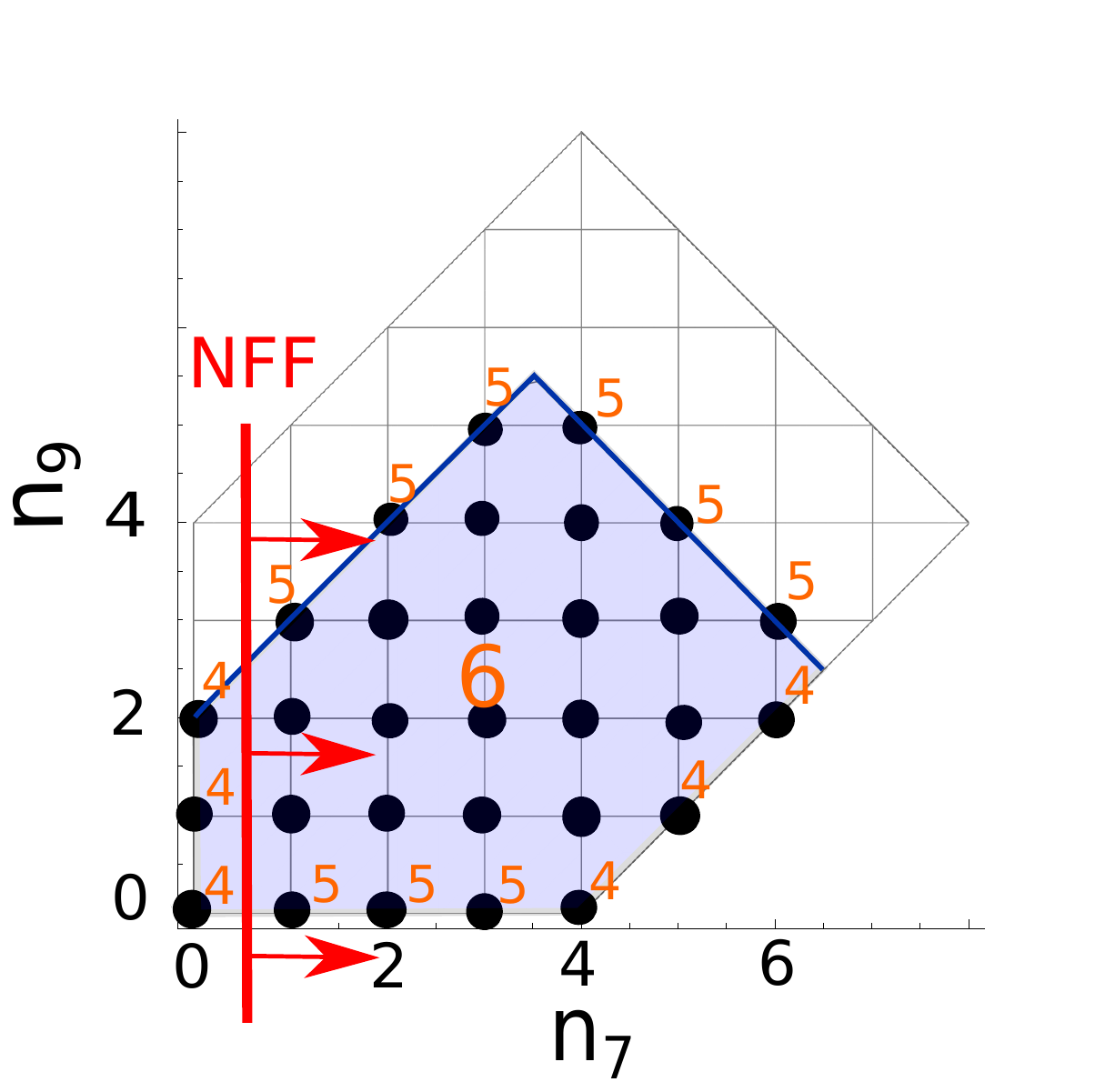} \qquad \qquad
 \caption{There are six independent constraints on the $G_4$-flux from
 the CS-terms. In order to obtain an anomaly free spectrum, for $n_7>0$,
 we force the vanishing condition \eqref{eq:nff-cond}, respectively,
 \eqref{eq:nff-cond2}. The small number in orange next to each point is
 the final number of parameters in $G_4$-flux. The maximal number is six
 in the bulk. }
 \label{fig:SU5n7n9-CF}
\end{figure}

In summary, for $n_7=0$  the vanishing of \eqref{eq:nff-cond} does
not impose any additional conditions on the $G_4$-flux, whereas for for
$n_7>0$ its vanishing implies the  additional condition
\beq \label{eq:nff-cond2}
 a_{12} \stackrel{!}{=} 0.
\eeq
Thus, the number of free parameters in the $G_4$-flux in
\eqref{eq:G4genSU5} is reduced to six.
Furthermore, as we demonstrate in the following imposing the condition
\eqref{eq:nff-cond} will yield a consistent 4D chiral spectrum with
all anomalies cancelled. For a better overview over the conditions we
impose, we summarize the number of independent elements in the $G_4$-flux graphically in figure \ref{fig:SU5n7n9-CF}.

Imposing now the condition \eqref{eq:nff-cond2} we solve the first line
of \eqref{eq:CSwithNFF}. The solutions are straightforwardly obtained,
but lengthy and can be found in appendix \ref{app:fluxesNChiralities}.
However, instead of parametrizing the chiralities in the parameters
$a_i$ of the $G_4$-flux in \eqref{eq:G4genSU5}, we can alternatively
express the chiralities in terms of the chiralities of the $\mathbf{5}$-
representations and one singlet. This will, on the one hand, reduce the
length of the expressions we obtain and, on the other hand, guarantee
that all chiralities are parametrized in terms of integers.\footnote{As
mentioned before, the determination of the integral basis of
$H^{(2,2)}_V(\Xsu)$ is a subtle and difficult problem, that requires the
application of more advanced techniques like mirror symmetry.}
We then obtain the following chiralities,
\begin{align} \label{eq:chi1SU5}
\chi\big(\mathbf{5}_{(-\tfrac{2}{5},0)}\big) &= c_1, &\qquad \chi\big(\mathbf {5}_{(-\tfrac{2}{5},1)}\big) &= c_2 ,&\qquad \chi\big(\mathbf{5}_{(\tfrac{3}{5},0)}\big) &= c_3, \nn \\ \qquad \chi\big(\mathbf{5}_{(\tfrac{3}{5},1)}\big)&=c_4, &\qquad  \chi\big(\mathbf{5}_{(-\tfrac{2}{5},-1)}\big)&=c_5, &\qquad \chi(\mathbf 1_{(-1,-2)})&=c_6,
\end{align}
with integral parameters $c_i$ in terms of which the other chiralities
are expressed as
\bea \label{eq:chisSU5}
\chi\big(\mathbf{10}_{(\tfrac{1}{5},0)}\big)\!\!\! &\!=\!&\!\!\! -c_1 - c_2 - c_3 - c_4 - c_5\,, \quad\qquad \chi(\mathbf 1_{(0,2)}) = c_6 + (c_2 + c_4 - c_5) n_9\,,\nn\\
\chi\big(\mathbf 1_{(1,0)}\big) \!\!\! &\!=\!&\!\!\! -22 c_2\! -\! c_3\! -\! 3 (2 (c_4 \!+\! c_5)\! +\! 3c_6) + 2 (c_2 \!+\! c_4\! -\! c_5) n_7  +
 2 c_1 (n_9\!-6) \!-\! 2 (c_4\! -\! 2 c_5) n_9 \,, \nn \\
\chi(\mathbf 1_{(1,1)}) \!\!\! &\!=\!&\!\!\! -14 c_5 + 3 c_6 + 2 c_5 n_7 - (c_2 + 3 c_5) n_9  + c_4 (5 - n_7 + n_9) +
 c_1 (n_7 - 2 (4 + n_9))\, ,\nn \\
\chi(\mathbf 1_{(0,1)}) \!\!\! &\!=\!&\!\!\! 14 c_4 - 5 c_5 - 2 c_6 + c_2 (21 - 2 n_7) - 2 c_5 n_7 - 2 c_4 n_9 + 4 c_5 n_9 + 
 c_1 (8 - 2 n_7 + 2 n_9)\,, \nn \\
\chi(\mathbf 1_{(-1,1)})\!\!\! &=& \!\!\! -2 c_2 - c_6 + c_1 n_7 + 2 c_2 n_7 + c_4 n_7 - (c_2 + c_4 - c_5) n_9\,.
\eea
We note that the chirality of the representation $\mathbf{10}_{(\tfrac{1}{5},0)}$ has the opposite sign of the $c_i$.  
For positive $c_i$ this would yield a negative chirality of left-handed 
Weyl fermions in 
$\mathbf{10}_{(\tfrac{1}{5},0)}$, which has to be understood as 
left-handed Weyl fermions in  the representation 
$\overline{\mathbf{10}}_{(-\tfrac{1}{5},0)}$. We adapt this convention 
in the following.

\subsubsection{Anomaly Cancellation}
\label{sec:anomaliesSU5}

We conclude this general discussion of the spectrum by
checking the cancellation of 4D anomalies. For a short
review, see section \ref{sec:anomalyCancellation}, or see
\cite{Cvetic:2012xn} for more details.

Having derived the spectrum in the previous
paragraph, all we have to calculate to check anomaly cancellation
are the gaugings $\Theta_{\alpha m}$ for $m=1,m=2$,
cf.~\eqref{eq:4Danomalies}. The results of this computations in terms of
the parameters $a_i$ of the $G_4$-flux can be found  in appendix
\ref{app:fluxesNChiralities}. As mentioned before, we can alternatively
use the chiralities \eqref{eq:chi1SU5} as parameters. Then we obtain
\beq \label{eq:ThetamalphaSU5}
\Theta_{\alpha,m=1}=-2 (c_1 + c_2 + c_5)\,,\qquad
\Theta_{\alpha,m=2}= 2 (c_2 + c_4 - c_5)\,.
\eeq
The coefficients $b_{mn}^\alpha$ of the GS-counter terms
change slightly compared to the U$(1)\times$U(1)-case of section
\ref{sec:anomalyCancellation} because of the altered Shioda map
\eqref{eq:ShiodaMapSQSRSU5}, with and $a^\alpha$ remaining the same.
We obtain
\bea \label{eq:bmnSU5}
	b_{mn}^\alpha&=&\begin{pmatrix}
		\frac{34}{5} & 4-n_7+n_9\\4-n_7+n_9 & 8+2 n_9
	\end{pmatrix}\,,\qquad\qquad  a^\alpha=-4\,.
\eea

We readily check that all anomalies in \eqref{eq:4Danomalies} cancel
beautifully. First, the purely non-Abelian anomaly vanishes trivially
\beq
\chi(\overline{\mathbf{10}}_{(-\tfrac{1}{5},0)})=
\sum_{(q_1,q_2)}\chi(\mathbf{5}_{(q_1,q_2)}),
\eeq
where the sum is over all different $\mathbf{5}$-representations, that
are labeled by their charges $(q_1,q_2)$, and where we used the representation $\overline{\mathbf{10}}_{(-\tfrac{1}{5},0)}$ 
as explained below \eqref{eq:chisSU5}.
The purely Abelian anomalies are
\bea
A^{\text{U(1)}}_{1,1,1} &:&\,\, -\tfrac{17}{5} (c_1 + c_2 + c_5), \nn \\
A^{\text{U(1)}}_{1,1,2}&:&\,\, \tfrac{1}{15} (17 c_4 - 37 c_5 + 5 c_5 n_7 + c_2 ( 5 n_7 - 5 n_9-3) +
 \,\,  5 c_1 (  n_7 - n_9-4) - 5 c_5 n_9), \nn \\
A^{\text{U(1)}}_{1,2,2}&:& \tfrac{1}{3} (-8 c_5 - c_2 n_7 + c_5 n_7 - 2 c_5 n_9 - c_1 (4 + n_9) + c_4 (4 - n_7 + n_9)), \nn \\
A^{\text{U(1)}}_{2,2,2}&:&\,\, (c_2 + c_4 - c_5) (4 + n_9),
\eea
The mixed non-Abelian-Abelian, and Abelian-gravitational anomalies
read
\beq
A^{\text{U(1)-SU(5)}}_{1}=A^{\text{U(1)-grav}}_{1}:\,\tfrac{1}{2} (-c_1 - c_2 - c_5)\,,\qquad
A^{\text{U(1)-SU(5)}}_{2}=A^{\text{U(1)-grav}}_{2}:\,\tfrac{1}{2} (c_2 + c_4 - c_5)\,.
\eeq
Using the gaugings \eqref{eq:ThetamalphaSU5} and the coefficients
\eqref{eq:bmnSU5} of the GS-terms, we confirm that the 4D anomaly
cancellation conditions \eqref{eq:4Danomalies} hold.

\subsection{A Toric Example}
\label{sec:ExamplesSU5U1U1}

In this concluding subsection we construct explicitly a toric model with
all characteristics described in this section, i.e.~an
$SU(5) \times U(1)^2$ 4D gauge group, the full spectrum computed in
sections \ref{sec:SingularitiesSU5U1U1}, \ref{sec:ChiralitiesU1xU1xSU5}
and with a non-flat fiber.

The concrete fourfold $\Xsu$ we construct has $n_7=n_9=4$ in figure
\ref{fig:XP3SU5n7n9region} and has $\Ssu=H_B$. It is engineered following
the general guideline of appendix \ref{app:tuning-s7-s9}. The vertices
specifying the reflexive polytope, along with their associated homogeneous coordinates and divisor classes, are as follows,
\beq \label{eq:ExampleTD3}
\text{
 \begin{tabular}{|c||c|c|c|c|c||c|} \hline
  variable  & \multicolumn{5}{c||}{vertices}	& divisor class   \rule{0pt}{13pt}\\  \hline
  $z_0$ & 1  & 1  & 1 & 0  &  0 	& $H_B$ 		 	\\
  $z_1$ & -1 & 0 & 0  & 0  &  1	& $H_B$ 		 	\\
  $z_3$ & 0& -1 & 0  & -1  &  -1	& $H_B$ 		 	\\
  $z_2$ & 0 & 0  & -1 & 1  &  0	& $H_B-D_1-D_2-D_3-D_4$ 			\\ \hline
  $d_1$ & 0& 0  & -1 & 0  &  0	& $D_1$		 	\\
  $d_2$ & 0&0  & -1 & -1  &  -1	& $D_2$		 	 \\
  $d_3$ & 0&0  & -1 & -1  &  -2 	& $D_3$			\\
  $d_4$ & 0&0  & -1 & 0  &  -1	& $D_4$		 	\\ \hline
  $u$  & 0&0  & 0  & 1  &  0 	& $-3H_B+H-D_2-D_3-E_1-E_2$ 		\\
  $v$ & 0&0  & 0  & 0  &  1 	& $H-D_1-D_2-D_3-D_4-E_2$ 			\\
  $w$ & 0&0  & 0  & -1 & -1		& $H-D_1-2D_2-3D_3-2D_4-E_1$ 		 	\\
  $e_1$ & 0&0  & 0  & 0  & -1 	& $E_1$			\\
  $e_2$ & 0&0  & 0  & 1  & 1 	& $E_2$ 			\\ \hline
 \end{tabular}
 }
\eeq
Here we follow the notation of section \ref{sec:SingularitiesSU5U1U1},
i.e.~$H_B$ is the hyperplane of $\mathbb{P}^3$, the $D_i$ are the Cartan
divisors of the SU(5) and $H$, $E_1$, $E_2$ are the classes of the
$dP_2$-fiber.

The Euler number is calculated as
\beq
\chi(\hat X_{SU(5)})=1110
\eeq
which agrees  perfectly with general formula \eqref{eq:EulerXSU5}, where
we have used $\chi(\hat X)=1632$ and $\cS_{SU(5)}=H_B$, $\cS_{7}=\cS_{9}=
4 H_B$. For the construction of the corresponding toric variety we have
to choose  the right triangulation so that the toric Stanley-Reissner
ideal is of the same from as the  Stanley-Reissner ideal in
\eqref{eq:SR-SU5}. The
toric divisors of the sections are given by \eqref{eq:SPSQSRP3} for
$n_9=4$ and  the Shioda maps \eqref{eq:ShiodaMapSQSRSU5} read
\bea
\sigma(\hat s_Q) &=& S_Q-S_P-4 H_B+\tfrac{1}{5}(2 D_1+4D_2+6D_3+3D_4) \\
\sigma(\hat s_R) &=& S_R-S_P-8H_B;
\eea

The full basis of  $H^{(2,2)}_V$ is given by the 13 elements in equation
\eqref{eq:basisH22Su5}.
The $G_4$-flux follows from \eqref{eq:G4genSU5} by inserting
$n_7=n_9=4$. Since we have $n_7\neq 0$, we have to impose the non-flat
fiber condition \eqref{eq:nff-cond2}. Our main interest lies in the
chiralities, which are then either computed torically or follow from
the general formulas in \ref{sec:ChiralitiesU1xU1xSU5} and appendix
\ref{app:fluxesNChiralities} as
\bea
\chi\big(\mathbf 5_{(-\tfrac{2}{5},0)}\big) &=& \tfrac{1}{2}a_{3} - \tfrac{2}{25} (40 a_{6} + 63 a_{7} + 75 a_{13})\,,\nn\\
\chi\big(\mathbf 5_{(-\tfrac{2}{5},1)}\big) &=&  \tfrac{1}{125} (50 a_{3} - 125 a_{4} + 80 a_{6} - 174 a_{7})\,,\nn \\
\chi\big(\mathbf 5_{(\tfrac{3}{5},0)}\big) &=& \tfrac{1}{250} (-225 a_{3} + 1640 a_{6} + 1888 a_{7})\,,\nn\\
\chi\big(\mathbf 5_{(\tfrac{3}{5},1)}\big) &=&  -\tfrac{2}{5} (-5 a_{3} + 5 a_{4} + 16 a_{6} + 38 a_{7})\,,\nn\\
\chi\big(\mathbf 5_{(-\tfrac{2}{5},-1)}\big) &=&  \tfrac{4}{5}a_{3} + 2 a_{4} + 8 a_{5} - \tfrac{468}{25} a_{6} - \tfrac{1608}{125} a_{7} +
 4 a_{13}\,,\nn \\
\chi(\mathbf 10_{(\tfrac{1}{5},0)}) &=&  -\tfrac{1}{5}a_{3} + \tfrac{32}{25} a_{6} + \tfrac{242}{125} a_{7} - 6 a_{13}\,,
\eea
for the non-trivial non-Abelian representations and
\bea
\chi(\mathbf 1_{(1,0)})&=&  -\tfrac{27}{2} a_{3} - 32 a_{5} + \tfrac{596}{5}a_{6} + \tfrac{3128}{25} a_{7} + 8 a_{13}\,, \nn \\
 \chi(\mathbf 1_{(1,1)})&=&  -13 a_{3} - 13 a_{4} - 88 a_{5} + 200 a_{6} + \tfrac{4226}{25} a_{7} + 16 a_{13}\,,  \nn \\
 \chi(\mathbf 1_{(0,1)})&=& -29 a_{4} - 88 a_{5} + \tfrac{836}{5} a_{6} + \tfrac{2006}{25} a_{7} + 28 a_{13}\,, \nn \\
\chi(\mathbf 1_{(-1,1)})&=& -\tfrac{2}{5} (-5 a_{3} + 5 a_{4} + 16 a_{6} + 38 a_{7})\,, \nn  \\
\chi(\mathbf 1_{(0,2)})&=& -8 a_{4} - 32 a_{5} + \tfrac{256}{5} a_{6} + \tfrac{648}{25} a_{7} + 8 a_{13}\,,  \nn \\
 \chi(\mathbf 1_{(-1,-2)})&=& 4 a_{3} + 8 a_{4} + 32 a_{5} - 80 a_{6} - \tfrac{296}{5} a_{7} - 8 a_{13}\,,
\eea
for the singlets.
As before, it is convenient to parametrize the chiralities
as in the previous subsections, in terms of the $c_i$.
We then obtain
\bea
\chi\big(\mathbf 5_{(-\tfrac{2}{5},0)}\big) \!&\!\!=\!\!&\! c_1\,, \qquad \chi\big(\mathbf 5_{(-\tfrac{2}{5},1)}\big) = c_2 \,,\qquad \chi\big(\mathbf 5_{(\tfrac{3}{5},0)}\big) = c_3\,,\!\!\!  \\ \qquad \chi\big(\mathbf 5_{(\tfrac{3}{5},1)}\big)\!&\!\!=\!\!&\!c_4\,,\,\,\, \quad  \chi\big(\mathbf 5_{(-\tfrac{2}{5},-1)}\big)=c_5\,,\,\quad \chi(\mathbf 1_{(-1,-2)})=c_6\,, \nn  \\
\chi(\mathbf 10_{(\tfrac{1}{5},0)}) \!&\!\!=\!\!&\! -c_1\! - c_2\! - c_3\! - c_4\! - c_5\,,\quad\,\,
\chi(\mathbf 1_{(1,0)}) = -4 c_1\! - 14 c_2\! - c_3\! - 6 c_4 + 2 c_5 - 3 c_6\,,\!\!\!\nn\\
\qquad \chi(\mathbf 1_{(0,1)}) &=& 8 c_1 + 13 c_2 + 6 c_4 + 3 c_5 - 2 c_6\,,\quad\,\,\chi(\mathbf 1_{(-1,1)}) = 4 c_1 + 2 c_2 + 4 c_5 - c_6\,, \nn \\
\chi(\mathbf 1_{(0,2)}) \!&\!\!=\!\!&\!  4 c_2 + 4 c_4 - 4 c_5 + c_6\,,\quad
\chi(\mathbf 1_{(1,1)}) = -12 c_1 - 4 c_2 + 5 c_4 - 18 c_5 + 3 c_6. \nn
\eea
It follows that all 4D anomalies are cancelled.

\section{Conclusions and Future Directions}
\label{sec:conclusions}

In this paper we have advanced the  program  on F-theory
compactifications on elliptic Calabi-Yau manifolds with rank two
Mordell-Weil group to four-dimensional chiral compactifications.

The analysis of resolved Calabi-Yau elliptic  fibrations with
$dP_2$-elliptic fiber has now been  performed for Calabi-Yau fourfolds
over a general three dimensional base $B$, extending earlier results
in six dimensions \cite{Borchmann:2013jwa,Cvetic:2013nia}.  We study general
compactifications with $U(1)\times U(1)$ and a specific
$SU(5)\times U(1)\times U(1)$ gauge
symmetry.  We determined the general matter representations associated
with the codimension two singularities of the fibration as well as its
Yukawa points.  However, at present the geometric techniques, relying
on the use prime ideals, that we employed allow us to determine all
matter curves and only a subset of matter surfaces.
As a next step, we  determine explicitly  the general  $G_4$-flux
of these resolved Calabi-Yau fourfolds. Since we considered  Calabi-Yau
fourfolds with a non-holomorphic zero section, we encountered
the novel problem of having to define $G_4$-flux over these manifolds.
For this purpose we had to derive the general F-theory conditions
on $G_4$-flux on Calabi-Yau manifolds with a non-holomorphic zero
section. We have formulated these conditions in terms of Chern-Simons
terms on the Coulomb branch of the effective theory obtained by
compactifying on a circle. The relevance of Kaluza-Klein states
generating new Chern-Simons terms in order to ensure a consistent
M-/F-theory duality mapping in 3D has been pointed out

After this interlude on $G_4$-fluxes in F-theory we constructed
the most general $G_4$-flux. We presented explicit calculations for the
three dimensional base $B=\mathbb{P}^3$. In both the U$(1)\times$U(1)
and SU$(5)\times$U$(1)\times$U(1) case we could determine certain
chiralities directly by evaluating the chiral index \eqref{eq:chi(R)}
directly while the remaining matter chiralities were determined by a
subset of the 3D Chern-Simons terms.  It turned  out that   Chern-Simons
terms are sufficient to determine chiralities of all the matter
multiplets if also the new, Kaluza-Klein generated Chern-Simons terms are
taken into account. The geometric techniques allowed an important,
independent consistency check.  The complications of non-flat fibers
could be successfully circumvented by imposing one additional condition
on the $G_4$-flux, namely the vanishing of the integral of the $G_4$-flux
over the non-flat fiber.  We also presented concrete explicit examples
using toric geometry. We emphasize that our techniques are in line,
but not restricted to toric geometry. In particular, we could calculate
the most general $G_4$-flux and chiralities, along with the proof of
anomaly cancellation, for the entire class of all elliptically fibered
Calabi-Yau fourfolds, cf.~Figure \ref{fig:XP3n7n9region} and Figure
\ref{fig:XP3SU5n7n9region},
with $dP_2$-elliptic fiber,   an additional SU(5)
GUT-sector and a fixed base $B$, here chosen to be $B=\mathbb{P}^3$
for simplicity.

There  paper leaves room for a number of further studies and
improvements:

$\bullet$ The geometric techniques developed  here apply to general
elliptic Calabi-Yau fourfolds, and they are {\it not}  restricted to
toric examples. In particular, we could derived closed formulas for 
e.g.~the Euler number, the $G_4$-flux and the 4D chiralities that explicitly
depend on the divisors $\cS_7$, $\cS_9$, introduced in Section \ref{sec:GeneraldP2fibrations}, which label 
the members in the family of
all inequivalent Calabi-Yau fourfolds obtained by varying the topologically data specifying 
their elliptic fibration. It would be desirable to obtain these formula   
for an arbitrary base $B$.
Our results are independent of the existence of toric realizations of the Calabi-Yau fourfolds. We emphasize that employing toric geometry techniques one 
can typically study only one polytope at a time, which obscures the dependence on $\cS_7$,
$\cS_9$. 

$\bullet$ The finite number of choices for the divisors $\cS_7$, $\cS_9$ on 
the F-theory side fix the topology of the fibration of $dP_2$ over the base $B$. 
It would be interesting to understand these new degrees of freedom from the point
of view of the heterotic string. In particular it would be important to
understand how the parameters in $\cS_7$, $\cS_9$ enter the heterotic vector bundle
and potentially modify the spectral cover construction \cite{workinprogress1}.
 
$\bullet$ Although all 
4D chiralities could still be derived by a combination of geometric and 
field theoretic techniques, we were at this point
unable to determine explicitly the matter surfaces for a subset of representations.  
It would be desirable to derive these missing 
homology classes purely geometrically by further extending the geometric techniques 
of Section \ref{sec:codim2U1xU1}. For this analysis important 
lessons might be learned from a geometrical interpretation of the field theoretically 
derived $G_4$-flux conditions in Section \ref{sec:G4constraintsF}.

$\bullet$ While explicit results were presented for a specific three-dimensional
base $B=\mathbb{P}^3$, it would be important to present
similar computations for  systematically classified three-dimensional
bases, extending a similar analysis in 6D \cite{Morrison:2012js,Morrison:2012np}. These studies 
are also motivated by a search for  $SU(5)$ GUT
models with  flat fibers.

$\bullet$ In this paper we have not  systematically performed a
classification of $SU(5)$  GUT models. We presented models for a specific
$SU(5)$ construction  with a specific base $B=\mathbb{P}^3$, only.  The
example does not have a flat fibration at codimension three, which may
lead  to an infinite tower of massless states (a so-called
tensionless string spectrum).  We removed chiral states in the tower by
further constraining the $G_4$-flux. It would be important to classify
SU$(5)$ constructions by carrying over the Tate classification to
$dP_2$-elliptic fibrations, and to construct a base $B$ where the
fibration can be engineered to be flat. For efforts in these directions,
primarily employing toric techniques, see
\cite{Borchmann:2013jwa,Braun:2013nqa}.  
We note that the classification of 
SU$(n)$-singularities for general $n$ should work similarly to the SU(5) 
case and thus will be facilitated by techniques developed  in this work.

$\bullet$ The techniques developed in this paper pave the way to
phenomenological studies of chiral four-dimensional models with rank-two
Abelian sectors. However in order to achieve these goals a number of
further details have to be addressed.  First, the quantization of
$G_4$-flux is not well understood in general. However, as pointed out in Section 4
we can choose the chiralities of our matter multiplets
to parameterize the spectrum which should in turn also yield integer
values for the flux-quanta in  the $G_4$-flux.  Second, the constraints imposed by
the self-duality condition on the $G_4$-flux  and  by D3-brane tadpole cancellation have to 
be investigated.  Those are important topics for future research.

\subsubsection*{Acknowledgments}
We would like to thank Jim Halverson, Albrecht Klemm, Wolfgang Lerche,
Song Peng and in particular Thomas W. Grimm  for discussions and
comments. M.C.~and D.K.~are grateful to  the Theory Division of CERN for
hospitality during completion of the project.  H.P is grateful to TASI and the University of Colorado, Boulder for hospitality. This research is supported in part by
the DOE grant DE-SC0007901 (M.C., H.P., D.K.), the NSF String Vacuum Project Grant No. NSF PHY05-51164 (H.P.), Dean's Funds for Faculty Working Group (M.C. and D.K.),
the Fay R. and Eugene L.Langberg Endowed Chair (M.C.) and the Slovenian Research Agency
(ARRS) (M.C.).


\appendix


\section{Chern Classes of $dP_2$-Elliptic Fibrations}
\label{app:CohomologyRing}

In this appendix we analyze the total Chern-class of the total space of $dP_2$-fibrations $dP_2^B(\cS_7,\cS_9)$
over a general base $B$. The following discussions hold in any complex dimension, although we specialize to Calabi-Yau
two-, three- and fourfolds in the following.

The final goal of this section is the derivation of general formulas
for the Euler number of smooth Calabi-Yau manifolds $\hat{X}$ of
complex dimension two, three and four in $dP^B_2(\cS_7,\cS_9)$.
For the reader only interested in these formulas, we
first state the results of the analysis and refer to the
remainder of this section for a detailed derivation. The Euler numbers
for Calabi-Yau two-, three- and fourfolds are expressed as the following
integrals over the one-, two- respectively, three-dimensional base $B$ of
the Calabi-Yau manifold $\hat{X}$,
\bea \label{eq:EulernumbergenX}
	&d=2\,:&\quad \chi(X)=12 \int_B c_1\nn\\
	&d=3\,:&\quad \chi(X)=\int_B(-24 c_1^2 + 8 c_1 \cS_7 - 4 \cS_7^2 + 8 c_1 \cS_9 + 2 \cS_7 \cS_9 - 4 \cS_9^2)\nn\\
	&d=4\,:&\quad \chi(X)=3\int_B\left[24 c_1^3 + 4c_1 c_2 - 16 c_1^2 (\cS_7+ \cS_9)+  c_1 (8\cS_7^2  + \cS_7 \cS_9 + 8 \cS_9^2)\right.\nn\\
	&&\qquad\qquad\,\, \left.-  \cS_7 \cS_9(\cS_7  +  \cS_9)\right]
\eea
Here we denoted the Poincar\'e dual $(1,1)$-forms  by abuse of notation by
the same symbol as their corresponding divisors $\cS_7$, $\cS_9$ in the
base $B$. Furthermore, wedge-products have been omitted for brevity of our
notation.
We emphasize that these formulas for $\chi(\hat{X})$ are a
direct generalization of the formulas in \cite{Klemm:1996ts} for
fibrations by the $E_6$ elliptic curve, that are obtained as the special
case $\cS_7=\cS_9\equiv 0$.

We prepare the derivation of the Euler numbers \eqref{eq:EulernumbergenX}
by a computation of the total Chern class of the ambient space
$dP_2^B(\cS_7,\cS_9)$.
The total space of the $dP_2$-fibration $dP_2^B(\cS_7,\cS_9)$ in \eqref{eq:dP2fibration} has the structure of a generalized
projective bundle: the two-dimensional fiber is $dP_2$ and it can be understood as the projectivization of a rank five
vector bundle over $B$ by three $\mathbb{C}^*$-actions. Consequently, the Chern class of the $dP_2^B(\cS_7,\cS_9)$ is
calculated analogous to the Chern-class of an ordinary projective bundle by adjunction. Denoting the projective
coordinates of the general $dP_2$ fiber by $[u:v:w:e_1:e_2]$ we follow the assignments \eqref{eq:dP2divs} of
line bundles over the base $B$ from the main text. Thus we obtain for the total Chern-class of $dP_2^B(\cS_7,\cS_9)$,
\beq \label{eq:c(dP2B)}
	c(dP_2^B)=
	c(B) (1 + H - E_1 - E_2 + \cS_9 -[K_B^{-1}]) (1 + H - E_2 + \cS_9 - \cS_7) (1 +  H - E_1)
	(1 +  E_1) (1 + E_2)\,.
\eeq
Here we suppressed the dependence on $\cS_7$, $\cS_9$ for brevity of our expression. By abuse of notation we denote
the divisors $E_i$ and $H$ and the first Chern-classes of their associated divisor bundles by the same symbol.
In addition, we used $c(B)=1+c_1(B)+c_2(B)+c_3(B)+\ldots$ to denote the total Chern class of $B$.  We readily expand
\eqref{eq:c(dP2B)} order by order to obtain the following Chern classes of $dP_2^B(\cS_7,\cS_9)$,
\bea \label{eq:cisdP2B}
	c_1(dP_2^B)\!\!\!&\!\!=\!\!\!&\!\! 3 H-E_1 - E_2+ 2 \cS_9- \cS_7\,,\\
    c_2(dP_2^B)\!\!\!&\!\!=\!\!\!&\!\! c_2-c_1^2 - 2 E_1^2 - 3 E_2^2 + 2 E_1 \cS_7 - 2 E_2 \cS_7 - E_1 \cS_9 + 4 E_2 \cS_9 -
                   H \cS_9 + \cS_7 \cS_9 - \cS_9^2 \nn\\
              \! \!&+\!&\!\! c_1 (4 H-2 E_1 - 3 E_2   - 2 \cS_7 + 3 \cS_9)\,,\\
    c_3(dP_2^B)\!\!\!&\!\!=\!\!\!&\!\!c_3 - c_2 (E_1 + E_2 - 3 H + \cS_7 - 2 \cS_9) +
 c_1^2 (H-E_1 - 2 E_2 - \cS_7 + \cS_9) \nn\\
 \!\!&-\!&\!\!  c_1 (c_2 + 2 E_1^2 + 3 E_2^2 +2\cS_7( E_2-  E_1) + \cS_9(H+E_1
 - 4 E_2 - \cS_7 + \cS_9))\,,\\
 c_4(dP_2^B)\!\!\!&\!\!=\!\!\!&\!\!c_3 (3 H-E_1 - E_2 ) +
 c_2 ( \cS_7(2 E_1- 2 E_2) + \cS_9(4 E_2 - E_1  - H ) -2 E_1^2 - 3 E_2^2  )\nn\\
 \!\!&+\!&\!\!  c_1 c_2 (H-E_1 - 2 E_2 )\,,\\
 c_5(dP_2^B)&=&-c_3 (2 E_1^2 + 3 E_2^2)=5c_3\vert_B\,.
\eea
Here we have used the notation $c_i\equiv c_i(B)$. All divisors in the
above expressions are understood as their Poincar\'e dual $(1,1)$-forms,
that are given as the first Chern class  of the
associated divisor line bundle, e.g.~$c_1(\mathcal{O}(E_1))$ for
$\mathcal{O}(E_1)$.  The expressions for the first to third Chern classes
are completely general, whereas we have assumed $dim_\mathbb{C}(B)=3$ for
the fourth and fifth Chern class to be able
to drop terms containing more than three vertical divisors.  In the last
line we have used the toric intersections $E_1^2=E_2^2=-1$ in
\eqref{eq:dP2ints} since $[c_3]\cong(\int_B c_3) pt$ with $pt$ denoting a
point on $B$. All expressions in \eqref{eq:cisdP2B} have been reduced
modulo the ideal $SR'$ in \eqref{eq:SRintsdP2B}.

Now we are in the position to calculate the total Chern class of the
Calabi-Yau fourfold $\hat{X}$. For this purpose we apply adjunction
to obtain
\beq
	c(\hat{X})=\frac{c(dP_2^B)}{1+3 H-E_1 - E_2+ 2 \cS_9- \cS_7}	
\eeq	
where the numerator is to be computed using \eqref{eq:c(dP2B)} and the denominator is the Chern class of the
anti-canonical bundle of $dP_2^B(\cS_7,\cS_9)$, cf.~the first line in \eqref{eq:cisdP2B}, employing that $\hat{X}$ is
the anti-canonical divisor. Expanding out this expression for the total Chern class of the fourfold $\hat{X}$ yields the following
individual Chern classes $c_i(\hat{X})$,
\footnotesize
\bea \label{eq:ChernClassesX}
	c_1(\hat{X})\!\!&\!\!=\!\!&\!\! 0\,,\nn\\ \nn \\
	c_2(\hat{X})\!\!&\!\!=\!\!&\!\! 3 c_1^2 + c_2 - 2 S_Q^2 - 3 S_P^2 +\cS_7(2 S_Q  - 2 S_P)+\cS_9(
 3 S_P- 2 S_Q -S_R +  \cS_7) \nn\\
\! \!&+\!&\!\!   c_1 (2 S_Q + S_P - 2 (\cS_7 + \cS_9 - 2 S_R)) \,,\nn\\ \nn\\
 c_3(\hat{X})\!\!&\!\!=\!\!&\!\! c_3 -8 c_1^3-c_1c_2 + \cS_7 (2 S_Q^2 - 2 S_P^2  - 2 S_Q \cS_7 - 2 S_P \cS_7 +
 2 S_P  \cS_9 + \cS_7 \cS_9 +  \cS_9^2) \nn\\
  \! \!&+\!&\!\!
  c_1 (  7 S_P^2 + 2 S_Q \cS_7 + 6 S_P \cS_7 - 2 \cS_7^2 + 8 S_Q \cS_9
 - 7 S_P \cS_9 - 6 \cS_7 \cS_9 - 2 \cS_9^2 + 7 \cS_9 S_R)\nn\\
\! \!&-\!&\!\! c_1^2 (8 S_Q + S_P - 8 (\cS_7 + \cS_9 - S_R))- 2 S_Q \cS_9^2 - 2 S_R\cS_9^2 \,, \nn\\ \nn \\
     c_4(\hat{X})\!\!&\!\!=\!\!&\!\! c_1^2 (8 S_Q^2 - 32 S_P^2 - \cS_7(22 S_Q  - 18 S_P ) + \cS_9(41 S_P - 41 S_Q   -
    40  S_R)) -2 S_Q^2 \cS_7^2 \nn\\
    \! \!&-\!&\!\!\! 2 S_P^2 \cS_7^2+ 2 S_Q \cS_7^3 - 2 S_P \cS_7^3 + 3 S_Q^2 \cS_7 \cS_9 -
 S_P^2 \cS_7 \cS_9 - 5 S_Q \cS_7^2 \cS_9 + 2 S_P \cS_7^2 \cS_9 + 2 S_Q^2 \cS_9^2  \nn\\
 \! \!&-\!&\!\! \!
 6 S_P^2 \cS_9^2- 2 S_Q \cS_7 \cS_9^2 - S_P \cS_7 \cS_9^2 - 2 S_Q \cS_9^3 + 6 S_P \cS_9^3 -
 4 \cS_9^3 S_R\! + 8 c_1^3 (4 S_Q\! - S_P\! + 3 S_R) \nn \\
 \! \!&-\!&\!\! \!
 c_2 (2 S_Q^2 + 3 S_P^2 - 2 S_Q \cS_7 + 2 S_P \cS_7 + 2 S_Q \cS_9 - 3 S_P \cS_9 +
    \cS_9 S_R) \nn\\
    \! \!&+\!&\!\! \!
 c_1 (\cS_7(14 S_P^2  + 12 S_P \cS_7- 11 S_P \cS_9) + 17 S_P^2 \cS_9  -
    19 S_P \cS_9^2 - 3 S_Q^2 (2 \cS_7 + 3 \cS_9)\nn \\
    \! \!&+\!&\!\! \!
    8 S_Q (\cS_7^2 + 2 \cS_7 \cS_9 + 2 \cS_9^2) + 23 \cS_9^2 S_R +
    c_2 (2 S_Q + S_P + 4 S_R))\,.
\eea
\normalsize
Here we have dropped terms quartic in divisors on the base $B$ and have reduced all expressions in the quotient
ring \eqref{eq:abstractCohomRing}. For the evaluation of \eqref{eq:ChernClassesX} it proves convenient to express all
the elements in terms of the divisor classes $S_P=E_2$, $S_Q=E_1$ and
$S_R=H-E_1-E_2+\cS_9+[K_B]$ of the sections of
$\hat{X}$. Then, we can make use of the intersection relations for the sections
\beq \label{eq:Sm^2}
	(S_m^2+[K^{-1}_B]\cdot S_m)\cdot D_\alpha\cdot D_\beta=0\,,
\eeq
where we collectively denote the divisor classes of the sections by $S_m$. We note that this relation is a
direct consequence of the property of the $S_m$ being the classes of sections, as noted before in \eqref{eq:SP^2}, but it is satisfying
that it can also be proven directly in the intersection ring \eqref{eq:abstractCohomRing}.

In the following we assume that $\hat{X}$ is either complex two-, three- and four-dimensional and for each case employ
\eqref{eq:Sm^2} and its lower dimensional analogs to simplify the Chern classes in \eqref{eq:ChernClassesX} even further.
We obtain for the top Chern class $c_d(\hat{X})$ on a $d$-dimensional Calabi-Yau manifold $\hat{X}$ the following results,
\footnotesize
\bea	 \label{eq:ChernClassesd=234}
	d=2\,:\quad  c_2(\hat{X})&=& 4 c_1 (S_Q + S_P + S_R)+  2\cS_7( S_Q- S_P ) + \cS_9 (3 S_P  -  S_R- 2 S_Q)\,,\nn\\
	d=3\,:\quad c_3(\hat{X})&=& -2\cS_7^2( S_Q + S_P) +2\cS_7\cS_9 S_P-2\cS_9^2( S_Q + S_R)
	-8 c_1^2(S_Q + S_P + S_R)\nn\\
	&+&	c_1(8 S_P \cS_7 + 8 S_Q \cS_9 - 7 S_P \cS_9 + 7 \cS_9 S_R)\,,\nn\\
	d=4\,:\quad c_4(\hat{X})&=&2\cS_7^3 ( S_Q -  S_P)+ \cS_7^2 \cS_9( 2 S_P-5 S_Q)-\cS_7 \cS_9^2(2 S_Q + S_P)-2\cS_9^3 (S_Q - 3 S_P + 2 S_R)\nn\\
	&+& c_1(14 S_P \cS_7^2 - 10 S_P \cS_7 \cS_9 - 13 S_P \cS_9^2 +
 S_Q (10 \cS_7^2 + 13 \cS_7 \cS_9 + 14\cS_9^2) + 23 \cS_9^2 S_R)\nn\\
 &+&4c_1  c_2 (S_Q + S_P + S_R)-8c_1^2 (4 S_P \cS_7 - 3 S_P \cS_9 + 2 S_Q (\cS_7 + 2 \cS_9) + 5 \cS_9 S_R)\nn\\
 &+&24 c_1^3 (S_Q + S_P + S_R)+c_2(2 S_Q (\cS_7 - \cS_9) -2 S_P \cS_7 + 3 S_P \cS_9 - \cS_9 S_R)
\eea
\normalsize
where we now in addition also dropped intersections of more than  $d-1$ vertical divisors.

Finally, we are in the position to calculate the Euler number for $\hat{X}$. We obtain the Euler number
for $\hat{X}$ being a twofold, i.e.~$K3$, a Calabi-Yau threefold and fourfold by integrating the above appropriate
Chern classes in \eqref{eq:ChernClassesd=234} over $\hat{X}$.
Using that $S_m\cdot D_\alpha\cdot D_\beta\cdot D_\gamma=(D_\alpha\cdot D_\beta\cdot D_\gamma)\vert_B$ for a fourfold $\hat{X}$,
respectively, analogous relations for the two- and threefold case,
we obtain that all integrals reduce to integrals over the base $B$.
By some algebra we immediately reproduces the
results anticipated in \eqref{eq:EulernumbergenX}.

As another application of the explicit formulas \eqref{eq:ChernClassesX}
for the fourfold $\hat{X}$ we calculate the Todd class
$\text{Td}_4(\hat{X})$ of $\hat{X}$. Upon integrating the  Todd class over
$\hat{X}$ we obtain the arithmetic genus $\chi_0(\hat{X})$ by means of an
Hirzebruch-Riemann-Roch index theorem as
\beq
	\chi_0(\hat{X}):=\sum_p (-1)^{p}h^{(p,0)}(\hat{X})=\int_{\hat{X}}\text{Td}_4(\hat{X})\,.
\eeq
In the case of a Calabi-Yau fourfold, see cf.~\cite{Klemm:1996ts} for more
details, the Todd class is given as
\beq
	\text{Td}_4(\hat{X})=\frac{1}{720}\left(3c_2(\hat{X})^2-c_4(\hat{X})\right)
\eeq
in terms of the Chern classes $c_2(\hat{X})$ and $c_4(\hat{X})$.
Thus, the arithmetic genus reads
\bea \label{eq:ToddX}
	\chi_0(\hat{X})=\int_{\hat{X}}\text{Td}_4(\hat{X})=\frac{1}{12}\int_B c_1 c_2=2\chi_0(B)\,.
\eea
Here we have evaluated the second equality equality using
the expression \eqref{eq:ChernClassesX} for the fourth Chern class
of $\hat{X}$ and  computed the square of the second Chern class
$c_2(\hat{X})^2$ as
\footnotesize
\bea \label{eq:c2X^2}
	c_2(\hat{X})^2&=&S_P\left[ 4 \left(2 c_1-\cS_7\right) \left(c_1^2+c_2-c_1 \cS_7\right)+\left(6 c_2+\left(2 c_1-\cS_7\right)
	\left(5 c_1+\cS_7\right)\right) \cS_9+\left(\cS_7-3 c_1\right) \cS_9^2+\cS_9^3\right]\nn\\
	&+&2S_Q\left[4 c_1^3+2 c_2 \left(\cS_7-\cS_9\right)-\cS_7 S_9^2-2 c_1^2 \left(\cS_7+3 \cS_9\right)+2 c_1 \left(2 c_2+\cS_7^2+\cS_7 \cS_9+\cS_9^2\right)\right]\nn\\
	&+&S_R\left[\left(4 c_1-\cS_9\right) \left(2 \left(c_1^2+c_2\right)-3 c_1 \cS_9+\cS_9^2\right)\right]\,,
\eea
\normalsize
after some algebra in the cohomology ring \eqref{eq:abstractCohomRing}.
In the last equality in \eqref{eq:ToddX} we employed the relation
\beq \label{eq:ToddB}
	\chi_0(B)=\frac{1}{24}\int_B c_1\, c_2
\eeq
for the arithmetic genus of the base $B$. We observe a surprising
cancellation of all terms involving $\cS_7$, $\cS_9$ in the fourth Chern
class in \eqref{eq:ChernClassesX} and in $c_2(\hat{X})^2$ in
\eqref{eq:c2X^2} so that the final result in \eqref{eq:ToddX} only
depends on the Chern classes of the base $B$.

We conclude that for a suitable base $B$ of a simply-connected
Calabi-Yau fourfold with $\chi_0(\hat{X})=2$, which follows from
$h^{(0,0)}(\hat{X})=h^{(4,0)}(\hat{X})=1$ and $h^{(p,0)}(\hat{X})=0$
otherwise, we have to require $\chi_0(B)=1$ or, using \eqref{eq:ToddB},
\beq
	\int_B c_1\, c_2=24\,.
\eeq
We even require the stronger conditions
$h^{(1,0)}(B)=h^{(2,0)}(B)=h^{(3,0)}(B)=0$ for $B$, since every
non-trivial element in these Hodge cohomology groups would give
rise via the pullback under $\pi:\,\hat{X}\rightarrow B$ to a
corresponding element in the Hodge cohomology of $\hat{X}$.
However, this is excluded by assumption of a simply-connected
fourfold.

\section{Intersection Ring on $\hat{X}$ with  $B=\mathbb{P}^3$}
\label{app:intsP3}

In this section we work out the intersection ring
$H^{(*,*)}_V(\hat{X})$ of the fourfold $\hat{X}$ over the
base $B=\mathbb{P}^3$. To this end,
we relate the intersections on $\hat{X}$  to intersections on
the ambient space $dP_2(n_7,n_9)$. On the ambient space
$dP_2(n_7,n_9)$ the intersections are determined completely
by the intersections of the fiber $dP_2$, the base $\mathbb{P}^3$
and basic properties of the fibration. This will allow us,
as demonstrated below and in the main text, to work out the basis of
$H^{(k,k)}_V(\hat{X})$ for all $k\leq 4$ on the one hand, and to
calculate the quartic intersections on $\hat{X}$.

We recall from the main text, cf.~\eqref{eq:divsdP2P3} and
\eqref{eq:SPSQSRP3}, that the basis $D_A$ of divisors on $\pi:\,\,
\hat{X}\rightarrow \mathbb{P}^3$ reads
\beq  \label{eq:XP2_H11}
	H^{(1,1)}(\hat{X})=\langle H_B,S_P,S_Q,S_R\rangle\,,
\eeq
where $S_P$, $S_Q$, $S_R$ denote the homology classes of the sections
$\hat{s}_P$, $\hat{s}_Q$ and $\hat{s}_R$, respectively.

\subsubsection*{Intersection relations and the quartic intersections}

Next we begin by calculating the intersections of two divisors
$D_A$, where we denote the intersection
pairing on $\hat{X}$ by a subscript. We employ the representation
\eqref{eq:abstractCohomRing} for $H^{(*,*)}(\hat{X})$ to relate these
to intersections on the ambient space $dP_2(n_7,n_9)$, that we denote
as `$\cdot$'. First we note that there are $\frac{5!}{2!3!}=10$ different
quadratic combinations $D_A\cdot_{\hat{X}} D_B$. We evaluate these
as
\bea \label{eq:XP3_quadInts}
	&S_P^2=S_P ( S_P[K_B^{-1}] - (\cS_7 - \cS_9) ( \cS_9-[K_B^{-1}]))\,,
 \quad S_P S_Q=0\,,&\nn\\
	 &S_P S_R=- S_P\cS_9(S_P -  \cS_9+[K_B^{-1}]) \,,\quad
	  S_Q^2=S_Q ( S_Q[K_B^{-1}] +  (\cS_7-[K_B^{-1}]) (\cS_7 - \cS_9))\,,& \nn\\	
	  &S_R^2= S_R \left(\left(\cS_7-[K_B^{-1}]\right) \left(\cS_9-[K_B^{-1}]\right)+ S_R[K_B^{-1}]\right)\,,\quad
 S_Q S_R=- S_Q\cS_7 (S_Q -   \cS_7 +[K_B^{-1}]) \,,&\nn\\
	 &S_P H_B=-S_P H_B (S_P + \cS_7 - 2 \cS_9)\,,\quad S_Q H_B=-S_Q H_B (S_Q + \cS_9-2 \cS_7 )\,,&\nn \\
	 &S_R H_B=S_R H_B \left( \cS_7+\cS_9-[K_B^{-1}]-S_R\right)\,,& \nn\\
 &H_B^2=H_B^2 \left(3 [K_B^{-1}]-\cS_7-\cS_9+2 S_P+2 S_Q+3 S_R\right)\,,&
\eea
where we used $\cS_7=n_7 H_B$, $\cS_9=n_9 H_B$ and $[K_B^{-1}]=4H_B$
for $B=\mathbb{P}^3$. Here we suppressed the intersection product
`$\cdot$' for brevity of our notation. The intersections on the left side
of the equations are carried out in the Calabi-Yau fourfold $\hat{X}$ and
related to the intersections on the ambient space $dP_2(n_7,n_9)$ on the
right side of the equations.

Next, we calculate the triple intersections of divisors on $\hat{X}$.
There are $\frac{6!}{3!3!}=20$ different cubic combinations
$D_A\cdot_{\hat{X}}D_B\cdot_{\hat{X}}D_C$ with intersections
\footnotesize
\begin{align} \label{eq:XP3_tripInts}
	&H_B^3=H_B^3 \left(2 S_P+2 S_Q+3 S_R\right)\,,\quad && H_B^2 S_P=S_PH_B^2( 2 \cS_9-\cS_7 -S_P)\,,\nn\\	
	&H_B^2S_Q=S_QH_B^2(2 \cS_7 - \cS_9-S_Q )\,,\quad && H_B^2S_R=S_R H_B^2 \left(\cS_7+\cS_9-[K_B^{-1}]-S_R\right)\,, \nn\\
&H_BS_P S_R=-S_P H_B \cS_9( S_P+ [K_B^{-1}]- \cS_9 )\,,	\quad &&
H_B S_Q S_R=- S_Q H_B \cS_7 (S_Q - \cS_7+[K_B^{-1}])\,,\nn
\end{align}
\begin{align}
 & H_BS_P^2
 =S_P H_B ( S_P[K_B^{-1}]+ (\cS_9 - \cS_7) ( \cS_9-[K_B^{-1}]))\,,\quad
 \nn\\
 & H_B S_Q^2=S_Q H_B ( S_Q[K_B^{-1}] + ( \cS_7-[K_B^{-1}])
 (\cS_7 - \cS_9))\,,  \nn\\
 &H_B S_R^2=S_R H_B  \left(\left(\cS_7-[K_B^{-1}]\right) \left(\cS_9-[K_B^{-1}]\right)+S_R[K_B^{-1}] \right)\,,\nn\\
 & S_P^3=-S_P  ( [K_B^{-1}] (\cS_9 - cS_7) (\cS_9-[K_B^{-1}])
 + S_P ([K_B^{-1}]^2 + (\cS_7-[K_B^{-1}] ) \cS_9 -\cS_9^2))\,,\nn\\
 & S_Q^3=S_Q  (S_Q (-[K_B^{-1}]^2 + \cS_7^2 - \cS_7 ( \cS_9-[K_B^{-1}]))
 - [K_B^{-1}]  (\cS_7-[K_B^{-1}]) (\cS_7 - \cS_9))\,,\nn\\
 & S_R^3=-S_R \left([K_B^{-1}] \left(\cS_7-[K_B^{-1}]\right) \left(\cS_9-[K_B^{-1}]\right)+\left([K_B^{-1}]^2-\cS_7 \cS_9\right) S_R\right),\,\nn
\end{align}
\begin{align}
 S_P^2 S_R&=S_P (S_P - \cS_9+[K_B^{-1}]) (\cS_7 - \cS_9) \cS_9\,,
 \quad &&S_P S_R^2=S_P ( [K_B^{-1}]-\cS_7) (S_P - \cS_9+[K_B^{-1}])) \cS_9\,,\nn\\
  S_Q^2 S_R&=-S_Q  (S_Q - \cS_7+[K_B^{-1}]) \cS_7 (\cS_7 - \cS_9)\,,\quad && S_Q S_R^2=-S_Q (S_Q - \cS_7+[K_B^{-1}]) \cS_7 ( \cS_9-[K_B^{-1}])\,.
\end{align}
\normalsize
We note that all other intersections vanish, since $S_P\cdot S_Q=0$ and
as before we suppressed the intersection product "$\cdot$"
of $\hat{X}$, respectively, $dP_2(n_7,n_9)$ on the left, respectively,
right side of the equations.

Finally we calculate the quartic intersections as
\footnotesize
\bea \label{eq:XP3_C0}
	\mathcal{C}_0&=&H_B^3 S_P-4 H_B^2 S_P^2
	+ \left(16+\left(n_7-4\right) n_9-n_9^2\right) H_B S_P^3
+H_B^3 S_Q-4 H_B^2 S_Q^2\nn\\
&+&\left(\left(32-4 n_7-n_7^2\right) n_9+3 n_7 n_9^2
-2 n_9^3-64\right) S_P^4+ \left(16-n_7^2+n_7 \left(n_9-4\right)\right) H_B S_Q^3
\nn\\
&-&\left(64+2 n_7^3-3 n_7^2 n_9-n_7
\left(32-4 n_9-n_9^2\right)\right)S_Q^4+H_B^3 S_R+n_9H_B^2S_P S_R\nn\\
&+&n_9\left(n_9-n_7\right) H_B  S_P^2 S_R
+\left(n_7-n_9\right)^2 n_9 S_P^3 S_R+n_7H_B^2 S_Q S_R+ n_7 \left(n_7-n_9\right)H_BS_Q^2 S_R\nn\\
&+&n_7 \left(n_7-n_9\right)^2 S_Q^3S_R
-4 H_B^2 S_R^2+ \left(n_7-4\right) n_9 H_BS_P S_R^2+\left(4-n_7\right) \left(n_7-n_9\right) n_9 S_P^2 S_R^2
\nn\\
&+& n_7 \left(n_9-4\right)H_B S_Q S_R^2
+n_7 \left(n_7-n_9\right) \left(n_9-4\right) S_Q^2 S_R^2+ \left(16-n_7 n_9\right) H_B S_R^3\nn\\
&+&\left(n_7-4\right)^2 n_9 S_PS_R^3+n_7 \left(n_9-4\right)^2 S_Q S_R^3
-\left(64+n_7^2 n_9+n_7 \left(n_9-12\right) n_9\right) S_R^4\,,
\eea
\normalsize
where the coefficient of the polynomial $\mathcal{C}_0$
of $D_A\cdot_{\hat{X}}D_B\cdot_{\hat{X}}D_C\cdot_{\hat{X}} D_D$
is the corresponding intersection number in $\hat{X}$.
In this computation we have used that the intersections
on $dP_2$ in \eqref{eq:dP2ints} are embedded into the intersections of
$dP_2(n_7,n_9)$ as the intersections containing $H_B^3\cdot D_A\cdot D_B$, in particular $H_B^3\cdot S_R^2=H_B^3\cdot S_P^2=H_B^3\cdot S_Q^2=-1$.

Since $\hat{X}$ is complex four-dimensional, intersections of more than
four divisors on $\hat{X}$ vanish, which is confirmed by the
concrete presentation \eqref{eq:abstractCohomRing} of the intersection
ring with $B=\mathbb{P}^3$.

\subsubsection*{The cohomology basis of $H^{(*,*)}_V(\hat{X})$}

The basis of $H_V^{(k,k)}(\hat{X})$ for each $k=0,\ldots,4$ is now
determined as follows.
First, we note that there is a canonical basis both for $k=0,1$.
Namely, the one-dimensional space $H^{(0,0)}(\hat{X})$ is spanned
by the generated by $1$, $H^{(0,0)}(\hat{X})=\langle 1\rangle$,
and four-dimensional $H^{(1,1)}(\hat{X})$ is generated by the $D_A$
as indicated in \eqref{eq:XP2_H11}.
Using the quartic intersections \eqref{eq:XP3_C0} we can
now calculate a basis of $H^{(2,2)}_V(\hat{X})$. For generic
$n_7$, $n_9$ this space is five-dimensional and a basis is given
by
\beq \label{eq:XP3_H22}
	H_V^{(2,2)}(\hat{X})=\langle H_B^2, H_B\cdot S_P,
	H_B\cdot \sigma(\hat{s}_Q), H_B \cdot\sigma(\hat{s}_R),S_P^2\rangle\,.
\eeq	
Here $\sigma(\hat{s}_Q)$, $\sigma(\hat{s}_R)$ denotes the Shioda map
\eqref{eq:ShiodaMapSQSR} of the rational sections, that takes in the
case at hand the form
\beq
	\sigma(\hat{s}_Q)=S_Q - S_P - 4 H_B\,,\quad \sigma(\hat{s}_R)=S_R - S_P  -(4 + n_9) H_B\,.
\eeq

The intersection matrix $\eta^{(2)}$ in this basis is readily
calculated from \eqref{eq:XP3_C0} as
\beq  \label{eq:XP3_eta2}
	 \eta^{(2)}=
\left(
\begin{array}{ccccc}
 0 & 1 & 0 & 0 & -4 \\
 1 & -4 & 0 & 0 & 16+\left(-4+n_7-n_9\right) n_9 \\
 0 & 0 & -8 & n_7-n_9-4 & n_9 \left(4-n_7+n_9\right) \\
 0 & 0 & \eta^{(2)}_{34} & -2 \left(4+n_9\right) & 2 n_9 \left(4-n_7+n_9\right) \\
 -4 & \eta^{(2)}_{25} & \eta^{(2)}_{35} & \eta^{(2)}_{45} & -64-\left(8+n_7-2 n_9\right) \left(-4+n_7-n_9\right) n_9 \\
\end{array}
\right)\,,
\eeq
where lengthy entries $\eta^{(2)}_{rs}$ that are determined by symmetry of
$\eta^{(2)}$ are omitted and denoted by $\eta^{(2)}_{sr}$.

From this
matrix it is apparent that if $(n_7,n_9)$ are on the boundary
of the allowed region in figure \ref{fig:XP3n7n9region}, the rank
is reduced to four. In this case, the five surfaces in \eqref{eq:XP3_H22}
become homologous.  A choice of basis in this case is given by
the first four surfaces in \eqref{eq:XP3_H22} with the surface $S_P^2$
dropped. Consequently, the intersection matrix $\eta^{(2)}$ in this case
is the $4\times 4$-submatrix of \eqref{eq:XP3_eta2} obtained by
deleting the fifth row and column. Let us discus these non-generic cases
in more detail, to illustrate the decrease from
five to four basis elements.

\underline{{$\mathbf{n_7=0:}$}} We see from \eqref{eq:sectionsFibration} that $s_7\neq 0$,
which implies that $\hat{s}_Q$ and $\hat{s}_R$ are holomorphic, cf.~the
discussion below \eqref{eq:defS_m}. we obtain the cohomology relation
\beq \label{eq:SP^2n7=0}
	S_P^2\cong -n_9(4+n_9) H_B^2-4H_B\cdot S_P-n_9H_B\cdot
	\sigma(\hat{s}_R)
\eeq
in $H^{(2,2)}_V(\hat{X})$ which follows from the relations
\bea
	H_B S_P&=&H_B (S_P ([K_B^{-1}]+\cS_9)
	- S_R (  S_R+[K_B^{-1}] - \cS_9))\,,\nn\\
	H_B S_R&=&H_B S_R (\cS_9-[K_B^{-1}]  - S_R)\nn \\
	S_P^2&=& \left([K_B^{-1}]+\cS_9\right) \left(\cS_9-[K_B^{-1}]\right) S_P
	+[K_B^{-1}] \left([K_B^{-1}]-\cS_9+S_R\right) S_R\,.
\eea
These relations can be proven for $n_7=0$ using the presentation
\eqref{eq:abstractCohomRing}.

\underline{{$\mathbf{n_9=4+n_7:}$}} In this situation, the section
$\hat{s}_Q$ is holomorphic. Analogously, as before, we compute the
homology relation on $\hat{X}$ reading
\beq\label{eq:SP^2n9_1}
	S_P^2\cong -8(4+n_7)H_B^2-4H_B \cdot S_P-n_7H_B\cdot
	\sigma(\hat{s}_Q)-4H_B\cdot \sigma(\hat{s}_R)  \,.
\eeq

\underline{{$\mathbf{n_9=12-n_7:}$}} In this case all sections are
rational,  but $s_1\neq 0$ in \eqref{eq:sectionsFibration}. As before,
we compute the homology relation
\beq\label{eq:SP^2n9_2}
	S_P^2\cong -2(n_7-12)(n_7-8)H_B^2-4H_B\cdot S_P
	+(n_7-8)H_B\cdot \sigma(\hat{s}_Q)+(n_7-8)H_B\cdot \sigma(s_R)\,.
\eeq

\underline{{$\mathbf{n_9=0:}$}} For this value the sections $\hat{s}_P$
and $\hat{s}_R$ are holomorphic.
Similarly, we compute for $n_9=0$ the following relation in homology,
\beq \label{eq:SP^2calculated}
	S_P^2+4H_B\cdot S_P=0\,.
\eeq
This is immediately clear since for a holomorphic zero section
the relation  \eqref{eq:holomorphicSec} holds in the homology of
$\hat{X}$, which is precisely the relation \eqref{eq:SP^2calculated}.
It is satisfying that this relation is reproduced by the intersection
calculation \eqref{eq:abstractCohomRing} by noting the relations
\bea \label{eq:SP^2calculated1}
	S_P^2&=& [K_B^{-1}]\left(\left(\cS_7-[K_B^{-1}]\right) S_P+ S_R\left([K_B^{-1}]+S_R\right) \right)\,,\nn\\
	S_P H_B&=&-H_B \left(\left(\cS_7-[K_B^{-1}]\right) S_P+S_R \left([K_B^{-1}]+S_R\right)\right)\,,
\eea
that are calculated for $n_9=0$. Combining these two relations precisely
yields \eqref{eq:SP^2calculated}.

\underline{{$\mathbf{n_9=n_7-4:}$}} For this value the sections
$\hat{s}_P$, but this time since $s_8\neq 0$ as is clear from
\eqref{eq:sectionsFibration}. Thus, the relation
\eqref{eq:SP^2calculated1} still holds.

Next, we note that the basis of $H^{(3,3)}(\hat{X})$ is determined
by Poincar\'e duality from $H^{(1,1)}(\hat{X})$. Thus, the group
$H^{(3,3)}(\hat{X})$ is generated by four elements, independently
of $n_7$ and $n_9$, which can
be checked by an explicit calculation. However, we will refrain
from presenting the details, since the group $H^{(3,3)}(\hat{X})$
is of no immediate relevance for F-theory, and leave this as an
exercise for the interested reader.  Finally, $H^{(4,4)}(\hat{X})$
is generated by a single element which is the volume form
normalized by the volume of $\hat{X}$. It is given, up
to combinatorical factors, by $\mathcal{C}_0$ in \eqref{eq:XP3_C0}.

\section{Total Chern Class of $\widehat{dP}^B_2(\cS_7,\cS_9)$}
\label{app:ChernOfdP2BU}

In this brief appendix we outline the calculation
of the total Chern class of the resolved
space $\widehat{dP}_2^B(\cS_7,\cS_9)$, which is straightforward but
tedious. We keep our discussion short, leaving some details of the
calculations for the interested reader.

The basic strategy to obtain the Chern class of $\widehat{dP}_2^B(\cS_7,
\cS_9)$ is to successively modify \eqref{eq:c(dP2B)} taking into
account all changes of divisor classes in the blow-up
\eqref{eq:resMapSU5}. First
we have to use the new divisor classes \eqref{eq:sectionsSU5} for
homogeneous coordinates $[\tilde{u}:\tilde{v}:\tilde{w}:e_1:e_2]$. This
alters the second to fourth factors in \eqref{eq:c(dP2B)}
accordingly. Then we have to include the classes of the Cartan
divisor, which is achieved by multiplying in \eqref{eq:c(dP2B)} by
$\prod_i(1+D_i)$. Finally we have to modify the first factor
in \eqref{eq:c(dP2B)} to incorporate the split and
shift  of the class of $z$ by the $D_i$ as in \eqref{eq:sectionsSU5}
appropriately. All we have to do is to formally replace the divisor
$\Ssu$ in the expression for the Chern class of $B$ as
\beq
	\Ssu \rightarrow \Ssu-\sum_i D_i\,,
\eeq
expand and then re-express the result we get in terms of the Chern class
of $B$ and the intersections of $\Ssu$. Taking all these changes into
account we obtain the following expression
\bea \label{eq:cdP2BU}
c(\widehat{dP}_2^B\!)\!\!\!&\!\!=\!\!&\!\!\! \left[1\! +\! c_1 \!+\! c_2\!-\!D_1^2 \!-\! D_1 D_2\! -\! D_2^2\! -\! D_2 D_3\! - \!D_3^2\! -\!
           D_2 D_4\! -\! D_3 D_4\! -\! D_4^2  \right.\nn\\
           &+\!\!& \!\!\!  \left(\Ssu\!-\!D_0\right) \Ssu \!+\!  c_3\!-\!c_1 \left(D_1^2\!+\!D_1 D_2\!+\!D_2^2\!+\!D_3^2\!+\!D_3 D_4\!+\!D_4^2\!+\!D_2D_3\!+\!D_2D_4\right)\nn\\
    &+\!\!&\!\!\Ssu \left(D_1^2+2 D_1 D_2+(D_2+D_3+D_4)^2\right)+\left(\Ssu-D_0\right)\Ssu\left(c_1-\Ssu\right)\nn\\
    &-\!\!&\!\! \left.D_1^2 D_2\!-\!D_1 D_2^2\!-\!(D_2\!+\!D_3) (D_2\!+\!D_4) (D_3\!+\!D_4)\right]\nn\\
    &\times\!\!&\!\!(1 \!+ \![u]) (1 \!+\!  [v]) (1\! +\! [w]) (1\! +\!  E_1) (1 \!+\!
       E_2)\,,
\eea
where we used the relation $\Ssu-D_0=D_1+D_2+D_3+D_4$ between the
extended node $D_0$ and $\Ssu$ as well as abbreviated the ambient space
as $\widehat{dP}_2^B$. We also denote the classes \eqref{eq:sectionsSU5}
of the homogeneous coordinates by brackets $[\cdot]$.  We emphasize that
the new factors  $(1+D_i)$ from the Cartan divisors $D_i$ have been
combined with the changes in the first factor in \eqref{eq:c(dP2B)}
containing the Chern class of $B$.

\section{Intersection Ring for $\Xsu$  with $B=\mathbb{P}^3$ }
\label{app:fluxP3SU5}

In this section we present the core cohomology calculations of
elliptically fibered Calabi-Yau fourfolds
$\pi:\,\hat{X}\rightarrow \mathbb{P}^3$  over $\mathbb{P}^3$ with
general elliptic fiber $\mathcal{E}$ in $dP_2$ and an additional
resolved  $\text{SU}(5)$-singularity  over codimension one in
$\mathbb{P}^3$. The geometry has been introduced in section
\ref{sec:FFEllipticFibSU5U12}, to which we refer for more details.
The following discussion is very similar to the one in
\ref{app:CohomologyRing} and \ref{app:intsP3}. Thus, we will keep
the exposition as short as possible and only present the key results
necessary for the construction of $G_4$-flux and 3D CS-terms in the main
text.

The starting point of the calculation of the cohomology ring
$H^{(*,*)}_V(\hat{X})$
of $\hat{X}$ is the representation \eqref{eq:abstractCohomRingSU5}
as a quotient ring.
We are working with a particular phase of the fourfold $\hat{X}$
with Stanley-Reissner ideal \eqref{eq:SR-SU5}.
The eight-dimensional basis of $H^{(1,1)}(\hat{X})$ is given in
\eqref{eq:basisH11XSU5} and the class of the anti-canonical
bundle $K_{\hat{dP}_2^B}^{-1}$  in \eqref{eq:KdP2SU5}.
Omitting lengthy details, that are best performed using a computer
algebra program, we immediately obtain the quartic intersections
on $\hat{X}$ as
\begin{footnotesize}
\bea  \label{eq:QuarticIntsSU5}
\mathcal{C}_0\!\!\!\!&\!\!\!=\!\!\!&\!\!\!\!
12 D_1^3 H_B+4 D_2 D_3 D_4 H_B-8 D_2 D_4^2 H_B+12 D_4^3 H_B
-2 D_1^2 H_B^2+D_1 D_2 H_B^2-2 D_2^2 H_B^2+D_2 D_3 H_B^2\nn\\
\!\!\!\!&\!\!\! -\!\!&\!\!\!\! 2 D_3^2 H_B^2+D_3 D_4 H_B^2-2 D_4^2 H_B^2+(n_7-7)D_2 D_3^2 H_B
-4(n_7-4)D_2^2 D_3 D_4+8(n_7-4) D_2^2 D_4^2\nn \\
\!\!\!\!&\!\!\! -\!\!&\!\!\!\!(n_7-4)D_2^2 D_3 H_B
-(n_7-7) (n_7-4)D_2^2 D_3^2+(n_7-4)^2D_2^3 D_3
-16(n_7-3) D_2 D_4^3-(n_9-7)D_1 D_2^2 H_B \nn\\
\!\!\!\!&\!\!\! -\!\!&\!\!\!\! ((n_7-5) n_7+(n_9-6) (n_9-5)) D_2^4
+(4+(n_7-n_9)^2-4 n_9)D_3 D_4^3+4(n_7-n_9-1) D_2 D_3^2 D_4\nn\\
\!\!\!\!&\!\!\! +\!\!&\!\!\!\!(n_7-n_9-1)D_3^2 D_4 H_B
-(n_7-n_9-1) (2+n_7-n_9)D_3^2 D_4^2+(n_9-10)D_1^2 D_2 H_B
+4(n_9-2) D_2 D_3 D_4^2
\nn\\
\!\!\!\!&\!\!\! +\!\!&\!\!\!\!(6-2 n_7+n_9)D_3^3 H_B +(n_9-n_7-2)D_3 D_4^2 H_B
+(1-n_7+n_9)^2D_3^3 D_4 +(n_7+n_9-9)D_2^3 H_B\nn\\
\!\!\!\!&\!\!\! +\!\!&\!\!\!\!(25+(n_7-14) n_7+4 n_9)D_2 D_3^3 +(16 n_7-12+(n_9-20) n_9)D_1^3 D_2
-(14+8 n_7+(n_9-17) n_9)D_1^2 D_2^2 \nn\\
\!\!\!\!&\!\!\! +\!\!&\!\!\!\!(21\!+\!4 n_7\!+\!(n_9\!-\!14) n_9)D_1 D_2^3 \!-\!2(16 n_7\!-\!36+(n_9-14) n_9) D_1^4
\!-\!2 (28+n_7^2+n_9(2+n_9)\!-\!2 n_7 (7\!+\!n_9))D_4^4 \nn\\
\!\!\!\!&\!\!\! -\!\!&\!\!\!\!(20+2 n_7^2+n_9 (5+n_9)-2 n_7 (7+n_9))D_3^4 +H_B^3 S_P
+(16+(n_7-n_9-4) n_9)H_B  S_P^3\nn\\
\!\!\!\!&\!\!\! -\!\!&\!\!\!\! 4 H_B^2 S_P^2+(-64-(8+n_7-2 n_9) (n_7-n_9-4) n_9) S_P^4+D_3^3 S_Q+D_3^2 H_B S_Q
+D_3 H_B^2 S_Q+H_B^3 S_Q\nn\\
\!\!\!\!&\!\!\! -\!\!&\!\!\!\! 4 D_3^2 S_Q^2\!-\!4 D_3 H_B S_Q^2\!-\!4 H_B^2 S_Q^2
+(16+n_7 (n_9\!-\!n_7\!-\!2))D_3  S_Q^3
+(16+n_7 (n_9-n_7-2))H_B  S_Q^3+H_B^3 S_R\nn\\
\!\!\!\!&\!\!\! -\!\!&\!\!\!\!(64+n_7 (2+n_7-n_9) (2 n_7-n_9-10)) S_Q^4\!-\!4n_7 D_4^3  S_R
-n_7 D_3^2 H_B S_R+D_3 D_4 H_B n_7 S_R-2n_7 D_4^2 H_B  S_R\nn\\
\!\!\!\!&\!\!\! +\!\!&\!\!\!\!n_7 (n_7\!-\!n_9\!-\!1)D_3^2 D_4  S_R
+n_7 (n_9\!-\!n_7)D_3^3  S_R
+n_7 (2\!-\!n_7\!+\!n_9)D_3 D_4^2  S_R+n_9H_B^2  S_P S_R+n_7D_3 H_B S_Q S_R
\nn\\
\!\!\!\!&\!\!\! +\!\!&\!\!\!\!n_9(n_9-n_7)H_BS_P^2 S_R+(n_7-n_9)^2 n_9 S_P^3 S_R+n_7D_3^2 S_Q S_R
+n_7H_B^2  S_Q S_R+n_7 (n_7-n_9-2)D_3  S_Q^2 S_R\nn\\
\!\!\!\!&\!\!\! +\!\!&\!\!\!\!n_7 (n_7-n_9-2)H_B  S_Q^2 S_R+n_7 (2-n_7+n_9)^2 S_Q^3 S_R-4 H_B^2 S_R^2-D_3 H_B n_7 S_R^2
+(n_7-4) n_7D_3 D_4  S_R^2\nn\\
\!\!\!\!&\!\!\! -\!\!&\!\!\!\!2 (n_7-4) n_7D_4^2  S_R^2-(n_7-3) n_7D_3^2 S_R^2+H_B (n_7-4) n_9 S_P S_R^2-(n_7-4) (n_7-n_9) n_9 S_P^2 S_R^2
\nn\\
\!\!\!\!&\!\!\! +\!\!&\!\!\!\!n_7 (n_9-2)D_3  S_Q S_R^2+n_7 (n_9-2)H_B  S_Q S_R^2+(n_7-4)^2 n_9 S_P S_R^3+n_7 (n_7-n_9-2) (n_9-2) S_Q^2 S_R^2\nn\\
\!\!\!\!&\!\!\! -\!\!&\!\!\!\!n_7 (n_7\!+\!n_9\!-\!6)D_3  S_R^3
+(16\!-\!n_7 (2+n_9))H_B  S_R^3+n_7 (n_9\!-\!2)^2 S_Q S_R^3\!-\!(64+n_7 (2+n_9) (n_7\!+\!n_9\!-\!10)) S_R^4\,.\nn\\
\eea
\end{footnotesize}
As before, all intersections are understood to be evaluated on
$\hat{X}$ and  the quartic  intersections
$D_A\cdot_{\hat{X}}D_B\cdot_{\hat{X}}D_C\cdot_{\hat{X}}D_D$
are read off as the coefficient of the appropriate monomial in $\mathcal{C}_0$.

As one immediate application we calculate the Cartan matrix $C_{IJ}$ of
the affine Lie algebra of SU$(5)$ as
\beq
	C_{IJ}=-D_I\cdot D_J\cdot H_B^2=\left(
\begin{array}{ccccc}
 2 & -1 & 0 & 0 & -1 \\
 -1 & 2 & -1 & 0 & 0 \\
 0 & -1 & 2 & -1 & 0 \\
 0 & 0 & -1 & 2 & -1 \\
 -1 & 0 & 0 & -1 & 2 \\
\end{array}
\right)\,,
\eeq
where we supplemented as before the Cartan divisors $D_i$,
$i=1,\ldots, 4,$ by the divisor $D_0=H_B-D_1-D_2-D_3-D_4$ corresponding
to the extended node $-\alpha_0$ as $D_I=(D_0,D_i)$.
In addition, we confirm the intersections \eqref{eq:intSecFiber},
\eqref{eq:SP^2} and \eqref{eq:S7S9} explicitly employing
\eqref{eq:QuarticIntsSU5}. Finally, we calculate the intersections
of the sections with the nodes of the Dynkin diagram of SU$(5)$ as
\beq
	S_P\cdot D_{I}\cdot H_B^2=S_R\cdot D_{I}\cdot H_B^2=(1,0,0,0,0)_I\,,\quad S_Q\cdot D_{I}\cdot H_B^2=(0,0,0,1,0)_I
\eeq
reproducing the independent findings in \eqref{eq:-alpha_IS_m} and,
consequently, the Shioda maps \eqref{eq:ShiodaMapSQSRSU5}.
Here we exploited \eqref{eq:c_alphaI} to represent the curve
$c_{-\alpha_I}$ corresponding to the simple root $\alpha_I$ as
\beq
	c_{-\alpha_I}=D_I\cdot H_B^2\,,
\eeq
employing $H_B^2\cdot \cS_{\text{SU}(5)}^b=H_B^3=1$. We note
that we can even make the stronger statement
\beq \label{eq:SPDi}
	S_P\cdot D_i=0
\eeq
in the homology of $\Xsu$, which will be important for the discussion
of section \ref{sec:ChiralitiesU1xU1xSU5}.

The cohomology group $H^{(2,2)}_V(\Xsu)$ is readily calculated.
We generically obtain a 13-dimensional vector space, with certain
jumps in the cohomology for non-generic values of $n_7$ and $n_9$,
see figure \ref{fig:XP3SU5n7n9region}. For the application of the
construction of $G_4$-flux, it proves useful to make a particular
choice of basis by hand. We present a thorough analysis of this basis in
the next appendix.

\section{Basis of $H^{(2,2)}_V(\Xsu)$  with $B=\mathbb{P}^3$ }
\label{app:H22BasisSU5}

The vector space $H^{(2,2)}_V$ is 13-dimensional for arbitrary values of
$n_7$ and $n_9$. A basis is \eqref{eq:basisH22Su5}, that we recall here
for convenience
\bea
H_{(V,13)}^{(2,2)} &=& \langle H_B^2,\, H_B \cdotp S_P,\, H_B \cdotp \sigma(S_Q),\, H_B \cdotp \sigma(S_R),\, S_P \cdotp \sigma(S_R),\,  \sigma(S_Q) \cdotp \sigma(S_R),\, \nn \\ && \sigma(S_Q)^2,\,  H_B \cdotp D_1,\, H_B \cdotp D_2,\,
H_B \cdotp D_3,\, H_B \cdotp D_4,\, D_2 \cdotp D_4,\, D_1^2 \rangle.
\eea
To obtain the smaller vector spaces of the boundaries we need to drop
some elements of this basis. The vectors that become linearly depended at
the boundary have to be re-expressed as a linear combination of the
smaller basis on the boundary. In the next subsections, we choose bases
for the different boundaries. We write the linearly dependent elements,
that we can drop from the above 13-dimensional basis, as linear
combinations of the rest of the vectors in the smaller basis.

\subsubsection*{12 Dimensional}
At the boundaries with a  twelve-dimensional basis we can take
\beq
H_{(V,12)}^{(2,2)}=H_{(V,13)}^{(2,2)} \backslash \lbrace S_P \cdotp \sigma(S_R)\rbrace,
\eeq
in which case the surface $S_P \cdotp \sigma(S_R)$ is re-expressed as:
\begin{itemize}
 \item At $n_9=0$
\beq
S_P \cdotp \sigma(S_R) \cong 0,
\eeq
\item At the boundary $n_9=n_7+2$
\bea
S_P \cdotp \sigma(S_R) &\cong&
-\frac{2}{35} \bigg(30 H_B^2 (26 + 3 n_7) + 25 \sigma(\hat s_Q) \cdot\sigma(\hat s_R) \nn \\ &&+
   H_B\cdot \big[4 D_1 n_7 + 8 D_2 n_7 + 12 D_3 n_7 + 6 D_4 n_7  \nn \\ && + 150 S_P  + 150 \sigma(\hat s_Q)+
      15 n_7 \sigma(\hat s_Q) + 130 \sigma(\hat s_R)\big]\bigg),
\eea
\item At the boundary $n_9=9-n_7$
\bea
S_P \cdotp \sigma(S_R) &\cong& \frac{2}{475} \bigg(18 D_1\cdot  H_B + 36 D_2\cdot H_B - 66 D_3 \cdot H_B - 33 D_4\cdot H_B \nn \\ && - 41100 H_B^2  +
   8400 n_7 H_B^2  - 400  n_7^2 H_B^2 - 4425 H_B\cdot S_P \nn \\ && + 550 n_7 H_B \cdot S_P  -
   5145 H_B \cdot \sigma(\hat s_Q)  + 450  n_7 H_B\cdot \sigma(\hat s_Q) \nn \\ && - 125 \sigma(\hat s_Q)^2 - 2790 H_B \cdot \sigma(\hat s_R)  +
   350 n_7 H_B \cdot \sigma(\hat s_R) \nn \\ && - 275 \sigma(\hat s_Q) \cdot \sigma(\hat s_R)\bigg).
\eea
\end{itemize}
\subsubsection*{11 Dimensional}
We obtain an eleven-dimensional basis by choosing
\beq
H_{(V,12)}^{(2,2)}=H_{(V,13)}^{(2,2)} \backslash \lbrace S_P \cdotp \sigma(S_R),\,  H_B \cdotp \sigma(S_Q)\rbrace,
\eeq
\begin{itemize}
 \item At $n_7=n_9-4$ we obtain the homology relation
\bea
 S_P \cdotp \sigma(S_R) &\cong& 0, \nn \\
H_B \cdotp \sigma(S_Q) &\cong& \frac{1}{1220 +
 150 n_7}\bigg(25 D_1^2 + 684 D_3 \cdot H_B + 342 D_4\cdot H_B - 3200 H_B^2 \nn \\ && +
  (438 - 45 n_7)  D_1 \cdot H_B  + 18 (32 - 5 n_7) D_2 \cdot H_B  - 120 n_7 D_3\cdot H_B  \nn \\ && -
  60 n_7 D_4\cdot H_B  - 800 H_B\cdot S_P - 125 \sigma(\hat s_Q)^2 \nn \\ && - 490 H_B\cdot \sigma(\hat s_R) - 275 \sigma(\hat s_Q)\cdot \sigma(\hat s_R)\bigg) .
\eea
\end{itemize}
\subsubsection*{10 Dimensional}
Finally a ten-dimensional basis is given by
\beq
H_{(V,12)}^{(2,2)}=H_{(V,13)}^{(2,2)} \backslash \lbrace S_P \cdotp \sigma(S_R),\,  H_B \cdotp \sigma(S_Q),\,D_2 \cdot D_4\rbrace.
\eeq
\begin{itemize}
 \item At $n_7=0$
\bea
 S_P \cdotp \sigma(S_R) &\cong& 2 n_9 \big[ (4 + n_9) H_B^2 + H_B \cdot \sigma(\hat s_R)\big], \nn \\
H_B \cdotp \sigma(S_Q) &\cong& -2  (2 + n_9)H_B^2  - 2  \left(\frac{2 + n_9}{4 + n_9}\right) \sigma(\hat s_R)\cdot H_B  \nn \\  && - \left(\frac{1}{4 + n_9}\right)\sigma(\hat s_Q)\cdot \sigma(\hat s_R) - H_B\cdot S_P , \nn \\
 D_2 \cdot D_4 &\cong& 4 D_4\cdot H_B.
\eea
\end{itemize}

\section{CS-Levels \& Chiralities for $\Xsu$ with $B=\mathbb{P}^3$}
\label{app:fluxesNChiralities}

In this appendix we present the 3D Chern-Simons terms on both
the M- and F-theory side and the chiralities, all of which
parametrized by the parameters $a_i$ of the general $G_4$-flux.

\subsection{Chern-Simons levels}

\subsubsection*{F-Theory Loop induced CS terms}
The one loop Chern-Simons levels read as follows. For the Cartan generators, we obtain $\Theta^F_{ij}$ as:
\normalsize
\bea
\Theta^F_{1,3}&=&\Theta^F_{1,4}=0, \nn \\
\Theta^F_{1,1}&=&- \frac{1}{2} \Big[6 \chi\big(\mathbf{10}_{(\tfrac{1}{5},0)}\big) + 2 \chi\big(\mathbf{5}_{(-\tfrac{2}{5},0)}\big) + 2 \chi\big(\mathbf{5}_{(-\tfrac{2}{5},1)}\big) + 2 \chi\big(\mathbf{5}_{(\tfrac{3}{5},0)}\big) + 2 \chi\big(\mathbf{5}_{(\tfrac{3}{5},1)}\big) - 2 \chi\big(\mathbf{5}_{(-\tfrac{2}{5},-1)}\big) \Big], \nn \\
\Theta^F_{1,2}&=&-\frac{1}{2} \Big[-3 \chi\big(\mathbf{10}_{(\tfrac{1}{5},0)}\big) - \chi\big(\mathbf{5}_{(-\tfrac{2}{5},0)}\big) - \chi\big(\mathbf{5}_{(-\tfrac{2}{5},1)}\big) - \chi\big(\mathbf{5}_{(\tfrac{3}{5},0)}\big) - \chi\big(\mathbf{5}_{(\tfrac{3}{5},1)}\big) + \chi\big(\mathbf{5}_{(-\tfrac{2}{5},-1)}\big)\Big], \nn \\
\Theta^F_{2,2}&=&-\frac{1}{2} \Big[2 \chi\big(\mathbf{10}_{(\tfrac{1}{5},0)}\big) + 2 \chi\big(\mathbf{5}_{(-\tfrac{2}{5},1)}\big) + 2 \chi\big(\mathbf{5}_{(\tfrac{3}{5},0)}\big) + 2 \chi\big(\mathbf{5}_{(\tfrac{3}{5},1)}\big) - 2 \chi\big(\mathbf{5}_{(-\tfrac{2}{5},-1)}\big)\Big], \nn \\
\Theta^F_{2,3}&=&-\frac{1}{2} \Big[-\chi\big(\mathbf{10}_{(\tfrac{1}{5},0)}\big) + \chi\big(\mathbf{5}_{(-\tfrac{2}{5},0)}\big) - \chi\big(\mathbf{5}_{(-\tfrac{2}{5},1)}\big) - \chi\big(\mathbf{5}_{(\tfrac{3}{5},0)}\big) - \chi\big(\mathbf{5}_{(\tfrac{3}{5},1)}\big) + \chi\big(\mathbf{5}_{(-\tfrac{2}{5},-1)}\big)\Big], \nn \\
\Theta^F_{3,3}&=&-\frac{1}{2} \Big[2 \chi\big(\mathbf{10}_{(\tfrac{1}{5},0)}\big) - 2 \chi\big(\mathbf{5}_{(-\tfrac{2}{5},0)}\big) + 2 \chi\big(\mathbf{5}_{(\tfrac{3}{5},1)}\big) - 2 \chi\big(\mathbf{5}_{(-\tfrac{2}{5},-1)}\big)\Big], \nn \\
\Theta^F_{3,4}&=&-\frac{1}{2} \Big[-\chi\big(\mathbf{10}_{(\tfrac{1}{5},0)}\big) + \chi\big(\mathbf{5}_{(-\tfrac{2}{5},0)}\big) + \chi\big(\mathbf{5}_{(-\tfrac{2}{5},1)}\big) + \chi\big(\mathbf{5}_{(\tfrac{3}{5},0)}\big) - \chi\big(\mathbf{5}_{(\tfrac{3}{5},1)}\big) + \chi\big(\mathbf{5}_{(-\tfrac{2}{5},-1)}\big)\Big], \nn \\
\Theta^F_{4,4}&=& -\frac{1}{2} \Big[2 \chi\big(\mathbf{10}_{(\tfrac{1}{5},0)}\big) - 2 \chi\big(\mathbf{5}_{(-\tfrac{2}{5},0)}\big) - 2 \chi\big(\mathbf{5}_{(-\tfrac{2}{5},1)}\big) - 2 \chi\big(\mathbf{5}_{(\tfrac{3}{5},0)}\big) + 2 \chi\big(\mathbf{5}_{(\tfrac{3}{5},1)}\big) - 2 \chi\big(\mathbf{5}_{(-\tfrac{2}{5},-1)}\big)\Big]. \nn \\
\eea
\normalsize
The mixed Abelian-non-Abelian $\Theta_{im}^F$ read
\bea
\Theta^F_{i=1,m=1}&=&\Theta^F_{i=4,m=1}=\Theta^F_{i=1,m=2}=\Theta^F_{i=2,m=2}=\Theta^F_{i=4,m=2}=0, \nn \\
\Theta^F_{i=2,m=1}&=& -\frac{1}{2} \left(-\frac{4}{5} \chi\big(\mathbf{10}_{(\tfrac{1}{5},0)}\big) + \frac{4}{5} \chi\big(\mathbf{5}_{(-\tfrac{2}{5},0)}\big)\right),\nn \\
\Theta^F_{i=3,m=1}&=&- \frac{1}{2} \left( \frac{4}{5} \chi\big(\mathbf{5}_{(-\tfrac{2}{5},1)}\big) - \frac{6}{5} \chi(\mathbf{5}_{(3/5,0)}\right), \nn \\
\Theta^F_{i=3,m=2}&=& \chi\big(\mathbf{5}_{(-\tfrac{2}{5},1)}\big).
\eea
The purely Abelian $\Theta_{mn}^F$ read
\normalsize
\bea
\Theta^F_{m=1,m=1}&=& -\frac{1}{2} \bigg(\frac{4}{25} \chi\big(\mathbf{10}_{(\tfrac{1}{5},0)}\big) - \frac{4}{25} \chi\big(\mathbf{5}_{(-\tfrac{2}{5},0)}\big) + \frac{4}{25} \chi\big(\mathbf{5}_{(-\tfrac{2}{5},1)}\big) \nn \\ && + \frac{9}{25} \chi\big(\mathbf{5}_{(\tfrac{3}{5},0)}\big) + \frac{9}{5} \chi\big(\mathbf{5}_{(\tfrac{3}{5},1)}\big) - \frac{4}{5} \chi\big(\mathbf{5}_{(-\tfrac{2}{5},-1)}\big) + \chi(\mathbf 1_{(1,0)})  \nn \\ && + \chi(\mathbf 1_{(1,1)}) + \chi(\mathbf 1_{(-1,1)}) - 3 \chi(\mathbf{1}_{(-1,-2)})\bigg), \nn \\
\Theta^F_{m=1,m=2}&=& -\frac{1}{2} \bigg(-\frac{2}{5} \chi\big(\mathbf{5}_{(-\tfrac{2}{5},1)}\big) + 3 \chi(\mathbf{5})_{(3/5,1)} - 2 \chi\big(\mathbf{5}_{(-\tfrac{2}{5},-1)}\big) \nn \\ && + \chi(\mathbf 1_{(1,1)}) -  \chi(\mathbf 1_{(-1,1)}) - 6 \chi(\mathbf{1}_{(-1,-2)}) \bigg), \nn \\
\Theta^F_{m=2,m=2}&=& -\frac{1}{4} \bigg(\chi\big(\mathbf{5}_{(-\tfrac{2}{5},1)}\big) + 5 \chi\big(\mathbf{5}_{(\tfrac{3}{5},1)}\big) - 5 \chi(\mathbf{5})_{(-2/5,-1)}  + \chi(\mathbf 1_{(1,1)})\nn \\ && + \chi(\mathbf 1_{(0,1)}) + \chi(\mathbf 1_{(-1,1)}) + 4 \chi(\mathbf{1}_{(0,2)}) -
   12 \chi(\mathbf{1}_{(-1,-2)}) \bigg).
\eea
\normalsize
\subsubsection*{M-Theory classical CS terms}
On the M-theory side we obtain for $\Theta_{ij}^M$:
\normalsize
\bea
\Theta^M_{1,3}&=&\Theta^M_{1,4}=0, \qquad \Theta^M_{2,4}=2 n_7 a_{12}, \nn \\
\Theta^M_{(1,1)}&=&\frac{1}{5} (6 + n_9) a_{3} + \frac{1}{2} (4 + n_9) a_{4} + n_9 (4 - n_7 + n_9) a_{5} \nn \\ && +
 \frac{1}{25} [-10 (4 + n_9) (8 + 3 n_9) + 3 n_7 (14 + 9 n_9)] a_{6} \nn \\ && +
 \frac{1}{250} [-2968 + n_7 (56 + 111 n_9) - n_9 (1148 + 115 n_9)] a_{7} \nn \\ && -
 \frac{1}{5} (-184 + 40 n_7 - 23 n_9 + n_9^2) a_{13}, \nn \\
\Theta^M_{1,2}&=&\frac{1}{500} \bigg(-50 (6 + n_9) a_{3} - 125 (4 + n_9) a_{4} \nn \\ &&+
   250 (-4 + n_7 - n_9) n_9 a_{5} \nn \\ && +
   10 [10 (4 + n_9) (8 + 3 n_9) - 3 n_7 (14 + 9 n_9)] a_{
     6} \nn \\ && + (2968 - n_7 (56 + 111 n_9) + n_9 (1148 + 115 n_9)) a_{7} \nn \\ && +
   50 (-184 + 40 n_7 + (-23 + n_9) n_9) a_{13}\bigg), 
   \eea
   \normalsize
   \bea
\Theta^M_{2,2}&=&\frac{1}{250} \bigg(25 (21 - n_7 + n_9) a_{3} + 125 (4 + n_9) a_{4} +
   250 n_9 (4 - n_7 + n_9) a_{5} \nn \\ && +
   5 [-2 n_7^2 - 5 (4 + n_9) (41 + 11 n_9) + n_7 (122 + 57 n_9)] a_{
     6} \nn \\ &&+ [-5524 + 7 n_7^2 - 3 n_9 (273 + 40 n_9) + n_7 (277 + 113 n_9)] a_{
     7} \nn \\ &&+ 25 [64 + n_7 (-24 + n_9) - (-13 + n_9) n_9] a_{13}\bigg), \nn \\
\Theta^M_{2,3}&=&\frac{1}{500} \bigg(50 (-17 + n_7) a_{3} - 125 (4 + n_9) a_{4} +
   250 (-4 + n_7 - n_9) n_9 a_{5} \nn \\ && +
   10 [2 n_7^2 - 6 n_7 (14 + 5 n_9) + 5 (4 + n_9) (27 + 5 n_9)] a_{
     6} \nn \\ &&+ [8616 - 14 n_7^2 - 5 n_7 (94 + 23 n_9) + 5 n_9 (114 + 25 n_9)] a_{
     7} \nn \\ &&- 500 n_7 a_{12} + 50 (-n_7 (-4 + n_9) + 2 (6 + n_9)) a_{13}\bigg), \nn  \\
\Theta^M_{3,3}&=&\frac{1}{500} \bigg(25 (50 - 5 n_7 + 3 n_9) a_{3} - 125 (n_7 - 2 (4 + n_9)) a_{4} \nn \\ && +
   500 n_9 (4 - n_7 + n_9) a_{5} +
   50 (-14 + n_9) n_9 a_{13} \nn \\ && +
   5 [-25 n_7^2 - 5 (4 + n_9) (90 + 23 n_9) + 2 n_7 (210 + 73 n_9)] a_{6} \nn \\ && +
   4 [-3030 - n_9 (483 + 65 n_9) + n_7 (310 + 66 n_9)] a_{7}  \bigg), \nn  \eea \bea
\Theta^M_{3,4}&=&\frac{1}{500} \bigg(25 (-16 + 3 n_7 - 3 n_9) a_{3} + 125 (-4 + n_7 - n_9) a_{4} +
   250 (-4 + n_7 - n_9) n_9 a_{5} \nn \\ && +
   5 [720 + 21 (-12 + n_7) n_7 + 440 n_9 - 86 n_7 n_9 + 65 n_9^2] a_{
     6} \nn \\ &&+ [14 (-55 + n_7) n_7 - 149 n_7 n_9 + 135 n_9^2 +
      6 (584 + 227 n_9)] a_{7} + 500 n_7 a_{12} \nn \\ && +
   50 [-12 + n_7 (-4 + n_9) - (-12 + n_9) n_9] a_{13}\bigg), \nn \\
\Theta^M_{4,4}&=&\frac{1}{10} (16 - 3 n_7 + 3 n_9) a_{3} + \frac{1}{2} (4 - n_7 + n_9) a_{4} +
 n_9 (4 - n_7 + n_9) a_{5} \nn \\ && +
 \frac{1}{50} [-21 n_7^2 - 5 (4 + n_9) (36 + 13 n_9) + n_7 (252 + 86 n_9)] a_{6} \nn \\ && +
 \frac{1}{250} [-14 n_7^2 + n_7 (770 + 149 n_9) - 3 (1168 + n_9 (454 + 45 n_9))] a_{
   7} \nn \\ && - 4 n_7 a_{12} + \frac{1}{5} [4 (3 + n_7) - (12 + n_7) n_9 + n_9^2] a_{13}.
\eea
\normalsize
For the mixed CS-levels $\Theta_{im}^M$ we obtain:
\normalsize
\bea
\Theta^M_{i=1,m=1}&=&\Theta^M_{i=1,m=2}=\Theta^M_{i=2,m=2}=0, \nn \\
\Theta^M_{i=4,m=1}&=&\frac{2}{5} n_7 a_{12}, \qquad
\Theta^M_{i=4,m=2}=2 n_7 a_{12}, \nn \\
\Theta^M_{i=2,m=1}&=& \frac{1}{625} \bigg(25 (-15 + n_7 + n_9) a_{3} +
   5 (-15 + n_7 + n_9) (2 n_7 - 5 (4 + n_9)) a_{
     6} \nn \\ &&- [-3360 + 7 n_7^2 + (209 - 5 n_9) n_9 + n_7 (179 + 2 n_9)] a_{7} \nn \\ &&-
   25 (-4 + n_9) (-5 + n_7 + n_9) a_{13} \bigg), \nn \\
\Theta^M_{i=3,m=1}&=& \frac{1}{2500}\bigg(-25 (54 + 13 n_7 - 9 n_9) a_{3} + 250 n_7 a_{4} \nn \\ && +
  5 [54 n_7^2 - 45 (-6 + n_9) (4 + n_9) + n_7 (52 + 3 n_9)] a_{
    6} \nn \\ &&+ [-639 n_7^2 - 6 (-6 + n_9) (426 + 5 n_9) +
     n_7 (1690 + 677 n_9)] a_{7} \nn \\ &&- 500 n_7 a_{12} -
  50 (18 + n_7 - 3 n_9) (-4 + n_9) a_{13}\bigg), \nn \\
\Theta^M_{i=4,m=1}&=&\frac{1}{500} n_7  \bigg(50 a_{3} - 125 a_{4} +
   5 (40 - 21 n_7 + 15 n_9) a_{6} \nn \\ &&+ (-158 + 111 n_7 - 115 n_9) a_{7} -
   500 a_{12} - 50 (-4 + n_9) a_{13} \bigg).
\eea
\normalsize
The CS-levels $\Theta_{\alpha m}^M$ for $D_{\alpha}=H_B$ read
\normalsize
\bea
\Theta^M_{\alpha,m=1} &=& -\frac{17}{5} a_{3} +
 \frac{1}{2} (-4 + n_7 - n_9) a_{4} + (-4 + n_7 - n_9) n_9 a_{
   5} \nn \\ &&+ \big[n_7^2/2 + \frac{1}{5} (4 + n_9) (27 + 5 n_9) - \frac{1}{50} n_7 (208 + 75 n_9)\big] a_{6} \nn \\ &&+ \frac{1}{250} \big[ 8616 - n_7 (312 + 125 n_7) + 5 n_9 (114 + 25 n_9)\big] a_{7} \nn \\ &&+
 \frac{2}{5} (6 + n_9) a_{13}, \nn \\
  \Theta^M_{\alpha,m=2} &=& \frac{1}{2} (-4 + n_7 - n_9) a_{3} - (4 + n_9) a_{4} +
 2 (-4 + n_7 - n_9) n_9 a_{5} \nn \\ &&+ \big[24 + 16 n_9 + \frac{5}{2} n_9^2 - \frac{1}{10} n_7 (56 + 25 n_9)\big] a_{6} \nn \\ &&+ \big[\frac{1}{2}n_7^2 + \frac{1}{5} (4 + n_9) (27 + 5 n_9) - \frac{1}{50} n_7 (208 + 75 n_9)\big] a_{7} \nn \\ &&+ (4 + n_9) a_{13}.
\eea

The purely Abelian CS-levels  $\Theta_{mn}^M$ read
\footnotesize
\bea
\Theta^M_{m=1,m=1}&=& \frac{1}{12500}\bigg(25 (8616 - n_7 (312 + 125 n_7) + 5 n_9 (114 + 25 n_9)) a_{3} \nn \\ && +
  125 (1080 - 208 n_7 + 25 n_7^2 + 470 n_9 - 75 n_7 n_9 + 50 n_9^2) a_{4} \nn \\ &&+
  1250 (-54 + 5 n_7 - 10 n_9) (-4 + n_7 - n_9) n_9 a_{5} \nn \\ &&+
  5 \Big[625 n_7^3   -
     5 (4 + n_9) (14016 + 5 n_9 (854 + 125 n_9)) \nn \\ &&- 4 n_7^2 (1411 + 625 n_9)+
     8 n_7 (6834 + 5 n_9 (893 + 125 n_9))\Big] a_{6} \nn \\ &&+
  2 \Big[-1090272 - 3125 n_7^3 -
     25 n_9 (6130 + n_9 (1852 + 125 n_9))\nn \\ && + n_7^2 (26777 + 3125 n_9)   +
     n_7 (82008 + 5 n_9 (2028 + 625 n_9))\Big] a_{7} \nn \\ &&+
  50 \big[-2968 + n_7 (56 + 111 n_9) - n_9 (1148 + 115 n_9)\big] a_{13}\bigg), \nn \\
\Theta^M_{m=1,m=2}&=& \frac{1}{100} (25 n_7^2 + 10 (4 + n_9) (27 + 5 n_9) - n_7 (208 + 75 n_9)) a_{
   3} \nn \\ &&+ [12 + 8 n_9 + \frac{5}{4} n_9^2 - \frac{1}{20} n_7 (56 + 25 n_9)] a_{4} \nn \\ &&+
 \frac{1}{2} (-12 + 3 n_7 - 5 n_9) (-4 + n_7 - n_9) n_9 a_{
   5} \nn \\ && + \Big[n_7^2 (-\frac{454}{125} -\frac{9}{4}  n_9 -
    \frac{1}{10} (4 + n_9) (228 + n_9 (157 + 30 n_9)) \nn \\ &&+
    \frac{3}{100} n_7 (1040 + n_9 (892 + 175 n_9))\Big] a_{
   6} \nn \\ &&+ \frac{1}{2500}\Big[625 n_7^3   -
    5 (4 + n_9) (14016 + 5 n_9 (854 + 125 n_9)) \nn \\ && +
    8 n_7 (6834 + 5 n_9 (893 + 125 n_9))- 4 n_7^2 (1411 + 625 n_9)\Big] a_{7} \nn \\ &&+
 \frac{1}{25} (-10 (4 + n_9) (8 + 3 n_9) + 3 n_7 (14 + 9 n_9)) a_{13}, \nn \\
\Theta^M_{m=2,m=2}&=& \big[12 + 8 n_9 + \frac{5}{4} n_9^2 - \frac{1}{20} n_7 (56 + 25 n_9)\big] a_{3} \nn \\ &&+
 \frac{1}{2} \big[-n_7 (1 + 4 n_9) + (4 + n_9) (12 + 5 n_9)\big] a_{
   4} \nn \\ &&+ (-12 + 2 n_7 - 5 n_9) (-4 + n_7 - n_9) n_9 a_{5} \nn \\ &&+
 \frac{1}{100} \Big[-25 (4 + n_9) (12 + 5 n_9)^2 - 6 n_7^2 (7 + 50 n_9) \nn \\ &&+
    5 n_7 (688 + n_9 (842 + 185 n_9))\Big] a_{6} \nn \\ &&+
 \frac{1}{500} \Big[-50 (4 + n_9) (12 + 5 n_9) (27 + 5 n_9)\nn \\ && - n_7^2 (1278 + 625 n_9) +
    n_7 (12164 + 5 n_9 (2542 + 375 n_9))\Big] a_{
   7} \nn \\ &&+ \big[-\frac{1}{2} (4 + n_9) (12 + 5 n_9) + \frac{1}{5} n_7 (4 + 9 n_9)\big] a_{13}.
\eea
\normalsize

\normalsize

\subsection{Chiralities}

Finally, we present the rest of the chiralities obtained by
matching the M-/F-theory CS-terms. The $G_4$-flux parameter supported
along the non-flat fiber is assumed to be zero by setting $a_{12}=0$. The
chiralities in terms of the $a_i$ in the $G_4$-flux are:
\bea
\chi\big(\mathbf{ 5}_{(-\tfrac{2}{5},0)}\big) &=& \frac{1}{250} \Big[-25 (-13 + n_7 + n_9) a_3 -
   5 (-13 + n_7 + n_9) (2 n_7 - 5 (4 + n_9)) a_6 \nn \\ &&+ [-3092 + 7 n_7^2 + (249 - 5 n_9) n_9 + n_7 (193 + 2 n_9)] a_{7} \nn \\ && +
   25 (-88 + (-17 + n_9) n_9 + n_7 (16 + n_9)) a_{13}\Big],
\nn \\
\chi\big(\mathbf{ 5}_{(\tfrac{3}{5},0)}\big) &=& \frac{1}{500} \Big[-75 (6 + n_7 - n_9) a_3 \nn \\ &&+
   5 \big[4 n_7^2 + 11 n_7 (4 + n_9) - 15 (-6 + n_9) (4 + n_9)\big] a_6 \nn \\ && + [-139 n_7^2 - 2 (-6 + n_9) (426 + 5 n_9) +
      n_7 (458 + 149 n_9)] a_7 \nn \\ && - 50 (6 + n_7 - n_9) (-4 + n_9) a_{13}\Big], \nn
\eea
\bea
\chi\big(\mathbf{ 5}_{(-\tfrac{2}{5},-1)}\big) &=& \frac{1}{500} \Big[50 (4 + n_9) a_3 + 125 (4 + n_9) a_{4} + 250 n_9 (4 - n_7 + n_9) a_{5} + \nn \\ &&
 10 (-10 (4 + n_9) (7 + 3 n_9) + n_7 (38 + 27 n_9)) a_{6} +\nn \\ && (3 n_7 (28 + 37 n_9) - (4 + n_9) (608 + 115 n_9)) a_{7} \nn \\ &&-
 50 (-76 + 20 n_7 - 15 n_9 + n_9^2) a_{13} \Big],
\eea
For the chiralities of the singlets we obtain
\bea
\chi(\mathbf{ 1}_{(1,0)}) &=& \frac{1}{100} \Big[ 25 [2 n_7^2 - (-9 + n_9) (-6 + n_9) - n_7 (15 + n_9)] a_{3} \nn \\ && +
 50 n_7 (-4 + n_7 - n_9) n_9 a_{5} + 200 n_7 a_{13} \nn \\ && +
 5 \big[5 (-9 + n_9) (-6 + n_9) (4 + n_9) \nn \\ && - 3 n_7^2 (12 + 5 n_9)  +
    n_7 (276 + n_9 (111 + 10 n_9))\big] a_{6} \nn \\ && + \big[50 n_7^3 + 244 (-9 + n_9) (-6 + n_9) \nn \\ && - 3 n_7^2 (309 + 25 n_9)  +
    n_7 (2694 + n_9 (883 + 25 n_9))\big] a_{7} \Big]\,, 
\eea
    \bea    
\chi(\mathbf{ 1}_{(1,1)}) &=& -\frac{1}{4} \big[76 + n_7^2 + n_7 (-17 + n_9) + 11 n_9 - 2 n_9^2\big] ( a_{3} + a_{4} ) \nn \\ && -
 \frac{1}{2} (-4 + n_7 - n_9) n_9 (-19 + 2 n_9) a_{5}+ [12 + \frac{1}{2} (2 + n_7 - n_9) n_9] a_{13} \nn \\ && +
 \frac{1}{4} \big[608 - n_7 (212 + (-25 + n_7) n_7) +
    316 n_9 \nn \\ &&+ (-56 + n_7) n_7 n_9 + (17 + 6 n_7) n_9^2 - 6 n_9^3 \big] a_{6} \nn \\ && +
 \frac{1}{100} \big[n_7^2 (262 - 25 n_9) +
    n_7 (-4454 + 3 n_9 (-91 + 25 n_9))  \nn \\ && - (4 + n_9) (-4918 +
       n_9 (39 + 50 n_9))\big] a_{7} , \nn \\
\chi(\mathbf{1}_{(0,1)}) &=& \frac{1}{4} \big[n_7 (-15 + 2 n_7) + (4 + n_9) (-19 + 2 n_9)\big] a_{4} \nn \\ && -
 \frac{1}{2} (-4 + n_7 - n_9) n_9 (-19 + 2 n_9) a_{5} \nn \\ && + \big[76 + 49 n_9 +
    \frac{1}{20} ((-12 + n_7) n_7 (1 + 10 n_7)  \nn \\ && - n_7 (79 + 10 n_7) n_9 +
       10 (7 + 2 n_7) n_9^2 - 20 n_9^3)\big] a_{6}  \nn \\ &&+
 \frac{1}{100} \big[-50 n_7^3 + n_7^2 (403 + 50 n_9) +
    2 n_7 (541 + n_9 (-488 + 25 n_9))  \nn \\ && - (4 + n_9) (-212 +
       n_9 (-449 + 50 n_9))\big] a_{7}  \nn \\ && + \frac{1}{2} [52 - n_9 (-9 + n_7 + n_9)] a_{13}.
\eea

\section{Toric Tuning of $\cS_7$ and $\cS_9$ for $B=\mathbb{P}^3$}
\label{app:tuning-s7-s9}

In this appendix we explain how to construct the toric polytopes of $dP_2^B(\cS_7,\cS_9)$ with base $B=\mathbb{P}^3$ for all the points in the allowed region of figure \ref{fig:XP3n7n9region}. In the following we present a list of vertices, each of which realizing one possible choice of $(n_7,n_9)$. 

In order to find this list of vertices we follow the algorithm presented in \cite{Braun:2013nqa} to construct all reflexive convex polytopes for a given toric base and a given toric fiber. This algorithm uses the $GL(\mathbb{Z},5)$ symmetry to
bring all the vertices of the polytope in the following form
\beq \label{eq:G1}
\text{
 \begin{tabular}{|c||ccccc|} \hline
  variable  & \multicolumn{5}{c|}{vertices}	  \rule{0pt}{13pt}\\  \hline
  $z_0$ & 1  & 1  & 1 & $p_1$  &  $p_2$ 		\\
  $z_1$ & -1 & 0 & 0  & 0  &  0	 	\\
  $z_3$ & 0& -1 & 0  &  0  &  0		\\
  $z_2$ & 0 & 0  & -1 & 0  &  0	 \\
\hline
  $u$  & 0&0  & 0  & 1  &  0   		\\
  $v$ & 0&0  & 0  & 0  &  1   				\\
  $w$ & 0&0  & 0  & -1 & -1			 	\\
  $e_1$ & 0&0  & 0  & 0  & -1 			\\
  $e_2$ & 0&0  & 0  & 1  & 1 				\\ \hline
 \end{tabular}
 }
\eeq
In the first four lines we see the polytope of the base, in the last five lines the polytope of the fiber. The degrees of freedom of the fibration are parametrized by the two integers $(p_1,p_2)$. The coordinates $(p_1,p_2)$ are fixed by the requirement of convexity of the polytope \eqref{eq:G1}. We obtain the following list of points $(p_1,p_2)$ each of which giving rise to a reflexive and convex polytope \eqref{eq:G1}, along with the corresponding values of $(n_7,n_9)$:

\beq \label{tab:app-n79}
\text{
\begin{tabular}{|c|c||c|c|} \hline
$p_1$ & $p_2$ & $n_7$ & $n_9$ \\ \hline
-4 & -4 & 8 & 4 \\
-3 & -4 & 7 & 3 \\
-3 & -3 & 7 & 4 \\
-3 & -2 & 7 & 5 \\
-2 & -4 & 6 & 2 \\
-2 & -3 & 6 & 3 \\
-2 & -2 & 6 & 4 \\
-2 & -1 & 6 & 5 \\
-2 & 0 & 6 & 6 \\
-1 & -4 & 5 & 1 \\
-1 & -3 & 5 & 2 \\
-1 & -2 & 5 & 3 \\
-1 & -1 & 5 & 4 \\
-1 & 0 & 5 & 5 \\
-1 & 1 & 5 & 6 \\
-1 & 2 & 5 & 7 \\ 
0 & -4 & 4 & 0 \\\hline
\end{tabular} \,\, \,\,  
\begin{tabular}{|c|c||c|c|} \hline
$p_1$ & $p_2$ & $n_7$ & $n_9$ \\ \hline
0 & -3 & 4 & 1 \\
0 & -2 & 4 & 2 \\
0 & -1 & 4 & 3 \\
0 & 0 & 4 & 4 \\
0 & 1 & 4 & 5 \\
0 & 2 & 4 & 6 \\
0 & 3 & 4 & 7 \\
0 & 4 & 4 & 8 \\
1 & -3 & 3 & 0 \\
1 & -2 & 3 & 1 \\
1 & -1 & 3 & 2 \\
1 & 0 & 3 & 3 \\
1 & 1 & 3 & 4 \\
1 & 2 & 3 & 5 \\
1 & 3 & 3 & 6 \\
1 & 4 & 3 & 7 \\
2 & -2 & 2 & 0 \\ \hline
\end{tabular}  \,\, \,\,
\begin{tabular}{|c|c||c|c|} \hline
$p_1$ & $p_2$ & $n_7$ & $n_9$ \\ \hline
2 & -1 & 2 & 1 \\
2 & 0 & 2 & 2 \\
2 & 1 & 2 & 3 \\
2 & 2 & 2 & 4 \\
2 & 3 & 2 & 5 \\
2 & 4 & 2 & 6 \\
3 & -1 & 1 & 0 \\
3 & 0 & 1 & 1 \\
3 & 1 & 1 & 2 \\
3 & 2 & 1 & 3 \\
3 & 3 & 1 & 4 \\
3 & 4 & 1 & 5 \\
4 & 0 & 0 & 0 \\
4 & 1 & 0 & 1 \\
4 & 2 & 0 & 2 \\
4 & 3 & 0 & 3 \\
4 & 4 & 0 & 4 \\ \hline
\end{tabular}
}
\eeq

\bibliographystyle{utphys}	
\bibliography{ref}

\end{document}